\documentclass[a4paper,11pt]{article}
\pdfoutput=1
\usepackage{jheppubMod}
\usepackage{amssymb,amsmath}
\usepackage{bm,bbm,bbm,bbold}
\usepackage{cancel}
\usepackage{datetime,dsfont}
\usepackage{graphicx}
\usepackage{epsfig,epstopdf}
\usepackage{lineno,lipsum}
\usepackage{subfigure}
\usepackage{textcomp}
\usepackage{times}
\usepackage{ulem}
\usepackage{verbatim}
\hypersetup{bookmarks=true,unicode=true,pdftoolbar=true,pdfmenubar=true,
  pdffitwindow=false,pdfstartview={FitH},
  pdftitle={Lepton Number Violation: Seesaw Models and Their Collider Tests},pdfauthor={Cai, Han, Li, Ruiz},
  pdfsubject={Subject},pdfcreator={Creator},pdfproducer={Producer},
  pdfkeywords={Lepton Number Violation}{Neutrino Mass Models}{Collider Physics},
  pdfnewwindow=true,colorlinks=true
}
\graphicspath{{./}{./plots/}}

\def\bl{~$B\mathrm{-}L$~}
\newcommand{\UX}[2]{{\rm U}(#1)_{#2}}
\newcommand{\SU}[2]{{\rm SU}(#1)_{#2}}
\def\lsim{\mathrel{\raise.3ex\hbox{$<$\kern-.75em\lower1ex\hbox{$\sim$}}}}
\def\gsim{\mathrel{\raise.3ex\hbox{$>$\kern-.75em\lower1ex\hbox{$\sim$}}}}

\def\ie{{\it i.e.}}
\def\eg{{\it e.g.}}
\def\etc{{\it etc}}

\def\tev{\,{\rm TeV}}
\def\gev{\,{\rm GeV}}

\def\kev{\,{\rm keV}}
\def\ev{\,{\rm eV}}
\newcommand{\MeV}{{\rm ~MeV}}
\newcommand{\GeV}{{\rm ~GeV}}

\newcommand{\invfb}{{\rm ~fb^{-1}}}
\newcommand{\invab}{{\rm ~ab^{-1}}}
\newcommand{\Dmq}{\Delta m^2}
\newcommand{\eVq}{\ensuremath{\text{eV}^2}}
\newcommand{\Tr}{\mathop{\rm Tr}}
\def\to{\rightarrow}

\def\beq{\begin{equation}}
\def\eeq{\end{equation}}
\def\Hc{\text{H.c.}}
\def\be{\begin{equation}}
\def\ee{\end{equation}}
\def\bea{\begin{eqnarray}}
\def\eea{\end{eqnarray}}
\newcommand{\minigraph}[5][0.25in]{\begin{minipage}{#2}\begin{center}\includegraphics[width=#2]{#5}\\\vspace{#3}\hspace{#1}{\footnotesize #4}\end{center}\end{minipage}}
\title{\textbf{Lepton Number Violation:\\ Seesaw Models and Their Collider Tests }}

\author[a]{Yi Cai,}
\author[b,c]{Tao Han,}
\author[d,e]{Tong Li,}
\author[f]{and Richard Ruiz}

\affiliation[a]{School of Physics, Sun Yat-sen University, Guangzhou, 510275, China}
\affiliation[b]{Department of Physics and Astronomy, University of Pittsburgh, Pittsburgh, PA 15260, USA}
\affiliation[c]{Department of Physics, Tsinghua University, and Collaborative Innovation Center of Quantum Matter, Beijing, 100086, China}
\affiliation[d]{School of Physics, Nankai University, Tianjin 300071, China}
\affiliation[e]{ARC Centre of Excellence for Particle Physics at the Terascale,
School of Physics and Astronomy,\\ Monash University, Melbourne, Victoria 3800, Australia}
\affiliation[f]{Institute for Particle Physics Phenomenology {\rm(IPPP)},\\
Department of Physics, Durham University, Durham, DH1 3LE, UK}

\emailAdd{caiy36@mail.sysu.edu.cn}
\emailAdd{than@pitt.edu}
\emailAdd{tong.li@monash.edu}
\emailAdd{richard.ruiz@durham.ac.uk}

\abstract{
The Majorana nature of neutrinos is strongly motivated from the theoretical and phenomenological point of view. A plethora of neutrino mass models, known collectively as Seesaw models, exist that could generate both a viable neutrino mass spectrum and mixing pattern. They can also lead to rich, new phenomenology,
including lepton number non-conservation as well as new particles, that may be observable at collider experiments.
It is therefore vital to search for such new phenomena and the mass scale associated with neutrino mass generation at high energy colliders.
In this review, we consider a number of representative Seesaw scenarios as phenomenological benchmarks,
including the characteristic Type I, II, and III Seesaw mechanisms, their extensions and hybridizations, as well as radiative constructions.
We present new and updated predictions for analyses featuring lepton number violation and expected coverage in the theory parameter space at current and future colliders.
We emphasize new production and decay channels, their phenomenological relevance and treatment across different facilities in $e^+e^-$, $e^-p$ and $pp$ collisions, as well as the available Monte Carlo tools available for studying Seesaw partners in collider environments.
}

\preprint{PITT-PACC-1712,~~~IPPP/17/74}
\arxivnumber{1711.02180}

\begin{document}
\maketitle
\flushbottom

\section{Introduction}
Neutrino flavor oscillation experiments from astrophysical and terrestrial sources
provide overwhelming evidence that neutrinos have small but nonzero masses.
Current observations paint a picture consistent with a mixing structure parameterized by the $3\times 3$ Pontecorvo-Maki-Nakagawa-Sakata (PMNS) matrix \cite{Pontecorvo:1957qd,Pontecorvo:1957cp,Maki:1962mu} with at least two massive neutrinos.
This is contrary to the Standard Model of particle physics (SM) \cite{Weinberg:1967tq}, which allows three massless neutrinos and hence no flavor oscillations.
Consequently, to accommodate these observations, the SM must~\cite{Ma:1998dn} be extended to a more complete theory by new degrees of freedom.

One could of course introduce right-handed (RH) neutrino states (${\nu_R}$) and construct Dirac mass terms, $m_D^{} \overline {\nu_L} \nu_R$,
in the same fashion as for all the other elementary fermions in the SM.
However, in this minimal construction, the new states do not carry any SM gauge charges, and thus these ``sterile neutrinos'' have the capacity to be Majorana fermions~\cite{Majorana:1937vz}.
The most significant consequence of this would be the existence of the RH Majorana mass term $M_R\overline{(\nu_R)^c} \nu_R$
and the explicit violation of lepton number ($L$).
In light of this prospect, 
a grand frontier opens for theoretical model-building with rich and new phenomenology at the energy scales accessible by collider experiments,
and which we will review in this article.

Generically, if we integrate out the new states, presumably much heavier than the electroweak (EW) scale,
the new physics may be parameterized at leading order through the dimension-5 lepton number violating operator~\cite{Weinberg:1979sa},
the so-called ``Weinberg operator,''
\begin{equation}
\mathcal{L}_5 = \frac{\alpha}{\Lambda} ~ (LH)(LH)
\xrightarrow{\rm EWSB}
\mathcal{L}_5 \ni \frac{\alpha v^2_0}{2\Lambda}\  \overline{(\nu_L)^c}\ \nu_L ,
\label{Weinberg}
\end{equation}
where $L$ and $H$ are, respectively, the SM left-handed (LH) lepton doublet and Higgs doublet, with vacuum expectation value (vev) $v_0 \approx 246$ GeV. 
After electroweak (EW) symmetry breaking (EWSB), $\mathcal{L}_5$ generates a Majorana mass term for neutrinos.
One significance of Eq.~(\ref{Weinberg}) is the fact that its ultraviolet (UV) completions are severely restricted. 
For example: extending the SM field content minimally, {\it i.e.}, by only a single SM multiplet, permits only three~\cite{Ma:1998dn} tree-level completions of
Eq.~(\ref{Weinberg}), a set of constructions famously known as 
the Type I~\cite{Minkowski:1977sc, Yanagida:1979as, GellMann:1980vs, Glashow:1979nm,Mohapatra:1979ia, Shrock:1980ct, Schechter:1980gr},
Type II~\cite{Konetschny:1977bn, Cheng:1980qt,Lazarides:1980nt,Schechter:1980gr, Mohapatra:1980yp}, and Type III~\cite{Foot:1988aq} Seesaw mechanisms.
These minimal mechanisms can be summarized with the following:

{\bf Minimal Type I Seesaw}~\cite{Minkowski:1977sc, Yanagida:1979as, GellMann:1980vs, Glashow:1979nm, Mohapatra:1979ia, Shrock:1980ct,Schechter:1980gr}:
In the minimal Type I Seesaw, one hypothesizes the existence of a right-handed (RH) neutrino $\nu_R$,
which transforms as a singlet, i.e., as $(1,1,0)$, under the SM gauge group SU(3)$_c \otimes$SU(2)$_L \otimes$U(1)$_Y$,
that possesses a RH Majorana mass $M_{\nu_R}$ and interacts with a single generation of SM leptons through a Yukawa coupling $y_\nu$.
After mass mixing and assuming $M_{\nu_R}\gg y_\nu v_0$, the light neutrino mass eigenvalue $m_\nu$ is given by $m_\nu \sim  y^2_\nu v_0^2/ M_{\nu_R}$.
If $y_\nu \simeq 1$, then to obtain a light neutrino mass of order an eV, $M_{\nu_R}$ is required to be of order $10^{14}-10^{15}$ GeV.
$M_{\nu_R}$ can be made much lower though by balancing against a correspondingly lower $y_\nu$.

{\bf Minimal  Type II Seesaw}~\cite{Konetschny:1977bn, Cheng:1980qt,Lazarides:1980nt,Schechter:1980gr, Mohapatra:1980yp}:
The minimal Type II Seesaw features the introduction of a Higgs field $\Delta$
with mass $M_{\Delta}$ in a triplet representation of SU$(2)_L$,
and hence transforms as $(1,3,2)$ under the SM gauge group.
In this mechanism, light neutrino masses are given by LH Majorana masses $m_\nu \approx Y_\nu  v_{\Delta}$,
where $v_{\Delta}$ is the vev of the neutral component of the new scalar triplet and $Y_\nu$ is the corresponding Yukawa coupling.
Due to mixing between the SM Higgs doublet and the new scalar triplet
via a dimensionful parameter $\mu$, EWSB leads to a relation $v_{\Delta} \sim \mu v_0^2/M_{\Delta}^2$.
In this case the new scale $\Lambda$ is replaced by $M_{\Delta}^2/\mu$.
With $Y_\nu \approx 1$ and $\mu \sim M_{\Delta}$, the scale is also $10^{14}- 10^{15}$ GeV.
Again, $M_\Delta$ can be of TeV scale if $Y_\nu$ is small or  $\mu \ll M_\Delta$.
It is noteworthy that in the Type II Seesaw no RH neutrinos are needed to explain the observed neutrino masses and mixing.

{\bf Minimal  Type III Seesaw}~\cite{Foot:1988aq}:
The minimal Type III Seesaw is similar to the other two cases  in that one introduces
the fermionic multiplet $\Sigma_L$ that is a triplet (adjoint representation) under SU$(2)_L$ and transforms as
$(1,3,0)$ under the SM gauge group.
The resulting mass matrix for neutrinos has the same form as in Type I Seesaw,
but in addition features heavy leptons that are electrically charged.
The new physics scale $\Lambda$ in Eq.~(\ref{Weinberg})
is replaced by the mass of the leptons $M_\Sigma$, which can also be as low as a TeV if balanced with a small Yukawa coupling.

However, to fully reproduce oscillation data, at least two of the three known neutrinos need nonzero masses.
This requires a nontrivial Yukawa coupling matrix for neutrinos if appealing to any of the aforementioned Seesaws mechanisms, and, 
if invoking the Type I or III Seesaws, extending the SM by at least two generations of multiplets~\cite{Wyler:1982dd}, 
which need not be in the same SM gauge representation.
In light of this, one sees that Weinberg's assumption of a high-scale Seesaw~\cite{Weinberg:1979sa}
is not necessary to generate tiny neutrino masses in connection with lepton $(L)$ number violation.
For example:
the so-called Inverse~\cite{Mohapatra:1986aw,Mohapatra:1986bd,Bernabeu:1987gr,Gavela:2009cd}
or Linear~\cite{Akhmedov:1995ip,Akhmedov:1995vm} variants of the Type I and III Seesaw models,
their generic extensions as well as hybridizations, {\it i.e.}, the combination of two or more Seesaw mechanisms,
can naturally lead to mass scales associated with neutrino mass-generation accessible at present-day experiments, 
and in particular, collider experiments. 
A qualitative feature of these low-scale Seesaws is that light neutrino masses are proportional to the scale of $L$ violation,
as opposed to inversely related as in high-scale Seesaws~\cite{Moffat:2017feq}. 

The Weinberg operator in Eq.~(\ref{Weinberg}) is the lowest order and simplest parameterization of neutrino mass generation using 
only the SM particle spectrum and its gauge symmetries.
Beyond its tree-level realizations, neutrino Majorana masses may alternatively be generated radiatively.
Suppression by loop factors may provide a partial explanation for the smallness of neutrino masses
and again allow much lower mass scales associated with neutrino mass-generation.
The first of such models was proposed at one-loop in Refs.~\cite{Zee:1980ai, Hall:1983id},
at two-loop order in Refs.~\cite{Cheng:1980qt, Zee:1985id, Babu:1988ki},
and at three-loop order in Ref.~\cite{Krauss:2002px}.
A key feature of radiative neutrino mass models is the absence of tree-level contributions to neutrino masses
either because there the necessary particles, such as SM singlet fermion as in Type I Seesaw, are not present
or because  relevant couplings are forbidden by additional symmetries.
Consequently, it is necessary that the new field multiplets run in the loop(s) that generate neutrino masses.

As observing lepton number violation would imply the existence of Majorana masses for neutrinos~\cite{Schechter:1981bd,Hirsch:2006yk,Duerr:2011zd},
confirming the existence of this new mass scale would, in addition, verify the presence of a Seesaw mechanism.
To this end, there have been on-going efforts in several directions, most notably the neutrinoless double beta $(0\nu\beta\beta)$-decay experiments, 
both current~\cite{KamLAND-Zen:2016pfg,Agostini:2017iyd,Alfonso:2015wka,Albert:2014awa} and upcoming~\cite{Arnaboldi:2002du,Arnold:2010tu,Alvarez:2012flf},
as well as proposed general purpose fixed-target facilities~\cite{Alekhin:2015byh,Anelli:2015pba}.
Complementary to this are on-going searches for lepton number violating processes at collider experiments,
which focus broadly on rare meson decays~\cite{Liventsev:2013zz,Aaij:2012zr,Aaij:2014aba},
heavy neutral leptons in Type I-like models~\cite{Sirunyan:2017yrk,Khachatryan:2016jqo,Khachatryan:2016olu,Khachatryan:2015gha,Aad:2015xaa},
heavy bosons in Type II-like models~\cite{ATLAS:2016pbt,ATLAS:2014kca,Chatrchyan:2012ya},
heavy charged leptons in Type III-like models \cite{Aad:2015cxa,CMS:2012ra,Sirunyan:2017qkz},
and lepton number violating contact interactions~\cite{CMS:2016blm,Sirunyan:2017xnz}.
Furthermore, accurate measurements of the PMNS matrix elements and stringent limits on the neutrino masses themselves provide crucial information and
knowledge of lepton flavor mixing that could shed light on the construction of Seesaw models.

In this context, we present a review of searches for lepton number violation at current and future collider experiments.
Along with the current bounds from the experiments at LEP, Belle, LHCb and ATLAS/CMS at 8 and 13 TeV, we present studies for the 13 and 14 TeV LHC.
Where available, we also include results for a future 100 TeV hadron collider, an $ep$ collider (LHeC), and a future high-energy $e^+e^-$ collider.
We consider a number of tree- and loop-level Seesaw models,
including, as phenomenological benchmarks, the canonical Type I, II, and III Seesaw mechanisms,
their extensions and hybridizations, and radiative Seesaw formulations in  $pp$, $ep$, and $ee$ collisions.
We note that the classification of collider signatures based on the canonical Seesaws is actually highly suitable,
as the same underlying extended and hybrid Seesaw mechanism can be molded to produce wildly varying collider predictions.

We do not attempt to cover the full aspects of UV-complete models for each type.
This review is only limited to a selective, but representative, presentation of tests of Seesaw models at collider experiments.
For complementary reviews, we refer readers to Refs.~\cite{Gluza:2002vs,Barger:2003qi,Mohapatra:2006gs,Rodejohann:2011mu,Chen:2011de,Atre:2009rg,Deppisch:2015qwa} and references therein.

This review is organized according to the following:
In Sec.~\ref{sec:nuparameters} we first show the PMNS matrix and summarize the mixing and mass-difference parameters from neutrino oscillation data. With those constraints, we also show the allowed mass spectra for the three massive neutrino scheme.
Our presentation is agnostic, phenomenological, and categorized according to collider signature, i.e.,
according to the presence of Majorana neutrinos (Type I) as in Sec.~\ref{sec:type1},
doubly charged scalars (Type II) as in Sec.~\ref{sec:type2},
new heavy charged/neutral leptons (Type III) as in Sec.~\ref{sec:type3}, and
new Higgs, diquarks and leptoquarks
in Sec.~\ref{sec:loop}.
Particular focus is given to state-of-the-art computations, newly available Monte Carlo tools,
and new collider signatures that offer expanded coverage of Seesaw parameter spaces at current and future colliders.
Finally in Sec.~\ref{sec:con} we summarize our main results.

\section{Neutrino Mass and Oscillation Parameters}
\label{sec:nuparameters}

In order to provide a general guidance for model construction and collider searches,
we first summarize the neutrino mass and mixing parameters in light of oscillation data.
Neutrino mixing can be parameterized by the PMNS matrix \cite{Pontecorvo:1957qd,Pontecorvo:1957cp,Maki:1962mu} as
\begin{equation}
U_{PMNS}  =
\begin{pmatrix}
1 & 0 & 0\\
0 & c_{23} & s_{23} \\
0 & -s_{23} & c_{23}
\end{pmatrix}
\begin{pmatrix}
c_{13} & 0 & e^{-i\delta}s_{13} \\
0 & 1 & 0 \\
-e^{i\delta}s_{13} & 0 & c_{13}
\end{pmatrix}
\begin{pmatrix}
c_{12} & s_{12}  & 0 \\
-s_{12} & c_{12} & 0 \\
0 & 0 & 1
\end{pmatrix}
\text{diag} (e^{i \Phi_1/2}, 1, e^{i \Phi_2/2})
\end{equation}
\beq = \left(
\begin{array}{lll}
 c_{12} c_{13} & c_{13} s_{12} & e^{-\text{i$\delta $}} s_{13}
   \\
 -c_{12} s_{13} s_{23} e^{\text{i$\delta $}}-c_{23} s_{12} &
   c_{12} c_{23}-e^{\text{i$\delta $}} s_{12} s_{13} s_{23} &
   c_{13} s_{23} \\
 s_{12} s_{23}-e^{\text{i$\delta $}} c_{12} c_{23} s_{13} &
   -c_{23} s_{12} s_{13} e^{\text{i$\delta $}}-c_{12} s_{23} &
   c_{13} c_{23}
\end{array}
\right)\times \text{diag} (e^{i \Phi_1/2}, 1, e^{i \Phi_2/2}),
\eeq
where $s_{ij}\equiv\sin{\theta_{ij}}$, $c_{ij}\equiv\cos{\theta_{ij}}$, $0 \le
\theta_{ij} \le \pi/2$, and $0 \le \delta, \Phi_i \le 2\pi$, with $\delta$ being the Dirac CP phase and $\Phi_i$ the Majorana phases.
While the PMNS is a well-defined $3\times3$ unitary matrix, throughout this review
we use the term generically to describe the $3\times3$ active-light mixing that may not, in general, be unitary.

\begin{table}\centering
  \begin{footnotesize}
    \begin{tabular}{l|cc|cc|c}
      \hline\hline
      & \multicolumn{2}{c|}{Normal Ordering (best fit)}
      & \multicolumn{2}{c|}{Inverted Ordering ($\Delta\chi^2=0.83$)}
      & Any Ordering
      \\
      \hline
      & bfp $\pm 1\sigma$ & $3\sigma$ range
      & bfp $\pm 1\sigma$ & $3\sigma$ range
      & $3\sigma$ range
      \\
      \hline
      \rule{0pt}{4mm}\ignorespaces
      $\sin^2\theta_{12}$
      & $0.306_{-0.012}^{+0.012}$ & $0.271 \to 0.345$
      & $0.306_{-0.012}^{+0.012}$ & $0.271 \to 0.345$
      & $0.271 \to 0.345$
      \\[1mm]
      $\theta_{12}/^\circ$
      & $33.56_{-0.75}^{+0.77}$ & $31.38 \to 35.99$
      & $33.56_{-0.75}^{+0.77}$ & $31.38 \to 35.99$
      & $31.38 \to 35.99$
      \\[3mm]
      $\sin^2\theta_{23}$
      & $0.441_{-0.021}^{+0.027}$ & $0.385 \to 0.635$
      & $0.587_{-0.024}^{+0.020}$ & $0.393 \to 0.640$
      & $0.385 \to 0.638$
      \\[1mm]
      $\theta_{23}/^\circ$
      & $41.6_{-1.2}^{+1.5}$ & $38.4 \to 52.8$
      & $50.0_{-1.4}^{+1.1}$ & $38.8 \to 53.1$
      & $38.4 \to 53.0$
      \\[3mm]
      $\sin^2\theta_{13}$
      & $0.02166_{-0.00075}^{+0.00075}$ & $0.01934 \to 0.02392$
      & $0.02179_{-0.00076}^{+0.00076}$ & $0.01953 \to 0.02408$
      & $0.01934 \to 0.02397$
      \\[1mm]
      $\theta_{13}/^\circ$
      & $8.46_{-0.15}^{+0.15}$ & $7.99 \to 8.90$
      & $8.49_{-0.15}^{+0.15}$ & $8.03 \to 8.93$
      & $7.99 \to 8.91$
      \\[3mm]
      $\delta_\text{CP}/^\circ$
      & $261_{-59}^{+51}$ & $\hphantom{00}0 \to 360$
      & $277_{-46}^{+40}$ & $145 \to 391$
      & $\hphantom{00}0 \to 360$
      \\[3mm]
      $\dfrac{\Dmq_{21}}{10^{-5}~\eVq}$
      & $7.50_{-0.17}^{+0.19}$ & $7.03 \to 8.09$
      & $7.50_{-0.17}^{+0.19}$ & $7.03 \to 8.09$
      & $7.03 \to 8.09$
      \\[3mm]
      $\dfrac{\Dmq_{3\ell}}{10^{-3}~\eVq}$
      & $+2.524_{-0.040}^{+0.039}$ & $+2.407 \to +2.643$
      & $-2.514_{-0.041}^{+0.038}$ & $-2.635 \to -2.399$
      & $\begin{bmatrix}
        +2.407 \to +2.643\\[-2pt]
        -2.629 \to -2.405
      \end{bmatrix}$
      \\[3mm]
      \hline\hline
    \end{tabular}
  \end{footnotesize}
  \caption{Three-neutrino oscillation fit based as obtained by the \texttt{NuFit} collaboration, taken from Ref.~\cite{Esteban:2016qun},
  where $\Delta m_{3\ell}^2=\Delta m_{31}^2>0$ for NO (or NH) and $\Delta m_{3\ell}^2=\Delta m_{32}^2<0$ for IO (or IH).
  }
  \label{tab:nufit}
\end{table}

The neutrino mixing matrix is very different from the quark-sector Cabbibo-Kobayashi-Maskawa (CKM) matrix,
in that most of the PMNS mixing angles are large whereas CKM angles are small-to-negligible.
In recent years, several reactor experiments, such as Daya Bay~\cite{An:2012eh}, Double Chooz~\cite{Abe:2011fz}, and RENO~\cite{Ahn:2012nd},
have reported non-zero measurements of $\theta_{13}$ by searching for the disappearance of antielectron neutrinos.
Among these reactor experiments, Daya Bay gives the most conclusive result with $\sin^22\theta_{13}\approx 0.084$ or $\theta_{13}\approx8.4^\circ$ \cite{An:2015rpe,Esteban:2016qun}, the smallest entry of the PMNS matrix.
More recently, there have been reports on indications of a non-zero Dirac CP phase, with $\delta \approx 3\pi/2$ \cite{Abe:2013hdq,Adamson:2016tbq,Abe:2017uxa}. However, it cannot presently be excluded that evidence for such a large Dirac phase may instead be evidence for sterile neutrinos or new neutral currents~\cite{Forero:2016cmb,Ge:2016dlx,deGouvea:2016pom,Miranda:2016wdr}.

\begin{figure}[tb]
\begin{center}
\subfigure[]{\includegraphics[scale=1,width=.45\textwidth]{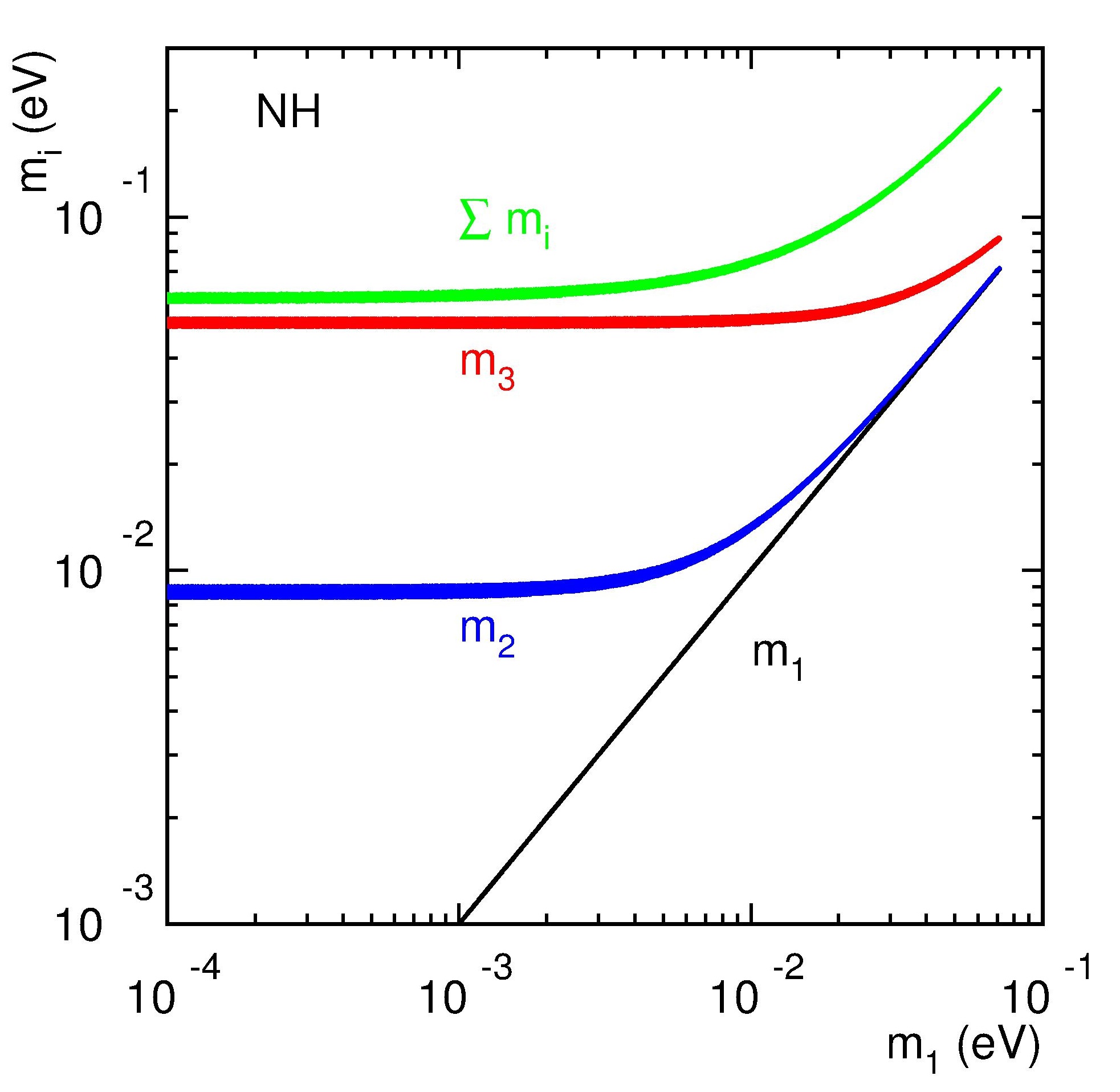}		}
\subfigure[]{\includegraphics[scale=1,width=.45\textwidth]{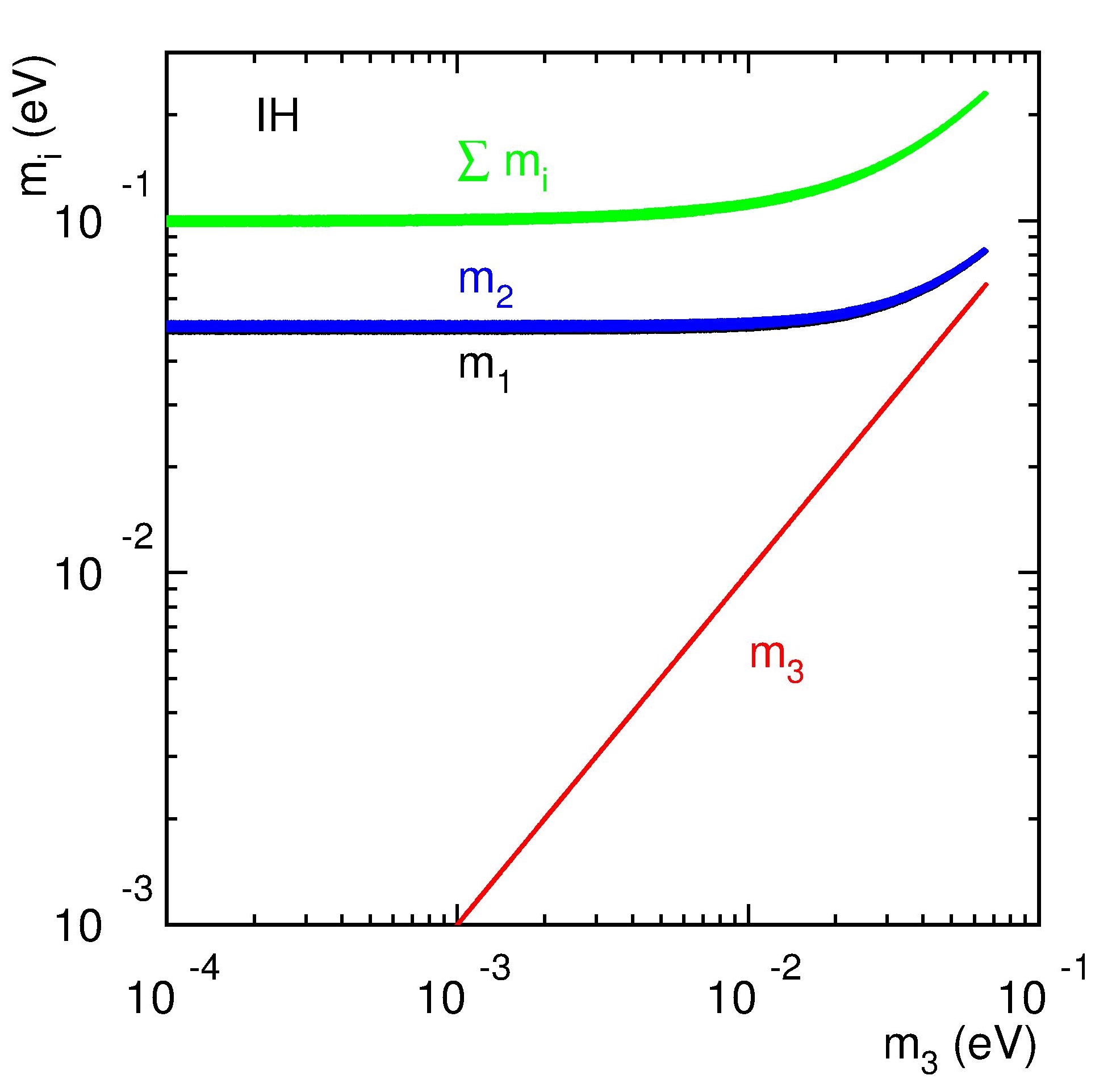}		}
\end{center}
\caption{The three active neutrino mass spectra versus the lowest neutrino mass for (a) NH, and (b) IH.}
\label{mabs}
\end{figure}

Neutrino oscillation experiments can help to extract the size of the mass-squared splitting between three neutrino mass eigenstates. 
The sign of $\Delta m_{31}^2 = m_{3}^2 - m_{1}^2$, however, still remains unknown at this time.
It can be either positive, commonly referred as the Normal Hierarchy (NH), or negative and referred to as the Inverted Hierarchy (IH).
The terms Normal Ordering (NO) and Inverted Ordering (IO) are also often used in the literature in lieu of NH and IH, respectively.
Taking into account the reactor data from the antineutrino disappearance experiments mentioned above together
with other disappearance and appearance measurement,
the latest global fit of the neutrino masses and mixing parameters from the \texttt{NuFit} collaboration~\cite{Esteban:2016qun}, 
are listed in Table \ref{tab:nufit} for NH (left) and IH (center).
The tightest constraint on the sum of neutrino masses comes from cosmological data.
Combining Planck+WMAP+highL+BAO data, this yields at $95\%$ confidence level (CL)~\cite{Ade:2015xua}
\beq
\sum_{i=1}^3 m_i < \ 0.230 \ \ev .
\eeq
Given this and the measured neutrino mass splittings, we show in Fig.~\ref{mabs}
the three active neutrino mass spectra as a function of the lowest neutrino mass in (a) NH and (b) IH.
With the potential sensitivity of the sum of neutrino masses being close to $0.1\ev$ in the near future ($5-7$ years) \cite{Lesgourgues:2014zoa},
upcoming cosmological probes will not be able to settle the issue of the neutrino mass hierarchy.
However, the improved measurement $\sim 0.01\ev$ over a longer term ($7-15$ years) \cite{Hamann:2012fe,Lesgourgues:2014zoa}
would be sensitive enough to determine the absolute mass scale of a heavier neutrino spectrum.
In addition, there are multiple proposed experiments aiming to determine the neutrino mass hierarchy.
The Deep Underground Neutrino Experiment (DUNE) will detect neutrino beams from the Long Baseline Neutrino Facility (LBNF),
and probe the Dirac CP-phase and mass hierarchy. With a baseline of 1300 km, DUNE is able to determine the mass hierarchy 
with at least $5\sigma$ significance~\cite{Acciarri:2016crz}.
The Jiangmen Underground Neutrino Observatory (JUNO) plans to precisely measure the reactor antielectron neutrinos and improve the accuracy of $\Delta m_{21}^2$, 
$\Delta m_{32}^2$ and $\sin^2\theta_{12}$ to 1\% level~\cite{An:2015jdp}.
The Hyper-Kamiokande (Hyper-K) experiment, an upgrade of the T2K experiment, can measure the precision of $\delta$ to be $7^\circ-21^\circ$ and 
reach $3~(5)\sigma$ significance for mass hierarchy determination after 5 (10) years exposure~\cite{Abe:2015zbg}.
Finally, the Karlsruhe Tritium Neutrino experiment (KATRIN), a tritium $\beta$ decay experiment, 
aims to measure the effective ``electron-neutrino mass'' with sub-eV sensitivity~\cite{Osipowicz:2001sq}.

\section{The Type I Seesaw and Lepton Number Violation at Colliders}\label{sec:type1}

We begin our presentation of collider searches for lepton number violation in the context of Type I Seesaw models.
After describing the canonical Type I mechanism~\cite{Minkowski:1977sc, Yanagida:1979as, GellMann:1980vs, Glashow:1979nm,Mohapatra:1979ia}
and its phenomenological decoupling at collider scales in Sec.~\ref{sec:type1Canon},
we discuss various representative, low-scale models that incorporate the Type I mechanism and its extensions.
We then present collider searches for lepton number violation mediated by Majorana neutrinos $(N)$,
which is the characteristic feature of Type I-based scenarios, in Sec.~\ref{sec:typeIhybrid}.
This is further categorized according to associated phenomena of increasing complexity:
$N$ production via massive Abelian gauge bosons is reviewed in Sec.~\ref{sec:type1Abelian},
via massive non-Abelian gauge bosons in Sec.~\ref{sec:lrsmCollider},
and via dimension-six operators in Sec.~\ref{sec:neftTests}.

\subsection{Type I Seesaw Models}

\subsubsection{The Canonical Type I Seesaw Mechanism}\label{sec:type1Canon}

In the canonical Type I Seesaw mechanism one hypothesizes a single RH neutral leptonic state, $N_R\sim (1,1,0)$,
in addition to the SM matter content.
However, reproducing neutrino oscillation data requires more degrees of freedom. Therefore, for our purposes,
we assume $i =1,\dots,3$ LH states and $j=1,\dots,n$ RH states.
Following the notation of Refs.~\cite{Atre:2009rg,Han:2012vk}, the full theory is
\begin{equation}
 \mathcal{L}_{\rm Type~I} = \mathcal{L}_{\rm SM} +  \mathcal{L}_{N~{\rm Kin}} +  \mathcal{L}_{N},
 \label{eq:type1Lag}
\end{equation}
where $\mathcal{L}_{\rm SM}$ is the SM Lagrangian, $\mathcal{L}_{N~{\rm Kin}}$ is $N_R$'s kinetic term,
and interactions and mass terms,
\begin{eqnarray}
{\cal L}_N &=& -\overline{L} \ Y_\nu^D \ \tilde{H} \ N_{R}
\ - \
{1\over 2} \overline{(N^c)}_{L} \ M_R \ N_{R} + \ \text{H.c.}
\end{eqnarray}
$L$ and $H$ are the SM LH lepton and Higgs doublets, respectively, and $\tilde{H}=i \sigma_2 H^*$.
Once $H$ settles on the vev $\langle H\rangle =v_0/\sqrt 2$, neutrinos acquire Dirac masses $m_D = Y_\nu^D \ v_0 / \sqrt{2}$ and we have
\begin{eqnarray}
{\cal L}_N \ni -{1\over 2} \ \left( \overline{\nu}_{L} \ m_D \ N_{R} \ + \ \overline{(N^c)}_{L} \ m_D^T \ (\nu^c)_{R}
\ + \ \overline{(N^c)}_{L} \ M_R \ N_{R} \right) \ + \ \text{H.c.}
\end{eqnarray}
After introducing a unitary transformation into $m~(m')$ light~(heavy) mass eigenstates,
\begin{eqnarray}
\left(
  \begin{array}{c}
    \nu \\
    N^c \\
  \end{array}
\right)_L = \mathbb{N}\left(
  \begin{array}{c}
    \nu_m \\
    N^c_{m'} \\
  \end{array}
\right)_L, \ \ \ \mathbb{N}= \left(
  \begin{array}{cc}
    U & V \\
    X & Y \\
  \end{array}
\right),
\label{eq:nuMixDefs}
\end{eqnarray}
one obtains the diagonalized mass matrix for neutrinos
\begin{eqnarray}
\mathbb{N}^\dagger
\left(
  \begin{array}{cc}
    0 & m_D \\
    m_{D}^{T} & M \\
  \end{array}
\right) \mathbb{N}^\ast &=&
\left(
  \begin{array}{cc}
    m_\nu & 0 \\
    0 & M_N  \\
  \end{array}
\right),
\label{eq:type1NuMixMatrix}
\end{eqnarray}
with mass eigenvalues $m_\nu=diag(m_1,m_2,m_3)$ and $M_N=diag(M_1,\cdots, M_{m'})$.
In the limit $m_D\ll M_R$, the light $(m_\nu)$ and heavy $(M_N)$ neutrino masses are respectively
\begin{eqnarray}
m_\nu\approx  -  m_D M^{-1}_R m_D^T \quad\text{and}\quad  M_N\approx M_R.
\end{eqnarray}
The mixing elements typically scale like
\begin{eqnarray}
UU^\dagger\approx I- m_\nu M_N^{-1}, \ \ \ VV^\dagger\approx m_\nu M_N^{-1},
\label{eq:massscale}
\end{eqnarray}
with the unitarity condition $UU^\dagger+VV^\dagger=I$.
With another matrix $U_\ell$ diagonalizing the charged lepton mass matrix,
we have the approximate neutrino mass mixing matrix $U_{PMNS}$ and the matrix $V_{\ell N}$,
which transits heavy neutrinos to charged leptons. These are given by
\begin{eqnarray}
U_\ell^\dagger U\equiv U_{PMNS}, \ \ \ U_\ell^\dagger V\equiv V_{\ell N}, \quad\text{and}\quad
U_{PMNS}U_{PMNS}^\dagger+V_{\ell N}V_{\ell N}^\dagger=I.
\end{eqnarray}
The decomposition of active neutrino states into a general number of massive eigenstates
is then given by~\cite{Atre:2009rg,Han:2012vk}, $\nu_{\ell} = \sum_{m=1}^{3} U_{\ell m}\nu_{m} + \sum_{m'=1}^{n}V_{\ell m'} N^{c}_{m'}.$
From this, the SM EW boson couplings to heavy mass eigenstates (in the mixed mass-flavor basis) are
\begin{eqnarray}
  \mathcal{L}_{\rm Int.} =
  &-& \frac{g}{\sqrt{2}}W^+_\mu \sum_{\ell=e}^\tau
  \left( \sum_{m=1}^3 ~\overline{\nu_m} ~U_{\ell m}^*  + \sum_{m'=1}^n \overline{N^c_{m'}} ~V_{\ell N_{m'}}^* 	\right)\gamma^\mu P_L\ell^-\nonumber\\
  &-& \frac{g}{2\cos\theta_W}Z_\mu	\sum_{\ell=e}^\tau
  \left( \sum_{m=1}^3	~\overline{\nu_m} ~U_{\ell m}^*	 + \sum_{m'=1}^n \overline{N^c_{m'}} ~V_{\ell N_{m'}}^* \right)\gamma^\mu P_L\nu_\ell\nonumber\\
  &-& \frac{g}{2M_W} h	\sum_{\ell=e}^\tau  \sum_{m'=1}^n	~m_{N_{m'}}\overline{N^c_{m'}} ~V_{\ell N_{m'}}^*	 P_L\nu_\ell + \text{H.c.}
  \label{eq:typeIEWLag}
\end{eqnarray}
There is a particular utility of using this mixed mass-flavor basis in collider searches for heavy neutrinos.
Empirically, $\vert V_{\ell N_{m'}}\vert \lesssim 10^{-2}$~\cite{delAguila:2008pw,Antusch:2014woa,deGouvea:2015euy,Fernandez-Martinez:2016lgt},
which means pair production of $N_{m'}$ via EW processes is suppressed by $\vert V_{\ell N_{m'}}\vert^2 \lesssim 10^{-4}$
relative to single production of $N_{m'}$.
Moreover, in collider processes involving $\nu_{m}-N_{m'}$ vertices,
one sums over $\nu_{m}$ either because it is an internal particle or an undetected external state.
This summation effectively undoes the decomposition of one neutrino interaction state for neutral current vertices,
resulting in the basis above.
In phenomenological analyses, it is common practice to consider only the lightest heavy neutrino mass eigenstate, i.e., $N_{m'=4}$,
to reduce the effective number of independent model parameters.
In such cases, the mass eigenstate is denoted simply as $N$ and one reports sensitivity on the associated mixing element,
labeled as $\vert V_{\ell N} \vert $ or $\vert V_{\ell 4} \vert$, and which are equivalent to $\vert V_{\ell N_{m'=4}} \vert.$
Throughout this text, the $\vert V_{\ell N} \vert$ notation is adopted where possible.

\begin{figure}[t]
\begin{center}
\subfigure[]{ \includegraphics[width=.45\textwidth]{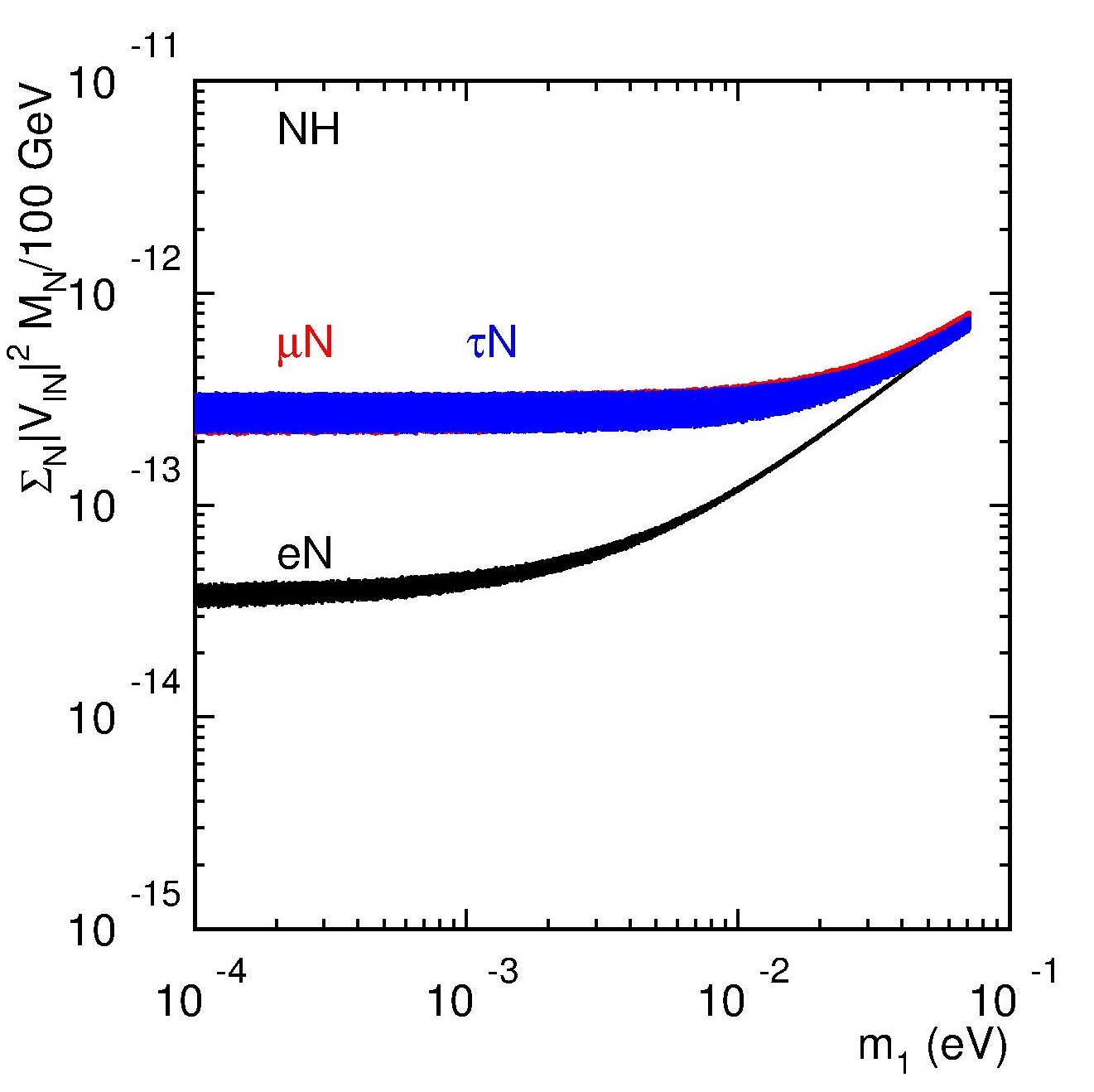}	}
\subfigure[]{ \includegraphics[width=.45\textwidth]{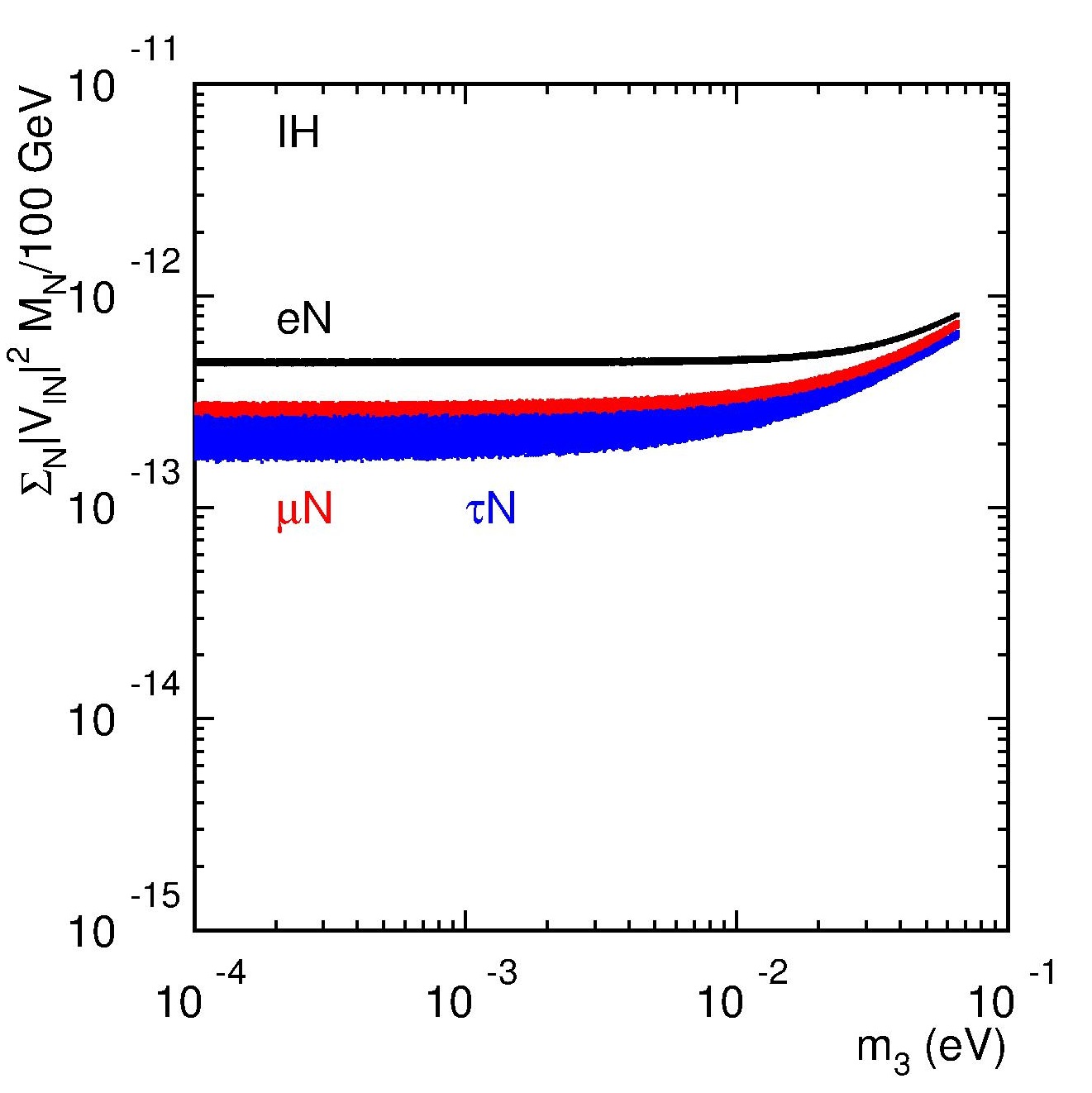}	}
\end{center}
\caption{ $\Sigma_N \left|V_{\ell N}\right|^2 M_N/100 \ \rm{GeV}$ versus the lightest neutrino mass for
(a) NH and (b) IH in the case of degenerate heavy neutrinos, assuming vanishing phases.}
\label{fig:vvmcase1}
\end{figure}

\begin{figure}[t]
\begin{center}
\subfigure[]{\includegraphics[width=.4\textwidth]{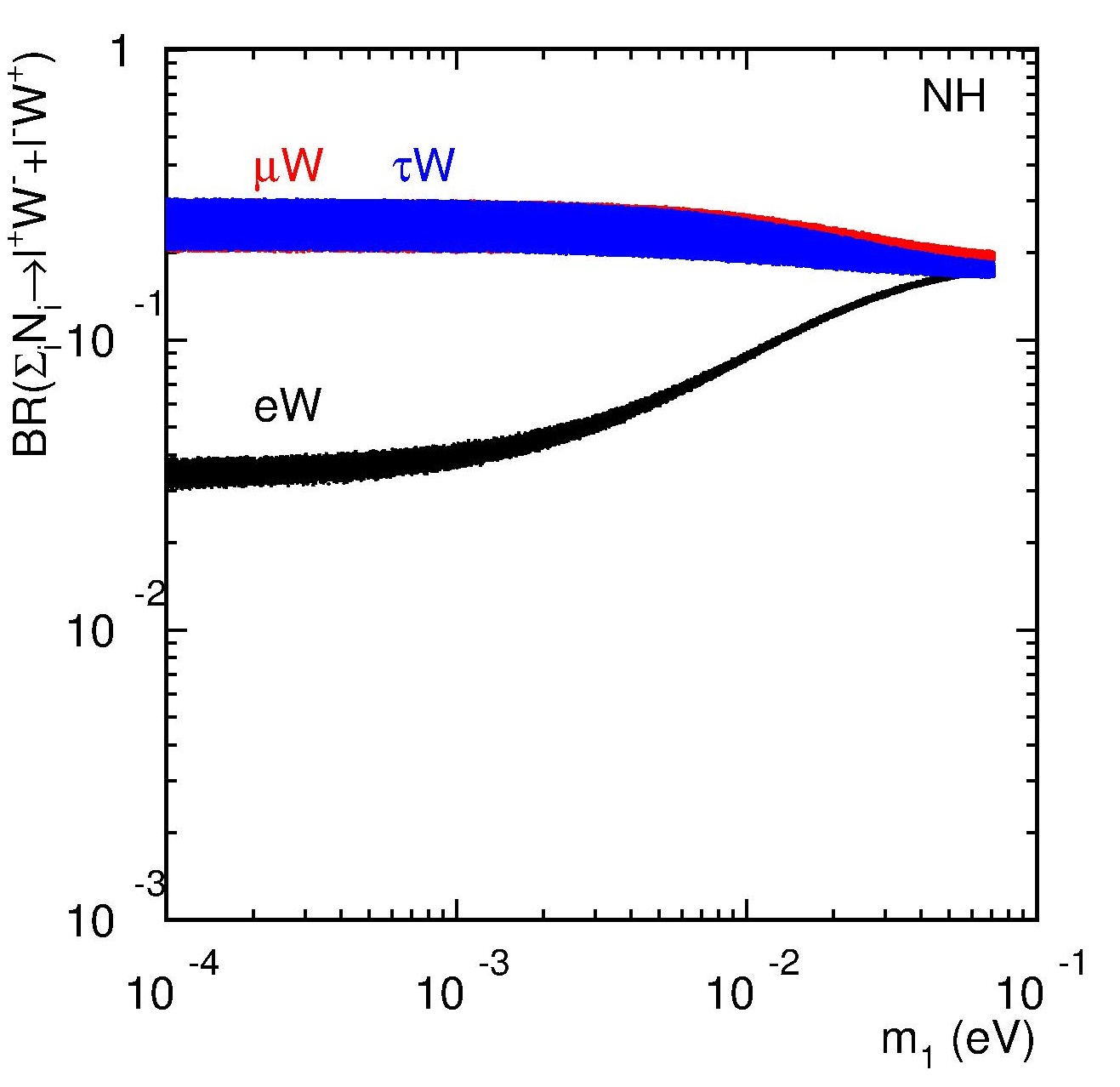}	}
\subfigure[]{\includegraphics[width=.4\textwidth]{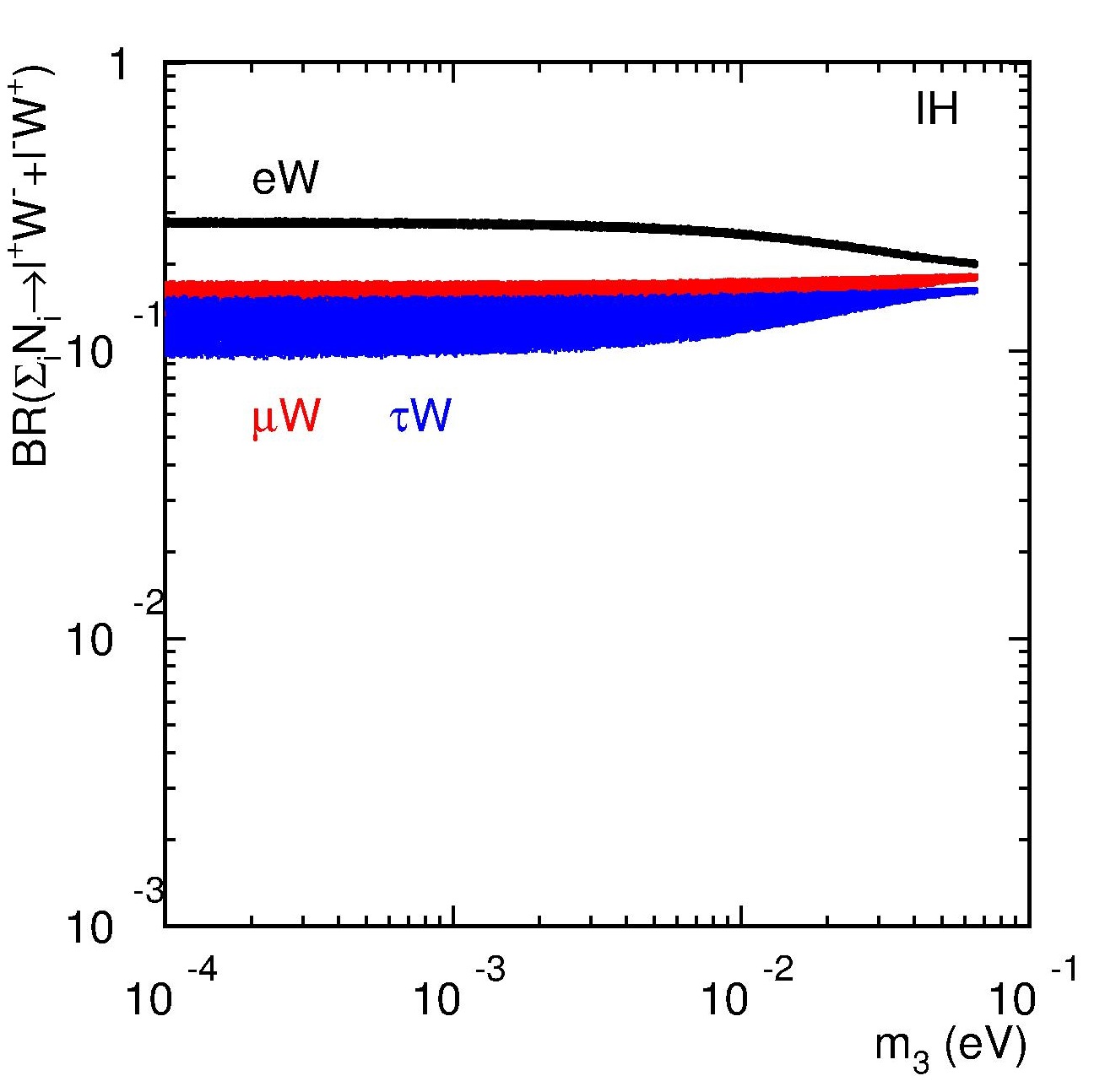}	}
\end{center}
\caption{Branching fractions of process $\sum_i N_i\rightarrow \ell^+W^- + \ell^-W^+$ versus the lightest neutrino mass for (a) NH and (b) IH 
in the degenerate case with $M_N=300 \ \rm{GeV}$ and $m_h=125 \ \rm{GeV}$, assuming vanishing phases.
} \label{fig:brnlwcase1}
\end{figure}

From Eq.~(\ref{eq:type1NuMixMatrix}), an important relation among neutrino masses can be derived. Namely, that
\begin{eqnarray}
U^\ast_{PMNS}m_\nu U^\dagger_{PMNS}+V^\ast_{\ell N}M_NV^\dagger_{\ell N}=0 \ .
\label{typei}
\end{eqnarray}
Here the masses and mixing of the light neutrinos in the first term are measurable from the oscillation experiments,
and the second term contains the masses and mixing of the new heavy neutrinos.
We now consider a simple case: degenerate heavy neutrinos with mass $M_N = diag(M_1,\cdots, M_{m'}) = M_N \mathds{I}_{m'}$.
Using this assumption, we obtain from Eq.~(\ref{typei}),
\begin{eqnarray}
M_N\sum_N (V_{\ell N}^\ast)^2=(U^\ast_{PMNS}m_\nu U^\dagger_{PMNS})_{\ell\ell} \ .
\label{dege}
\end{eqnarray}
Using the oscillation data in Table \ref{tab:nufit} as
inputs\footnote{This is done for simplicity since $U_{PMNS}$ in Table \ref{tab:nufit} is unitary whereas here it is not;
for more details, see~\cite{Esteban:2016qun,Parke:2015goa}.},
we display in Fig.~\ref{fig:vvmcase1} the normalized mixing of each lepton flavor in this 
scenario\footnote{$\sum_N (V_{\ell N}^\ast)^2=\sum_N |V_{\ell N}|^2$ only when all phases on the right-hand side of Eq.~(\ref{dege}) vanish~\cite{Perez:2009mu}.}.
Interestingly, one can see the characteristic features:
\begin{eqnarray}
\sum_N |V_{eN}|^2 & \ll \sum_N |V_{\mu N}|^2,	\sum_N |V_{\tau N}|^2  	&  \quad {\rm  for \ NH}, \\
\sum_N |V_{eN}|^2 & > 	\sum_N |V_{\mu N}|^2,	\sum_N |V_{\tau N}|^2  	&  \quad {\rm  for  \ IH}.
\end{eqnarray}
As shown in Fig.~\ref{fig:brnlwcase1}, a corresponding pattern also emerges in the branching
fraction\footnote{Where $\text{BR}(A\to X)\equiv \Gamma(A\to X)/\sum_Y\Gamma(A\to Y)$ for partial width $\Gamma(A\to Y)$.}
of the degenerate neutrinos decaying into charged leptons plus a $W$ boson,
\begin{align}
&{\rm BR}(\mu^\pm W^\mp), {\rm BR}(\tau^\pm W^\mp)\sim (20-30)\%\gg {\rm BR}(e^\pm W^\mp)\sim (3-4)\% \  & {\rm for} \ {\rm NH},\\
&{\rm BR}(e^\pm W^\mp)\sim 27\% > {\rm BR}(\mu^\pm W^\mp), {\rm BR}(\tau^\pm W^\mp)\sim (10-20)\% \  &{\rm for} \ {\rm IH},
\end{align}
with ${\rm BR}(\ell^\pm W^\mp)={\rm BR}(N_i\to \ell^+W^-+\ell^-W^+)$.
These patterns show a rather general feature that ratios of Seesaw partner observables, \eg, cross sections and branching fractions,
encode information on light neutrinos, such as their mass hierarchy~\cite{Perez:2009mu,Ibarra:2011xn}.
Hence, one can distinguish between competing light neutrino mass and mixing patterns with high energy observables.

\begin{figure}[t]
\begin{center}
\includegraphics[scale=1,width=11cm]{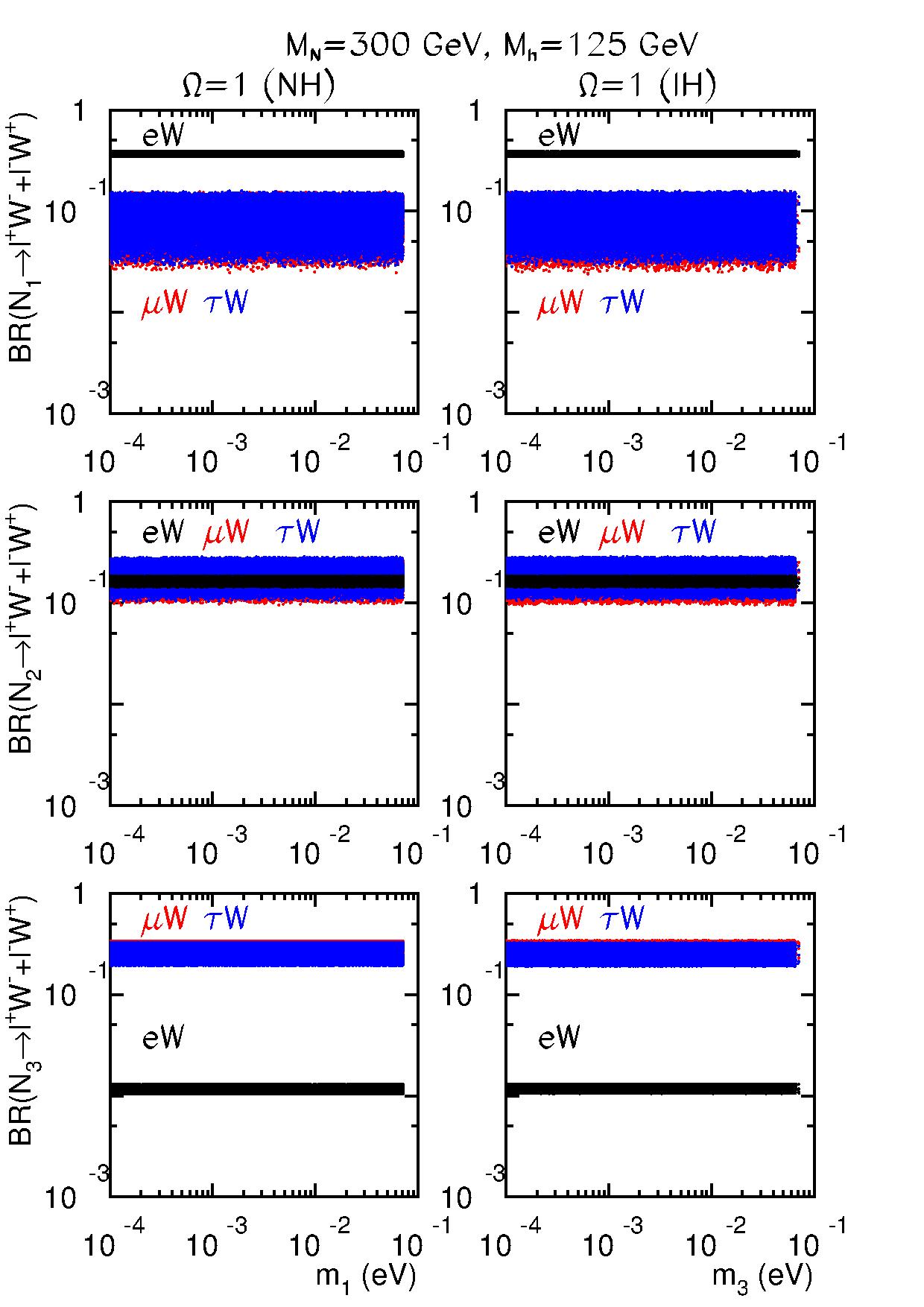}
\end{center}
\caption{Branching fractions of process $N_i\rightarrow \ell^+W^- + \ell^-W^+$ versus
the lightest neutrino mass for NH and IH in the case $\Omega=I$ with $M_i=300 \ \rm{GeV}$ and $m_h=125 \ \rm{GeV}$,
assuming vanishing Majorana phases.
} \label{fig:nibr}
\end{figure}

More generally, the $V_{\ell N}$ in Eq.~(\ref{typei}) can be formally solved in terms of an arbitrary orthogonal complex matrix $\Omega$,
known as the Casas-Ibarra parametrization~\cite{Casas:2001sr}, using the ansatz
\begin{eqnarray}
V_{\ell N}=U_{PMNS} ~m_\nu^{1/2}\Omega M_N^{-1/2},
\label{omega}
\end{eqnarray}
with the orthogonality condition $\Omega \Omega^T=I$.
For the simplest incarnation of a unity matrix $\Omega=I$, 
the $|V_{\ell N_{m'}}|^2$ are proportional to one and only one light neutrino mass, and
thus the branching ratio of $N_{m'}\to \ell^\pm W^\mp$ for each lepton flavor is independent of neutrino mass and universal for both NH and IH~\cite{Perez:2009mu}.
Nevertheless, one can still differentiate between the three heavy neutrinos according to the decay rates to their leading decay channels.
As shown in Fig.~\ref{fig:nibr} for $\Omega=I$, one sees
\begin{align}
&{\rm BR}(e^\pm W^\mp)\sim 40\%>{\rm BR}(\mu^\pm W^\mp),{\rm BR}(\tau^\pm W^\mp)\sim (4-15)\% \  & {\rm for} \ N_1,\\
&{\rm BR}(e^\pm W^\mp)\sim 20\%\approx {\rm BR}(\mu^\pm W^\mp)\approx {\rm BR}(\tau^\pm W^\mp)\sim (10-30)\% & \ {\rm for} \ N_2,\\
&{\rm BR}(\mu^\pm W^\mp),{\rm BR}(\tau^\pm W^\mp)\sim (15-40)\%\gg {\rm BR}(e^\pm W^\mp)\sim 1\% \  & {\rm for} \ N_3.
\end{align}
A realistic Dirac mass matrix can be quite arbitrary with three complex angles parameterizing the orthogonal matrix $\Omega$. 
However, the arbitrariness of the Dirac mass matrix is not a universal feature of Seesaw models;
the neutrino Yukawa matrix in the Type II Seesaw, for example, is much more constrained. 

Beyond this, Fig.~\ref{fig:vvmcase1} also shows another general feature of minimal, high-scale Seesaw constructions,
namely that the active-sterile mixing $\vert V_{\ell N}\vert$ is vanishingly small.
For a heavy neutrino mass of $M_N \sim 100$ GeV, Eq.~(\ref{dege}) implies $\vert V_{\ell N}\vert^2 \sim 10^{-14} - 10^{-12}$.
This leads to the well-known decoupling of observable lepton number violation in the minimal, 
high-scale Type I Seesaw scenario at colliders experiments~\cite{Pilaftsis:1991ug,Kersten:2007vk,Moffat:2017feq}.
For low-scale Type I Seesaws, such decoupling of  observable lepton number violation also occurs:
Due to the allowed arbitrariness of the matrix $\Omega$ in Eq.~(\ref{omega}),
it is possible to construct $\Omega$ and $M_N$ with particular entry patterns or symmetry structures, also known as ``textures'' in the literature,
such that $V_{\ell N}$ is nonzero but $m_\nu$ vanishes.
Light neutrino masses can then be generated as perturbations from these textures.
In Ref.~\cite{Moffat:2017feq} it was proved that such delicate
(and potentially fine-tuned~\cite{AristizabalSierra:2011mn,Lopez-Pavon:2015cga,Fernandez-Martinez:2015hxa})
constructions result in small neutrino masses being proportional to small $L$-violating parameters,
instead of being inversely proportional as in the high-scale case.
Subsequently, in low-scale Seesaw scenarios that assume only fermionic gauge singlets,
tiny neutrino masses is equivalent to an approximate conservation of lepton number,
and leads to the suppression of observable $L$ violation in high energy processes.
Hence, any observation of lepton number violation (and Seesaw partners in general)
at collider experiments implies a much richer neutrino mass-generation scheme than just the canonical, high-scale Type I Seesaw.

\subsubsection{Type I+II Hybrid Seesaw Mechanism}\label{sec:typeIHybrid}

While the discovery of lepton number violation in, say, $0\nu\beta\beta$ or hadron collisions 
would imply the Majorana nature of neutrinos~\cite{Schechter:1981bd,Hirsch:2006yk,Duerr:2011zd},
it would be less clear which mechanism or mechanisms are driving light neutrino masses to their sub-eV values.
This is because in the most general case neutrinos possess both LH and RH Majorana masses in addition to Dirac masses.
In such hybrid Seesaw models, two or more ``canonical'' tree- and loop-level mechanisms
are combined and, so to speak, may give rise to phenomenology that is greater than the sum of its parts.

A well-studied hybrid model is the Type I+II Seesaw mechanism, wherein the light neutrino mass matrix $M_{\nu}$,
when $M_D M_R^{-1}\ll1$, is given by~\cite{Chen:2005jm,Akhmedov:2006de,Akhmedov:2006yp,Chao:2007mz,Chao:2009ef,Gu:2008yj,Chao:2008mq}
\begin{equation}
 M_{\nu}^{light} = M_L - M_D M_R^{-1} M_D^T.
 \label{eq:hybridNuMass}
\end{equation}
Here, the Dirac and Majorana mass terms, $M_D$, $M_R$, have their respective origins according to the Type I model,
whereas $M_L$ originates from the Type II mechanism; see Sec.~\ref{sec:type2} for details. 
In this scenario, sub-eV neutrino masses can arise not only from parametrically small Type I and II masses 
but additionally from an incomplete cancellation of the two terms~\cite{Akhmedov:2006de,Akhmedov:2006yp,Chao:2007mz}. 
While a significant or even moderate cancellation requires a high-degree of fine tuning and is radiatively instable~\cite{Chao:2008mq}, 
this situation cannot theoretically be ruled out \textit{a priori}. 
For a one-generation mechanism, the relative minus sign in Eq.~(\ref{eq:hybridNuMass}) is paramount for such a cancellation; however, 
in a multi-generation scheme, it is not as crucial as $M_D$ is, in general, complex and can absorb the sign through a phase rotation. 
Moreover, this fine-tuning scenario is a caveat of the aforementioned
decoupling of $L$-violation in a minimal Type I Seesaw from LHC phenomenology~\cite{Pilaftsis:1991ug,Kersten:2007vk,Moffat:2017feq}.
As we will discuss shortly, regardless of its providence, if such a situation were to be realized in nature, then vibrant and rich collider signatures emerges.


\subsubsection{Type I Seesaw in U$(1)_X$ Gauge Extensions of the Standard Model}\label{sec:type1U1X}

Another manner in which the decoupling of heavy Majorana neutrinos $N$ from collider experiments can be avoided is through the introduction of new gauge symmetries,
under which $N$ is charged.
One such example is the well-studied U$(1)_{X}$ Abelian gauge extension of the 
SM~\cite{Langacker:1980js,Hewett:1988xc,Faraggi:1990ita,Accomando:2010fz,Faraggi:2015iaa},
where U$(1)_X$ is a linear combination of U$(1)_Y$ and U$(1)_{B-L}$
after the spontaneous breaking of electroweak symmetry and $B-L$ (baryon minus lepton number) symmetries.
In this class of models, RH neutrinos are introduced to cancel gauge anomalies and realize a Type I Seesaw mechanism.

Generally, such a theory can be described by modifying the SM covariant derivatives by~\cite{Salvioni:2009mt}
\begin{eqnarray}
D_\mu \ni ig_1YB_\mu \quad\to\quad D_\mu \ni ig_1YB_\mu + i (\tilde{g}Y + g_1' Y_{BL}) B_\mu',
\label{U1}
\end{eqnarray}
where $B_\mu (Y)$ and $B_\mu' (Y_{BL})$ are the gauge fields (quantum numbers) of U$(1)_Y$ and U$(1)_{B-L}$, respectively.
The most economical extension with vanishing mixing between U$(1)_Y$ and U$(1)_{B-L}$, \ie, U$(1)_X={\rm U}(1)_{B-L}$ and $\tilde{g}=0$ in Eq.~(\ref{U1}),
introduces three RH neutrinos and  a new complex scalar $S$ that are all charged under the new gauge group
but remain singlets under the SM symmetries~\cite{Carlson:1986cu,Buchmuller:1991ce,Abbas:2007ag}.
In this extension one can then construct the neutrino Yukawa interactions
\begin{eqnarray}
{\cal L}_I^Y  &  =  & -   \bar{L}_L \ Y_\nu^D \ \tilde{H} \ N_R
\ -\frac{1}{2}Y_\nu^M \ \overline{(N^c)_L}  \ N_R \ S+ \ \rm{H.c.}
\label{Y}
\end{eqnarray}
Once the Higgs $S$ acquires the vacuum expectation value $\langle S\rangle = v_S/\sqrt{2}$,
$B-L$ is broken, spontaneously generating the RH Majorana mass matrix $M_N=Y_\nu^M v_S/\sqrt{2}$ from Eq.~(\ref{Y}).

It is interesting to note that the scalar vev provides a dynamical mechanism for the heavy,
RH Majorana mass generation, \ie, a Type I Seesaw via a Type II mechanism; see Sec.~\ref{sec:type2} for more details.
The Seesaw formula and the mixing between the SM charged leptons and heavy neutrinos here are exactly the same as those in the canonical Type I Seesaw.
The mass of neutral gauge field $B_\mu'$, $M_{Z'}=M_{Z_{B-L}}=2g_{BL}v_S$, is generated from $S$'
kinetic term, $\left( D_\mu S \right)^\dagger \left( D^\mu S \right)$
with $D_\mu S = \partial_\mu S \ + \ i 2 g_{BL} B_\mu' S$.
Note that in the minimal model, $g_{BL}=g_1'$.
As in other extended scalar scenarios, the quadratic term $H^\dagger H S^\dagger S$ in the scalar potential
results in the SM Higgs $H$ and $S$ interaction states mixing into two CP-even mass eigenstates, $H_1$ and $H_2$.

\subsubsection{Type I+II Hybrid Seesaw in Left-Right Symmetric Model}\label{sec:hybrid}
As discussed in Sec.~\ref{sec:typeIHybrid}, it may be the case that light neutrino masses result from an interplay of multiple Seesaw mechanisms.
For example: the Type I+II hybrid mechanism with light neutrino masses given by Eq.~(\ref{eq:hybridNuMass}).
It is also worth observing two facts:
First, in the absence of Majorana masses, the minimum fermionic field content for a Type I+II Seesaw automatically obeys an accidental global U$(1)_{B-L}$ symmetry.
Second, with three RH neutrinos, all fermions can be sorted into either SU$(2)_L$ doublets (as in the SM) or SU$(2)_R$ doublets, its RH analogue.
As the hallmark of the Type II model (see Sec.~\ref{sec:type2}) is the spontaneous generation of LH Majorana masses from a scalar SU$(2)_L$ triplet $\Delta_L$,
it is conceivable that RH neutrino Majorana masses could also be generated spontaneously, but from a scalar SU$(2)_R$ triplet $\Delta_R$.
(This is similar to the spontaneous breaking of U$(1)_{B-L}$ in Sec.~\ref{sec:type1U1X}.)
This realization of the Type I+II Seesaw is known as the Left-Right Symmetric Model
(LRSM)~\cite{Pati:1974yy,Mohapatra:1974hk,Mohapatra:1974gc,Senjanovic:1975rk,Senjanovic:1978ev},
and remains one of the best-motivated and well-studied extensions of the SM.
For recent, dedicated reviews, see Ref.~\cite{Duka:1999uc,Mohapatra:2006gs,Senjanovic:2016bya}.

The high energy symmetries of the LRSM is based on the extended gauge group
\begin{equation}
\mathcal{G}_{\rm LRSM} = \SU{3}{c}\otimes\SU{2}{L}\otimes\SU{2}{R}\otimes\UX{1}{B-L},
\label{eq:lrsmGaugeGrp}
\end{equation}
or its embeddings, and conjectures that elementary states, in the UV limit, participate in LH and RH chiral currents with equal strength.
While the original formulation of model supposes a generalized parity $\mathcal{P}_X=\mathcal{P}$
that enforces an exchange symmetry between fields charged under $\SU{2}{L}$ and $\SU{2}{R}$,
it is also possible to achieve this symmetry via a generalized charge conjugation $\mathcal{P}_X=\mathcal{C}$~\cite{Maiezza:2010ic}.
For fermionic and scalar multiplets $Q_{L,R}$ and $\Phi$, the exchange relationships are~\cite{Maiezza:2010ic},
\begin{equation}
 \mathcal{P}:\left\{\begin{matrix}
Q_L \leftrightarrow Q_R
\\
\Phi \leftrightarrow \Phi^\dagger
\end{matrix}\right.,
\quad\text{and}\quad
\mathcal{C}:\left\{\begin{matrix}
Q_L \leftrightarrow (Q_R)^c
\\
\Phi \leftrightarrow \Phi^T
\end{matrix}\right.,
\quad\text{where}\quad (Q_R)^c = C\gamma^0Q_R^*.
\end{equation}
A non-trivial, low-energy consequence of these complementary formulations of the LRSM is the
relationship between the LH CKM matrix in the SM, $V_{ij}^{\rm L}$, and its RH analogue, $V_{ij}^{\rm R}$.
For generalized conjugation, one has $\vert V_{ij}^{\rm R} \vert = \vert V_{ij}^{\rm L}\vert$,
whereas $\vert V_{ij}^{\rm R} \vert \approx \vert V_{ij}^{\rm L}\vert+ \mathcal{O}(m_b/m_t)$ for generalized
parity~\cite{Zhang:2007fn,Zhang:2007da,Maiezza:2010ic,Senjanovic:2014pva,Senjanovic:2015yea}.
Moreover, LR parity also establishes a connection between the Dirac and Majorana masses in the leptonic sector~\cite{Nemevsek:2012iq,Senjanovic:2016vxw}.
Under generalized parity, for example, the Dirac $(Y^D_{1,2})$ and Majorana $(Y_{L,R})$ Yukawa matrices must satisfy~\cite{Senjanovic:2016vxw},
\begin{equation}
 Y_{1,2}^D = Y_{1,2}^{D\dagger} \quad\text{and}\quad Y_L = Y_R.
\end{equation}
Such relationships in the LRSM remove the arbitrariness of neutrino Dirac mass matrices, as discussed in Sec.~\ref{sec:type1Canon},
and permits one to calculate $\Omega$, even for nonzero $\Delta_L$ vev~\cite{Akhmedov:2008tb,Nemevsek:2012iq}.
However, the potential cancellation between Type I and II Seesaw masses in Eq.~\ref{eq:hybridNuMass} still remains. 

In addition to the canonical formulation of the LRSM are several alternatives. For example:
It is possible to instead generate LH and RH Majorana neutrino masses radiatively
in the absence of triplet scalars~\cite{FileviezPerez:2017zwm,FileviezPerez:2016erl}.
One can gauge baryon number and lepton number independently, which, for an anomaly-free theory, gives rise to vector-like leptons and
a Type III Seesaw mechanism~\cite{FileviezPerez:2008sr,Duerr:2013opa}~(see Sec.~\ref{sec:type3}),
as well as embed the model into an $R$-parity-violating Supersymmetric framework~\cite{FileviezPerez:2008sx,Everett:2009vy}.

Despite the large scalar sector of the LRSM
(two complex triplets and one complex bidoublet), and hence a litany of neutral and charged Higgses,
the symmetry structure in Eq.~(\ref{eq:lrsmGaugeGrp}) confines the number in independent degrees of freedom to 18~\cite{Deshpande:1990ip,Duka:1999uc}.
These consist of three mass scales  $\mu_{1,\dots,3}$, 14 dimensionless couplings $\lambda_{1,\dots,4}$,
$\rho_{1,\dots,4}$, $\alpha_{1,\dots,3}$, $\beta_{1,\dots,3}$, and one CP-violating phase, $\delta_2$.
For further discussions on the spontaneous breakdown of CP in LR scenarios, see also~Refs.~\cite{Senjanovic:1978ev,Basecq:1985sx,Kiers:2002cz}. 
With explicit CP conservation, the minimization conditions on the scalar potential give rise to
the so-called LRSM vev Seesaw relationship~\cite{Deshpande:1990ip},
\begin{equation}
v_L = \frac{\beta_2 k_1^2  + \beta_1 k_1 k_2 + \beta_3 k_2^2}{(2\rho_1 - \rho_3)v_R},
\label{eq:vevSeesaw}
\end{equation}
where, $v_{L,R}$ and $k_{1,2}$ are the vevs of $\Delta_{L,R}$ and the Higgs bidoublet $\Phi$, 
respectively, with $v_L^2 \ll k_1^2 + k_2^2 \approx (246\gev)^2 \ll v_R$.

In the LRSM, the bidoublet $\Phi$ fulfills the role of the SM Higgs to generate the known Dirac masses of elementary fermions and 
permits a neutral scalar $h_i$ with mass $m_{h_i}\approx125$ GeV and SM-like couplings. 
In the absence of egregious fine-tuning, \ie, $\rho_3 \not\approx 2\rho_1$, Eq.~(\ref{eq:vevSeesaw}) suggests that $v_L$ in the LRSM is inherently small because, 
in addition to $k_{1},k_{2}\ll v_R$, custodial symmetry is respected (up to hypercharge corrections) when all $\beta_i$ are identically zero~\cite{Mitra:2016kov}.
Consistent application of such naturalness arguments reveals a lower bound on the  scalar potential parameters~\cite{Mitra:2016kov},
\begin{eqnarray}
 \rho_{1,2,4} > \frac{g^2_R}{4}\left(\frac{m_{\rm FCNH}}{M_{W_R}}\right)^2,
 &\quad&
 \rho_{3} > g^2_R\left(\frac{m_{\rm FCNH}}{M_{W_R}}\right)^2 + 2\rho_1 \sim 6\rho_1,
 \\
 \alpha_{1,\dots,3} > g^2_R\left(\frac{m_{\rm FCNH}}{M_{W_R}}\right)^2,
 &\quad&
 \mu_{1,2}^2 > (m_{\rm FCNH})^2, \quad \mu_3^2 > \frac{1}{2}(m_{\rm FCNH})^2,
\end{eqnarray}
where $M_{W_R}$ and $g_R$ are the mass and coupling of the $W_R^\pm$ gauge boson associated with $\SU{2}{R}$, 
and $m_{\rm FCNH}$ is the mass scale of the LRSM scalar sector participating in flavor-changing neutral transitions.
Present searches for neutron EDMs~\cite{Zhang:2007da,Zhang:2007fn,Xu:2009nt,Maiezza:2014ala} and
FCNCs~\cite{Chakrabortty:2012pp,Bambhaniya:2013wza,Maiezza:2016bzp,Bertolini:2014sua,Maiezza:2014ala} require $m_{\rm FCNH} > 10-20$ TeV at 90\% CL.
Subsequently, in the absence of FCNC-suppressing mechanisms, $\rho_{i} > 1$ for LHC-scale $W_R$. 
Thus, discovering LRSM at the LHC may suggest a strongly coupled scalar sector. Conversely, for $\rho_i<1$ and $m_{\rm FCNH}\sim15~(20)\tev$, 
one finds $M_{W_R}\gtrsim10~(12)\tev$, scales that are within the reach of future hadron colliders~\cite{Arkani-Hamed:2015vfh,Golling:2016gvc,Mitra:2016kov}. 
For more detailed discussions on the perturbativity and stability of the LRSM scalar section, 
see Refs.~\cite{Bertolini:2009qj,Bertolini:2009es,Bertolini:2012im,Maiezza:2016bzp,Mitra:2016kov,Mohapatra:2014qva,Maiezza:2016ybz} and references therein.

After $\Delta_R$ acquires a vev and LR symmetry is broken spontaneously, the neutral component of $\SU{2}{R}$, \ie, $W_R^{3}$, and the $\UX{1}{B-L}$ boson, 
\ie, $X_{B-L}$, mix into the massive eigenstate $Z'_{\rm LRSM}$ (sometimes labeled $Z_R$) and the orthogonal, massless vector boson $B$. 
$B$ is recognized as the gauge field associated with weak hypercharge in the SM, 
the generators of which are built from the remnants of $\SU{2}{R}$ and $\UX{1}{B-L}$. 
The relation between electric charge $Q$, weak left/right isospin $T_{L/R}^3$, baryon minus lepton number $B$-$L$, and weak hypercharge $Y$ is given by
\begin{equation}
 Q = T_{L}^3 + T_{R}^3 + \frac{(B-L)}{2} \equiv T_{L}^3 + \frac{Y}{2}, \quad\text{with}\quad Y = 2T_{R}^3 + (B-L).
\end{equation}
This in turn implies that the remaining components of $\SU{2}{R}$, $W_R^{1}$ and $W_R^{2}$, 
combine into the state $W_R^\pm$ with electric charge $Q^{W_R}=\pm1$ and mass $M_{W_R}=g_R v_R/\sqrt{2}$. 
After EWSB, it is possible for the massive $W_R$ and $W_L$ gauge fields to mix, 
with the mixing angle $\xi_{\rm LR}$ given by $\tan2\xi_{\rm LR} = 2k_1 k_2/(v_R^2 - v_L^2)\lesssim 2 v_{\rm SM}^2/v_R^2$.
Neutral meson mass splittings~\cite{Beall:1981ze,Langacker:1989xa,Bertolini:2014sua,Maiezza:2010ic,Bernard:2015boz,Buras:2013ooa}
coupled with improved lattice calculations, e.g.~\cite{Mescia:2012fg,Garron:2016mva},
Weak CPV~\cite{Maiezza:2010ic,Cirigliano:2016yhc,Buras:2013ooa},
EDMs~\cite{Maiezza:2010ic,Zhang:2007da,Zhang:2007fn,Buras:2013ooa}, and CP violation in the electron EDM~\cite{Nemevsek:2012iq},
are particularly sensitive to this mixing, implying the competitive bound of $M_{W_R} \gtrsim 3$ TeV at 95\% CL~\cite{Bertolini:2014sua}. 
This forces $W_L-W_R$ mixing to be, $\tan2\xi_{\rm LR}/2 \approx  \xi_{\rm LR} \lesssim M_W^2/M_{W_R}^2 < 7-7.5\times10^{-4}.$ 
A similar conclusion can be reached on $Z- Z'_{LRSM}$ mixing. 
Subsequently, the light and heavy mass eigenstates of LRSM gauge bosons, $W_1^\pm,~W_2^\pm,~Z_1,~Z_2$, where $M_{V_1}<M_{V_2}$, 
are closely aligned with their gauge states. 
In other words, to a very good approximation, $W_1\approx W_{\rm SM}$, $Z_1\approx Z_{\rm SM}$, 
$W_2\approx W_R$ and $Z'\approx Z'_{LRSM}$~(or sometimes $Z'\approx Z_R$).
The mass relation between the LR gauge bosons is $M_{Z_R} = \sqrt{2\cos^2\theta_W / \cos2\theta_W}M_{W_R}\approx (1.7)\times M_{W_R}$,
and implies that bounds on one mass results in indirect bounds on the second mass; see, for example, Ref.~\cite{Lindner:2016lpp}.

\subsubsection{Heavy Neutrino Effective Field Theory}\label{sec:neft}
It is possible that the coupling of TeV-scale Majorana neutrinos to the SM sector is dominated
by new states with masses that are hierarchically larger than the heavy neutrino mass or the reach of present-day collider experiments.
For example: Scalar SU$(2)_R$ triplets in the Left-Right Symmetric Model may acquire vevs $\mathcal{O}(10)$ TeV,
resulting in new gauge bosons that are kinematically accessible at the LHC but,
due to $\mathcal{O}(10^{-3}-10^{-2})$ triplet Yukawa couplings, give rise to EW-scale RH Majorana neutrino masses.
In such a pathological but realistic scenario, the LHC phenomenology appears as a canonical Type I Seesaw mechanism
despite originating from a different Seesaw mechanism~\cite{Ruiz:2017nip}.
While it is generally accepted that such mimicry can occur among Seesaws, few explicit examples exist in the literature and further investigation is encouraged.

For such situations, it is possible to parameterize the effects of super-heavy degrees of freedom
using the Heavy Neutrino Effective Field Theory (NEFT) framework~\cite{delAguila:2008ir}.
NEFT is an extension of the usual SM Effective Field Theory (SMEFT)~\cite{Burges:1983zg,Leung:1984ni,Buchmuller:1985jz,Grzadkowski:2010es},
whereby instead of augmenting the SM Lagrangian with higher dimension operators
one starts from the Type I Seesaw Lagrangian in Eq.~(\ref{eq:type1Lag}) and builds operators using that field content.
Including all SU$(3)$ $\otimes$ SU$(2)_L$ $\otimes$ U$(1)_Y$-invariant, operators of mass dimension $d>4$, the NEFT Lagrangian before EWSB is given by
\begin{eqnarray}
 \mathcal{L}_{\rm NEFT} &=& \mathcal{L}_{\rm Type~I} + \sum_{d=5}\sum_{i} \frac{\alpha_i^{(d)}}{\Lambda^{(d-4)}}\mathcal{O}_{i}^{(d)}.
\end{eqnarray}
Here, $\mathcal{O}_{i}^{(d)}$ are dimension $d$, Lorentz and gauge invariant permutations of Type I fields,
and $\alpha_i^{(d)}\ll4\pi$ are the corresponding Wilson coefficients.
The list of $\mathcal{O}_{i}^{(d)}$ are known explicitly for $d=5$~\cite{Aparici:2009fh,Elgaard-Clausen:2017xkq},
6~\cite{delAguila:2008ir,Elgaard-Clausen:2017xkq},
and 7~\cite{Bhattacharya:2015vja,Liao:2016qyd,Elgaard-Clausen:2017xkq}, and can be built for larger $d$  
following Refs.~\cite{Henning:2015alf,Kobach:2016ami,Liao:2017amb}.

After EWSB, fermions should then be decomposed into their mass eigenstates via quark and lepton mixing.
For example: among the $d=6$, four-fermion contact operations $\mathcal{O}_i^{(6)}$ that contribute to
heavy $N$ production in hadron colliders (see Eq.~(\ref{eq:heavyNDYCC})) in the interaction/gauge basis are~\cite{delAguila:2008ir}
\begin{eqnarray}
 \mathcal{O}_V^{(6)} 	= \left(\overline{d}\gamma^\mu P_R u\right)\left(\overline{e}\gamma_\mu P_R N_R\right)
 \quad\text{and}\quad
 \mathcal{O}_{S3}^{(6)} = \left(\overline{Q}\gamma^\mu P_R N_R\right)\varepsilon\left(\overline{L}\gamma_\mu P_R d\right).
\end{eqnarray}
In terms of light $(\nu_m)$ and heavy $(N_{m'})$ mass eigenstates and using Eq.~(\ref{eq:nuMixDefs}),
one can generically~\cite{Atre:2009rg,Han:2012vk}
decompose the heavy neutrino interaction state $N_\ell$ as
$N_{\ell} = \sum_{m=1}^{3} X_{\ell m}\nu_{m}^c + \sum_{m'=1}^{n}Y_{\ell N_{m'}} N_{m'},$
with $\vert Y_{\ell N_{m'}} \vert$ of order the elements of $U_{PMNS}$.
Inserting this into the preceding operators gives quantities in terms of leptonic mass eigenstates:
\begin{eqnarray}
  \mathcal{O}_V^{(6)} 	&=&
  \sum_{m=1}^3\left(\overline{d}\gamma^\mu P_R u\right)\left(\overline{\ell}\gamma_\mu P_R~X_{\ell m}~\nu_{m}^c\right) ~+~
  \sum_{m'=1}\left(\overline{d}\gamma^\mu P_R u\right)\left(\overline{\ell}\gamma_\mu P_R~Y_{\ell N_{m'}}~N_{m'}\right),
 \quad\text{and}\quad\nonumber\\
 \mathcal{O}_{S3}^{(6)} &=&
 \sum_{m=1}^3\left(\overline{Q}\gamma^\mu P_R ~X_{\ell m} \nu_{m}^c \right)\left(\overline{\ell}\gamma_\mu P_R d\right) ~+~
 \sum_{m'=1}\left(\overline{Q}\gamma^\mu P_R ~Y_{\ell N_{m'}} N_{m'} \right)\left(\overline{\ell}\gamma_\mu P_R d\right).
\end{eqnarray}
After EWSB, a similar decomposition for quarks gauge states in terms of CKM matrix elements and mass eigenstates should be applied.
For more information on such decompositions, see, \eg,~\cite{Ruiz:2017nip} and references therein.
It should be noted that after integrating out the heavy $N$ field, the marginal operators at $d>5$ generated from the Type I Lagrangian
are not the same operators generated by integrating
the analogous Seesaw partner in the Type II and III scenarios~\cite{Abada:2007ux,delAguila:2012nu}.


\subsection{Heavy Neutrinos at Colliders}
\label{sec:typeIhybrid}
The connection between low-scale Seesaw models and colliders is made no clearer than in searches for heavy neutrinos,
both Majorana and (pseudo-)Dirac, in the context of Type I-based scenarios.
While extensive, the topic's body of literature is still progressing in several directions.
This is particularly true for the development of collider signatures, Monte Carlo tools, and high-order perturbative corrections.
Together, these advancements greatly improve sensitivity to neutrinos and their mixing structures at collider experiments.

We now review the various searches for $L$-violating collider processes facilitated by Majorana neutrinos $N$.
We start with low-mass (Sec.~\ref{sec:type1LowMass}) and high-mass (Secs.~\ref{sec:type1HighMassPP} and~\ref{sec:type1HighMassEX}) neutrinos
in the context of Type I-based hybrid scenarios, before moving onto Abelian (Sec.~\ref{sec:type1Abelian}) and non-Abelian (Sec.~\ref{sec:lrsmCollider})
gauge extensions, and finally the semi-model independent NEFT framework (Sec.~\ref{sec:neftTests}).
Lepton number violating collider processes involving pseudo-Dirac neutrinos are,
by construction, suppressed~\cite{Wolfenstein:1981kw,Petcov:1982ya,Leung:1983ti,Valle:1983dk,Moffat:2017feq}.
Thus, a discussion of their phenomenology is outside the scope of this review and we refer readers to
thorough reviews such as Refs.~\cite{Ibarra:2011xn,Weiland:2013wha,Antusch:2016ejd}.

\subsubsection{Low-Mass Heavy Neutrinos at $pp$ and $ee$ Colliders}\label{sec:type1LowMass}

\begin{figure}[!t]
\begin{center}
\subfigure[]{\includegraphics[scale=1,width=.48\textwidth]{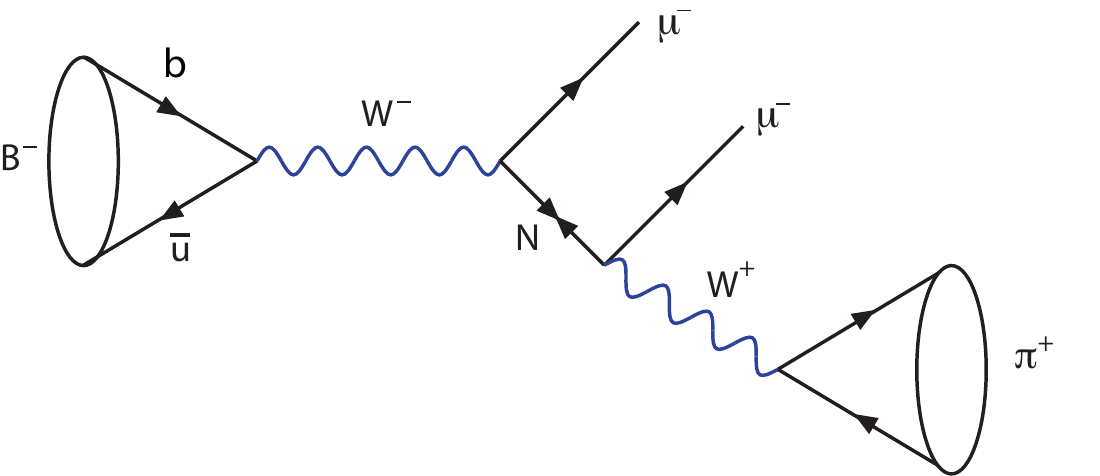}	\label{fig:lnvBDecayDiagram}	}
\subfigure[]{\includegraphics[scale=1,width=.48\textwidth]{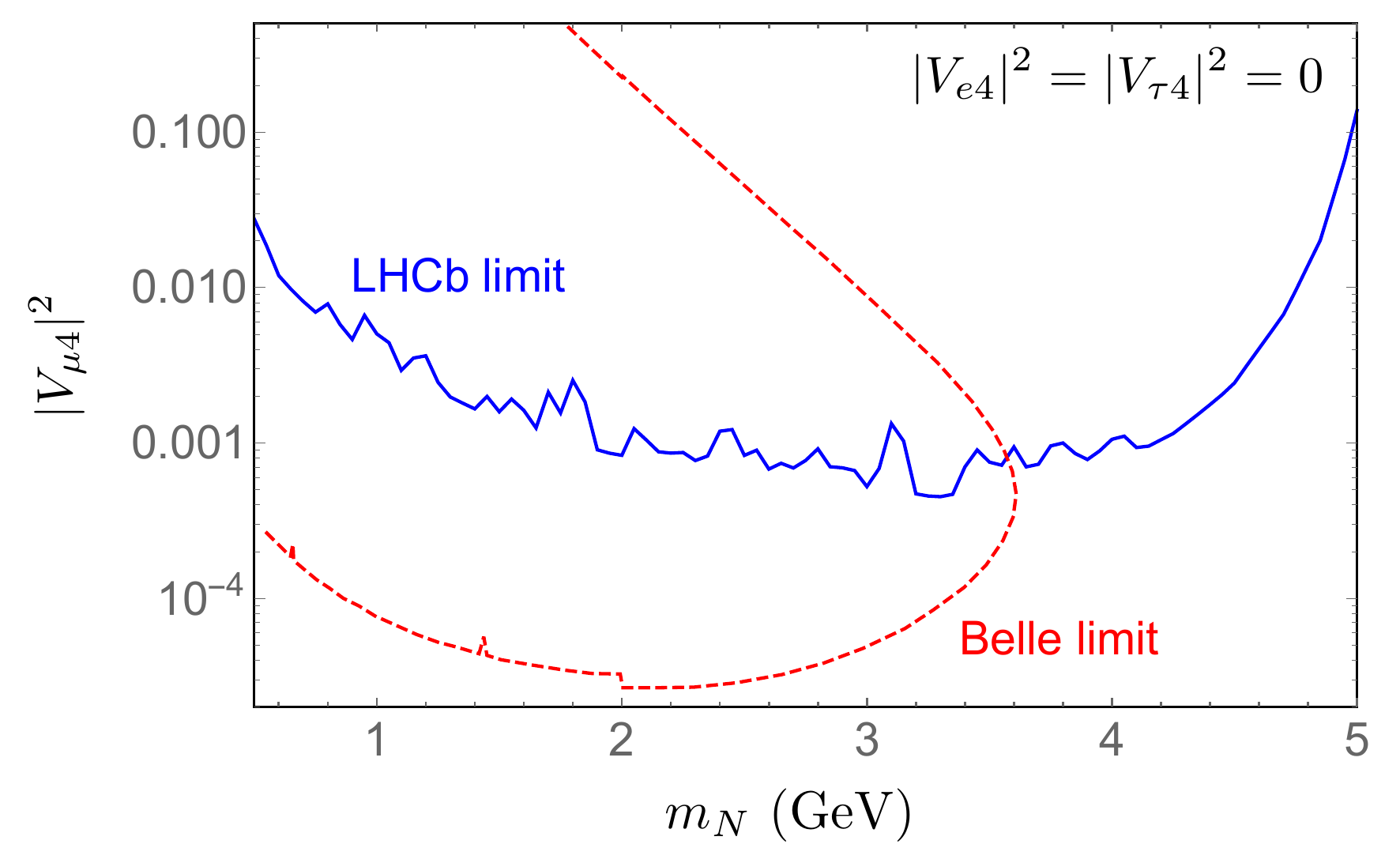}		\label{fig:lnvBDecayLimits}	}
\end{center}
\caption{
(a) $B^-$ meson decay to $L$-violating final state via heavy Majorana $N$~\cite{Aaij:2014aba}.
(b) LHCb and Belle I limits on $\vert V_{\mu N}\vert^2$ (labeled $\vert V_{\mu 4}\vert^2$ in the figure)
as a function of $N$ mass after $\mathcal{L}=3$ fb$^{-1}$ at 7-8 TeV LHC~\cite{Shuve:2016muy}.
}
\label{fig:lhcbLimits}
\end{figure}

For Majorana neutrinos below the $M_W$ mass scale, lepton number violating processes may manifest in numerous ways, including
rare decays of mesons, baryons, $\mu$ and $\tau$ leptons, and even SM electroweak bosons.
Specifically, one may discover $L$ violation in
three-body meson decays to lighter
mesons $M_1^\pm \to M_2^\mp \ell_1^\pm \ell_2^\pm$~\cite{Littenberg:1991ek,Littenberg:2000fg,Dib:2000wm,Atre:2005eb,Atre:2009rg,Cvetic:2010rw,Dib:2014iga,Cvetic:2014nla,Cvetic:2015ura,Cvetic:2015naa,
Cvetic:2016fbv,Dib:2014pga,Quintero:2016iwi,Milanes:2016rzr,Wang:2014lda,Dong:2013raa,Asaka:2016rwd},
such as that shown in Fig.~\ref{fig:lnvBDecayDiagram};
four-body meson decays to lighter mesons $M_1^\pm \to  M_2^\mp M_3^0 \ell_1^\pm \ell_2^\pm $
~\cite{Quintero:2016iwi,Milanes:2016rzr,Castro:2013jsn,Yuan:2013yba,Cvetic:2017vwl};
four-body meson decays to leptons $M^\pm \to \ell_1^\pm \ell_1^\pm \ell_2^\mp \nu$
~\cite{Cvetic:2015naa,Cvetic:2016fbv,Cvetic:2017vwl,Cvetic:2012hd,Cvetic:2013eza};
five-body meson decays~\cite{Cvetic:2017vwl};
four-body baryon decays to mesons, $B \to M \ell_1^\pm \ell_2^\pm$~\cite{Mejia-Guisao:2017nzx};
three-body $\tau$ decay to mesons, $\tau^\pm \to \ell^\mp M_1^\pm M_2^\pm$~\cite{Quintero:2016iwi,Kobach:2014hea,Zamora-Saa:2016ito};
four-body $\tau$ decays to mesons, $\tau^\pm \to \ell_1^\pm \ell_1^\pm M^\mp \nu$~\cite{Quintero:2016iwi,Kobach:2014hea,Yuan:2017xdp,Castro:2012gi,Mandal:2016hpr};
four-body $W$ boson decays, $W^\pm \to \ell_1^\pm \ell_1^\pm \ell_2^\mp \nu$~\cite{BarShalom:2006bv,Dib:2017iva,Dib:2017vux,Dib:2016wge,Dib:2015oka};
Higgs boson decays, $h \to NN \to \ell_1^\pm \ell_2^\pm + X$~\cite{BhupalDev:2012zg,Gago:2015vma,Caputo:2017pit,Das:2017zjc}.
and even top quark decays, $t \to b W^{+*} \to b \ell_1^+ N \to b \ell_1^+ \ell_2^\pm q \overline{q'}$~\cite{Weinberg:1979sa,BarShalom:2006bv,Si:2008jd,Quintero:2011yh}.
The $W$ boson case is notable as azimuthal and polar distributions~\cite{Han:2012vk} and endpoint kinematics~\cite{Dib:2016wge}
can differentiate between $L$ conservation and non-conservation.
Of the various collider searches for GeV-scale $N$, great complementarity is afforded by $B$-factories.
As shown in \ref{fig:lnvBDecayLimits}, an analysis of Belle I~\cite{Liventsev:2013zz} and LHCb Run I~\cite{Aaij:2012zr,Aaij:2014aba}
searches for $L$-violating final states from meson decays excluded~\cite{Shuve:2016muy} $\vert V_{\mu N}\vert^2\gtrsim3\times10^{-5}$ for $M_N = 1-5$ GeV.
Along these same lines, the observability of
displaced decays of heavy neutrinos~\cite{Helo:2013esa,Blondel:2014bra,Izaguirre:2015pga,Gago:2015vma,Antusch:2017hhu,Antusch:2016vyf}
and so-called ``neutrino-antineutrino oscillations''~\cite{Anamiati:2016uxp,Antusch:2017ebe,Das:2017hmg,Antusch:2017pkq}
(in analogy to $\mathcal{B}-\overline{\mathcal{B}}$ oscillations) and  have also been discussed.

Indirectly, the presence of heavy Majorana neutrinos can appear in precision EW measurements as deviations from lepton flavor unitarity and universality,
and is ideally suited for $e^+e^-$
colliders~\cite{delAguila:2008pw,Gomez-Ceballos:2013zzn,Antusch:2014woa,deGouvea:2015euy,Antusch:2015mia,Fernandez-Martinez:2016lgt,Antusch:2016ejd},
such as the International Linear Collider~(ILC) ~\cite{Baer:2013cma,Fujii:2015jha},
Circular $e^-e^+$ Collider (CepC)~\cite{CEPC-SPPCStudyGroup:2015csa},
and Future Circular Collider-$ee$ (FCC-ee) \cite{Gomez-Ceballos:2013zzn}.
An especially famous example of this is the number of active, light neutrino flavors $N_\nu$,
which can be inferred from the $Z$ boson's invisible width $\Gamma^Z_{\rm Inv}$.
At lepton colliders, $\Gamma^Z_{\rm Inv}$ can be determined in two different ways:
The first is from line-shape measurements of the $Z$ resonance as a function of $\sqrt{s}$,
and is measured to be $N_\nu^{\rm Line}=2.9840\pm0.0082$~\cite{ALEPH:2005ab}.
The second is from searches for invisible $Z$ decays, \ie, $e^+e^- \to Z\gamma$,
and is found to be $N_\nu^{\rm Inv}=2.92\pm0.05$~\cite{Beringer:1900zz}.
Provocatively, both measurements deviate from the SM prediction of $N_\nu^{\rm SM}=3$ at the $2\sigma$ level.
It is unclear if deviations from $N_\nu^{\rm SM}$ are the result of experimental uncertainty or indicate the presence of, for example,
heavy sterile neutrinos~\cite{Jarlskog:1990kt,Blondel:2014bra}. 
Nonetheless, a future $Z$-pole machine can potentially clarify this discrepancy~\cite{Blondel:2014bra}.
For investigations into EW constraints on heavy neutrinos, see Refs.~\cite{delAguila:2008pw,Antusch:2014woa,deGouvea:2015euy,Fernandez-Martinez:2016lgt}.

\subsubsection{High-Mass Heavy Neutrinos at $pp$ Colliders}\label{sec:type1HighMassPP}

Collider searches for heavy Majorana neutrinos with masses above $M_W$ have long been of interest
to the community~\cite{Keung:1983uu,Gronau:1984ct,Willenbrock:1985tj,Petcov:1984nf},
with exceptionally notable works appearing in the early 1990s~\cite{Pilaftsis:1991ug,Dicus:1991fk,Dicus:1991wj,Datta:1991mf,Datta:1993nm}
and late-2000s~\cite{Kersten:2007vk,Atre:2009rg,Han:2006ip,Bray:2007ru,delAguila:2006bda,delAguila:2007qnc,delAguila:2008cj,delAguila:2009bb}.
In the past decade, among the biggest advancements in Seesaw phenomenology
is the treatment of collider signatures for such hefty $N$ in Type I-based models.
While coupled to concurrent developments in Monte Carlo simulation packages,
the progression has been driven by attempts to reconcile conflicting reports of heavy neutrino production cross sections for the LHC.
This was at last resolved in Refs.~\cite{Alva:2014gxa,Degrande:2016aje},
wherein new, infrared- and collinear- (IRC-)safe definitions for inclusive and semi-inclusive\footnote{
A note on terminology:
High-$p_T$ hadron collider observables, \eg, fiducial distributions, are inherently inclusive with respect to jets with arbitrarily low $p_T$.
In this sense, we refer to hadronic-level processes with a fixed multiplicity of jets satisfying kinematical requirements
(and with an arbitrary number of additional jets that do not) as \textit{exclusive,} \eg, $pp\to W^\pm + 3j + X$;
those with a minimum multiplicity meeting these requirements are labeled \textit{semi-inclusive,} \eg, $pp\to W^\pm + \geq 3j + X$;
and those with an arbitrary number of jets are labeled \textit{inclusive,} \eg, $pp\to W^\pm + X$.
Due to DGLAP-evolution, exclusive, partonic amplitudes convolved with PDFs are semi-inclusive at the hadronic level.
}
production channels were introduced.
The significance of such collider signatures is that they are well-defined at all orders in $\alpha_s$, and hence correspond to physical observables.
We now summarize this extensive body of literature, emphasizing recent results.

\begin{figure}[t!]
\includegraphics[scale=1,width=0.96\textwidth]{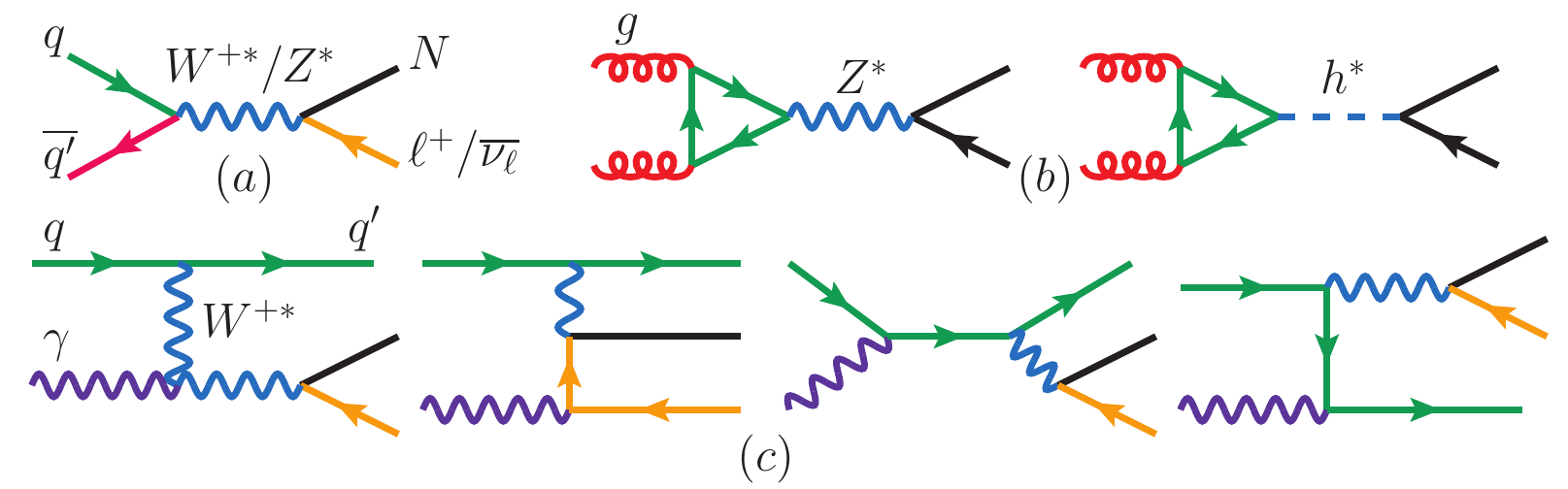}
\caption{Born diagrams for heavy neutrino  $(N)$ production via (a) Drell-Yan, (b) gluon fusion, and (c) electroweak vector boson fusion;
from Ref.~\cite{Ruiz:2017yyf} and drawn using \texttt{JaxoDraw}~\cite{Binosi:2008ig}.
} \label{fig:heavyNDiagrams}
\end{figure}

For Majorana neutrinos with $M_N > M_W$, the most extensively
studied ~\cite{Keung:1983uu,Gronau:1984ct,Datta:1991mf,Han:2006ip,delAguila:2006bda,delAguila:2007qnc,Bray:2007ru,Chao:2009ef,
delAguila:2008cj,delAguila:2009bb,Atre:2009rg,Antusch:2016ejd,Das:2017hmg,Chen:2011hc}
collider production mechanism is the $L$-violating, charged current (CC) Drell-Yan (DY) process~\cite{Keung:1983uu},
shown in Fig.~\ref{fig:heavyNDiagrams}(a), and given by
\begin{equation}
 q_1 ~\overline{q}_2 \rightarrow W^{\pm *} \rightarrow N ~\ell^\pm_1, ~\quad\text{with}\quad N \to \ell_2^\pm W^{\mp } \to \ell_2^\pm q_1' ~\overline{q'}_2.
 \label{eq:heavyNDYCC}
\end{equation}
A comparison of Fig.~\ref{fig:heavyNDiagrams}(a) to the meson decay diagram of Fig.~\ref{fig:lnvBDecayDiagram}
immediately reveals that Eq.~(\ref{eq:heavyNDYCC}) is the former's high momentum transfer completion.
Subsequently, much of the aforementioned kinematical properties related
to $L$-violating meson decays also hold for the CC DY channel~\cite{Han:2012vk,Dev:2015kca}.
Among the earliest studies are those likewise focusing on
neutral current (NC) DY production~\cite{Gronau:1984ct,Willenbrock:1985tj,Dicus:1991wj,Datta:1991mf,Datta:1993nm},
again shown in Fig.~\ref{fig:heavyNDiagrams}(a), and given by
\begin{eqnarray}
 q ~\overline{q} \rightarrow Z^* \rightarrow N ~\overset{(-)}{\nu_\ell},
 \label{eq:heavyNDYNC}
\end{eqnarray}
as well as the gluon fusion mechanism~\cite{Willenbrock:1985tj,Dicus:1991wj}, shown in Fig.~\ref{fig:heavyNDiagrams}(b), and given by
\begin{equation}
 g ~g \rightarrow Z^*/h^* \rightarrow N ~ \overset{(-)}{\nu_\ell}.
 \label{eq:heavyNGF}
\end{equation}
Interestingly, despite gluon fusion being formally an $\mathcal{O}(\alpha_s^2)$ correction to Eq.~(\ref{eq:heavyNDYNC}), it is non-interfering,
separately gauge invariant, and the subject of renewed interest \cite{Hessler:2014ssa,Degrande:2016aje,Ruiz:2017yyf}.
Moreover, in accordance to the Goldstone Equivalence Theorem~\cite{Chanowitz:1985hj,Lee:1977yc},
the $ggZ^*$ contribution has been shown~\cite{Hessler:2014ssa,Ruiz:2017yyf}
to be as large as the $ggh^*$ contribution, and therefore should not be neglected.
Pair production of $N$ via $s$-channel scattering~\cite{Willenbrock:1985tj,Datta:1991mf}, \eg, $gg \to  N N$,
or weak boson scattering~\cite{Dicus:1991fk,Datta:1993nm,Han:2006ip}, \eg, $W^\pm W^\mp \to NN$, have also been discussed,
but are relatively suppressed compared to single production by an additional mixing factor of $\vert V_{\ell N_{m'}}\vert^2 \lesssim 10^{-4}$.

\begin{figure}[t!]
\includegraphics[scale=1,width=0.96\textwidth]{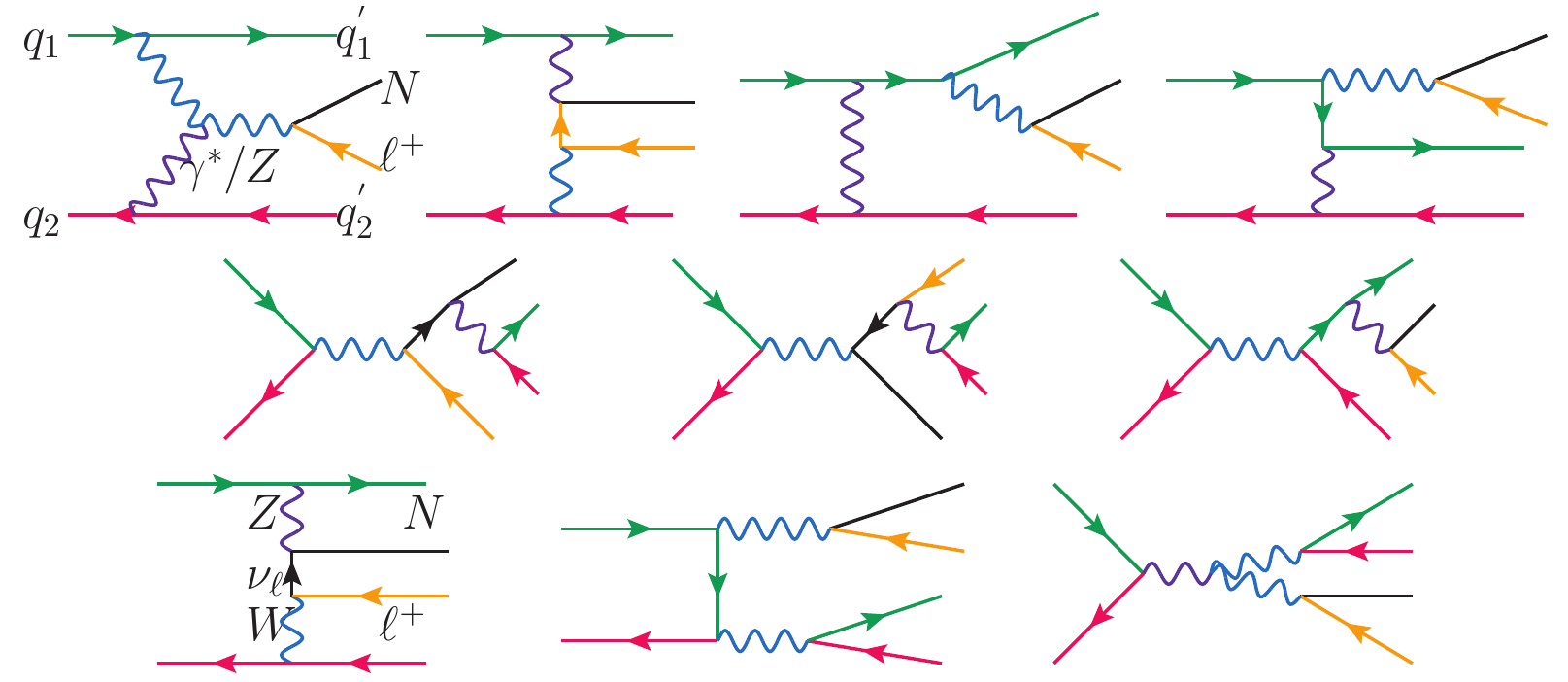}
\caption{Born diagrams for the $\mathcal{O}(\alpha^4)$ heavy neutrino $(N)$ production process $q_1q_2\to N\ell^\pm q_1' q_2'$~\cite{Alva:2014gxa}.}
\label{fig:heavyNVBF2}
\end{figure}

A recent, noteworthy development is the interest in semi-inclusive and exclusive
production of heavy neutrinos at hadron colliders, \ie, $N$ production in association with jets.
In particular, several studies have investigated the semi-inclusive, photon-initiated vector boson fusion (VBF)
process~\cite{Datta:1993nm,Dev:2013wba,Alva:2014gxa,Degrande:2016aje}, shown in Fig.~\ref{fig:heavyNDiagrams}(c), and given by
\begin{equation}
 q ~\gamma \rightarrow N ~\ell^\pm ~q',
 \label{eq:heavyNvbfIncl}
\end{equation}
and its deeply inelastic, $\mathcal{O}(\alpha)$ radiative
correction~\cite{Datta:1993nm,Dev:2013wba,Bambhaniya:2014hla,Alva:2014gxa,Ng:2015hba,Arganda:2015ija,Degrande:2016aje,Andres:2017daw},
\begin{equation}
 q_1 ~q_2 \xrightarrow{W\gamma+WZ \to N\ell^\pm} N ~\ell^\pm ~q'_1 ~q'_2.
 \label{eq:heavyNvbfExcl}
\end{equation}
At $\mathcal{O}(\alpha^4)$ (here we do not distinguish between $\alpha$ and $\alpha_W$), the full, gauge invariant set of diagrams,
which includes the sub-leading $W^\pm Z\to N\ell^\pm$ scattering, is given in Fig.~\ref{fig:heavyNVBF2}.

Treatment of the VBF channel is somewhat subtle in that it receives contributions from collinear QED radiation off the proton~\cite{Dev:2013wba},
collinear QED radiation off initial-states quarks~\cite{Alva:2014gxa},
and QED radiation in the deeply inelastic/high momentum transfer limit~\cite{Datta:1993nm}.
For example: In the top line of diagrams in  Fig.~\ref{fig:heavyNVBF2}, one sees that in the collinear limit of the $q_2 \to \gamma^* q_2'$ splitting,
the virtual $\gamma^*$ goes on-shell and the splitting factorizes into a photon parton distribution function (PDF),
recovering the process in Eq.~(\ref{eq:heavyNvbfIncl})~\cite{Alva:2014gxa,Degrande:2016aje}.
As these sub-channels are different kinematic limits of the same process, care is needed when combining channels so as to not double count regions of phase space.
While ingredients to the VBF channel have been known for some time,
consistent schemes to combine/match the processes are more recent~\cite{Alva:2014gxa,Degrande:2016aje}.
Moreover, for inclusive studies, Ref.~\cite{Degrande:2016aje} showed that the use of Eq.~(\ref{eq:heavyNvbfIncl})
in conjunction with a $\gamma$-PDF containing both elastic and inelastic contributions~\cite{Martin:2014nqa}
can reproduce the fully matched calculation of Ref.~\cite{Alva:2014gxa}
within the $\mathcal{O}(20\%)$ uncertainty resulting from missing NLO in QED terms.
Neglecting the collinear $q_2 \to \gamma^* q_2'$ splitting accounts for the unphysical cross sections reported in Refs.~\cite{Dev:2013wba,Deppisch:2015qwa}.
Presently, recommended PDF sets containing such $\gamma$-PDFs include:	
MMHT QED (no available \texttt{lhaid}) ~\cite{Martin:2014nqa,Harland-Lang:2016kog},
NNPDF 3.1+LUXqed (\texttt{lhaid=324900}) ~\cite{Bertone:2017bme},
LUXqed17+PDF4LHC15 (\texttt{lhaid=82200}) ~\cite{Manohar:2017eqh,Manohar:2016nzj},
and CT14 QED Inclusive (\texttt{lhaid = 13300}) ~\cite{Schmidt:2015zda}.
Qualitatively, the MMHT~\cite{Martin:2014nqa} and LUXqed~\cite{Manohar:2017eqh,Manohar:2016nzj} treatments of photon PDFs are the most rigorous.
In analogy to the gluon fusion and NC DY, Eq.~(\ref{eq:heavyNvbfIncl}) (and hence Eq.~(\ref{eq:heavyNvbfExcl})) is a non-interfering, $\mathcal{O}(\alpha)$ correction to the CC DY process.
Thus, the CC DY and VBF channels can be summed coherently.

In addition to these channels, the semi-inclusive, associated $n$-jet production mode,
\begin{equation}
p ~p ~\to W^* ~+~ \geq nj ~+~ X \quad\to\quad N ~\ell^\pm ~+~ \geq n j ~+~ X, \quad\text{for}\quad  n\in \mathbb{N},
\label{eq:heavyNnj}
\end{equation}
has also appeared in the recent literature~\cite{Dev:2013wba,Das:2014jxa,Degrande:2016aje}.
As with VBF, much care is needed to correctly model Eq.~(\ref{eq:heavyNnj}).
As reported in Refs.~\cite{Ruiz:2015zca,Degrande:2016aje}, the production of heavy leptons in association with QCD jets
is nuanced due to the presence of additional $t$-channel propagators that can lead to artificially large cross sections
if matrix element poles are not sufficiently regulated. (It is not enough to simply remove the divergences with phase space cuts.)
After phase space integration, these propagators give rise to logarithmic dependence on the various process scales.
Generically~\cite{Collins:1984kg,Ruiz:2015zca}, the cross section for heavy lepton and jets in Eq.~(\ref{eq:heavyNnj}) scales as:
\begin{equation}
 \sigma(pp\rightarrow N\ell^\pm+nj +X) \sim \sum^n_{k=1} \alpha_s^k(Q^2)\log^{(2k-1)}\left(\frac{Q^2}{q_T^2}\right).
 \label{eq:heavyNnjetXSec}
\end{equation}
Here, $Q\sim M_N$ is the scale of the hard scattering process, $q_T = \sqrt{\vert \vec{q}_T \vert^2}$, and  $\vec{q}_T \equiv \sum_k^n \vec{p}_{T,k}^{~j}$,
is the $(N\ell)$-system's transverse momentum, which recoils against the vector sum of all jet $\vec{p}_T$.
It is clear for  a fixed $M_N$ that too low jet $p_T$ cuts can lead to too small $q_T$ and cause
numerically large (collinear) logarithms such that $\log(M_N^2/q_T^2) \gg 1/\alpha_s(M_N)$,
spoiling the perturbative convergence of Eq.~(\ref{eq:heavyNnjetXSec}).
Similarly, for a fixed $q_T$, arbitrarily large $M_N$ can again spoil perturbative convergence.
As noted in Refs.~\cite{Alva:2014gxa,Degrande:2016aje}, neglecting this fact has led to conflicting predictions in several studies
on heavy neutrino production in $pp$ collisions.

\begin{figure}[tb]
\subfigure[]{\includegraphics[scale=1,width=0.48\textwidth]{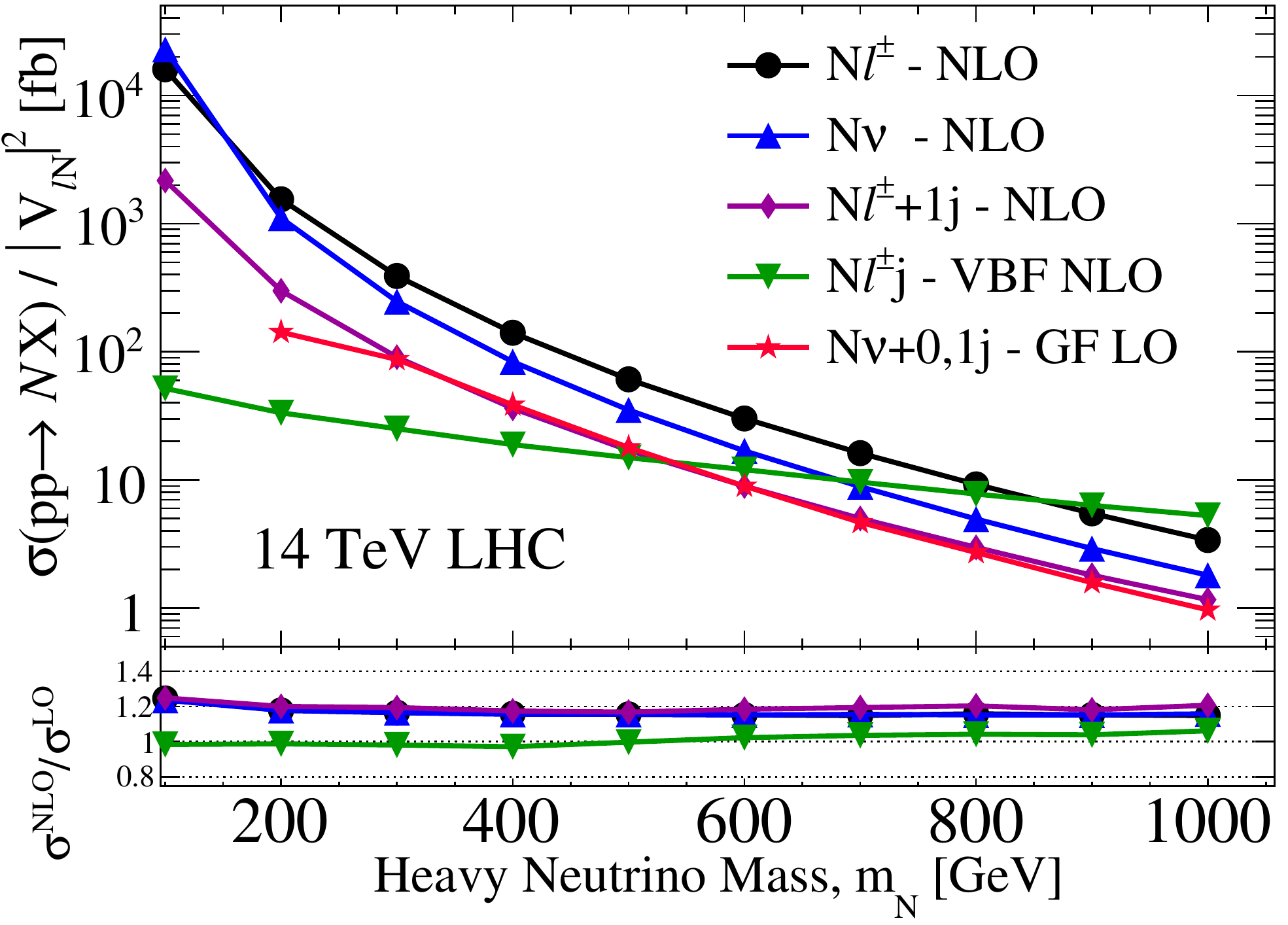}	\label{fig:heavyNXSecVsM}	}
\subfigure[]{\includegraphics[scale=1,width=0.48\textwidth]{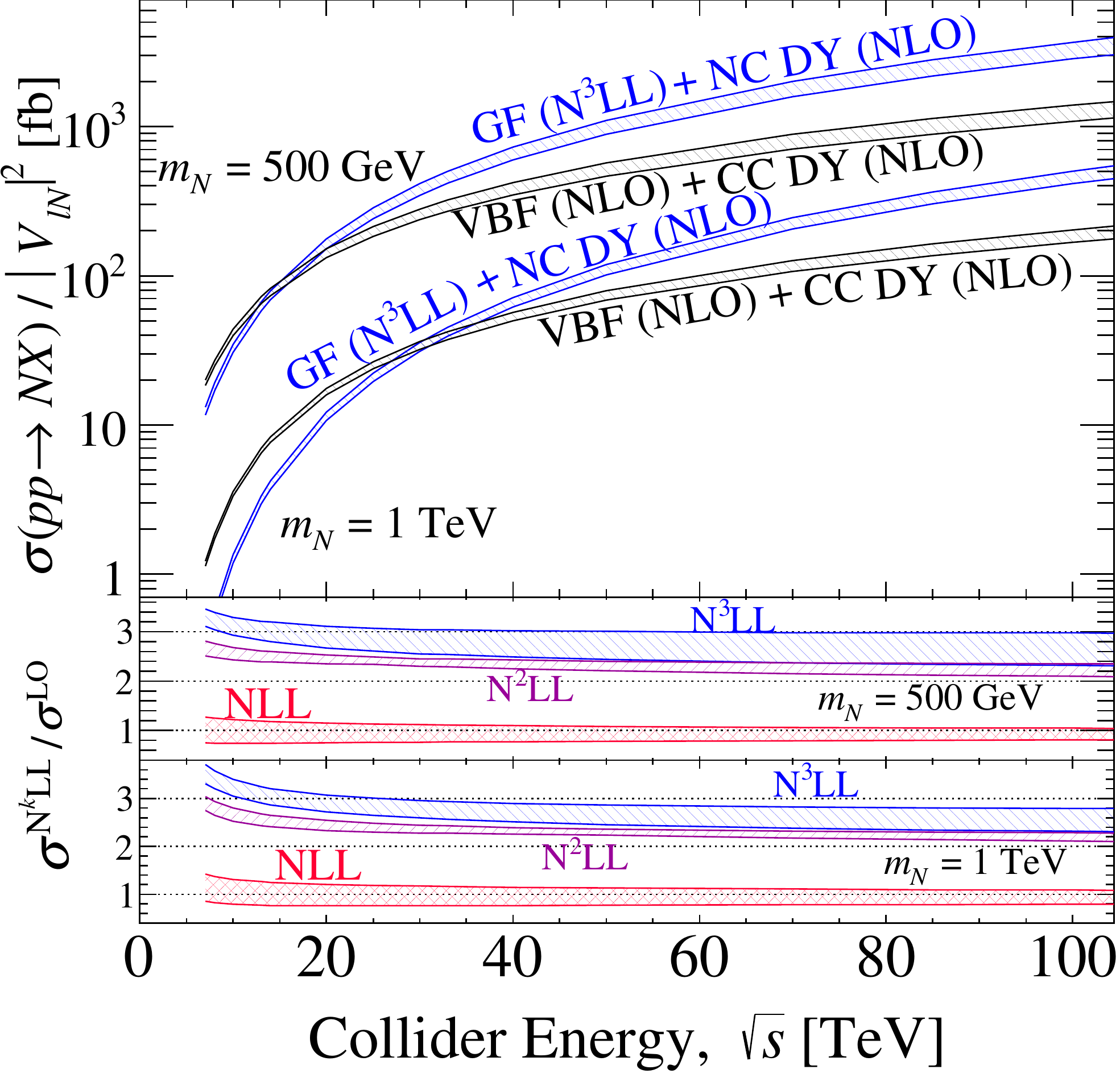}	\label{fig:heavyNXSecVsS}	}
\caption{Heavy neutrino $(N)$ hadron collider production cross sections, divided by active-heavy mixing $\vert V_{\ell N}\vert^2$,
for various production modes as a function of (a) $N$ mass at $\sqrt{s}=14$  ~\cite{Degrande:2016aje}
and (b) collider energy for representative $M_N$ (band thickness corresponds to residual scale uncertainty)~\cite{Ruiz:2017yyf}.
} \label{fig:heavyNXSec}
\end{figure}

It is possible~\cite{Degrande:2016aje}, however, to tune $p_T$ cuts on jets with varying $M_N$ to enforce the validity of Eq.~(\ref{eq:heavyNnjetXSec}).
Within the Collins-Soper-Sterman (CSS) resummation formalism~\cite{Collins:1984kg}, Eq.~(\ref{eq:heavyNnjetXSec}) is trustworthy when
$\alpha_s(Q^2)$ is perturbative and $q_T\sim Q$, \ie,
\begin{equation}
 \log (Q/\Lambda_{\rm QCD}) \gg1 \quad\text{and}\quad \alpha_s(Q)\log^2 (Q^2/q_T^2) \lesssim 1.
\end{equation}
Noting that at 1-loop $\alpha_s(Q)$ can be written as $1/\alpha_s(Q) \approx (\beta_0/2\pi)\log (Q/\Lambda_{\rm QCD})$, and setting $Q=M_N$,
one can invert the second CSS condition and obtain a consistency relationship~\cite{Degrande:2016aje}:
\begin{equation}
  q_T = \vert \vec{q}_T \vert = \left\vert \sum_{k=1}^n \vec{p}_{T,k}^{~j}
  \right\vert \gtrsim M_N \times e^{-(1/2)\sqrt{(\beta_0/2\pi)\log(M_N/\Lambda_{\rm QCD})}}.
  \label{eq:cssConsistency}
\end{equation}
This stipulates a minimum $q_T$ needed for semi-inclusive processes like Eq.~(\ref{eq:heavyNnjetXSec}) to be valid in perturbation theory.
When $q_T$ of the $(N\ell)$-system is dominated by a single, hard radiation, Eq.~(\ref{eq:cssConsistency}) is consequential:
In this approximation, $q_T \approx \vert \vec{p}_{T,1}^{~j} \vert$ and Eq.~(\ref{eq:cssConsistency}) suggests a \textit{scale-dependent},
minimum  jet $p_T$ cut to ensure that specifically the semi-inclusive $pp\to N\ell +\geq 1j+X$ cross section is well-defined in perturbation theory.
Numerically, this is sizable: for $M_N = 30~(300)~[3000]~{\rm GeV}$, one requires that $\vert \vec{p}_{T,1}^{~j} \vert \gtrsim 9~(65)~[540]$ GeV,
or alternatively $\vert \vec{p}_{T,1}^{~j} \vert \gtrsim 0.3~(0.22)~[0.18]\times M_N$,
and indicates that na\"ive application of fiducial $p_T^j$ cuts for the LHC do not readily apply for $\sqrt{s} = $ 27-100 TeV
scenarios, where one can probe much larger $M_N$.
The perturbative stability of this approach is demonstrated by the (roughly) flat $K$-factor of $K^{\rm NLO}\approx1.2$
for the semi-inclusive $pp\to N\ell^\pm+1j$ process, shown in the lower panel of Fig.~\ref{fig:heavyNXSecVsM}.
Hence, the artificially large $N$ production cross sections reported in Refs.~\cite{Dev:2013wba,Das:2014jxa,Deppisch:2015qwa}
can be attributed to a loss of perturbative control over their calculation,
not the presence of an enhancement mechanism.
Upon the appropriate replacement of $M_N$, Eq.~(\ref{eq:cssConsistency}) holds for other color-singlet processes~\cite{Degrande:2016aje},
including mono-jet searches, and is consistent with explicit $p_T$ resummations
of high-mass lepton~\cite{Ruiz:2015zca} and slepton~\cite{Dreiner:2006sv,Chen:2006ep} production.

A characteristic of heavy neutrino production cross sections is that the active-sterile mixing, $\vert V_{\ell N}\vert$,
factorizes out of the partonic and hadronic scattering expressions.
Exploiting this one can define~\cite{Han:2006ip} a ``bare'' cross section $\sigma_{0}$, given by
\begin{equation}
\sigma_{0}(pp\rightarrow N+X) ~\equiv~  \sigma(pp\rightarrow N+X) /  \vert V_{\ell N}\vert^2.
\end{equation}
Assuming resonant production of $N$, a similar expression can be extracted at the $N$ decay level,
\begin{equation}
 \sigma_{0}(pp \to \ell^\pm_1 \ell^\pm_2 + X) ~\equiv~
     \sigma(pp \to \ell^\pm_1 \ell^\pm_2 + X) / S_{\ell_1 \ell_2}, \quad
 S_{\ell_1 \ell_2} = \cfrac{\vert V_{\ell_1 N}\vert^2\vert V_{\ell_2 N}\vert^2 }{\sum_{\ell=e}^\tau \vert V_{\ell N}\vert^2}.
\end{equation}
These definitions, which hold at higher orders in $\alpha_s$~\cite{Ruiz:2015zca,Degrande:2016aje},
allow one to make cross section predictions and comparisons independent of a particular flavor model,
including those that largely conserve lepton number, such as the inverse and linear Seesaws.
It also allows for a straightforward reinterpretation of limits on collider cross sections as limits
on $S_{\ell_1 \ell_2}$, or $\vert V_{\ell N}\vert$ with additional but generic assumptions.
An exception to this factorizablity is the case of
nearly degenerate neutrinos with total widths that are comparable to their mass splitting~\cite{Bray:2007ru,Fuchs:2014ola,Fuchs:2016swt,Anamiati:2016uxp}.

\begin{figure}[!t]
\begin{center}
\subfigure[]{\includegraphics[scale=1,width=.48\textwidth]{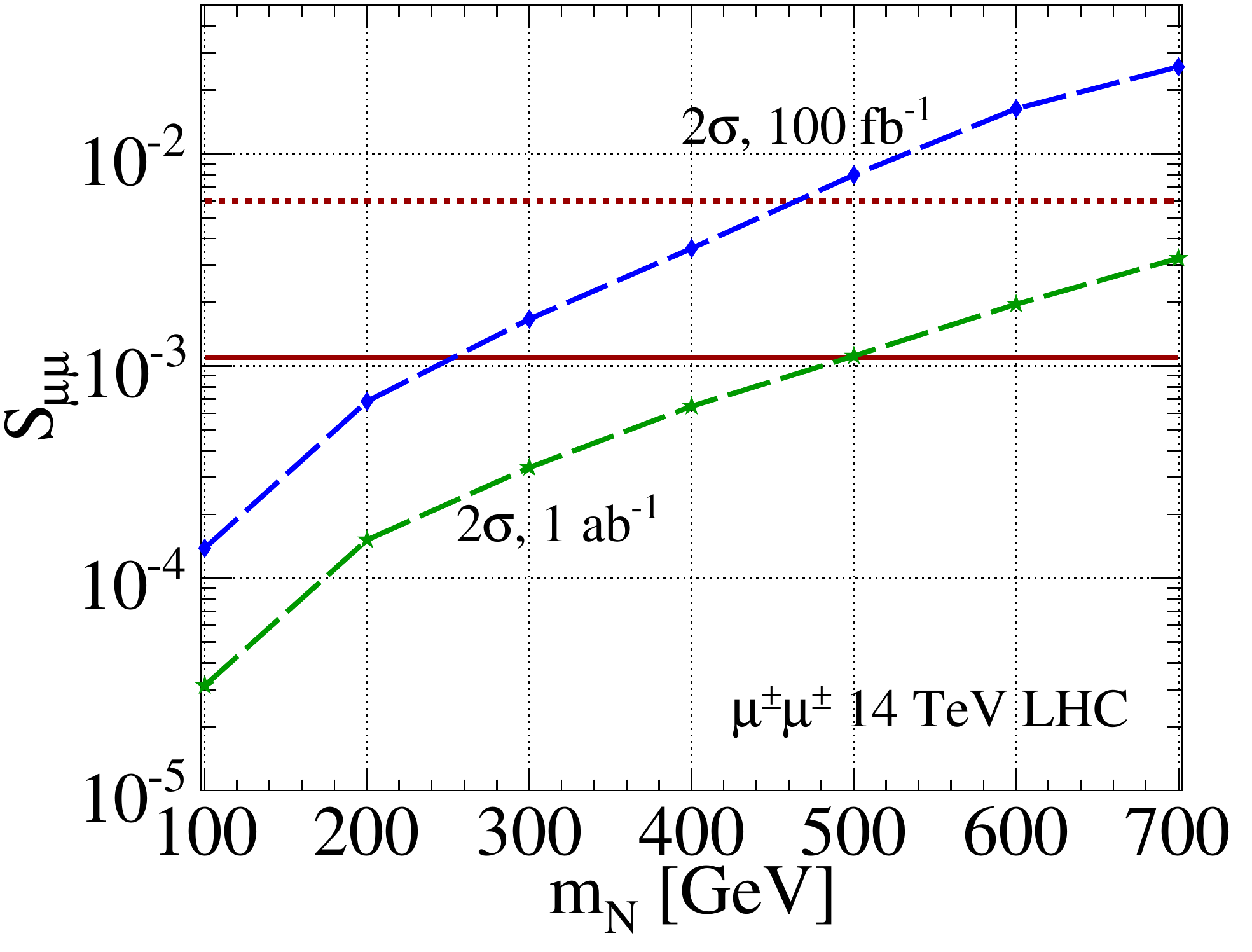}		}
\subfigure[]{\includegraphics[scale=1,width=.48\textwidth]{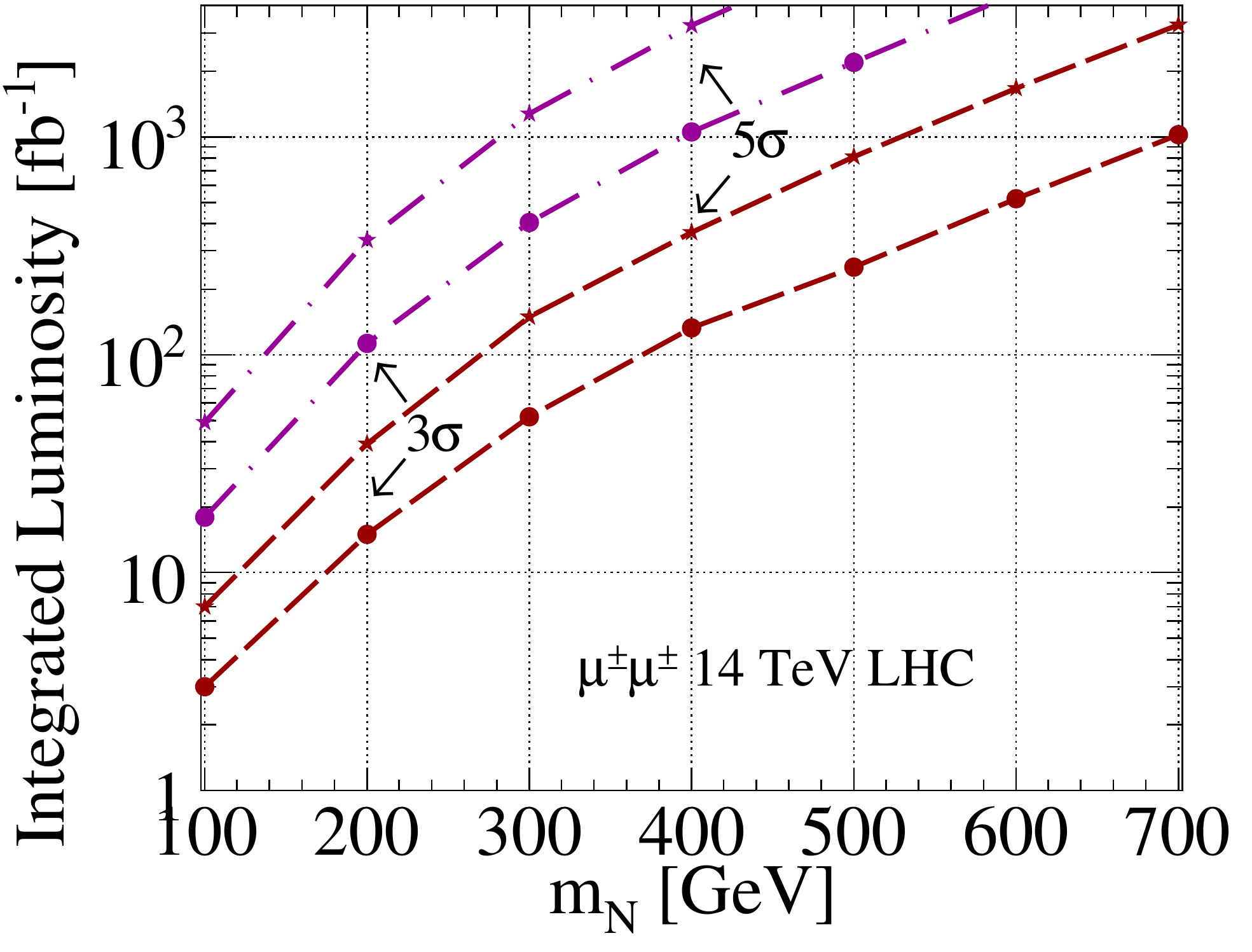}		}
\vspace{.2in}\\
\subfigure[]{\includegraphics[scale=1,width=.48\textwidth]{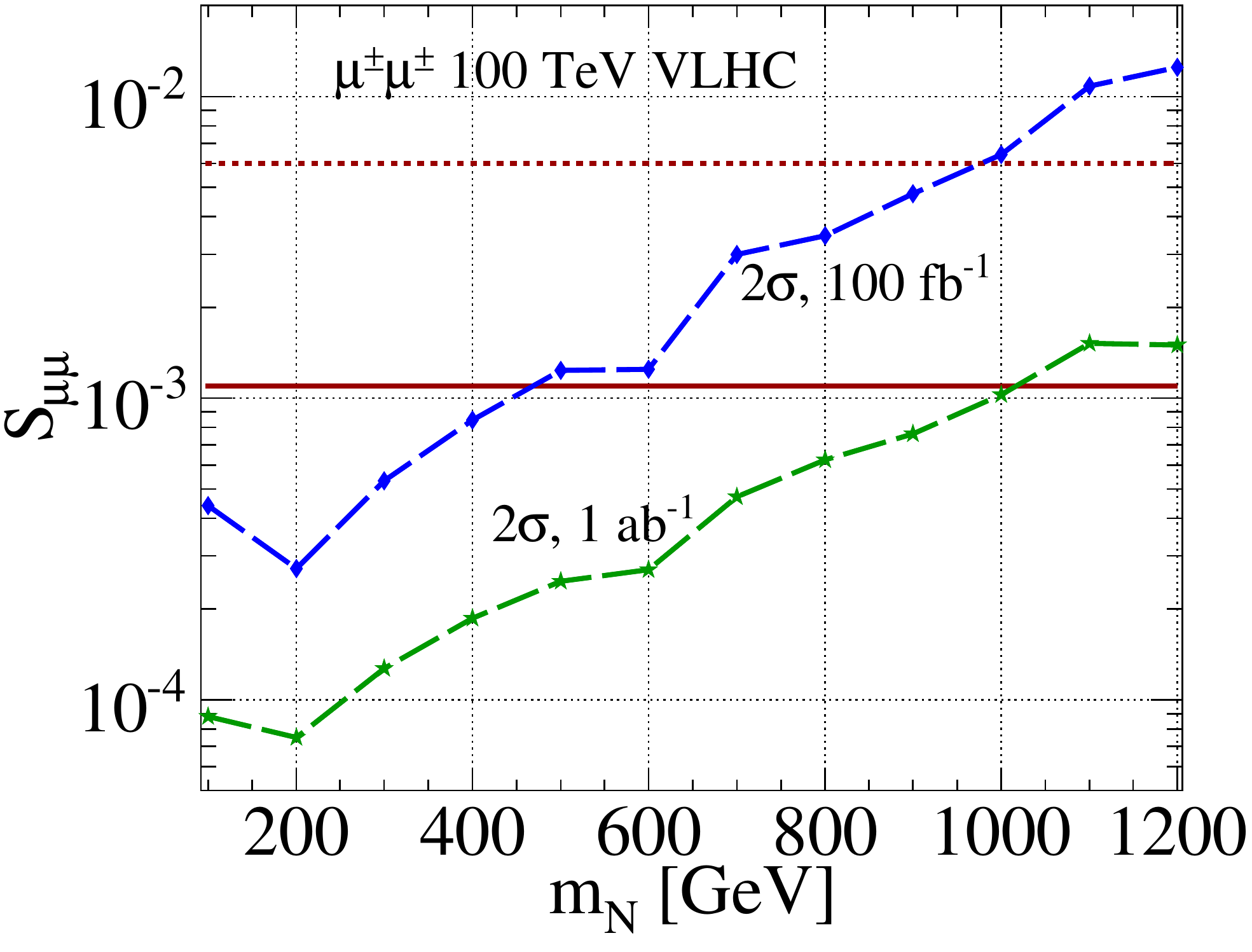}		}
\subfigure[]{\includegraphics[scale=1,width=.48\textwidth]{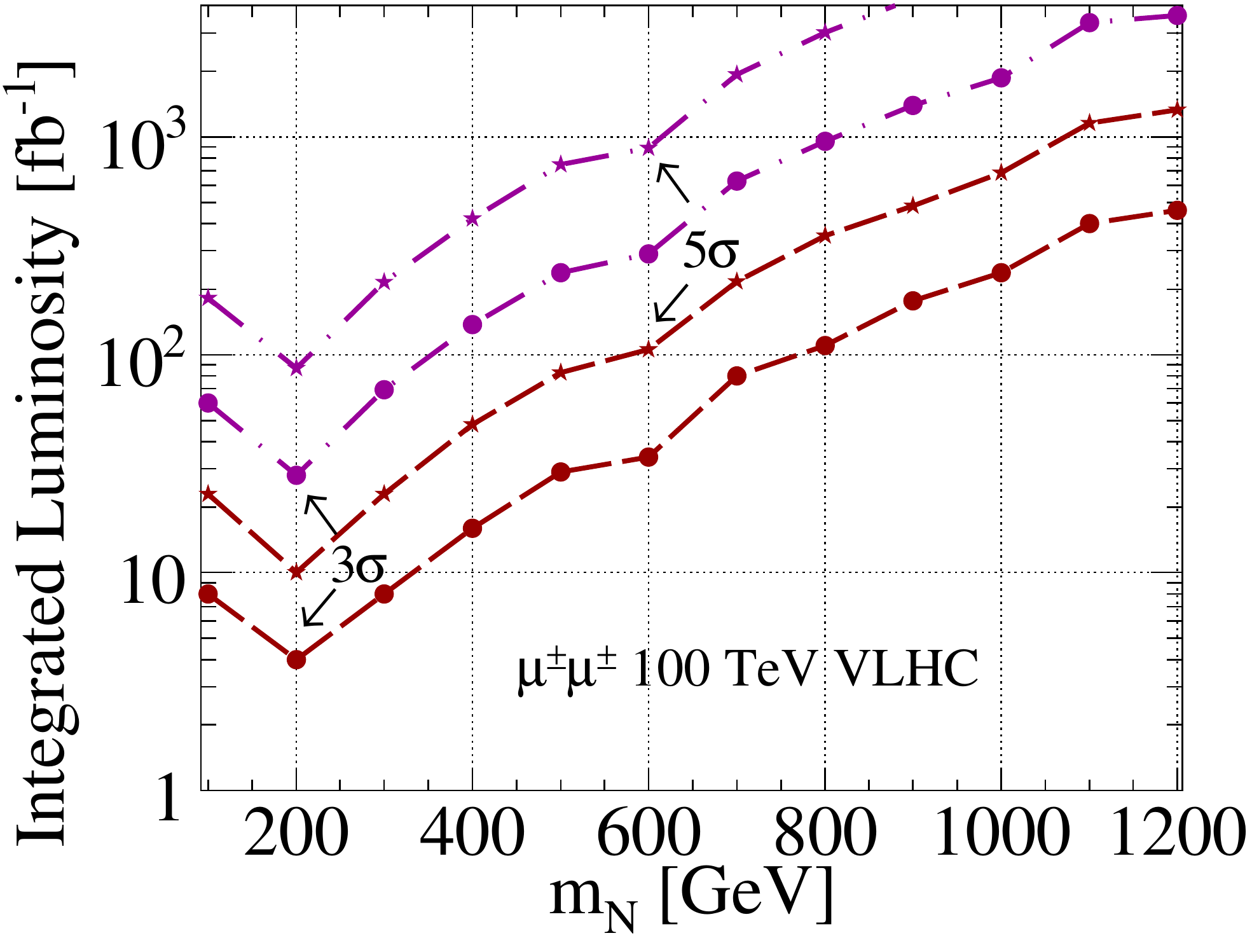}		}
\end{center}
\caption{
At 14 TeV and as a function of $M_N$, (a) the $2\sigma$ sensitivity to $S_{\ell\ell^{'}}$ for the $pp\to\mu^{\pm}\mu^{\pm}+X$ process.
(b) The required luminosity for a 3 (dash-circle) and 5$\sigma$ (dash-star) discovery in the same channel
(c,d) Same as (a,b) but for 100 TeV~\cite{Alva:2014gxa}.
}
\label{100TeVdiscovery.fig}
\end{figure}

Figure~\ref{fig:heavyNXSec} shows a comparison of the leading, single $N$ hadronic production cross sections,
divided by active-heavy mixing $\vert V_{\ell N}\vert^2$, as a function of
(a) heavy neutrino mass $M_N$ at $\sqrt{s}=14$~\cite{Degrande:2016aje}
and
(b) collider energy $\sqrt{s}$ up to 100 TeV for $M_N = 500,~1000$ GeV~\cite{Ruiz:2017yyf}.
The various accuracies reported reflect the maturity of modern Seesaw calculations.
Presently, state-of-the-art predictions for single $N$ production modes are automated up to NLO+PS in QCD
for the Drell-Yan and VBF channels~\cite{Ruiz:2015gsa,Degrande:2016aje}, amongst others,
and known up to N$^3$LL(threshold) for the gluon fusion channel~\cite{Ruiz:2017yyf}.
With Monte Carlo packages,
predictions are available at LO with multi-leg merging (MLM)~\cite{delAguila:2007qnc,Alpgen:HeavyN,Degrande:2016aje,FeynRules:HeavyNnlo}
as well as up to NLO with parton shower matching and merging~\cite{Degrande:2016aje,FeynRules:HeavyNnlo}.
The NLO in QCD-accurate~\cite{Degrande:2014vpa}, \texttt{HeavyNnlo} universal FeynRules object (UFO)~\cite{Degrande:2011ua}
model file is available from Refs.~\cite{Degrande:2016aje,FeynRules:HeavyNnlo}. 
Model files built using \texttt{FeynRules}~\cite{Christensen:2008py,Degrande:2011ua,Alloul:2013bka}
construct and evaluate $L$-violating currents following the Feynman rules convention of Ref.~\cite{Denner:1992me}. 
A brief comment is needed regarding choosing MLM+PS or NLO+PS computations:
To produce MLM Monte Carlo samples, one must sum semi-inclusive channels with successively higher leg multiplicities in accordance with
Eqs.~(\ref{eq:heavyNnjetXSec})-(\ref{eq:cssConsistency}) and correct for phase space double-counting.
However, such MLM samples are formally LO in $\mathcal{O}(\alpha_s)$ because of missing virtual corrections.
NLO+PS is formally more accurate, under better perturbative control (due to explicit cancellation of infrared singularities), 
and thus is recommended for modeling heavy $N$ at colliders.
Such computations are possible with modern, general-purpose event generators,
such as Herwig~\cite{Bellm:2015jjp}, MadGraph5\_aMC@NLO~\cite{Alwall:2014hca}, and Sherpa~\cite{Gleisberg:2008ta}.

\begin{figure}[!t]
\begin{center}
\subfigure[]{\includegraphics[scale=1,width=.48\textwidth]{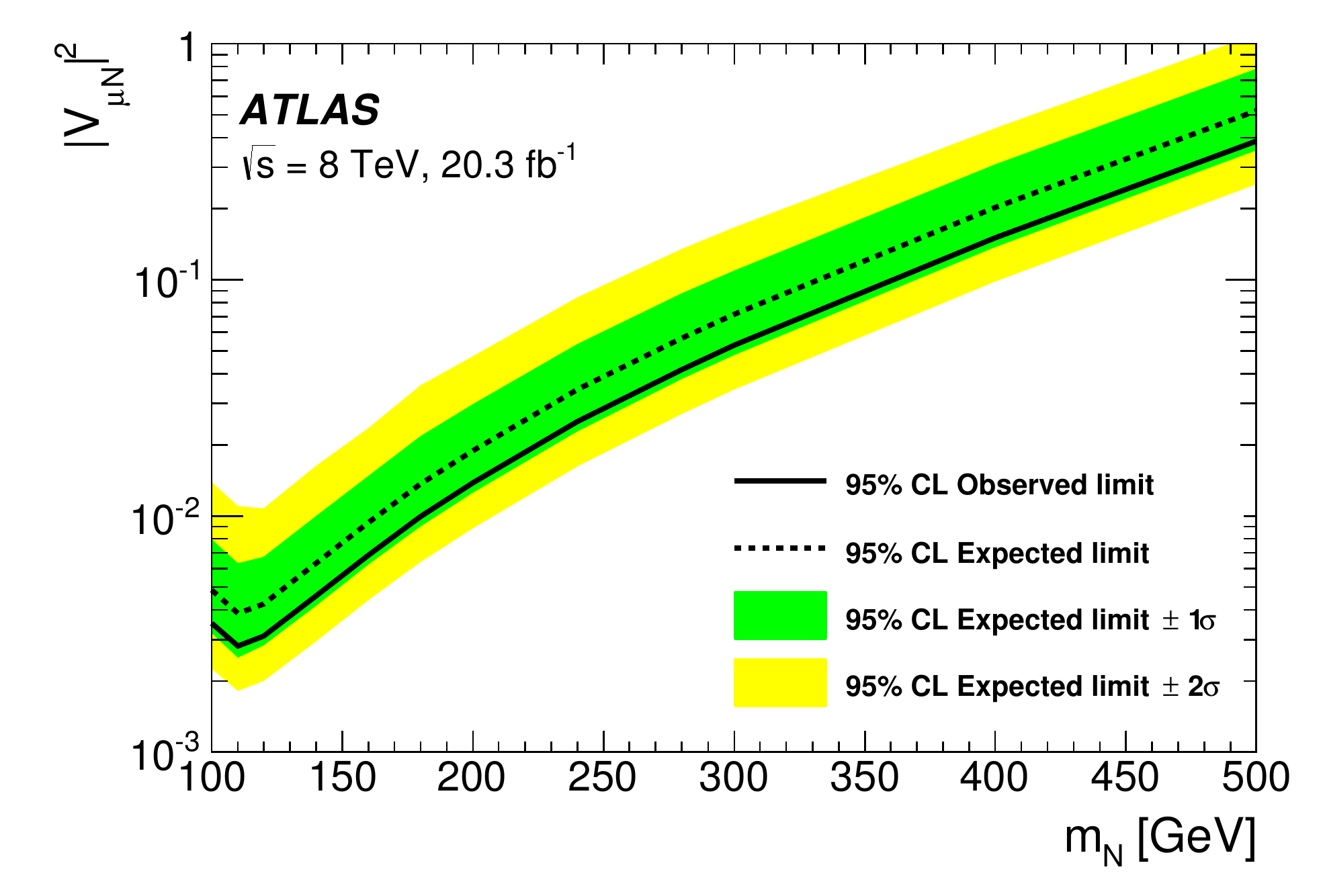}	}
\subfigure[]{\includegraphics[scale=1,width=.48\textwidth]{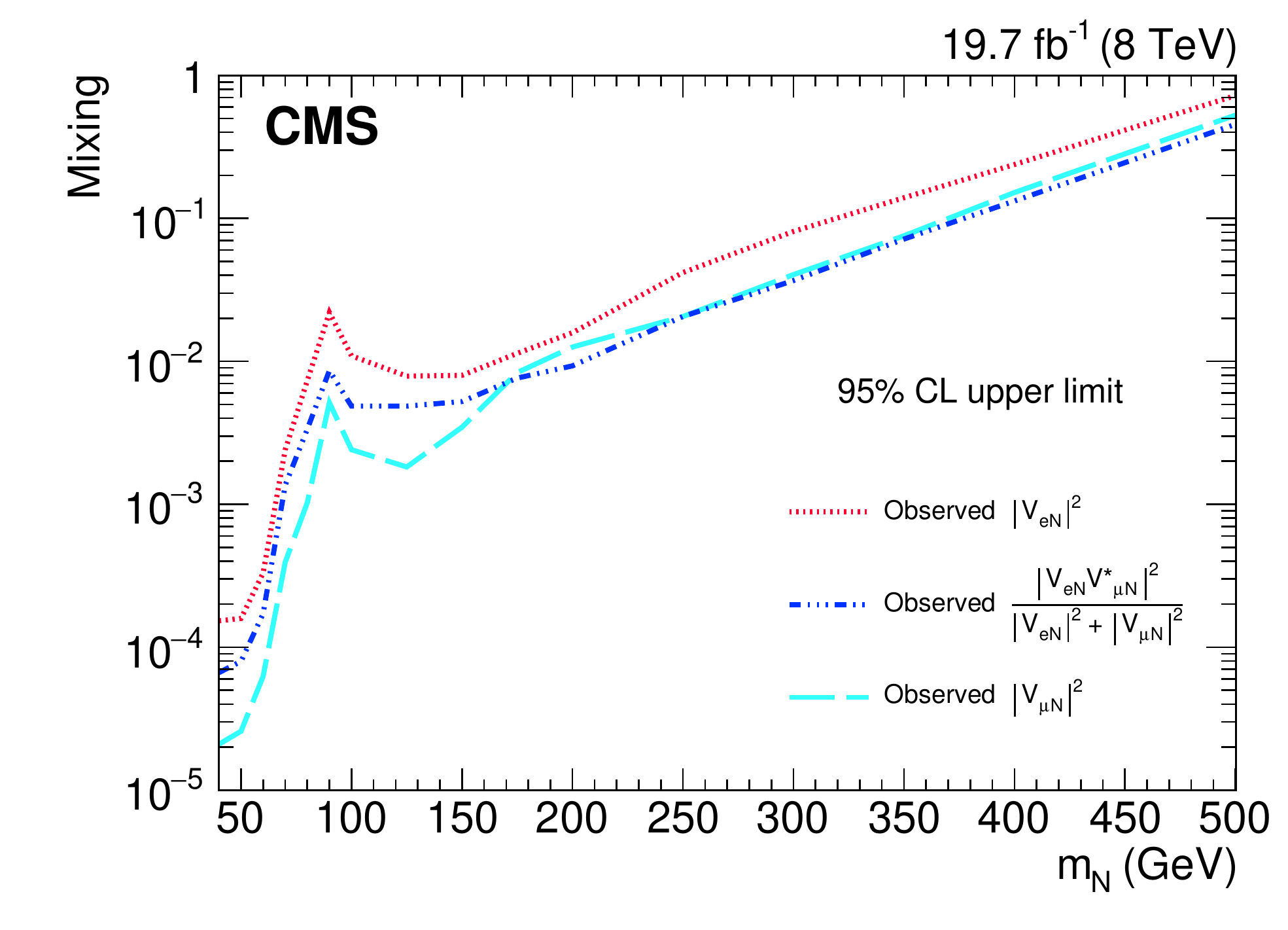}	}
\end{center}
\caption{
8 TeV LHC limits on neutrino mixing $\vert V_{\ell N}\vert^2$ from searches for
$pp\to \ell^\pm_1 \ell^\pm_2 +nj$ at (a) ATLAS~\cite{Aad:2015xaa} and (b) CMS~\cite{Khachatryan:2016olu}
with $\mathcal{L}\approx20\invfb$ of data.
}
\label{fig:lhcTypeILimits}
\end{figure}

At the 13 and 14 TeV LHC, heavy $N$ production is dominated by charged-current mechanisms for
phenomenologically relevant mass scales, \ie, $M_N \lesssim 700$ GeV~\cite{Alva:2014gxa}.
At more energetic colliders, however,
the growth in the gluon-gluon luminosity increases the $gg\to N\nu$ cross section faster than the CC DY channel.
In particular, at $\sqrt{s} = 20-30$ TeV, neutral-current mechanisms surpass charged-current modes
for heavy $N$ production with $M_N = 500-1000$ GeV~\cite{Ruiz:2017yyf}.
As seen in the sub-panel of Fig.~\ref{fig:heavyNXSecVsM},
NLO in QCD contributions only modify inclusive, DY-type cross section normalizations by $+20$-to-$+30\%$ and VBF negligibly, indicating that
the prescriptions of Ref.~\cite{Degrande:2016aje} are sufficient to ensure perturbative control over a wide-range of scales.
One should emphasize that while VBF normalizations do not appreciably change under QCD corrections~\cite{Han:1992hr},
VBF kinematics do change considerably~\cite{Degrande:2016aje,Degrande:2015xnm,Cacciari:2015jma,Dreyer:2016oyx}.
The numerical impact, however, is observable-dependent and can be large if new kinematic channels are opened at higher orders of $\alpha_s$.
In comparison to this, the sub-panel of Fig.~\ref{fig:heavyNXSecVsS} shows that QCD corrections to gluon fusion  are huge ($+150$-to-$+200\%$),
but convergent and consistent with SM Higgs, heavy Higgs, and heavy pseudoscalar production~\cite{Bonvini:2014qga,Anastasiou:2016cez,Anastasiou:2015ema};
for additional details, see Ref.~\cite{Ruiz:2017yyf}.

With these computational advancements, considerable collider sensitivity to $L$-violating processes in the Type I Seesaw has been reached.
In Fig.~\ref{100TeVdiscovery.fig} is the expected sensitivity to active-sterile neutrino mixing
via the combined CC DY+VBF channels and in same-sign $\mu^\pm\mu^\pm+X$ final-state.
With $\mathcal{L}=1$ ab$^{-1}$ of data for $M_N > M_W$ at $\sqrt{s} = $ 14 (100) TeV,
one can exclude at $2\sigma$ $S_{\mu\mu} \approx \vert V_{\mu N}\vert^2 \gtrsim 10^{-4}~(10^{-5})$~\cite{Alva:2014gxa}.
This is assuming the 2013 Snowmass benchmark detector configuration for $\sqrt{s}=100$ TeV~\cite{Avetisyan:2013onh}.
Sensitivity to the $e^\pm e^\pm$ and $e^\pm\mu^\pm$ channels is comparable, up to detector (in)efficiencies for electrons and muons.
As shown in Fig.~\ref{fig:lhcTypeILimits}, with $\mathcal{L}\approx20$ fb$^{-1}$ at 8 TeV,
the ATLAS and CMS experiments have excluded at 95\% CLs $\vert V_{\ell N}\vert^2 \gtrsim 10^{-3} - 10^{-1}$
for $M_N = 100 - 450$ GeV~\cite{Sirunyan:2017yrk,Khachatryan:2016jqo,Khachatryan:2016olu,Khachatryan:2015gha,Aad:2015xaa}.
For heavier $M_N$, quarks from the on-shell $W$ boson decay can form a single jet instead of the usual two-jet configuration.
In such cases, well-known ``fat jet'' techniques can be used~\cite{ATLAS:2012ak,Cox:2017eme}.
Upon discovery of $L$-violating processes involving heavy neutrinos,
among the most pressing quantities to measure are $N$'s chiral couplings to other fields~\cite{Han:2012vk,Dev:2015kca},
its flavor structure~\cite{Chen:2011hc,Nemevsek:2012iq,Anamiati:2016uxp,Das:2017hmg},
and a potential determination if the signal is actually made of multiple, nearly degenerate $N$~\cite{Chao:2009ef,Antusch:2017ebe}.

\begin{figure}[!t]
\begin{center}
\subfigure[]{\includegraphics[scale=1,width=.40\textwidth]{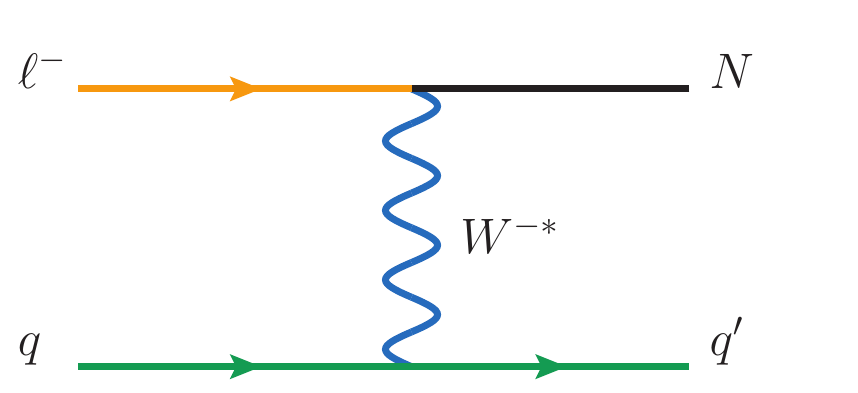}	}
\subfigure[]{\includegraphics[scale=1,width=.56\textwidth]{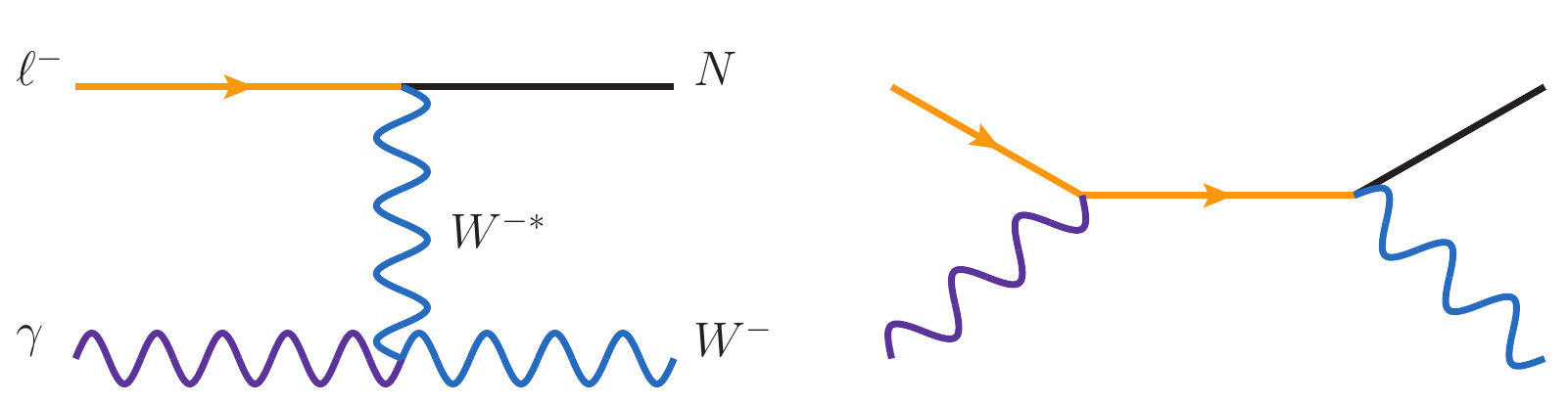}	}
\\
\subfigure[]{\includegraphics[scale=1,width=.48\textwidth]{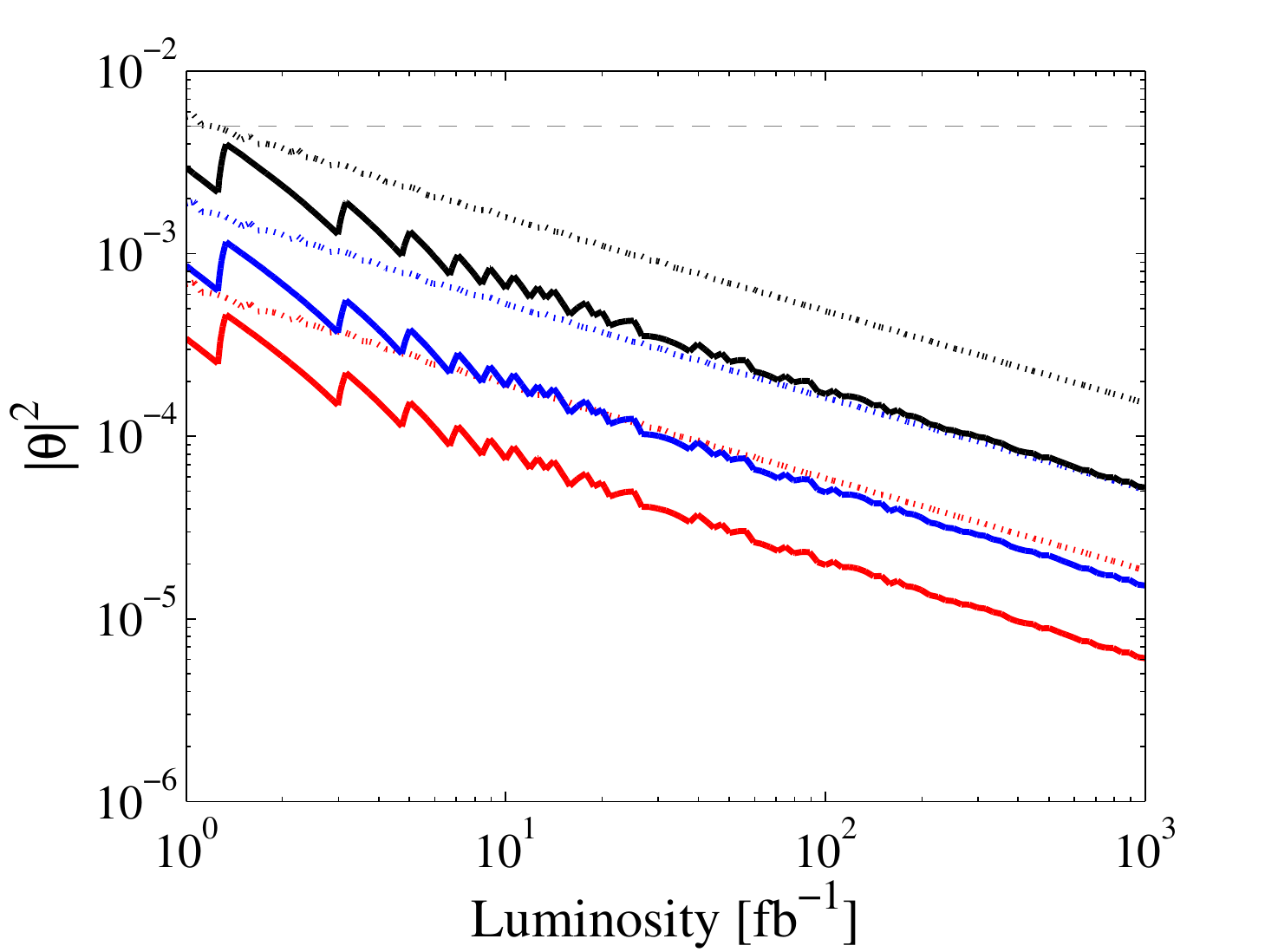}	}
\subfigure[]{\includegraphics[scale=1,width=.48\textwidth]{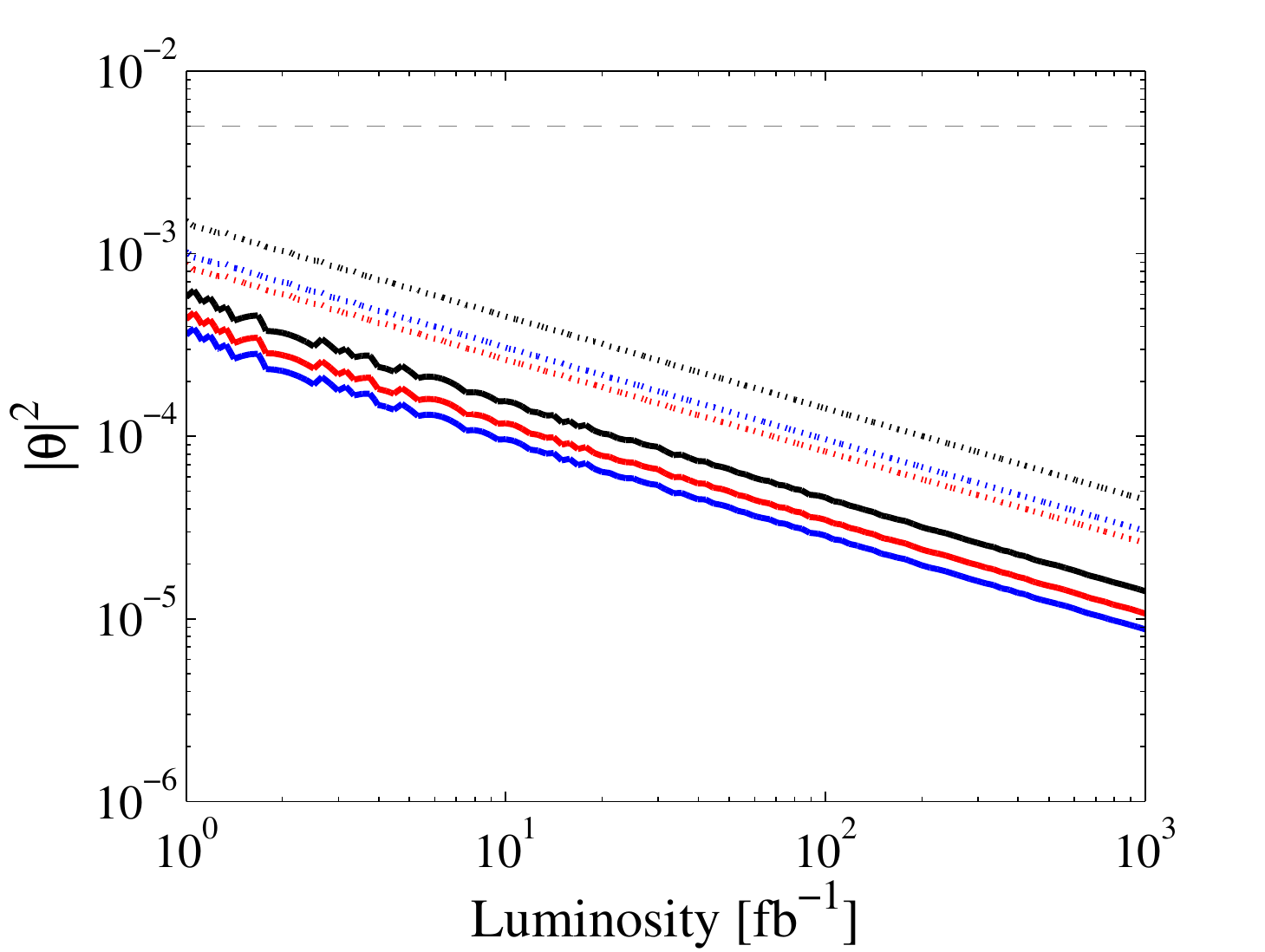}	}
\end{center}
\caption{
Born diagrams for DIS heavy neutrino $(N)$ production via (a) $W$-exchange and (b) $W\gamma$ fusion.
90\% CL active-sterile mixing $\vert\theta\vert^2$~(or $\vert V_{\ell N}\vert^2$) sensitivity versus integrated luminosity
at DIS experiment assuming (c) $ep$ configuration with $E_e = 150$ GeV and (d) $\mu p$ configuration with $E_\mu = 2$ TeV;
red~(blue)~[black] line  in (c,d) correspond to $M_N = 250~(500)~[750]$~GeV, whereas
the solid/dotted lines are the sensitivities with/without cuts~\cite{Blaksley:2011ey}.
}
\label{fig:heavyNDISproduction}
\end{figure}

\subsubsection{High-Mass Heavy Neutrinos at $ep$ Colliders}\label{sec:type1HighMassEX}

Complementary to searches for $L$ violation in $pp$ collisions
are the prospects for heavy $N$ production at $ep$ deeply inelastic scattering (DIS)
colliders~\cite{Buchmuller:1990vh,Buchmuller:1991tu,Blaksley:2011ey,Duarte:2014zea,Banerjee:2015gca,Mondal:2016kof,Antusch:2016ejd,Ingelman:1993ve,Liang:2010gm},
such as  proposed Large Hadron-electron Collider (LHeC) ~\cite{AbelleiraFernandez:2012cc}, or a $\mu p$ analogue~\cite{Blaksley:2011ey}.
As shown in Fig.~\ref{fig:heavyNDISproduction}, DIS production of Majorana neutrinos can occur in multiple ways,
including (a) $W$ exchange and (b) $W\gamma$ fusion.
For treatment of initial-state photons from electron beams, see Ref.~\cite{Frixione:1993yw}.
Search strategies for Majorana neutrinos at DIS experiments typically rely on production via the former
since  $e\gamma\to NW$ associated production can suffer from large phase space suppression, especially at lower beam energies.
On the other hand, at higher beam energies, the latter process can provide additional polarization information on $N$
and its decays~\cite{Antusch:2016ejd}.

At DIS facilities, one usually searches for $L$ violation by requiring that $N$
decays to a charged lepton of opposite sign from the original beam configuration, \ie,
\begin{equation}
 \ell_1^\pm ~q_i ~\to ~N ~q_f, \quad\text{with}\quad N \to \ell_2^\mp W^\pm \to ~\ell_2^\mp ~q ~\overline{q'},
 \label{eq:heavyNdis}
\end{equation}
which is only possible of $N$ is Majorana and is relatively free of SM backgrounds.
As in the $pp$ case, the existence of a high-$p_T$ charged lepton without accompanying MET (at the partonic level) greatly reduces SM backgrounds.
At the hadronic level, this translates to requiring one charged lepton and three high-$p_T$ jets:
two that arise from the decay of $N$, which scale as $p_T^j \sim M_N/4$, and the third from the $W$ exchange, which scales as $p_T^j \sim M_W/2$.
However, it was recently noted~\cite{Mattelaer:2016ynf} that tagging this third jet is not necessary to reconstruct and identify the heavy neutrino, and that a more inclusive search may prove more sensitive. 
Although Eq.~(\ref{eq:heavyNdis}) represents the so-called ``golden channel'', searches for $N\to Z/h+\nu$ decays,
but which do not manifestly violate lepton number, have also been proposed~\cite{Ingelman:1993ve}.

While the lower beam energies translate to a lower mass reach for $M_N$, large luminosity targets and relative cleaner hadronic environment
result in a better sensitivity than the LHC to smaller active-sterile mixing for smaller neutrino Majorana masses.
In Fig.~\ref{fig:heavyNDISproduction}, one sees the expected 90\% CL active-sterile mixing $\vert\theta\vert^2$~(or $\vert V_{\ell N}\vert^2$) sensitivity
assuming (c) $ep$ configuration with $E_e = 150$ GeV and (d) $\mu p$ configuration with $E_\mu = 2$ TeV.
For $\mathcal{L}\sim \mathcal{O}(100)$ fb$^{-1}$, one can probe $\vert V_{\ell N}\vert^2 \sim 10^{-5}-10^{-3}$ for $M_N = 250-750$ GeV~\cite{Blaksley:2011ey}.

\subsubsection{Heavy Neutrinos and U$(1)_{X}$ Gauge Extensions at Colliders}\label{sec:type1Abelian}

\begin{figure}[t!]
\begin{center}
\subfigure[]{\includegraphics[scale=1,width=.48\textwidth,height=7cm]{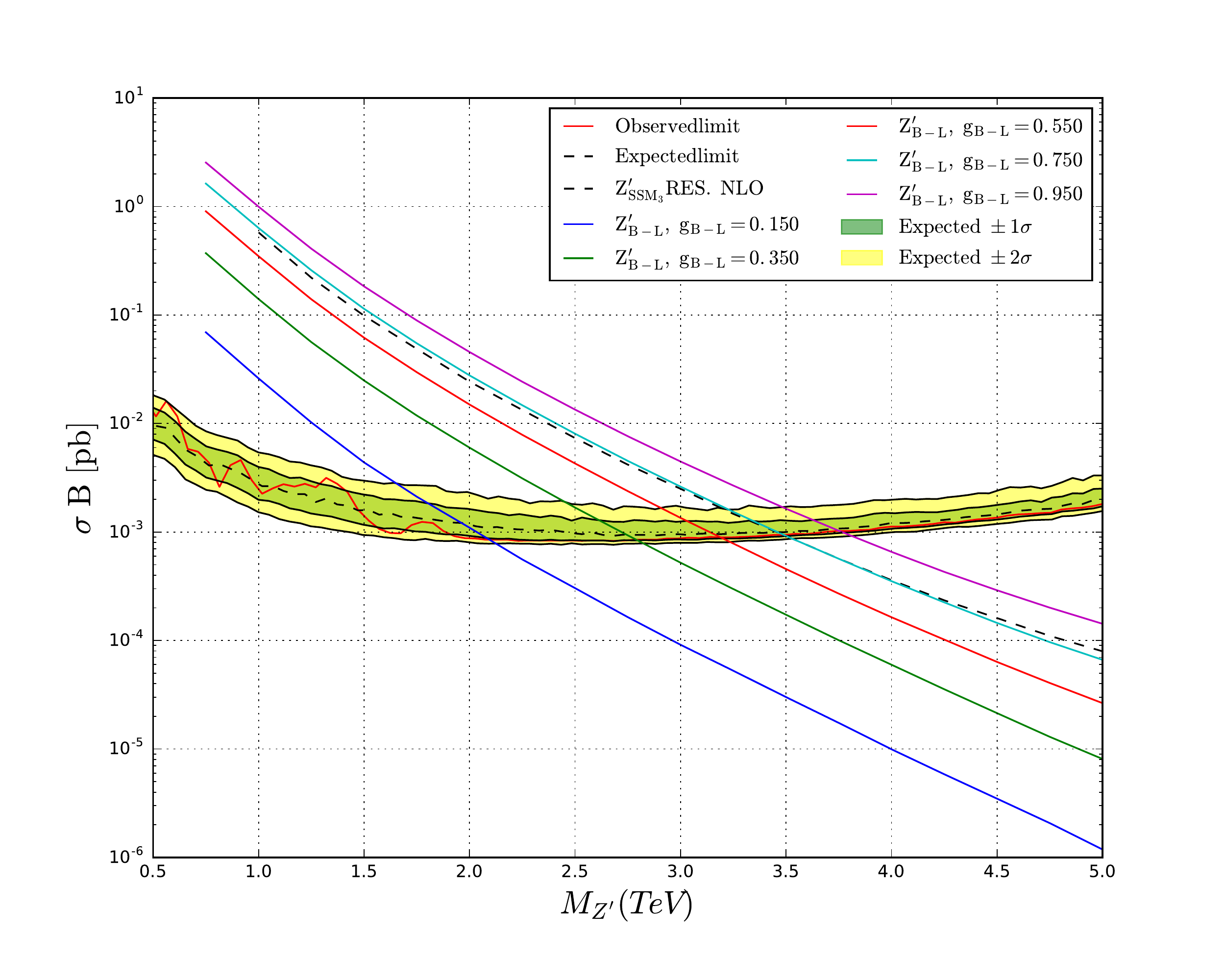}	\label{fig:type1BLxSecNLONNLL}}
\subfigure[]{\includegraphics[scale=1,width=.48\textwidth]{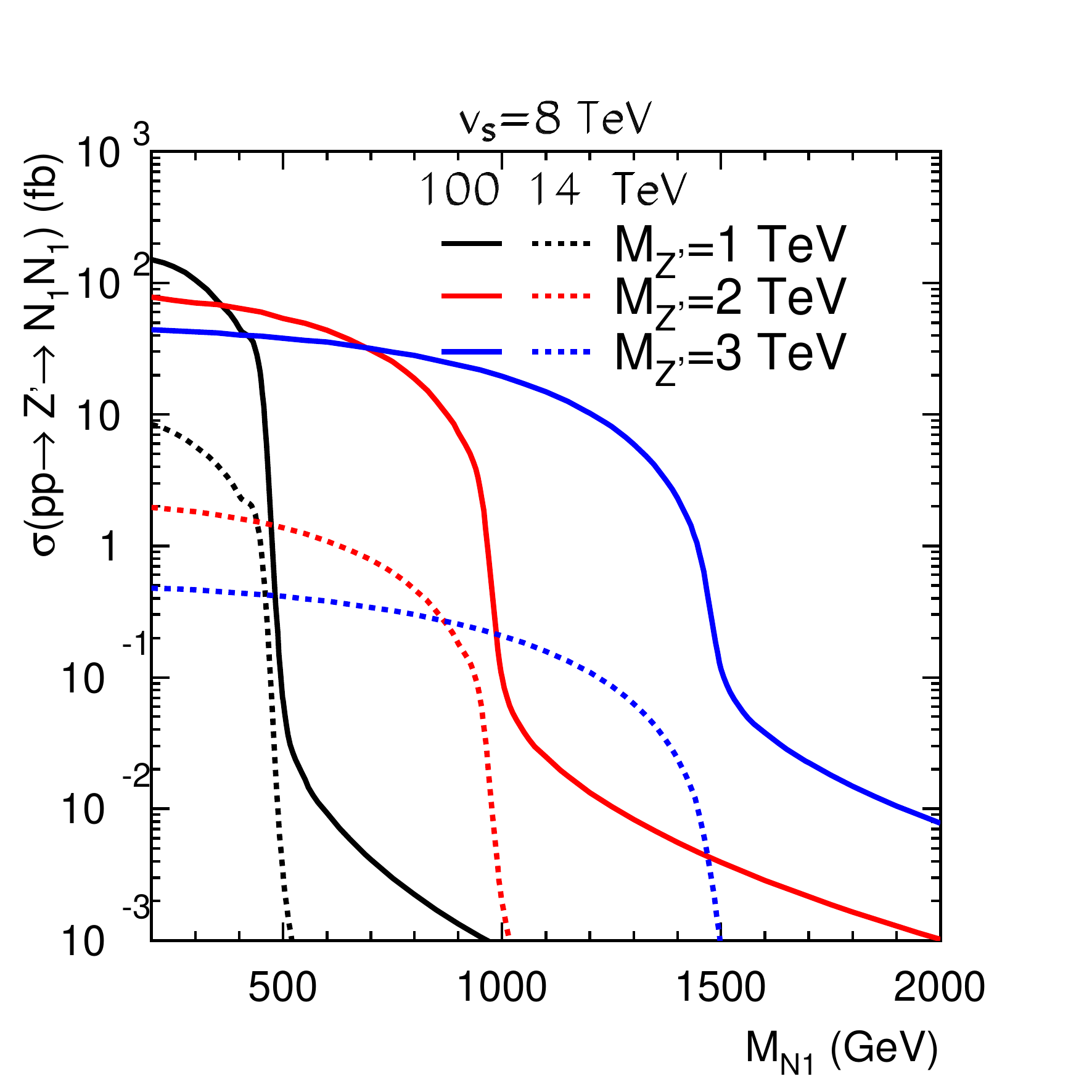}				\label{fig:type1BLxSecLO}}
\end{center}
\caption{
(a) The total cross section of  $pp\to Z_{BL} \to \ell^+\ell^-$
as a function of for various representative values of $g_{BL}$ at NLO+NLL(thresh.) for $\sqrt{s}=13$ TeV~\cite{Klasen:2016qux}.
(b) The total cross section of $pp\to Z'\to NN$ as a function of $M_N$ for $M_{Z'}=1, 2, 3$ TeV, $v_S=8$ TeV, with $\sqrt{s}=14$ TeV and 100 TeV.
}
\label{typei-xsec}
\end{figure}

\begin{figure}[t]
\begin{center}
\subfigure[]{\includegraphics[scale=1,width=.48\textwidth]{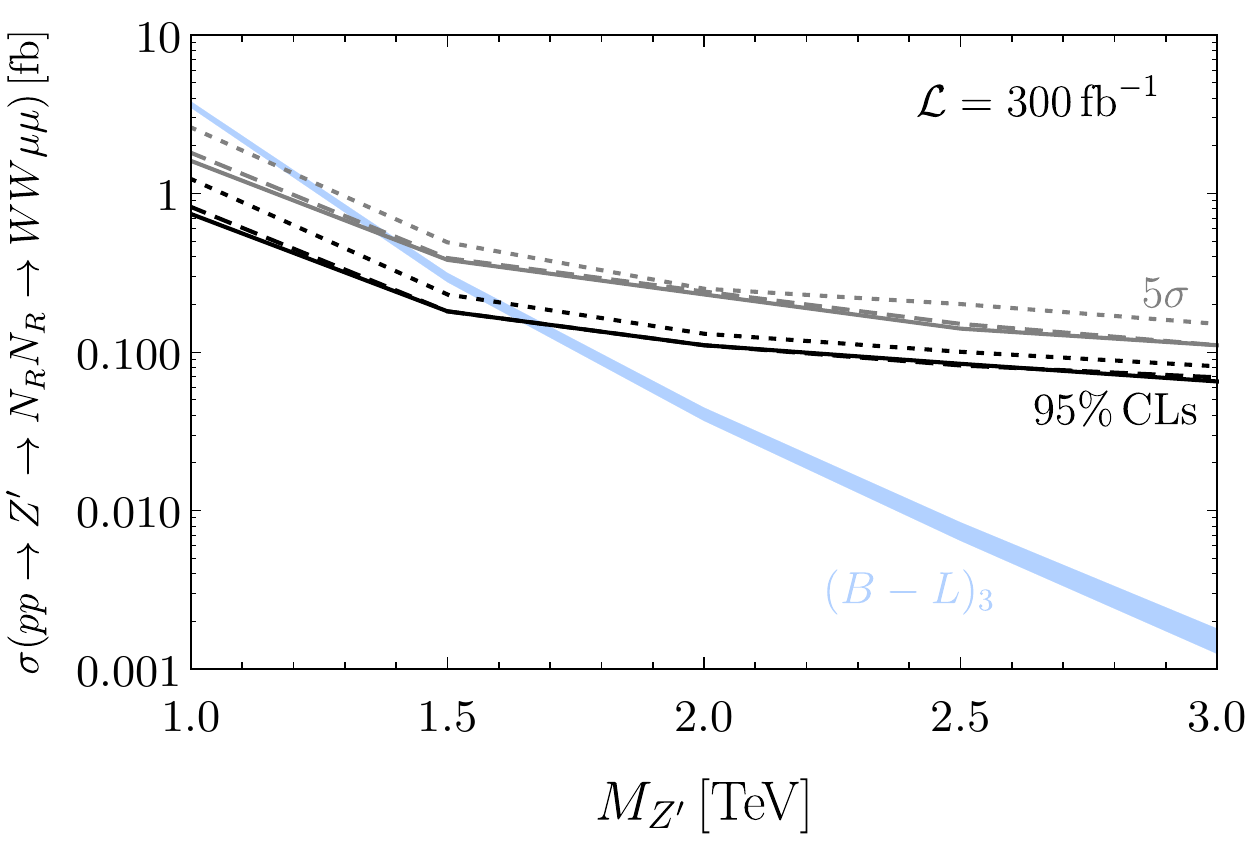}}
\subfigure[]{\includegraphics[scale=1,width=.48\textwidth]{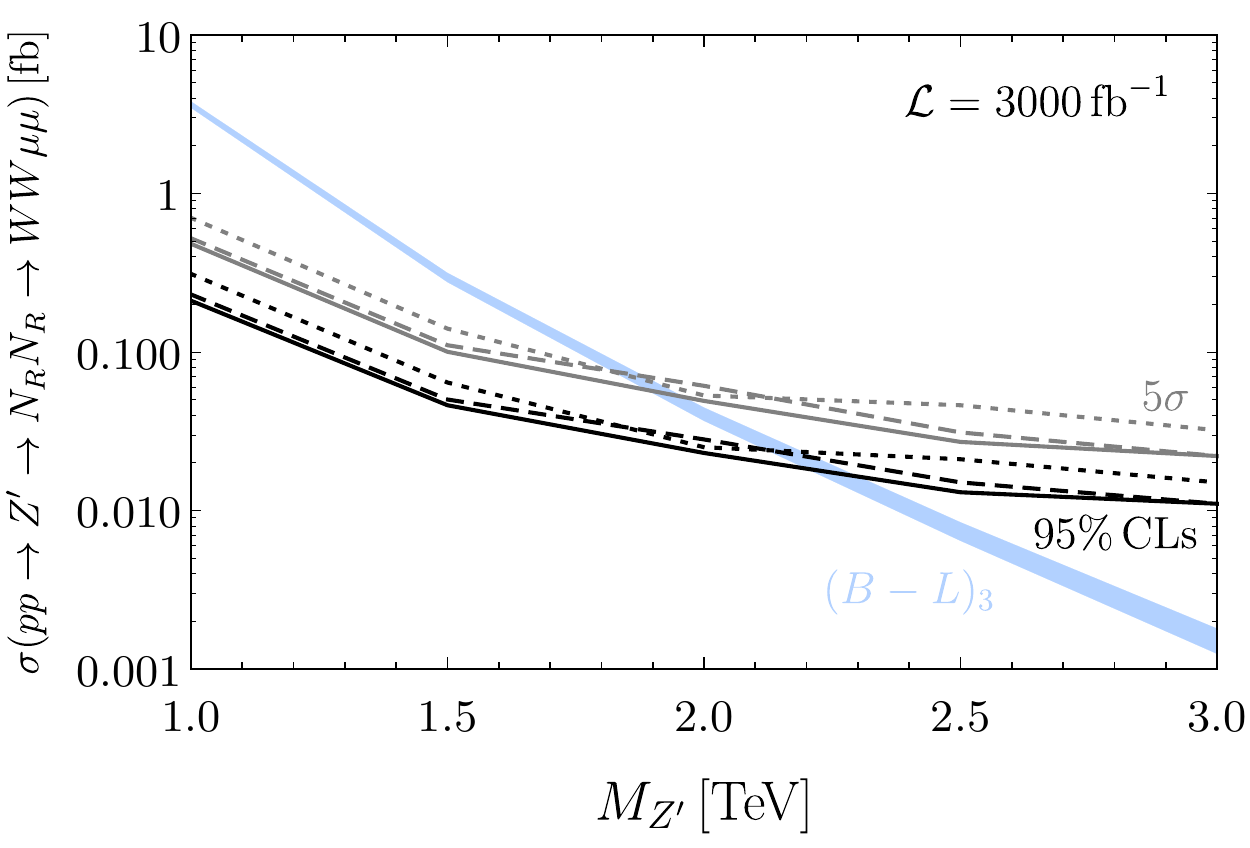}}
\end{center}
\caption{HL-LHC sensitivity for $pp\to Z'\to NN$ with $\sqrt{s}=14$ TeV for (a) $\mathcal{L}=300$ fb$^{-1}$ and for (b) $\mathcal{L}=3000$ fb$^{-1}$, assuming $M_N=M_{Z'}/4$ and $g_1'=0.6$~\cite{Cox:2017eme}.
}
\label{typei-sensitivity}
\end{figure}

Due to the small mixing between the heavy neutrinos and the SM leptons in minimal Type I Seesaw scenarios,
typically of the order $|V_{\ell N}|^2\sim \mathcal{O}(m_\nu/M_N)$,
the predicted rates for collider-scale lepton number violation is prohibitively small.
With a new gauge interaction, say, from U$(1)_{B-L}$, the gauge boson $Z'=Z_{BL}$ can be produced copiously in
$pp$ and $p\bar{p}$ collisions via gauge interactions in quark annihilation~\cite{Salvioni:2009mt,Jenkins:1987ue,Carena:2004xs,Emam:2007dy,Langacker:2008yv,Iso:2009nw,Basso:2010pe,Deppisch:2013cya}
and at Linear Colliders in $e^+e^-$ annihilation~\cite{Freitas:2004hq,Basso:2009hf,Iso:2009nw,Ramirez-Sanchez:2016ugz},
\begin{equation}
q \bar{q} \to Z' \to N N \quad\text{and}\quad
e^+e^- \to Z' \to N N.
\end{equation}
$Z_{BL}$'s subsequent decay to a pair of heavy Majorana neutrinos may lead to a large
sample of events without involving the suppression from a small active-sterile mixing angles
~\cite{delAguila:2007ua,Huitu:2008gf,Basso:2008iv,AguilarSaavedra:2009ik,Perez:2009mu,Li:2010rb,AguilarSaavedra:2012gf,Accomando:2016sge,Accomando:2017qcs}.
As a function of $M_{Z_{BL}}$, Fig.~\ref{fig:type1BLxSecNLONNLL} shows the NLO+NLL(Thresh.) $pp\to Z_{BL}\to \ell^+\ell^-$ production and decay rate
for $\sqrt{s}=13$ TeV and representative values of coupling $g_{BL}$.
As a function of Majorana neutrino mass $M_{N_1}$, Fig.~\ref{fig:type1BLxSecLO} shows the LO $pp\to Z_{BL}\to NN$ production and decay rate
for $\sqrt{s}=14$ TeV and 100 TeV and representative $M_{Z_{BL}}$.
As $N$ is Majorana, the mixing-induced decays modes $N\to \ell^\pm W^\mp, \nu Z, \nu h$ open for $M_{N_1}>M_W,M_Z,M_h$, respectively.
Taking these into account, followed by the leptonic and/or hadronic decays of $W$, $Z$ and $h$,
the detectable signatures include
the lepton number violating, same-sign dileptons, $NN\to \ell^\pm\ell^\pm W^\mp W^\mp \to \ell^\pm\ell^\pm+nj$~\cite{Perez:2009mu,Cox:2017eme};
final states with three charged leptons, $\ell^\pm\ell^\pm\ell^\mp+nj+$MET~\cite{Basso:2008iv,Kang:2015uoc,Accomando:2017qcs};
and four-charged lepton, $\ell^\pm\ell^\pm\ell^\mp\ell^\mp+$MET~\cite{Huitu:2008gf,Abdelalim:2014cxa}.
Assuming that only the third generation fermions are charged under $B-L$ symmetry, the HL-LHC can probe $Z'$ mass up to 2.2 TeV
and heavy neutrino mass in the range of $0.2-1.1$ TeV as shown in Fig.~\ref{typei-sensitivity}~\cite{Cox:2017eme}.

For super-heavy $Z_{BL}$, {\it e.g.}, $M_{Z_{BL}}\gtrsim5$ TeV $\gg M_{N}$,
one should note that at the 13 TeV LHC, a nontrivial contribution of the total $pp\to Z_{BL} \to NN$ cross section
comes from the kinematical threshold region, where the $(NN)$ system's invariant mass is near $m_{NN}\sim 2M_N$ and $Z_{BL}^*$ is far off-shell.
This implies that the $L$-violating process $pp\to NN \to \ell^\pm\ell^\pm+nj$
can still proceed despite $Z_{BL}$ being kinematically inaccessible~\cite{Ruiz:2017nip}. For more details, see Sec.~\ref{sec:neftTests}.
Additionally, for such heavy $Z_{BL}$ that are resonantly produced, the emergent $N$ are highly boosted with Lorentz factors of $\gamma \sim M_{Z_{BL}}/2M_N$.
For $M_N\ll M_{Z_{BL}}$, this leads to highly collimated decay products, with separations scaling as $\Delta R\sim 2/\gamma \sim 4M_N/M_{Z_{BL}}$,
and eventually the formation of lepton jets~\cite{Izaguirre:2015pga,Dube:2017jgo},
\ie, collimated clusters of light, charged leptons and electromagnetic radiation,
and neutrino jets~\cite{Mitra:2016kov,Mattelaer:2016ynf,Cox:2017eme,Nemevsek:2018bbt}, \ie, collimated clusters of electromagnetic and hadronic activity
from decays of high-$p_T$ heavy neutrinos.

Leading Order-accurate Monte Carlo simulations for tree-level processes involving $Z'$ bosons and heavy neutrinos in $U(1)_X$ theories are possible
using the \texttt{SM+B-L} FeynRules UFO model~\cite{FeynRules:BL,Basso:2011na,Basso:2008iv}.
At NLO+PS accuracy, Monte Carlo simulations can be performed using the \texttt{Effective LRSM at NLO in QCD} UFO model\cite{Mattelaer:2016ynf,FeynRules:LRSMnlo},
and, for light, long-lived neutrinos and arbitrary $Z'$ boson couplings,
the \texttt{SM~+~W'~and~Z'~at~NLO~in~QCD} UFO model~\cite{Fuks:2017vtl,FeynRules:WZPrimeAtNLO}.

\begin{figure}[h!]
\begin{center}
\subfigure[]{\includegraphics[scale=1,width=.48\textwidth,height=6cm]{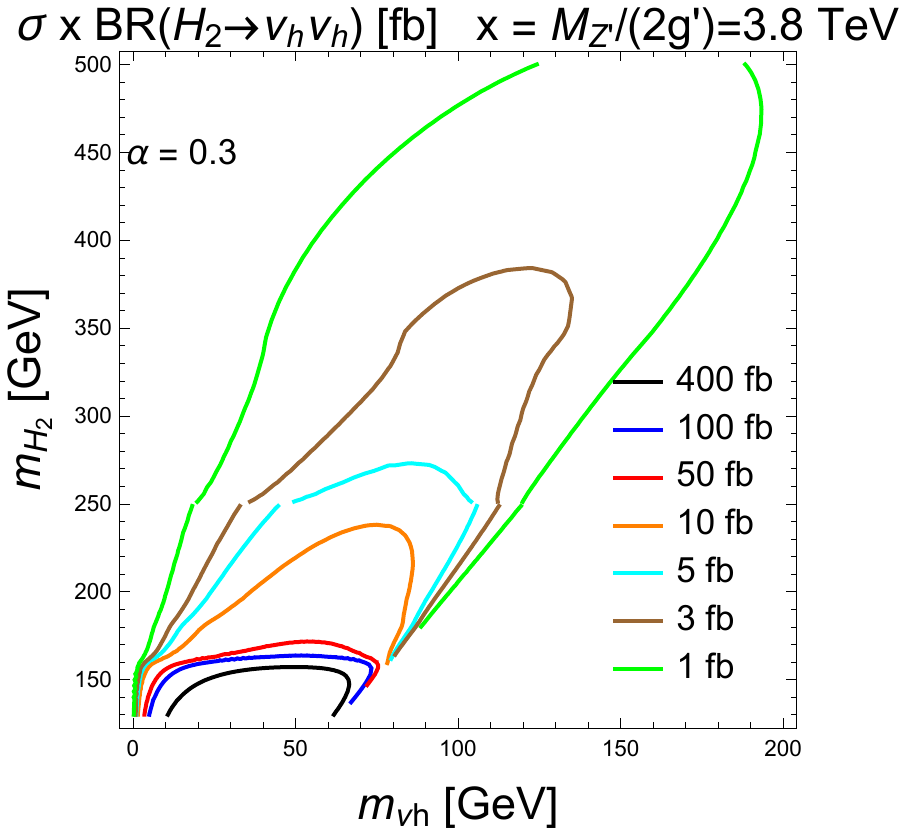}}
\subfigure[]{\includegraphics[scale=1,width=.48\textwidth,height=6cm]{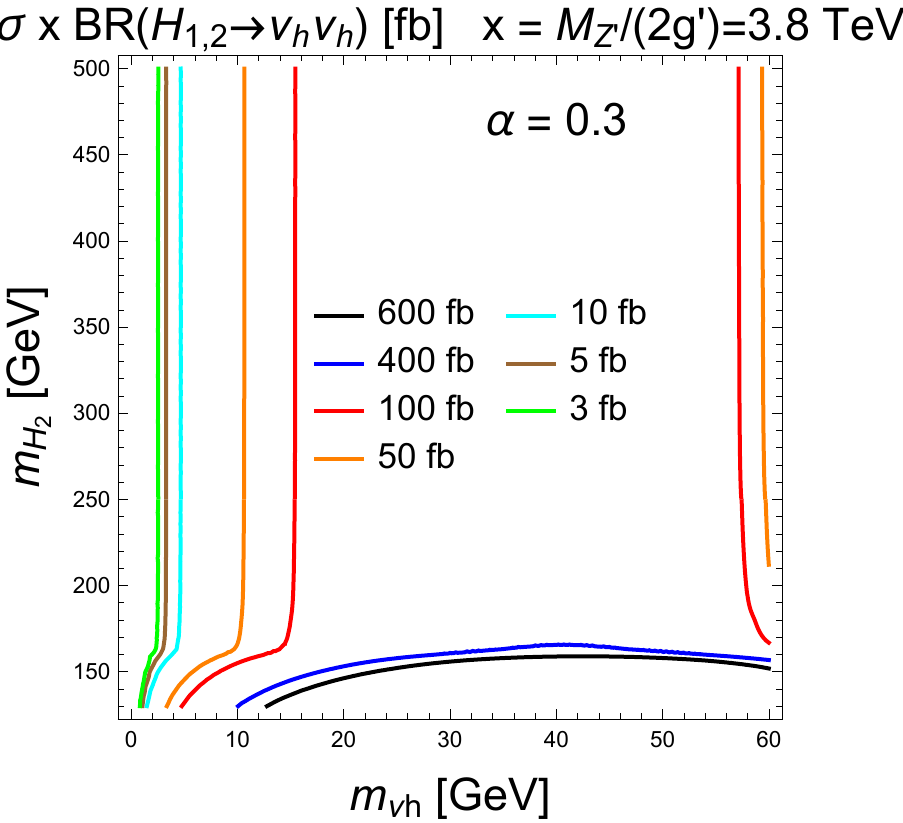}}
\end{center}
\caption{
(a) Contour of the cross section for   $pp\to H_2\to NN$ with $\sqrt{s}=13$ TeV in the plane of $M_{H_2}$ vs. $M_N$ for $M_{Z'}=5$ TeV and $g_1'=0.65$;
(b) the same but for $pp\to H_1,H_2\to NN$ with $\sqrt{s}=13$ TeV and $M_N<M_W$ \cite{Accomando:2017qcs}.
}
\label{typei-xsechiggs}
\end{figure}

In $B-L$ models, heavy neutrino pairs can also be produced through the gluon fusion process
mediated by the two $H_1$ and $H_2$~\cite{Basso:2010yz,Pruna:2011me,Accomando:2016rpc,Accomando:2017qcs}, and given by
\begin{eqnarray}
gg\to H_1, H_2\to N N.
\end{eqnarray}
For long-lived heavy neutrinos with $M_N\lesssim 200$ GeV, this process becomes important compared to the channel mediated by $Z'$. Fig.~\ref{typei-xsechiggs} (a) shows that for $M_{H_2}<500$ GeV, $M_N<200$ GeV, and $M_{Z'}=5$ TeV,
the cross section $\sigma(pp\to H_2\to NN)$ can be above 1 fb at the $\sqrt{s}=13$ TeV LHC.
For $M_N<60$ GeV, decays of the SM-like Higgs $H_1$ also contributes to neutrino pair production.
Summing over the contributions via $H_1$ and $H_2$ the total cross section can reach about 700 fb for $M_{H_2}<150$ GeV as shown in Fig.~\ref{typei-xsechiggs} (b).

\begin{figure}[!t]
\begin{center}
\subfigure[]{\includegraphics[scale=1,width=.48\textwidth,height=6cm]{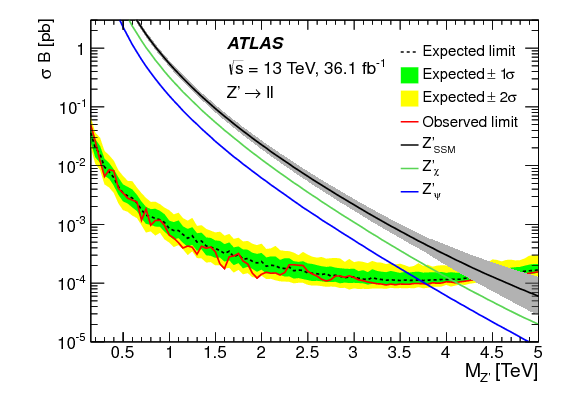}}
\subfigure[]{\includegraphics[scale=1,width=.48\textwidth,height=6cm]{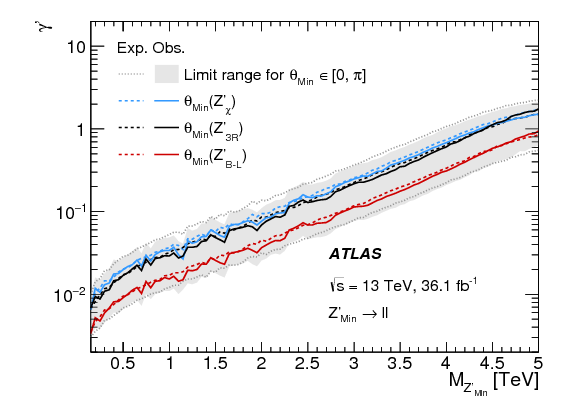}}
\end{center}
\caption{(a) Exclusion limit on $pp\to Z'\to \ell^+\ell^-$ by ATLAS at $\sqrt{s}=13$ TeV with 36.1 fb$^{-1}$;
(b) 13 TeV upper limit at 95\% CL on the coupling strength $\gamma'=g_{BL}/g_Z$ as a function of $M_{Z'}$~\cite{Aaboud:2017buh}.
}
\label{B-LcouplingmassATLAS}
\end{figure}

Owing to this extensive phenomenology, collider experiments are broadly sensitive to $Z'$ bosons from U$(1)_{BL}$ gauge theories.
For example: Searches at LEP-II have set the lower bound of $M_{Z'}/g_{BL}\gtrsim 6$ TeV~\cite{Carena:2004xs}. For more generic $Z'$ (including $Z_R$ in LRSM models),
comparable limits from combined LEP+EW precision data have been derived in Ref.~\cite{delAguila:2010mx,delAguila:2011yd}.
Direct searches for a $Z'$ with SM-like couplings to fermions exclude $M_{Z'}<2.9$ TeV at 95\% CLs
by ATLAS~\cite{Aad:2014cka} and CMS~\cite{Khachatryan:2014fba} at $\sqrt{s}=8$ TeV.
$Z_{BL}$ gauge bosons with the benchmark coupling $g_1'=g_{BL}$ are stringently constrained by searches for dilepton resonances at the LHC, with $M_{Z'}\lesssim 2.1-3.75$ TeV excluded at 95\% CLs for $g_{BL}=0.15-0.95$, as seen in Fig.~\ref{fig:type1BLxSecNLONNLL}~\cite{Klasen:2016qux}.
Searches for $Z'$ decays to dijets at the LHC have exclude $M_{Z'}<1.5-3.5$ TeV for $g_{BL}=0.07-0.27$~\cite{Sirunyan:2016iap,Aaboud:2017yvp}.
Fig.~\ref{B-LcouplingmassATLAS} (a) shows that ATLAS excludes $M_{Z'}<4.5$ TeV at $\sqrt{s}=13$ TeV.
Further constraints are given in the plane of coupling strength $\gamma'=g_{BL}/g_Z$ vs. $M_{Z'}$ by ATLAS at $\sqrt{s}=13$ TeV
with 36.1 fb$^{-1}$~\cite{Aaboud:2017buh} as shown in the lower curve of Fig.~\ref{B-LcouplingmassATLAS}(b).
For $\sqrt{s}=27$ TeV, early projections show that with $\mathcal{L}=1~(3)$ ab$^{-1}$, $M_{Z'}\lesssim19~(20)$ TeV can be probed in the dijet channel~\cite{Chekanov:2017pnx}.

\subsubsection{Heavy Neutrinos and the Left-Right Symmetric Model at Colliders}\label{sec:lrsmCollider}

In addition to the broad triplet scalar phenomenology discussed later in Sec.~\ref{sec:type2Collider}, the LRSM predicts at low scales massive $W_R^{\pm}$ and $Z_R$ gauge bosons that couple appreciably to SM fields as well as to heavy Majorana neutrinos $N$. The existence of these exotic states leads to a rich collider phenomenology that we now address, focusing, of course, on lepton number violating final states.
The collider phenomenology for $Z_R$ searches is very comparable to that for $Z'$ gauge bosons in U$(1)_X$ theories
~\cite{delAguila:2007ua,Huitu:2008gf,Basso:2008iv,AguilarSaavedra:2009ik,Perez:2009mu,Li:2010rb,AguilarSaavedra:2012gf,Accomando:2016sge,Accomando:2017qcs},
and thus we refer readers to Sec.~\ref{sec:type1Abelian} for more generic collider phenomenology.

\begin{figure}[!t]
\begin{center}
\subfigure[]{\includegraphics[scale=1,width=.48\textwidth]{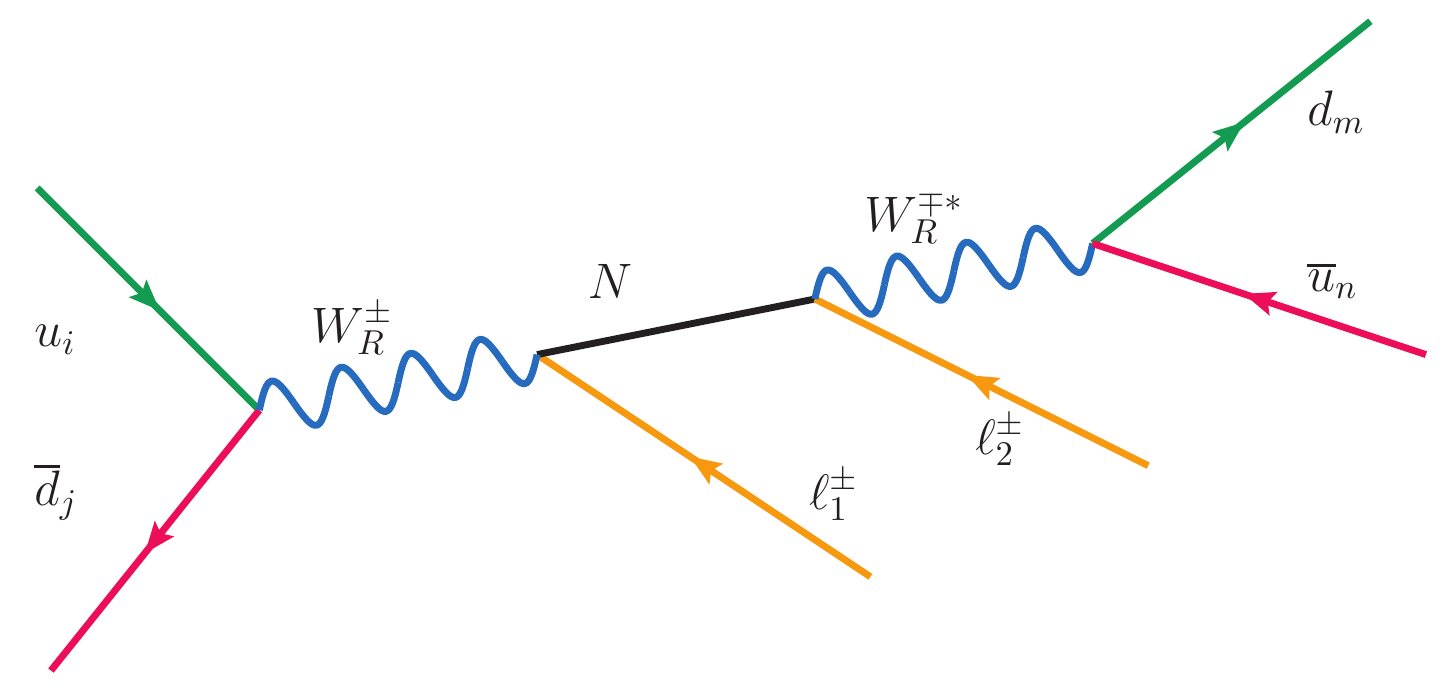} \label{fig:mLRSM} }
\subfigure[]{\includegraphics[scale=1,width=.48\textwidth]{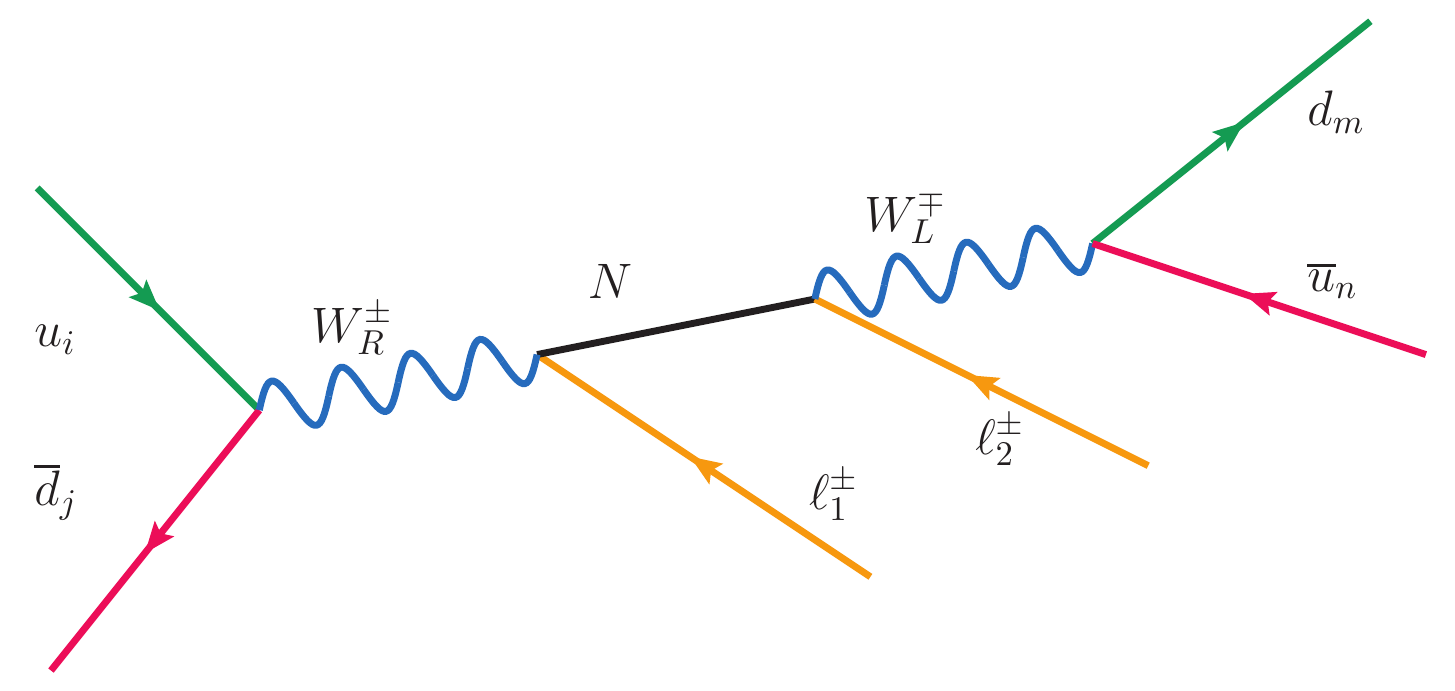} \label{fig:aLRSM} }
\end{center}
\caption{
Born-level diagrams depicting resonant $W_R, N$ production and decay to same-sign leptons in LRSM via
(a) successive right-handed currents and
(b) mixed right- and left-handed currents.
}
\label{fig:lrsmFeynman}
\end{figure}

In the LRSM, for $M_N<M_{W_R}$ or $M_N<M_{Z_R}/2$,
the most remarkable collider processes are the single and pair production of heavy Majorana neutrinos $N$ through resonant charged and neutral $\SU{2}{R}$ currents,
\begin{eqnarray}
 q \overline{q'} \to W_R^\pm \to ~N_i ~\ell^\pm  \quad\text{and}\quad q \overline{q'} \to Z_R \to ~N_i ~N_j.
\end{eqnarray}
As first observed in Ref.~\cite{Keung:1983uu}, $N_i$ can decay into $L$-violating final-states, giving rise to the collider signatures,
\begin{equation}
  pp \to W_R^\pm \to ~N_i ~\ell^\pm \to \ell_1^\pm ~\ell_2^\pm + nj  \quad\text{and}\quad   pp \to Z_R \to ~N_i ~N_j \to \ell_1^\pm ~\ell_2^\pm + nj.
  \label{eq:lrsmppssllnj}
\end{equation}
In the minimal/manifest LRSM, the decay of $N_i$ proceeds primarily via off-shell three-body right-handed currents, as shown in Fig.~\ref{fig:mLRSM},
due to mixing suppression to left-handed currents.
In a generic LRSM scenario, the na\"ive mixing suppression of $\vert V_{\ell N}\vert^2 \sim\mathcal{O}(m_\nu / M_N)$ is not guaranteed
due to the interplay between the Types I and II Seesaws, \eg, as in Refs~\cite{Anamiati:2016uxp,Das:2017hmg}.
(However, heavy-light neutrino mixing in the LRSM is much less free than in pure Type I scenarios
due to constraints on Dirac and RH masses from LR parity; see Sec.~\ref{sec:hybrid} for more details.)
Subsequently, if $\vert V_{\ell N}\vert$ is not too far from present  bounds (see, \eg, ~\cite{Fernandez-Martinez:2016lgt}), then decays of $N_i$ to on-shell EW bosons, as shown in Fig.~\ref{fig:aLRSM}, can occur with rates comparable to decays via off-shell $W_R^*$~\cite{Han:2012vk}. The inverse process~\cite{Chen:2013fna}, \ie, $N_i$ production via off-shell EW currents and decay via off-shell RH currents
as well as vector boson scattering involving $t$-channel $W_R$ and $Z_R$ bosons~\cite{Dev:2016dja}
are in theory also possible but insatiably phase space-suppressed.
For $M_N>M_{W_R},M_{Z_R}$, resonant $N$ production via off-shell $\SU{2}{R}$ currents is also possible,
and is analogous to the production through off-shell, $\SU{2}{L}$ currents in Eqs.~(\ref{eq:heavyNDYCC})-(\ref{eq:heavyNDYNC}).
As $M_{W_R},M_{Z_R}$ are bound to be above a few-to-several TeV, the relevant collider phenomenology is largely
the same as when $M_N<M_{W_R},M_{Z_R}$~\cite{Chakrabortty:2012pp},
and hence will not be individually discussed.

\begin{figure}[!t]
\begin{center}
\subfigure[]{\includegraphics[scale=1,width=.48\textwidth]{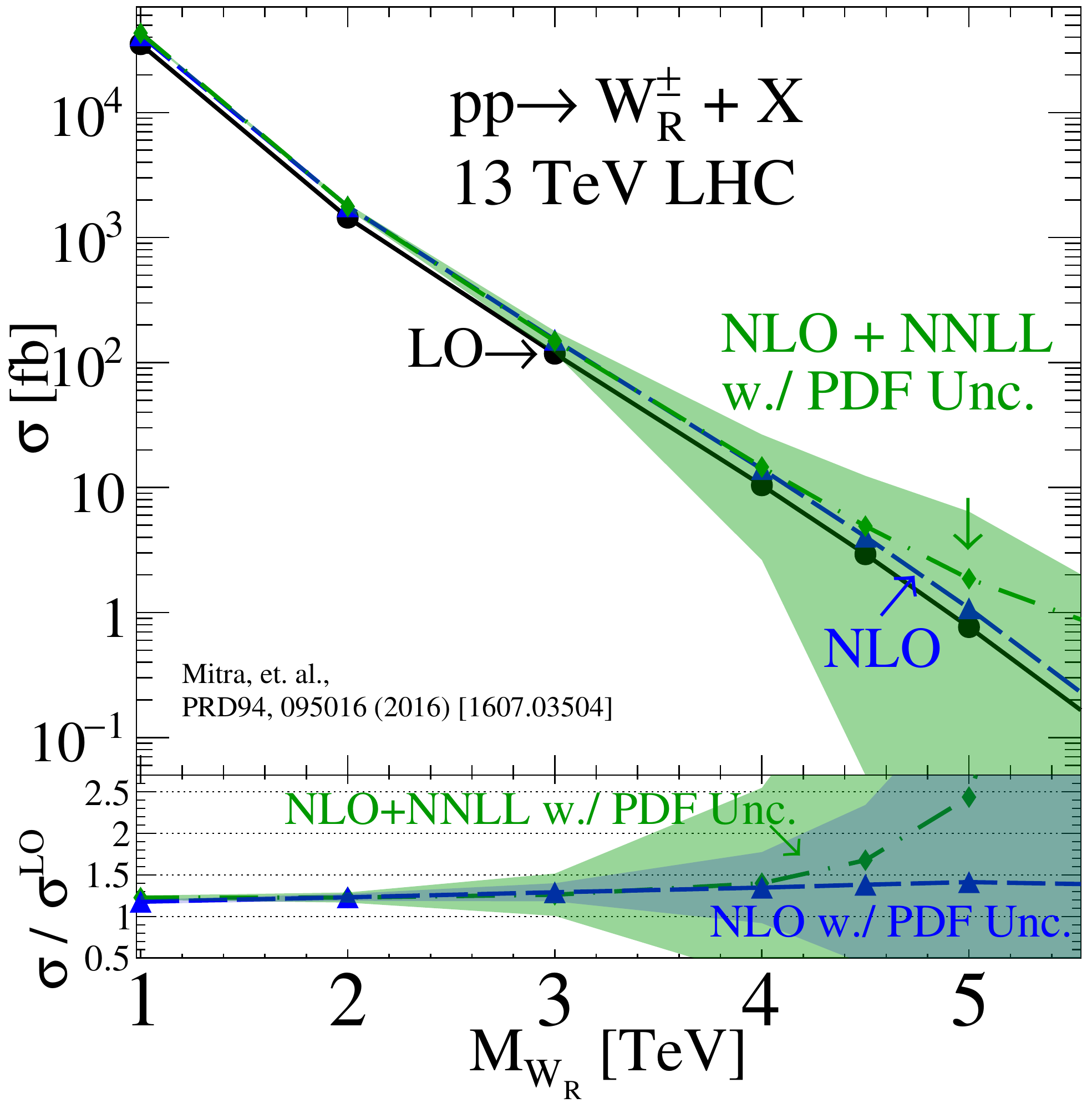}	\label{fig:wrNLONNLLxsec13TeV}	}
\subfigure[]{\includegraphics[scale=1,width=.48\textwidth]{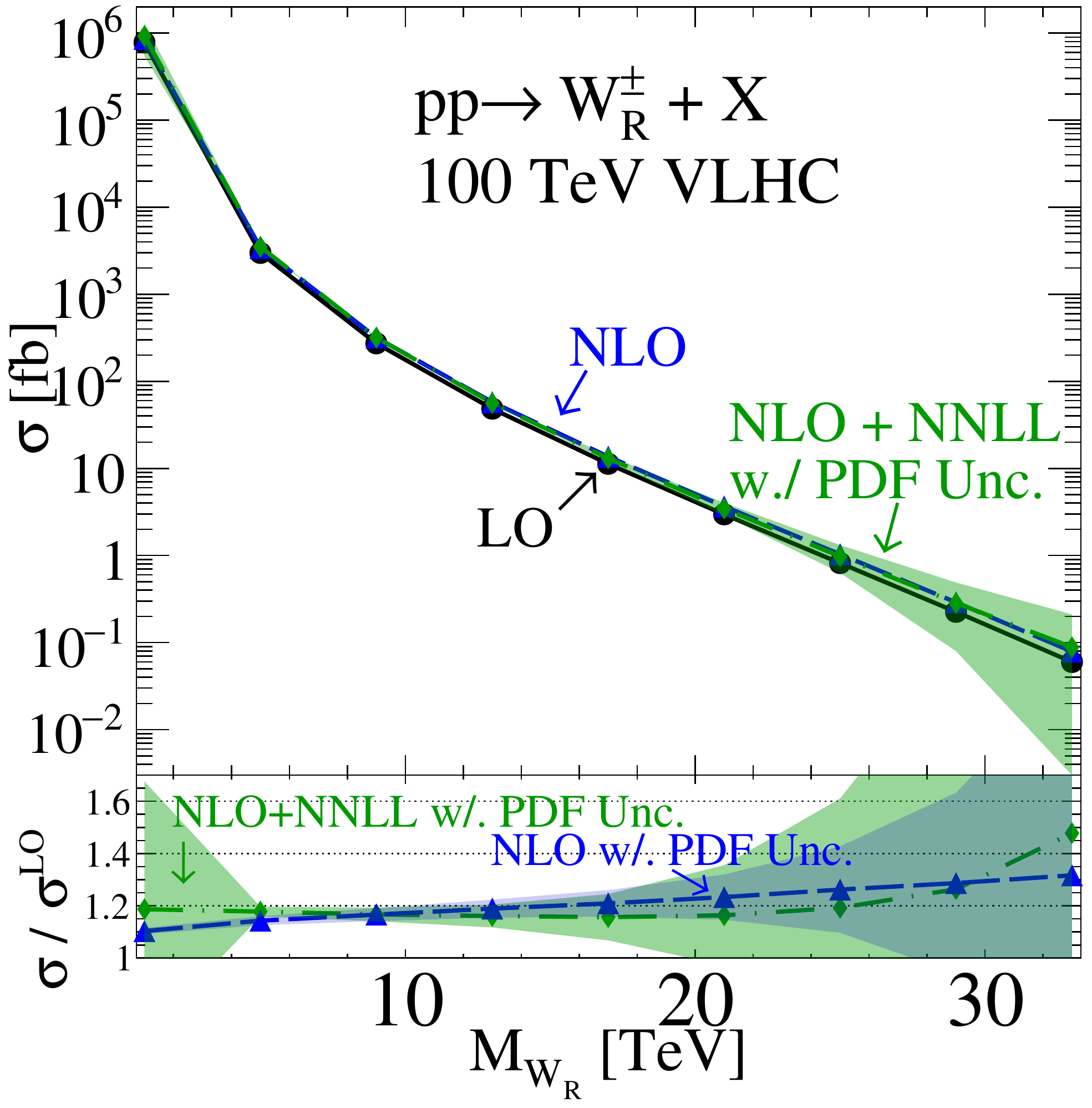}	\label{fig:wrNLONNLLxsec100TeV} }
\end{center}
\caption{
Upper panel: As a function of $M_{W_R}$, $pp\rightarrow W_R$ production cross section for $\sqrt{s} =$ (a) 13 and (b) 100 TeV,
at LO (solid), NLO (dash), and NLO+NNLL (dash-dot) with $1\sigma$ PDF uncertainty (shaded);
Lower: NLO (dash) and NLO+NNLL (dash-dot) $K$-factors and PDF uncertainties~\cite{Mitra:2016kov}.
}
\label{fig:lrsmXSec}
\end{figure}

Aside from the mere possibility of $L$ violation, what makes these channels so exceptional, if they exist, are their production rates. Up to symmetry-breaking corrections, the RH gauge coupling is $g_R \approx g_L \approx 0.65$, which is not a small number. In Fig.~\ref{fig:lrsmXSec}, we show for $\sqrt{s}=13$ and 100 TeV the production rate for resonant $W_R$ at various accuracies as a function of mass~\cite{Mitra:2016kov};
rates for $Z_R$ are marginally smaller due to slight coupling suppression. As in other Seesaw scenarios, much recent progress has gone into advancing the precision of integrated and differential predictions for the LRSM:
The inclusive production of $W_R$ and $Z_R$ are now known up to NLO+NNLL(Thresh)~\cite{Mitra:2016kov}, automated at NLO+NLL(Thresh+$k_T$)~\cite{Fuks:2013vua,Jezo:2014wra},
automated at NNLO ~\cite{Gavin:2010az,Gavin:2012sy},
and differentially has been automated at NLO with parton shower matching for Monte Carlo simulations~\cite{Mattelaer:2016ynf}.
For $\sqrt{\tau_0} = M_{W_R/Z_R}/\sqrt{s} \gtrsim 0.3$, threshold corrections become as large as (N)NLO corrections, which span roughly $+20\%$ to $+30\%$,
and have an important impact cross section normalizations~\cite{Mitra:2016kov,Appell:1988ie}.
For example: The inclusive $W_R$ cross section at LO~(NLO+NNLL) for $M_{W_R}=5$ TeV is $\sigma\sim0.7~(1.7)$ fb. After $\mathcal{L}=1\invab$ and assuming a combined branching-detection efficiency-selection acceptance of
BR$\times\varepsilon\times\mathcal{A}=2\%$, the number of observed events is $N\sim14~(34)$. For simple Gaussian statistics with a zero background hypothesis, this is the difference between a $6\sigma$ ``discovery'' and $4\sigma$ ``evidence''. Clearly, the HL-LHC program is much more sensitive to ultra-high-mass resonances than previously argued.

\begin{figure}[!t]
\begin{center}
\subfigure[]{\includegraphics[scale=1,width=.45\textwidth]{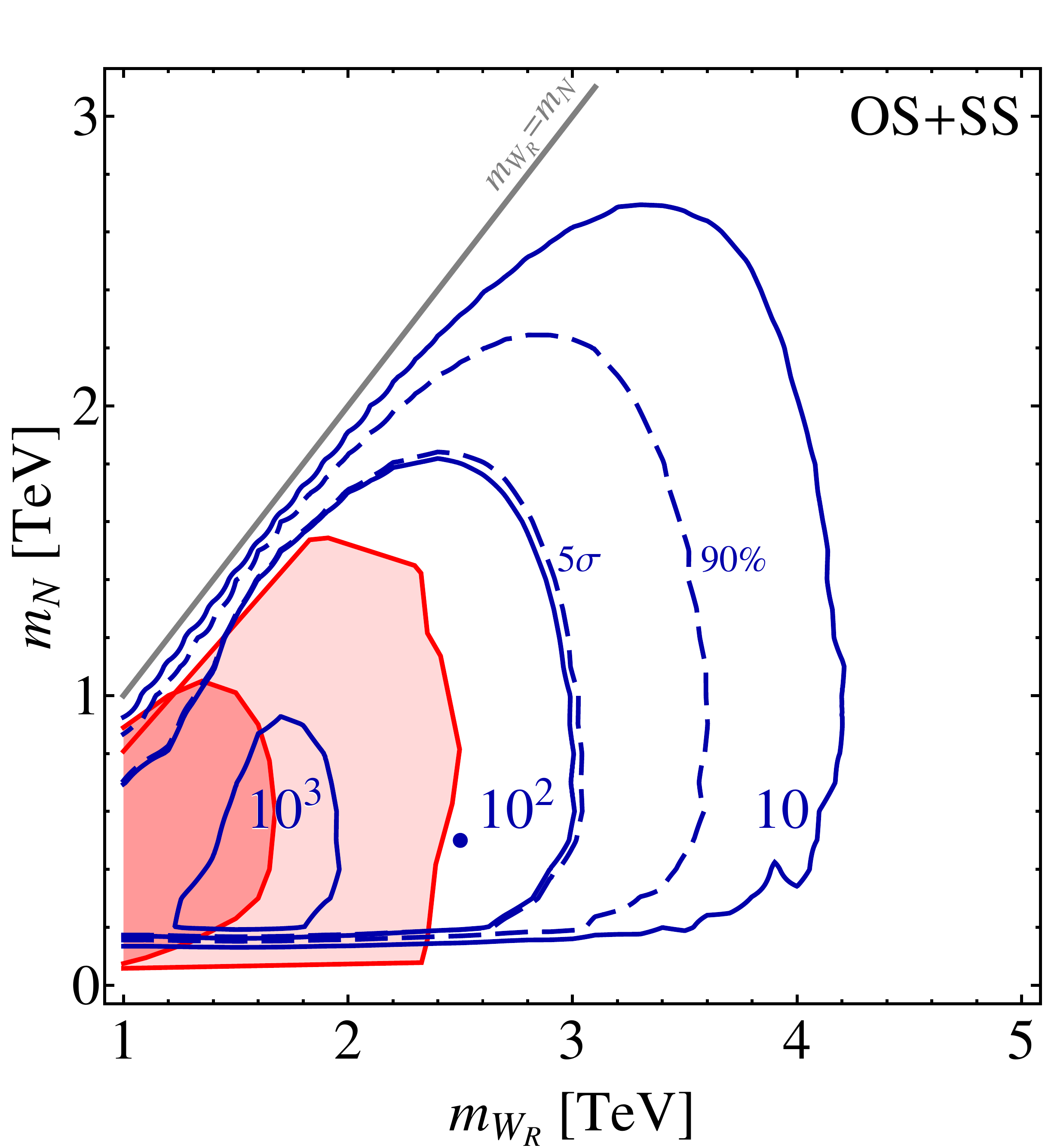}	}
\subfigure[]{\includegraphics[scale=1,width=.45\textwidth]{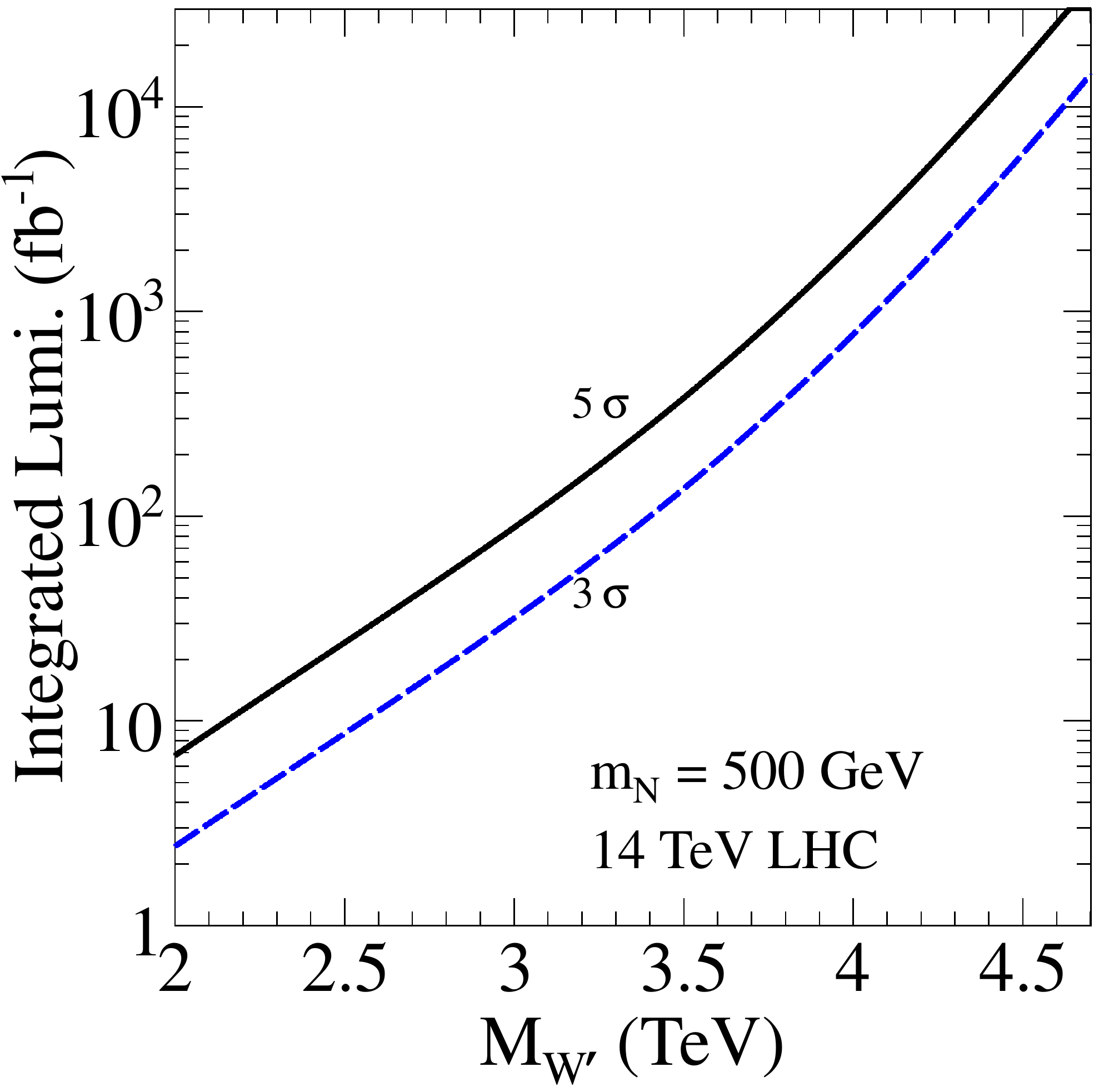} }
\end{center}
\caption{
Discovery potential at 14 TeV LHC of $W_R$ and $N$ in
(a) the minimal LRSM as in Fig.~\ref{fig:mLRSM} after $\mathcal{L}=30\invfb$~\cite{Das:2012ii}
and
(b) the agnostic mixing scenario as in Fig.~\ref{fig:aLRSM}~\cite{Han:2012vk}.
}
\label{fig:lrsmDisc}
\end{figure}

For the collider processes in Eq.~(\ref{eq:lrsmppssllnj}), such estimations of branching, acceptance/selection,
and background rates resemble actual rates;
see, \eg,~\cite{Keung:1983uu,Ferrari:2000sp,Das:2012ii,Han:2012vk,Mitra:2016kov,Chen:2013fna,Dev:2016dja,Goh:2009fm}.
For $M_{W_R},~M_{Z_R}\gg M_{N}$, one finds generically that BR$(W_R \to \ell^\pm N_i)\sim 1/(1+3N_c)\sim\mathcal{O}(10\%)$, 
BR$(Z_R \to N_i N_j)\sim\mathcal{O}(10\%)$, and, for the lightest heavy $N_i$ in this limit,  BR$(N_1 \to \ell^\pm X)\sim\mathcal{O}(100\%)$.
Trigger rates for multi-TeV, stable charged leptons $(e,\mu)$ at ATLAS and CMS exceed $80\%-95\%$, but
conversely, the momentum resolution for such energetic muons is severely degraded;
for additional information, see~\cite{Aad:2015xaa,ATLAS:2016ecs,Khachatryan:2014dka,Khachatryan:2016jww} and references therein.
As in searches for Majorana neutrinos in the previous Type I-based scenarios, the final-states in Eq.~(\ref{eq:lrsmppssllnj})
possess same-sign, high-$p_T$ charged leptons without accompanying MET at the partonic level~\cite{Keung:1983uu,Han:2006ip,Ferrari:2000sp}.
For the LRSM, this is particularly distinct since the kinematics of the signal process scale with the TeV-scale $W_R$ and $Z_R$ masses.
Accordingly, top quark and EW background processes that can mimic the fiducial collider definition correspondingly 
must carry \textit{multi}-TeV system invariant masses, and are inherently more phase space suppressed than the signal processes at the LHC~\cite{Ferrari:2000sp}. 
Consequently, so long as $M_N \lesssim M_{W_R},~M_{Z_R} \ll \sqrt{s}$, 
$s$-channel production of $W_R$ and $Z_R$ remains the most promising mechanism for discovering $L$ violation in the LRSM at hadron colliders. 
In Fig.~\ref{fig:lrsmDisc} we show the discovery potential at 14 TeV LHC of $W_R$ and $N$ in (a) the minimal LRSM as in Fig.~\ref{fig:mLRSM} 
after $\mathcal{L}=30\invfb$~\cite{Das:2012ii} and (b) the agnostic mixing scenario as in Fig.~\ref{fig:aLRSM}~\cite{Han:2012vk}.
Final-states involving $\tau$ leptons are also possible, but inherently suffer from the difficult signal event reconstruction and
larger backgrounds due to partonic-level MET induced by $\tau$ decays~\cite{AguilarSaavedra:2012fu}.

\begin{figure*}[!t]
\begin{center}
\subfigure[]{\includegraphics[width=.48\textwidth]{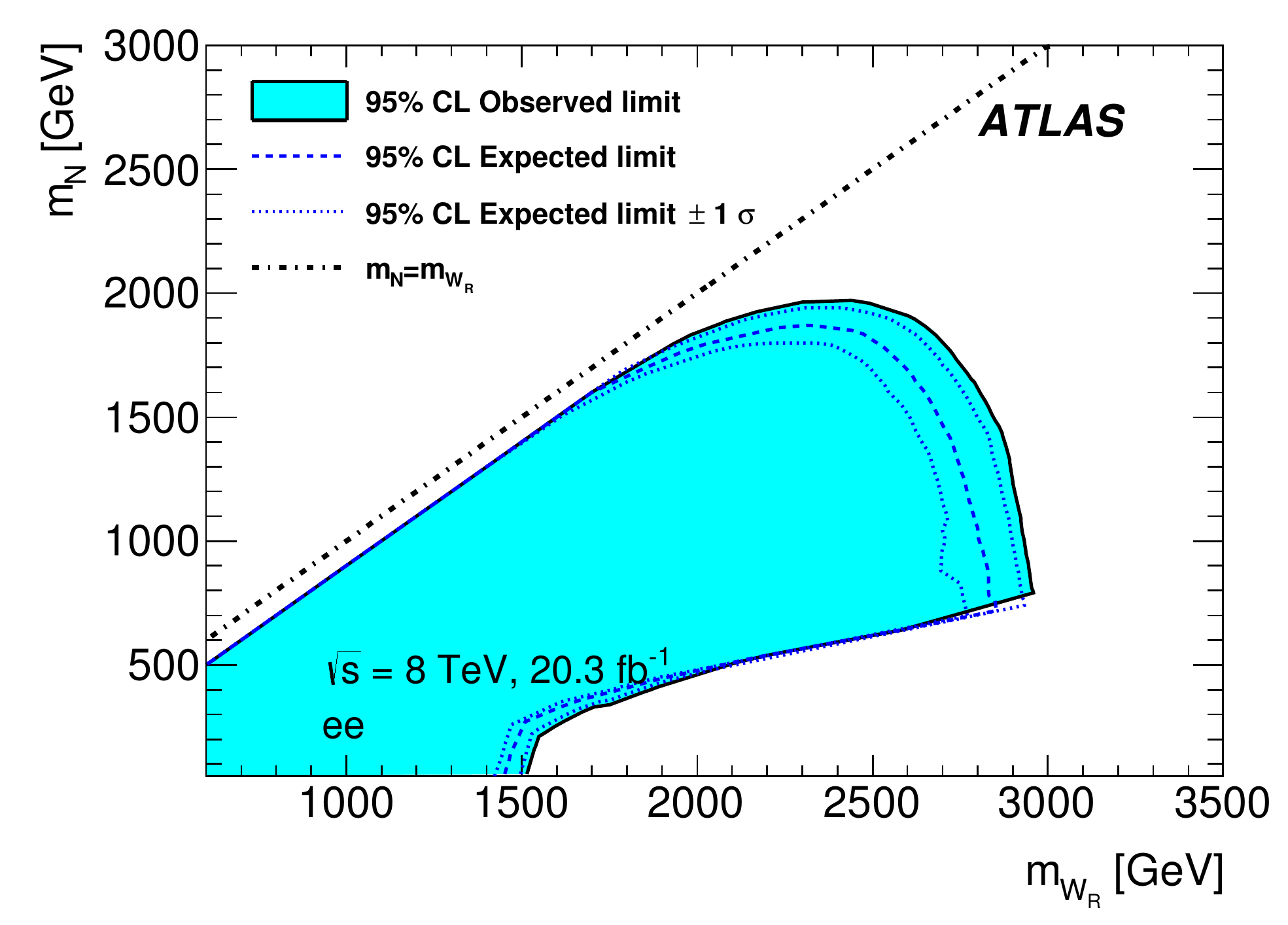}	}
\subfigure[]{\includegraphics[width=.48\textwidth]{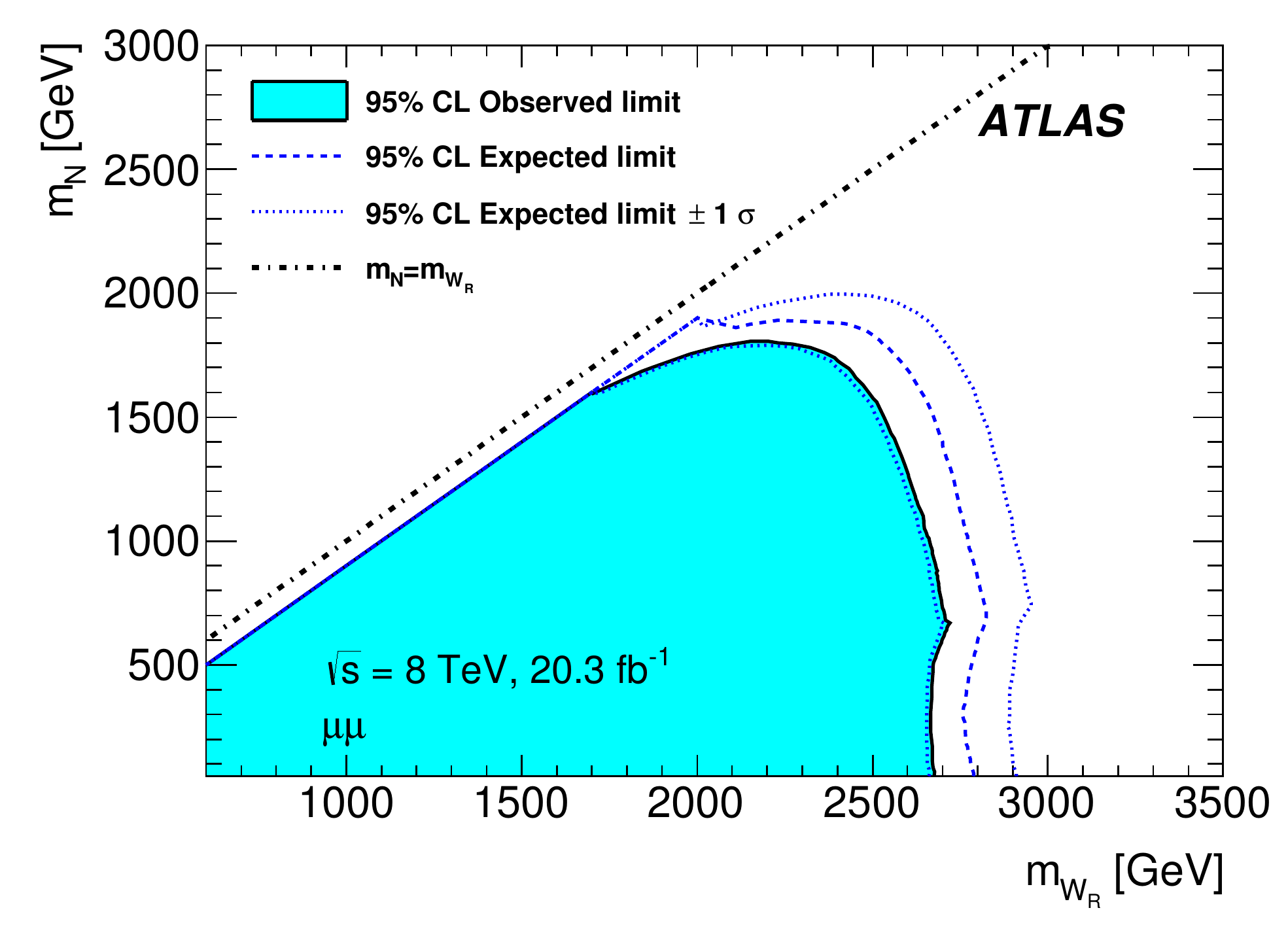}	}
\\
\subfigure[]{\includegraphics[width=.48\textwidth]{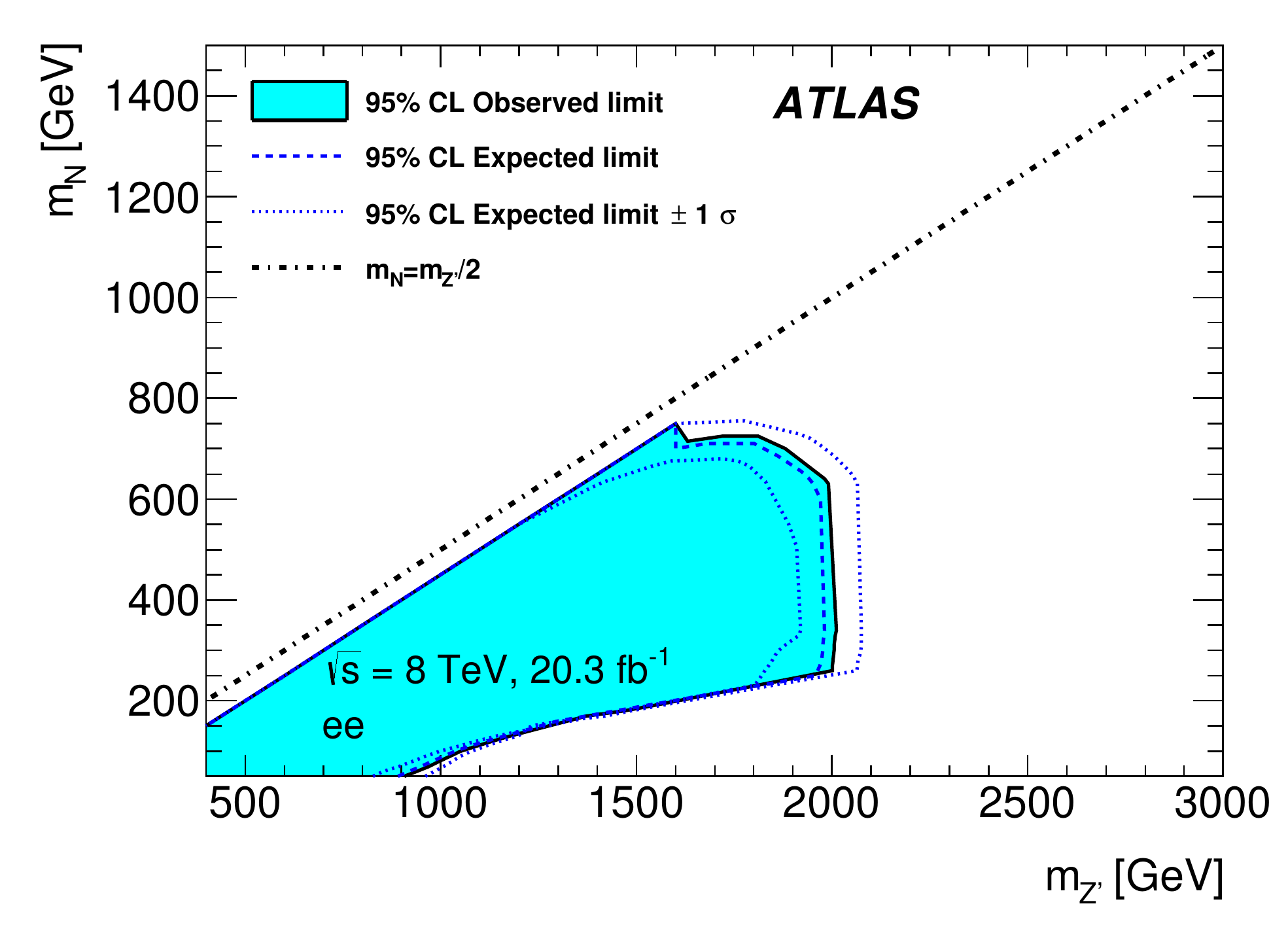}	}
\subfigure[]{\includegraphics[width=.48\textwidth]{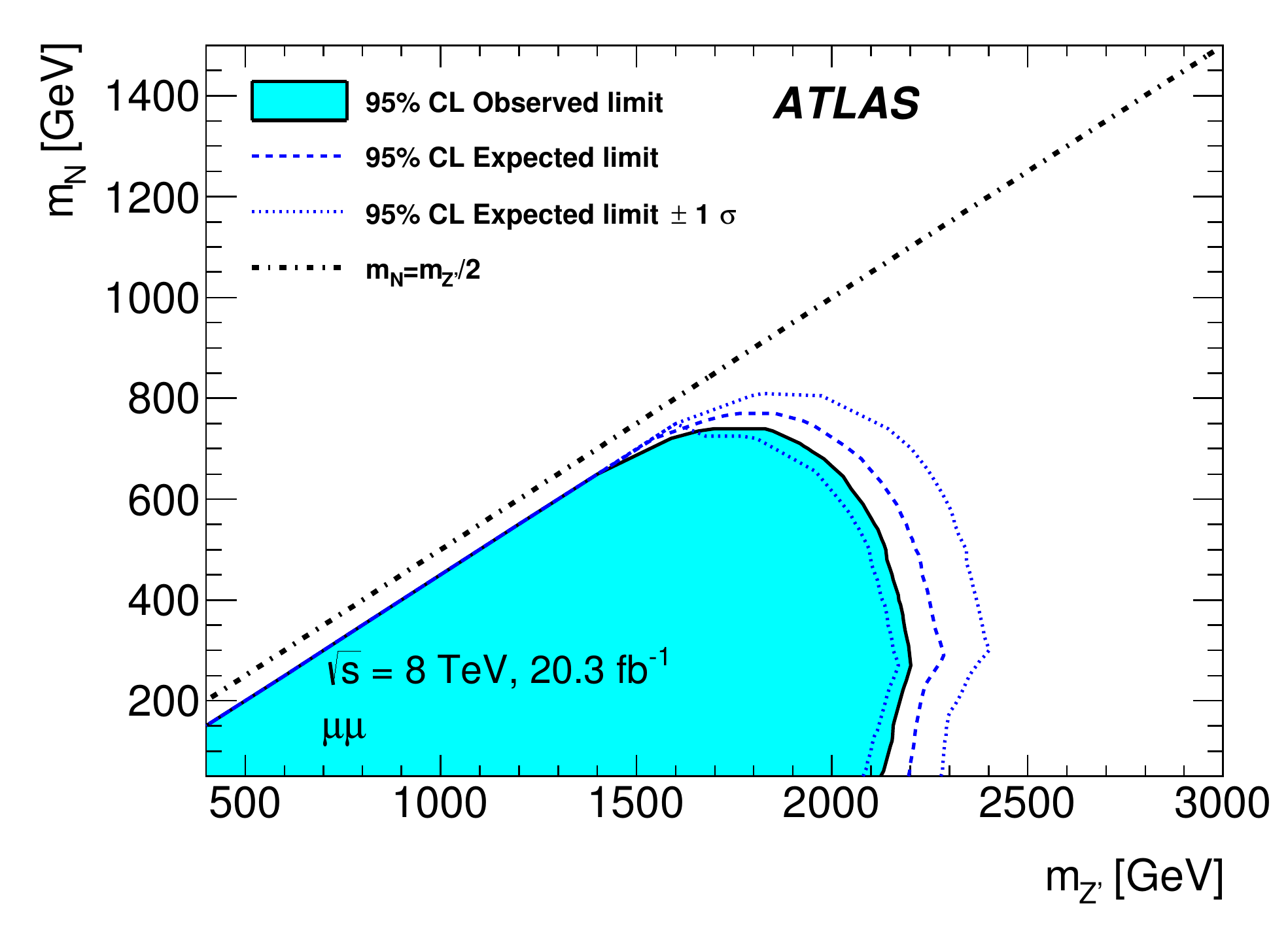}	}
\end{center}
\caption{
95\% CL exclusion of the $(M_V,M_N)$ parameter space
by the ATLAS experiment at $\sqrt{s}=8$ for $V=W_R$ (top) and $V=Z_R$ (bottom) production
in the (L) $e^\pm e^\pm+nj$ and (R) $\mu^\pm\mu^\pm+nj$ final state~\cite{Aad:2015xaa}.
}
\label{fig:atlasLRSMExcl}
\end{figure*}

Unfortunately, direct searches at the $\sqrt{s}=7/8$ TeV LHC via the DY channels have yielded no evidence for lepton number violating processes mediated
by $W_R$ and $Z_R$ gauge bosons from the LRSM~\cite{ATLAS:2012ak,Aad:2015xaa,Khachatryan:2014dka,CMS:2012zv}.
As shown in Fig.~\ref{fig:atlasLRSMExcl}, searches for $W_R/Z_R$ in the $e^\pm e^\pm+nj$ and $\mu^\pm\mu^\pm + nj$ final state
have excluded, approximately, $M_{W_R/Z_R}\lesssim1.5-2.5$ TeV and $M_N\lesssim2$ TeV.
However, sensitivity to the $e^\pm e^\pm+nj$ greatly diminishes for $M_N\ll M_{W_R/Z_R}$.

\begin{figure}[!t]
\begin{center}
\subfigure[]{\includegraphics[scale=1,width=.48\textwidth]{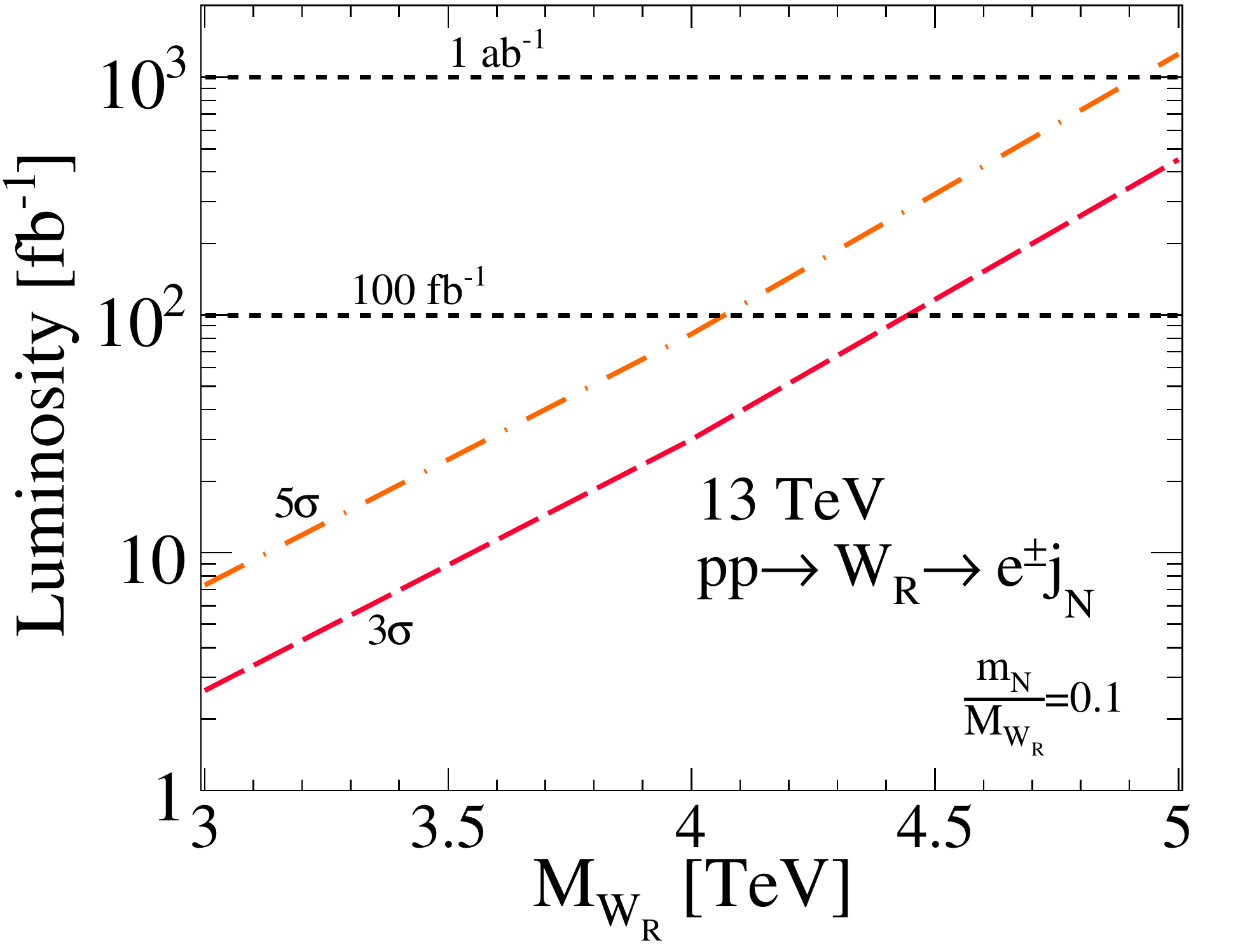}	\label{fig:disc13}		}
\subfigure[]{\includegraphics[scale=1,width=.48\textwidth]{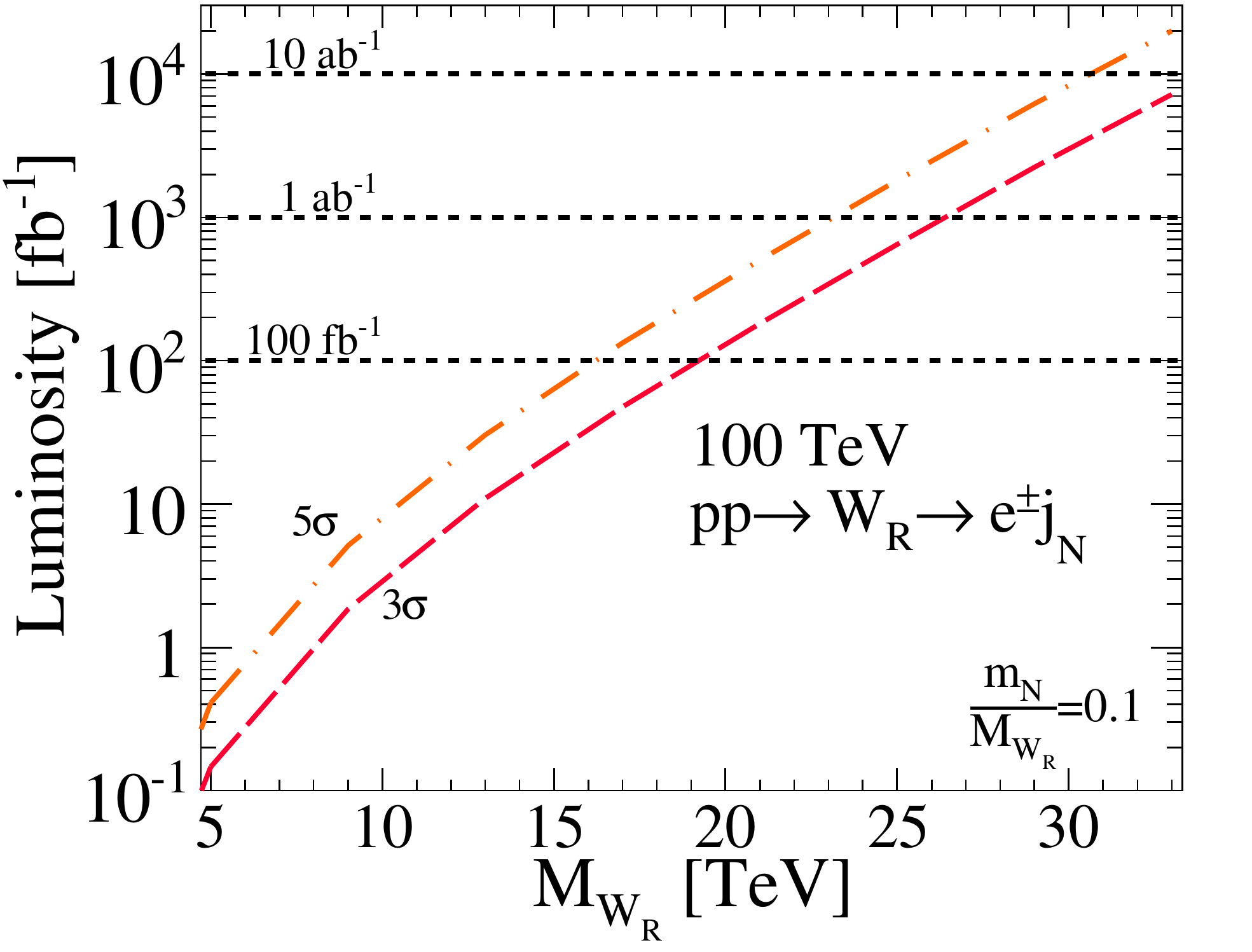}	\label{fig:disc100}		}
\\
\subfigure[]{\includegraphics[scale=1,width=.48\textwidth]{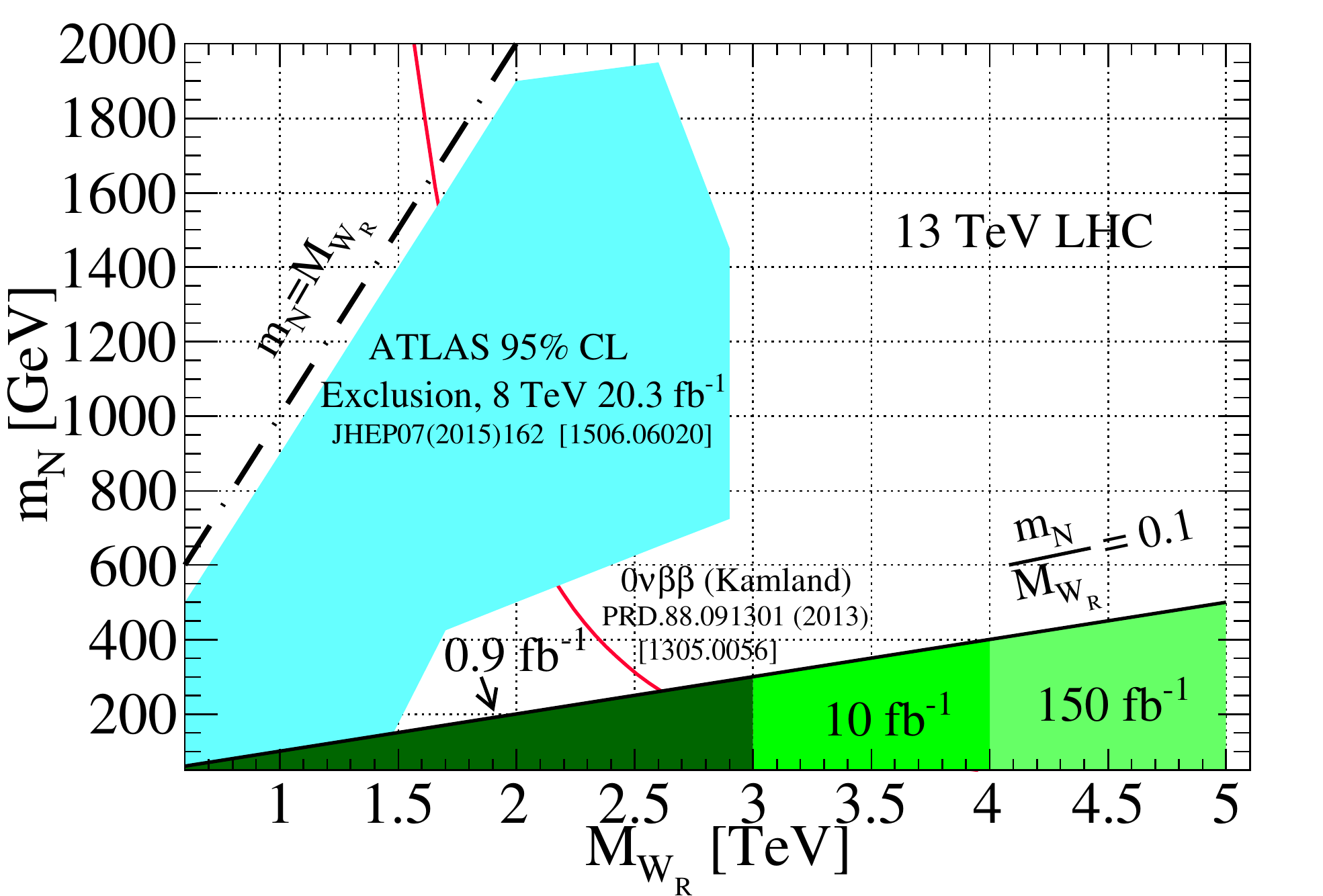}	\label{fig:exclusion13}		}
\subfigure[]{\includegraphics[scale=1,width=.48\textwidth]{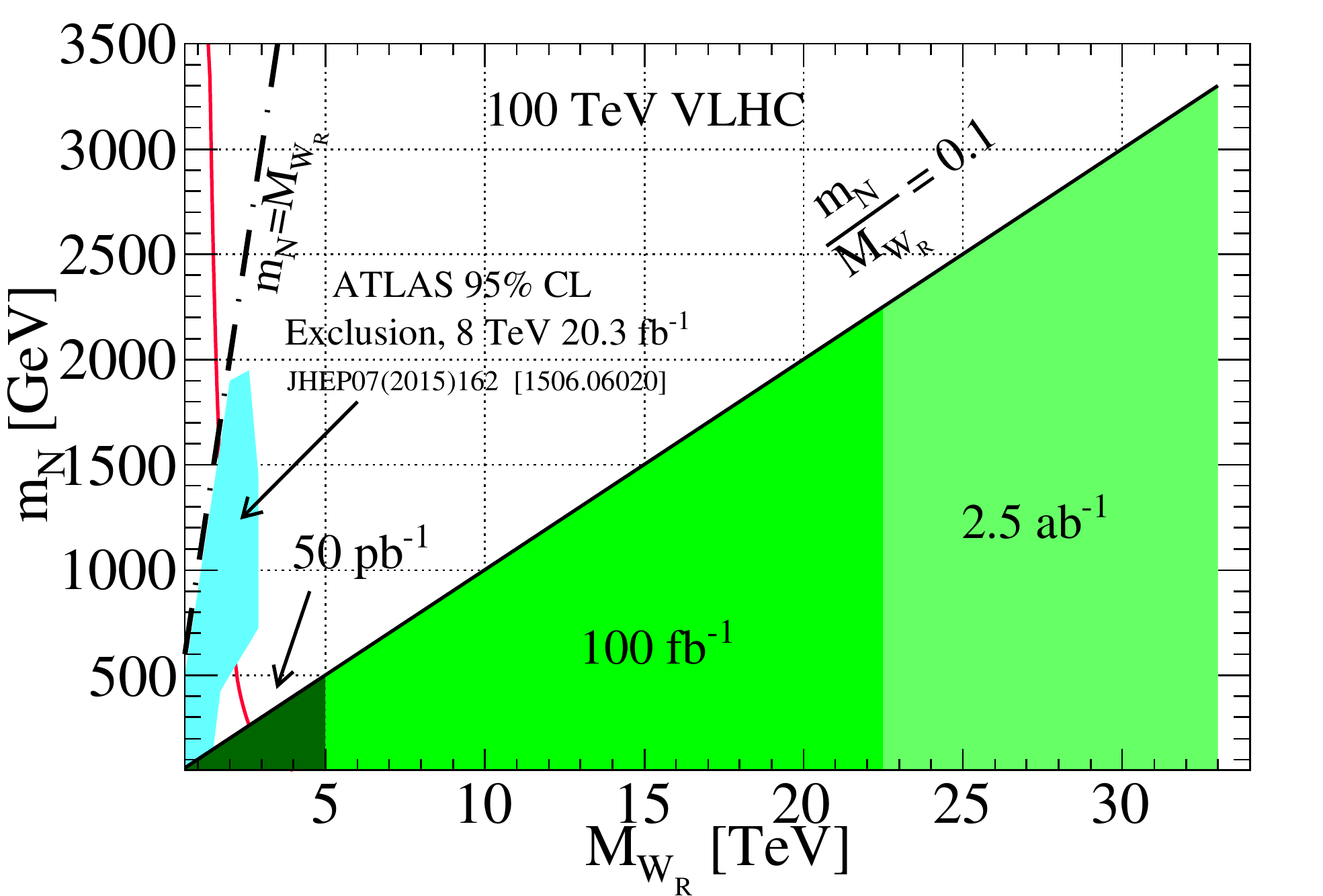}	\label{fig:exclusion100}	}
\end{center}
\caption{Discovery (a,b) and 95\% CL exclusion (c,d) potential of neutrino jet searches, \ie, $pp\rightarrow W_R\rightarrow e^\pm j_N$,
at (a,c) $\sqrt{s} =$ 13 and (b,d) 100 TeV.
Also shown in (c,d), ATLAS experiment's 8 TeV 95\% CL~\cite{Aad:2015xaa}
and KamLAND-Zen 90\% CL~\cite{Dev:2013vxa,KamLAND-Zen:2016pfg} exclusion limits. Figure from Ref.~\cite{Mitra:2016kov}.
}
\label{fig:lrsmNjetDiscov}
\end{figure}

Interestingly, for $M_N\ll M_{W_R}, M_{Z_R}$, decays of $N$ become highly boosted and its decay products, \ie, $\ell^\pm_2 q \overline{q'}$,
become highly collimated.
In such cases, the isolation criterion for electrons (and some muons) in detector experiments fail, particularly when
$\sqrt{r_N} = M_N / M_{W_R} < 0.1$~\cite{Ferrari:2000sp,Han:2012vk,Aad:2015xaa,Mitra:2016kov}.
Instead of requiring the identification of two well-isolated charged leptons for the processes given in Eq.~(\ref{eq:lrsmppssllnj}),
one can instead consider the $N$-decay system as a single, high-$p_T$ \textit{neutrino jet}~\cite{Mitra:2016kov,Mattelaer:2016ynf}.
The hadronic-level collider signature is then
\begin{eqnarray}
 p p \to W_R \to \ell^\pm ~N \to \ell^\pm ~j_N,
\end{eqnarray}
where the neutrino jet $j_N$ is comprised of three ``partons'', $(\ell_2,q,\overline{q'})$, with an invariant mass of $m_j \sim M_N$.
(Neutrino jets are distinct from so-called ``lepton jets''~\cite{Izaguirre:2015pga}, 
which are built from collimated charged leptons and largely absent of hadrons.)
This alternative topology for $M_N\ll M_{W_R}$ recovers the lost sensitivity of the same-sign dilepton final state,
as seen in Fig.~\ref{fig:lrsmNjetDiscov}.
Inevitably, for $N$ masses below the EW scale, rare $L$-violating decay modes also of SM particles open.
In particular, for $M_N$ below the top quark mass $m_t$, one has the rare decay
mode, $t \to b W_R^{+*} \to b \ell_1^+ N \to b \ell_1^+ \ell_2^\pm q \overline{q'}$~\cite{Si:2008jd}.
Such processes, however, can be especially difficult to distinguish from rare SM processes, \eg, $t\to W b \ell^+\ell^-$~\cite{Quintero:2014lqa},
particularly due to the large jet combinatorics.

For too small $M_N/M_{W_R}$ ratio, the lifetime for $N$, which scales as $\tau_N \sim M_{W_R}^4/M_N^5$, can become quite long.
In such instances, the decays of $N$ are no longer prompt and searches for $pp\to W_R\to N\ell$ map onto searches for
Sequential Standard Model $W'$ bosons~\cite{Maiezza:2015lza,Fuks:2017vtl}.
Likewise, searches for $L$-violating top quark decays become searches for RH currents in $t\to b\ell\not\!p_T$ decays.
For intermediate lifetimes, displaced vertex searches become relevant~\cite{Helo:2013esa,Anamiati:2016uxp,Das:2017hmg,Cottin:2018kmq,Nemevsek:2018bbt}.

\begin{figure*}[!t]
\begin{center}
\subfigure[]{\includegraphics[width=.48\textwidth]{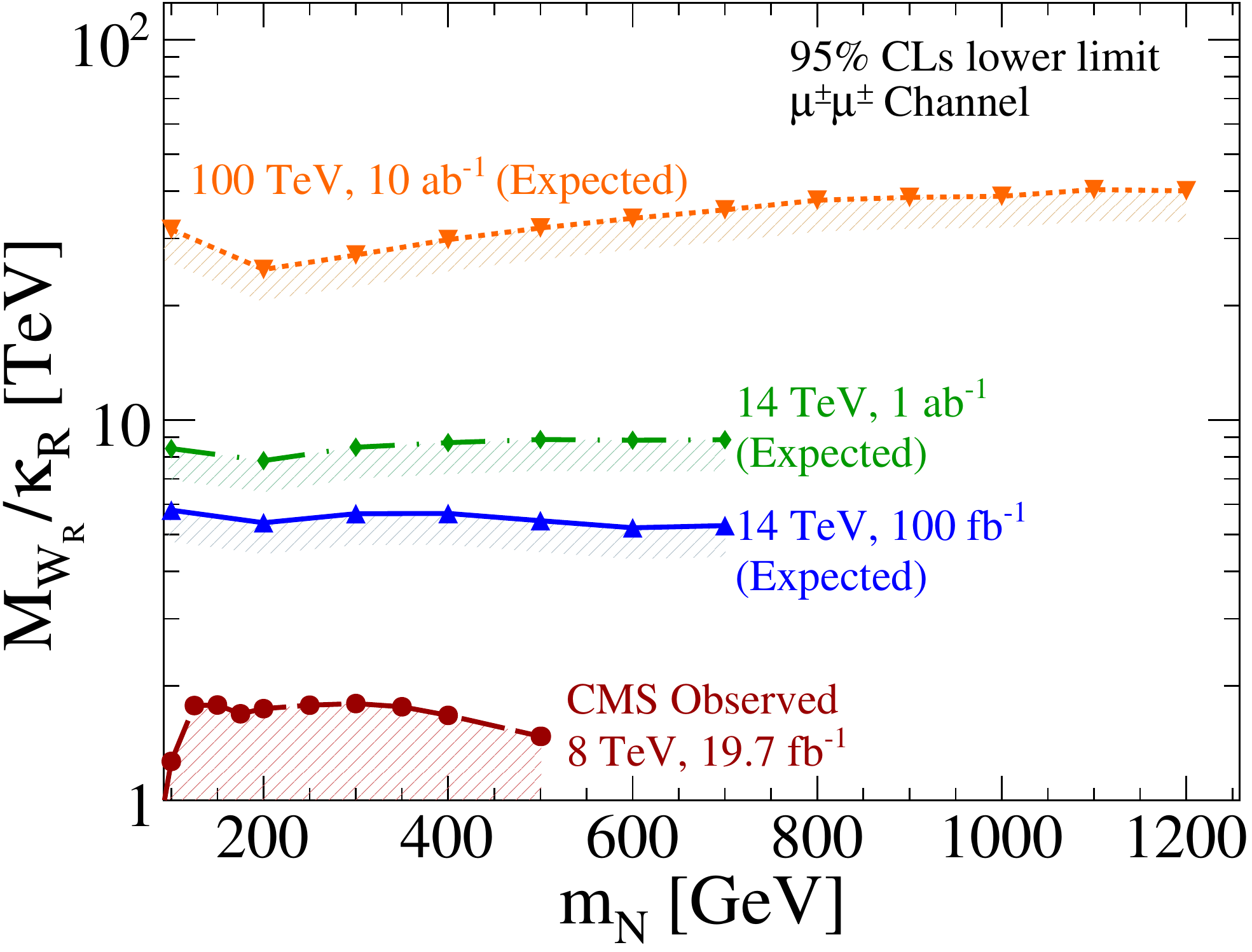}	\label{fig:lnvWithoutVR_mwrKapR_dimuon} }
\subfigure[]{\includegraphics[width=.48\textwidth]{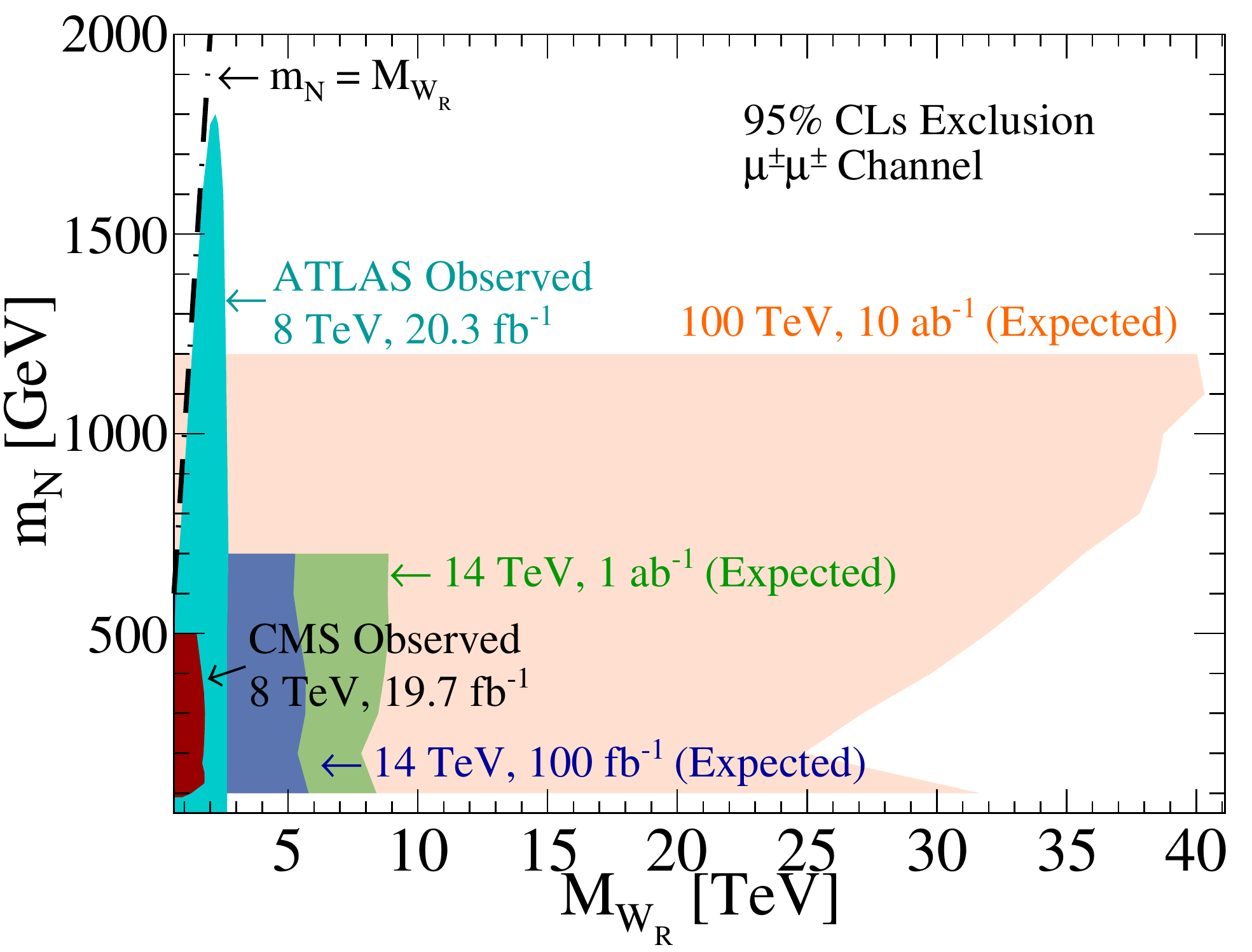}		\label{fig:lnvWithoutVR_MWRmNExcl}	}
\end{center}
\caption{
(a) As a function of $M_N$ and for right-left coupling ratio $\kappa_R = g_R/g_L$,
the observed 8 TeV LHC 95\% CLs lower limit on $(M_{W_R}/\kappa_R)$ (dash-dot),
expected 14 TeV sensitivity with $\mathcal{L}=100$ fb$^{-1}$ (solid-triangle) and $1$ ab$^{-1}$ (dash-dot-diamond),
and expected 100 TeV VLHC sensitivity with $10$ ab$^{-1}$ (dot-star).
(b) Observed and expected 95\% CL$_s$ sensitivities to the $(M_{W_R},M_N)$ parameter space for various collider configurations via
direct and indirect searches in the $\mu^\pm\mu^\pm$ final state~\cite{Ruiz:2017nip}.}
\label{fig:lrsmRecast}
\end{figure*}

Another recent avenue of exploration is the reassessment for resonant production of $W_R$ and $Z_R$ in Eq.~(\ref{eq:lrsmppssllnj}).
In the limit where $M_{W_R}\gtrsim \sqrt{s}$ but $M_N\ll \sqrt{s}$, resonant production of $N$, and hence a lepton number violating final state,
is still possible despite $W_R$ being kinematically inaccessible~\cite{Ruiz:2017nip}.
In such cases, $N$ is produced near mass threshold with $p_T^N \sim M_N$ instead of the usual $p_T^N \sim M_{W_R}/2$.
The same-sign leptons discovery channel is then kinematically and topologically identical to Type I Seesaw searches,
and hence is actively searched for at the LHC, despite this kinematic regime not being well-studied in the literature.
Reinterpretation of observed and expected sensitivities at the 14 and 100 TeV LHC are shown in Fig.~\ref{fig:lrsmRecast}.
One sees that with the anticipated cache of LHC data, $M_{W_R}\lesssim 9$ TeV can be excluded for $M_N\lesssim 1$ TeV.

In addition to the aforementioned DY and VBF channels,
there has been recent attention~\cite{Chen:2014qda,Frank:2010cj,Dev:2016dja,Mattelaer:2016ynf} given to the production of LRSM scalar and vector bosons in association with heavy flavor quarks, \eg,
\begin{equation}
 g \overset{(-)}{b} \to \overset{(-)}{t} W_R^\pm ~\text{or}~ \overset{(-)}{t}H^\pm_R
 \quad\text{and}\quad
 g g \to t\overline{t}Z_R ~\text{or}~~t\overline{t}H_R^0.
\end{equation}
As in the SM, such processes are critical in measuring the couplings of gauge bosons to quarks as well as determining heavy flavor PDFs.
However, also as in the SM, care is needed in calculating the rates of these processes when $M_R \gg m_b,~m_t$.
Here, $M_R$ is generically the mass of the RH scalar or vector boson.
As discussed just after Eq.~(\ref{eq:heavyNnj}), it has been noted recently in Ref.~\cite{Mattelaer:2016ynf} that such associated processes possess logarithmic dependence on the outgoing top quarks' kinematics, \ie, that the inclusive cross section scales as $\sigma\sim \alpha_s^k\log^{2k-1}\left(M_R^2/(m_t^2+p_T^{t~2})\right)$. Subsequently, for $M_R \gtrsim 1-2$ TeV, these logarithms grow numerically large since $\log^2(M_R^2/m_t^2)\gtrsim 1/\alpha_s$ and can spoil the perturbativity convergence of fixed order predictions.
For example, the (N)NLO $K$-factor of $K^{\rm (N)NLO}\gtrsim 1.6-2.0$ claimed in Ref.~\cite{Dev:2016dja} indicate a loss of perturbative control, not an enhancement, and leads to a significant overestimation of their cross sections. As in the case of EW boson production in association with heavy flavors~\cite{Dicus:1988cx,Maltoni:2012pa}, the correct treatment requires either a matching/subtraction scheme with top quark PDFs to remove double counting of phase space configurations~\cite{Dawson:2014pea,Han:2014nja} or kinematic requirements on the associated top quarks/heavy quark jets,
\eg, Eq.~(\ref{eq:cssConsistency})~\cite{Degrande:2016aje}.

In all of these various estimates for discovery potential,
it is important to also keep in mind what can be learned from observing $L$ violation and LR symmetry at the LHC or a future collider,
including $ep$ machines~\cite{Kaya:2015tia,Mattelaer:2016ynf,Lindner:2016lxq,Buchmuller:1992wm,Mondal:2016czu,Mondal:2015zba,Sarmiento-Alvarado:2014eha,Dev:2015vra}.
Primary goals post-discovery include:
determination of $W_R$ and $Z_R$ chiral coupling to fermions~\cite{Han:2012vk,Gopalakrishna:2010xm,Nemevsek:2012iq}, which can be quantified for quarks and leptons independently~\cite{Han:2012vk},
determination of the leptonic and quark mixing~\cite{Vasquez:2014mxa,Senjanovic:2016vxw,Nemevsek:2012iq,Tello:2010am,Gluza:2015goa,Gluza:2016qqv,Anamiati:2016uxp,Das:2017hmg},
as well as potential CP violation~\cite{Bajc:2009ft,Anamiati:2016uxp,Gluza:2015goa,Gluza:2016qqv,Das:2017hmg}.
We emphasize that the discovery of TeV-scale LRSM could have profound implications on high-scale baryo- and leptogenesis~\cite{GellMann:1980vs,Frere:2008ct,Dhuria:2015cfa,Deppisch:2013jxa,Harz:2015fwa}
as well as searches for $0\nu\beta\beta$ ~\cite{Tello:2010am,Barry:2013xxa,Nemevsek:2012iq,Peng:2015haa,Lindner:2016lpp}.
The latter instance is particularly noteworthy as the relationship between $m^{ee}_\nu$ and $m_{\nu_1}$ in the LRSM is different because of the new mediating fields~\cite{Tello:2010am}.

We finish this section by noting our many omissions, in particular: supersymmetric extensions of the LRSM, {\it e.g.},~\cite{Frank:1995dh,Demir:2009nq};
embeddings into larger internal symmetry structures, {\it e.g.},~\cite{Goh:2009fm,Appelquist:2003uu};
as well as generic extensions with additional vector-like or mirror quarks, {\it e.g.},~\cite{Goh:2009fm,deAlmeida:2010qb}.
While each of these extensions have their phenomenological uniquenesses, their collider signatures are broadly indistinguishable from the minimal LRSM scenario.
With regard to Type I-based Seesaws in extra dimensional frameworks, it is worthwhile to note that it has
recently~\cite{Agashe:2015izu,Agashe:2016ttz,Agashe:2017ann} been observed that in warped five-dimensional models,
a more careful organization of Kaluza-Klein states and basis decomposition results in an inverse Seesaw mechanism as opposed to a canonical
Type I-like Seesaw mechanism, as conventionally believed. Again, this leads to greatly suppressed $L$ violation at collider experiments.

\subsubsection{Heavy Neutrino Effective Field Theory at Colliders}\label{sec:neftTests}
\begin{figure}[!t]
\begin{center}
\subfigure[]{\includegraphics[scale=1,width=.48\textwidth,height=6cm]{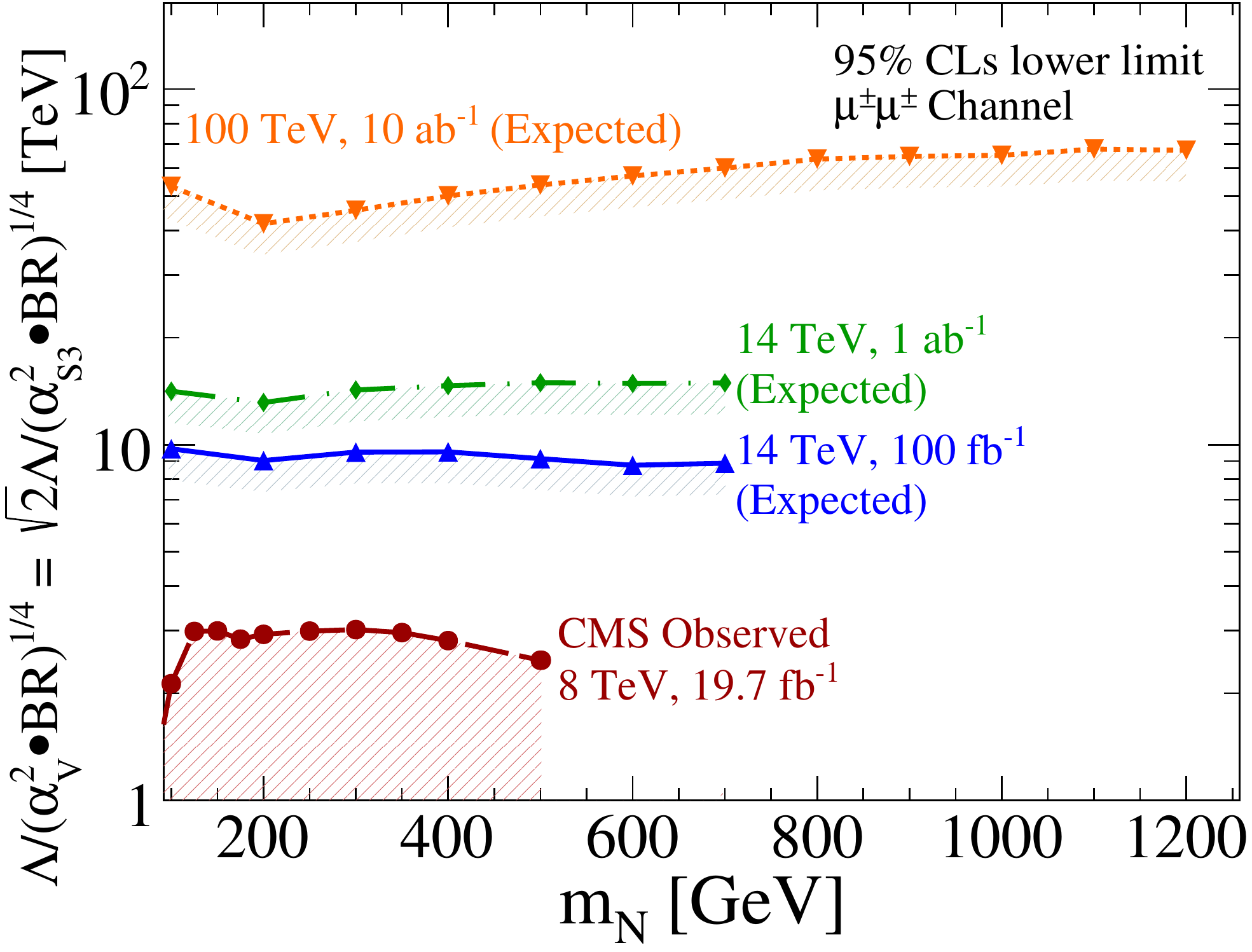}		}
\subfigure[]{\includegraphics[scale=1,width=.48\textwidth,height=6.5cm]{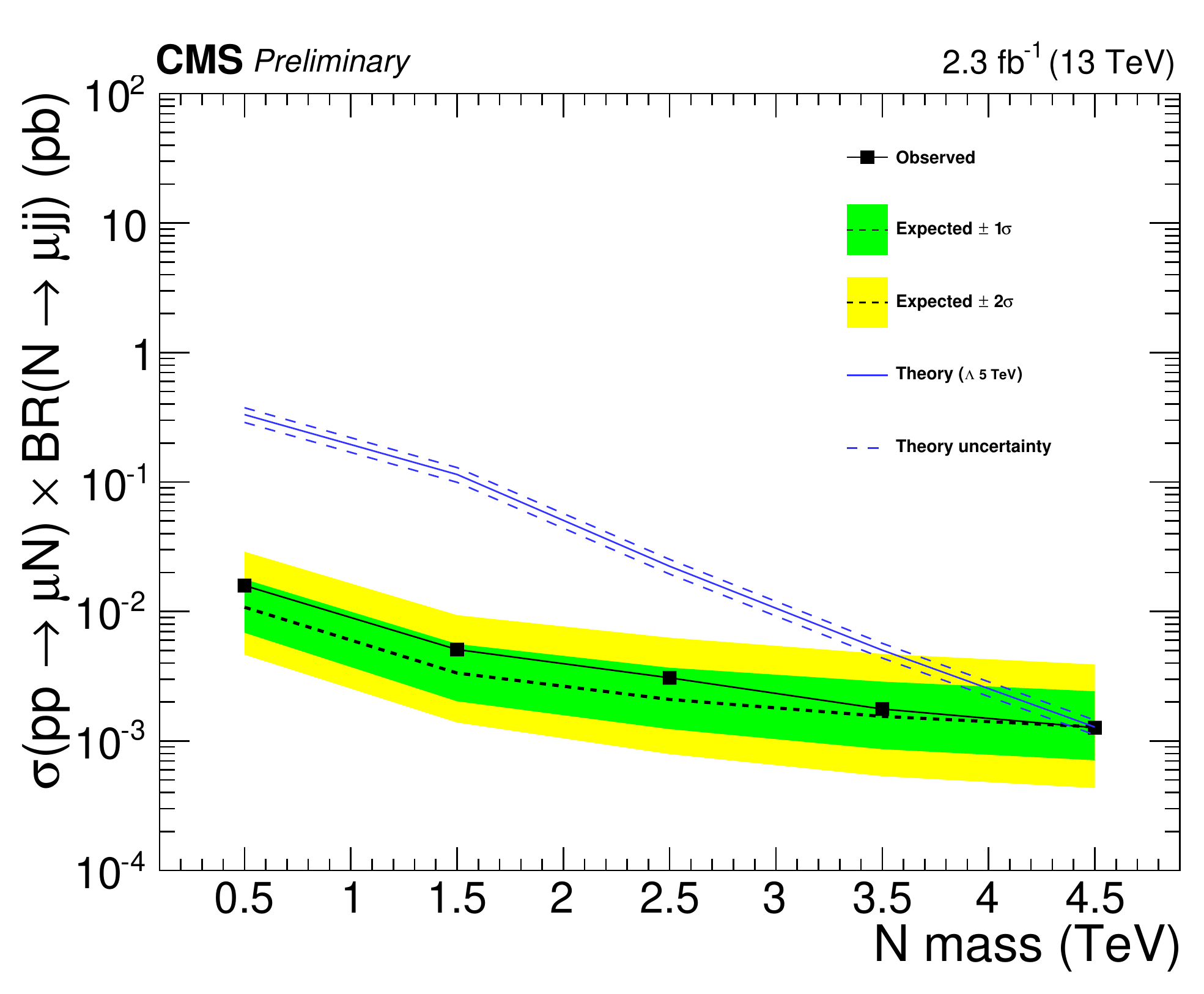}	}
\end{center}
\caption{
Observed limits and expected sensitivities at current and future hadron collider experiments
on NEFT mass scale $\Lambda$ for low-mass~\cite{Ruiz:2017nip} and high-mass~\cite{CMS:2016blm} Majorana neutrinos $N$
via the $L$-violating $pp\to \ell^\pm_1 \ell^\pm_2+X$.
}
\label{fig:neftLimits}
\end{figure}

As discussed in Sec.~\ref{sec:neft}, the production and decay of Majorana neutrinos in colliders may occur through contact interactions
if mediating degrees of freedom are much heavier than the hard scattering process scale.
Such scenarios have recently become a popular
topic~\cite{Bhattacharya:2015vja,Liao:2016qyd,Ruiz:2017nip,Leonardi:2015qna,Duarte:2016miz,Duarte:2016caz,Duarte:2015iba,Duarte:2014zea,Caputo:2017pit},
in part because of the considerable sensitivity afforded by collider experiments.
This is particularly true for $L$-violating final-states in $pp$ collisions, which naturally have small experimental backgrounds.
As shown in Fig.~\ref{fig:neftLimits}, for various operators, searches for $L$-violating process $pp\to N\ell^\pm_1 \to \ell^\pm_1 \ell^\pm_2+X$
by the ATLAS and CMS experiments have set wide limits on the effective mass scale
of $\Lambda > 1-5$ TeV for $M_N = 100$ GeV$ -$4.5 TeV~\cite{Leonardi:2015qna,Ruiz:2017nip,CMS:2016blm}.
Projections for $\sqrt{s}=14~(100)$ TeV after $\mathcal{L}=1~(10)$ ab$^{-1}$ show that $\Lambda\lesssim9~(40)$ TeV can be achieved \cite{Ruiz:2017nip}.
These search strategies are also applicable for the more general situation where $L$ violation
is mediated entirely via SMEFT operators~\cite{Abada:2007ux,delAguila:2012nu} as introduced
in Sec.~\ref{sec:neft}.

\section{The Type II Seesaw and Lepton Number Violation at Colliders}\label{sec:type2}

In this section we review lepton number violating collider signatures associated with the Type II Seesaw
mechanism~\cite{Konetschny:1977bn,Gelmini:1980re, Cheng:1980qt,Lazarides:1980nt,Schechter:1980gr,Mohapatra:1980yp} and its extensions.
The Type II model is unique among the original tree-level realizations of the Weinberg operator in that lepton number is spontaneously broken;
in the original formulations of the Type I and III Seesaws, lepton number violation is explicit by means of a Majorana mass allowed by gauge invariance.
In Sec.~\ref{sec:type2Models}, we summarize the main highlights of the canonical Type II Seesaw
and other Type II-based scenarios.
We then review in Sec.~\ref{sec:type2Collider} collider searches for lepton number violation mediated by exotically charged scalars $(H^\pm,H^{\pm\pm})$,
which is the characteristic feature of Type II-based scenarios.

\begin{figure}[tb]
\begin{center}
\subfigure[]{\includegraphics[scale=1,width=0.48\textwidth]{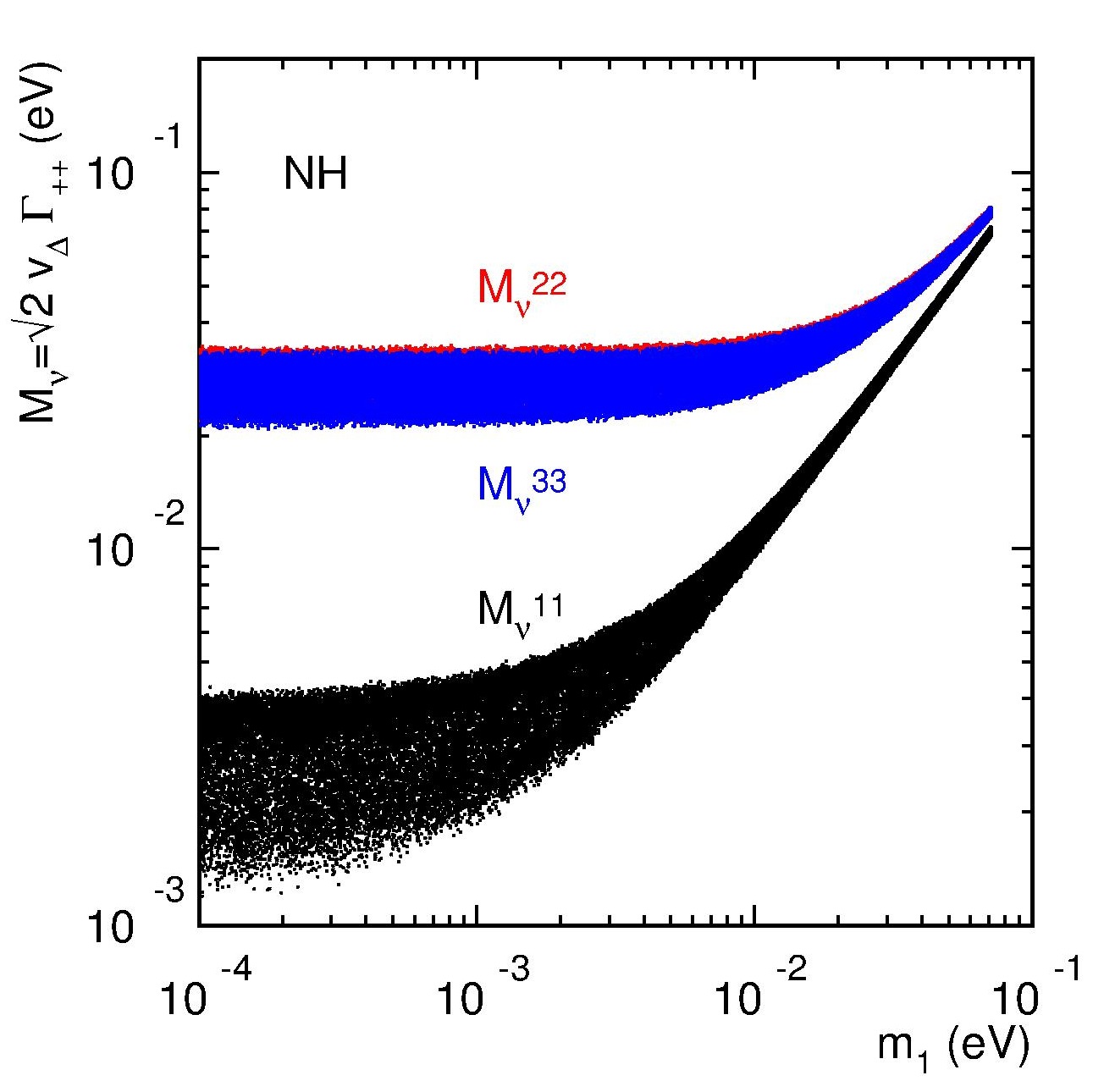} }
\subfigure[]{\includegraphics[scale=1,width=0.48\textwidth]{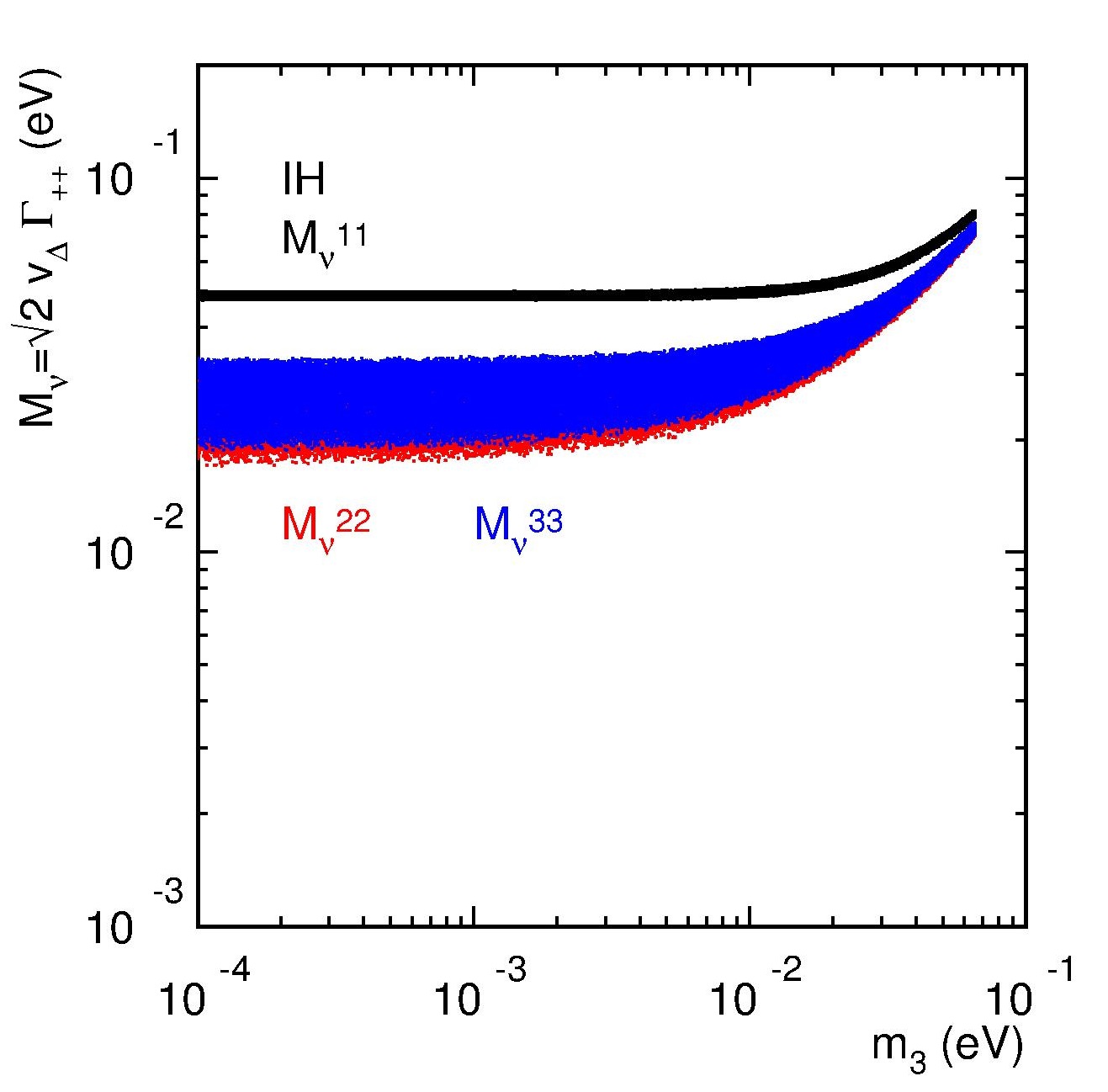}	}
\\
\subfigure[]{\includegraphics[scale=1,width=0.48\textwidth]{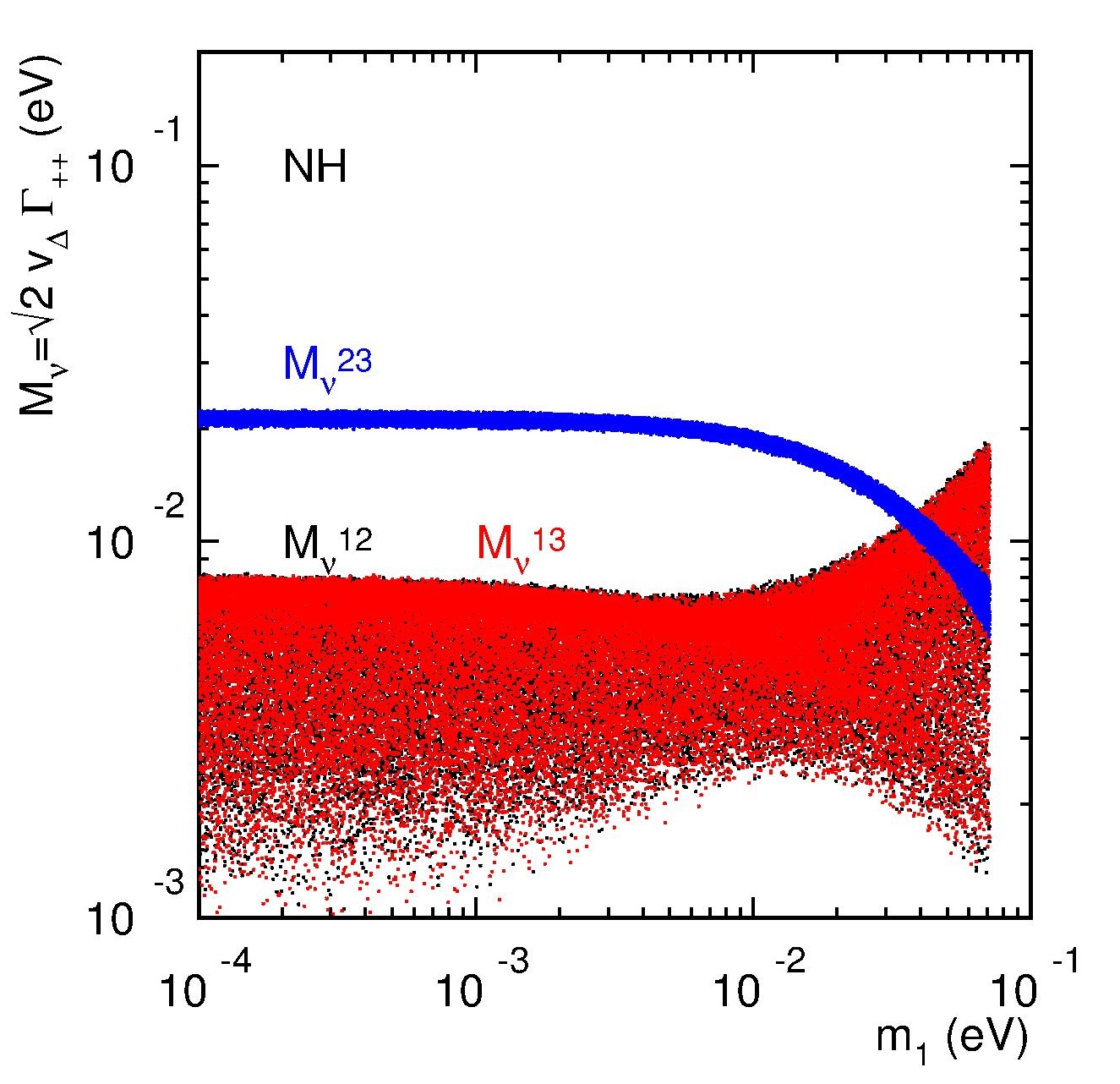}	}
\subfigure[]{\includegraphics[scale=1,width=0.48\textwidth]{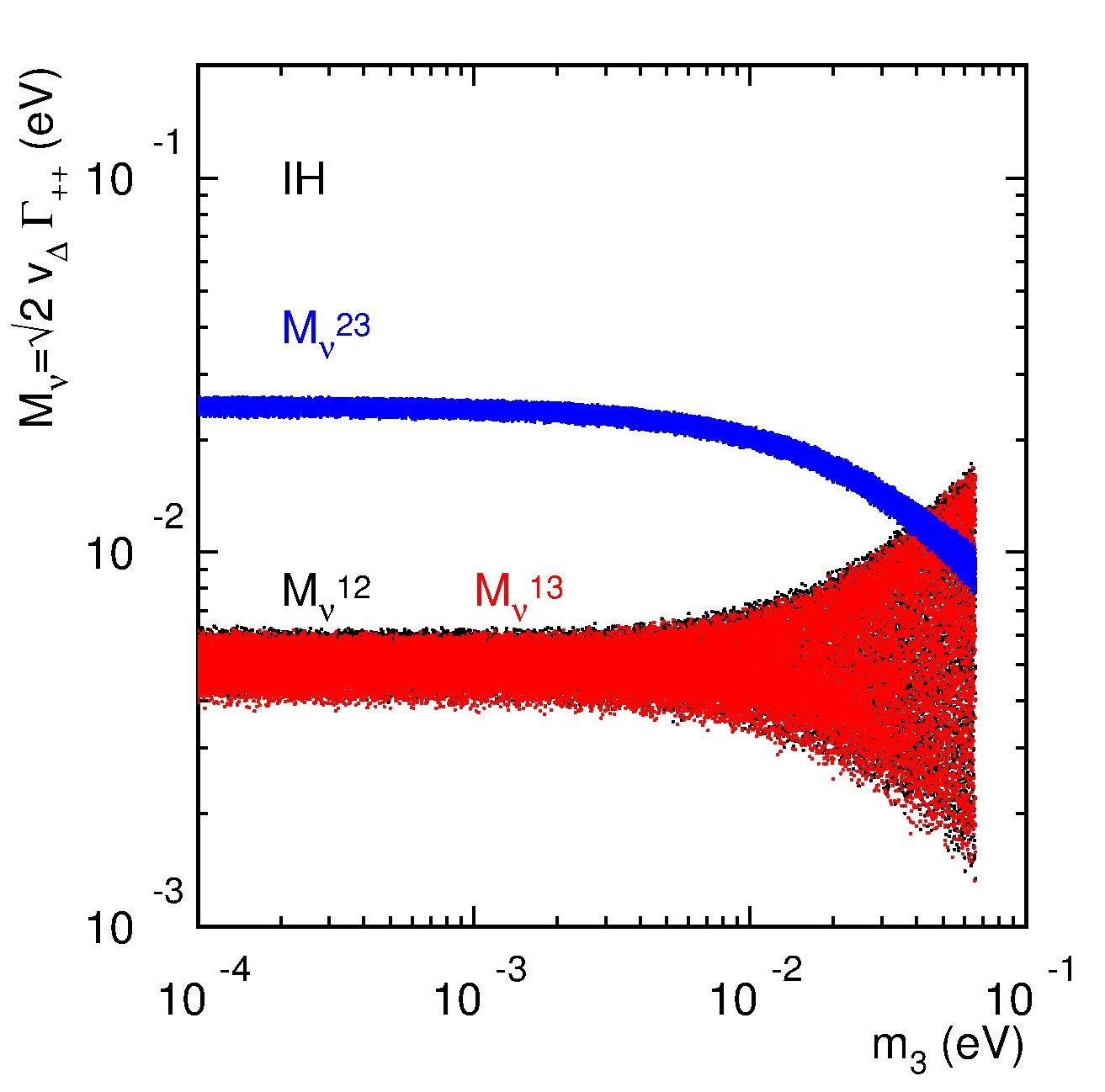}	}
\end{center}
\caption{Constraints on the diagonal (a,b) and off-diagonal (c,d)
elements of the neutrino mass matrix $M_\nu\equiv\sqrt{2}v_\Delta \Gamma_{++}$ versus the lowest
neutrino mass for NH (a,c) and IH (b,d) when $\Phi_1 =0$ and
$\Phi_2 = 0$.}
\label{mii}
\end{figure}

\subsection{Type II Seesaw Models}\label{sec:type2Models}

In the Type II mechanism~\cite{Konetschny:1977bn,Gelmini:1980re, Cheng:1980qt,Lazarides:1980nt,Schechter:1980gr,Mohapatra:1980yp},
tiny neutrino masses arise through the Yukawa interaction,
\begin{eqnarray}
\Delta\mathcal{L}_{II}^m  = - \overline{L^c} ~Y_\nu~ i\sigma_2 \ \Delta_L \ L+~\Hc,
\label{eq:typeIIYukawa}
\end{eqnarray}
between the SM LH lepton doublet $L$, its charge conjugate, and an SU$(2)_L$ scalar triplet (adjoint representation) $\Delta_L$ with mass $M_\Delta$
and Yukawa coupling $Y_\nu$.
More precisely, the new scalar transforms as $(1,3,1)$ under the full SM gauge symmetry and possesses lepton number $L=-2$,
thereby ensuring that Eq.~(\ref{eq:typeIIYukawa}) conserves lepton number before EWSB.
Due to its hypercharge and $L$ assignments,  $\Delta_L$ does not couple to quarks at tree-level.
It does, however, couple to the SM Higgs doublet, particularly through the doublet-triplet mixing operator
\begin{equation}
\Delta\mathcal{L}_{H\Delta_L} \ni \mu H^T \ i\sigma_2 \ \Delta_L^\dagger H+~\Hc
\end{equation}
The importance of this term is that after minimizing the full Type II scalar potential $V_{\rm Type~II}$,
$\Delta_L$ acquires a small vev $v_\Delta$ that in turn induces a LH Majorana mass for SM neutrinos, given by
\begin{eqnarray}
M_\nu=\sqrt{2}Y_\nu v_\Delta
\quad\text{with}\quad
v_\Delta= \langle\Delta_L\rangle = {\mu v_0^2\over \sqrt{2} M_\Delta^2}.
\end{eqnarray}
In the above, $v_0 = \sqrt{2}\langle H \rangle$ is the vev of the SM Higgs and $v_0^2+v_\Delta^2 = (\sqrt{2}G_F)^{-1}\approx (246 \ {\rm GeV})^2$.
As a result of \bl being spontaneously broken by $\Delta_L$,
tiny $0.1\ev$ neutrino masses follow from the combination of three scales: $\mu$, $v_0$, and $M_\Delta$.
In addition, after EWSB, there are seven physical Higgses, including the singly and doubly electrically charged $H^\pm$ and $H^{\pm\pm}$
with masses $M_{H^\pm,H^{\pm\pm}} \sim M_\Delta$.
As $v_\Delta$ contributes to EWSB at tree-level, and hence the EW $\rho/T$-parameter, $v_\Delta$ is constrained by precision EW observables,
with present limits placing $v_\Delta\lesssim\mathcal{O}(1$ GeV) \cite{Perez:2008ha,Dutta:2014dba,Han:2005nk,Melfo:2011nx,Arhrib:2011uy,Kanemura:2012rs,Chen:2008jg,Chen:2005jx,Das:2016bir}.
The impact of triplet scalars on the naturalness of the SM-like Higgs at 125 GeV has also been studied~\cite{Dev:2013ff,Arhrib:2011uy,Dev:2017ouk}.
The simultaneous sensitivity of $M_\nu$ to collider, neutrino mass measurement, and neutrino oscillation experiments is one of the clearest
examples of their complementarity and necessity to understanding neutrinos physics.

\begin{figure}[tb]
\begin{center}
\minigraph{6cm}{-0.15in}{(a)}{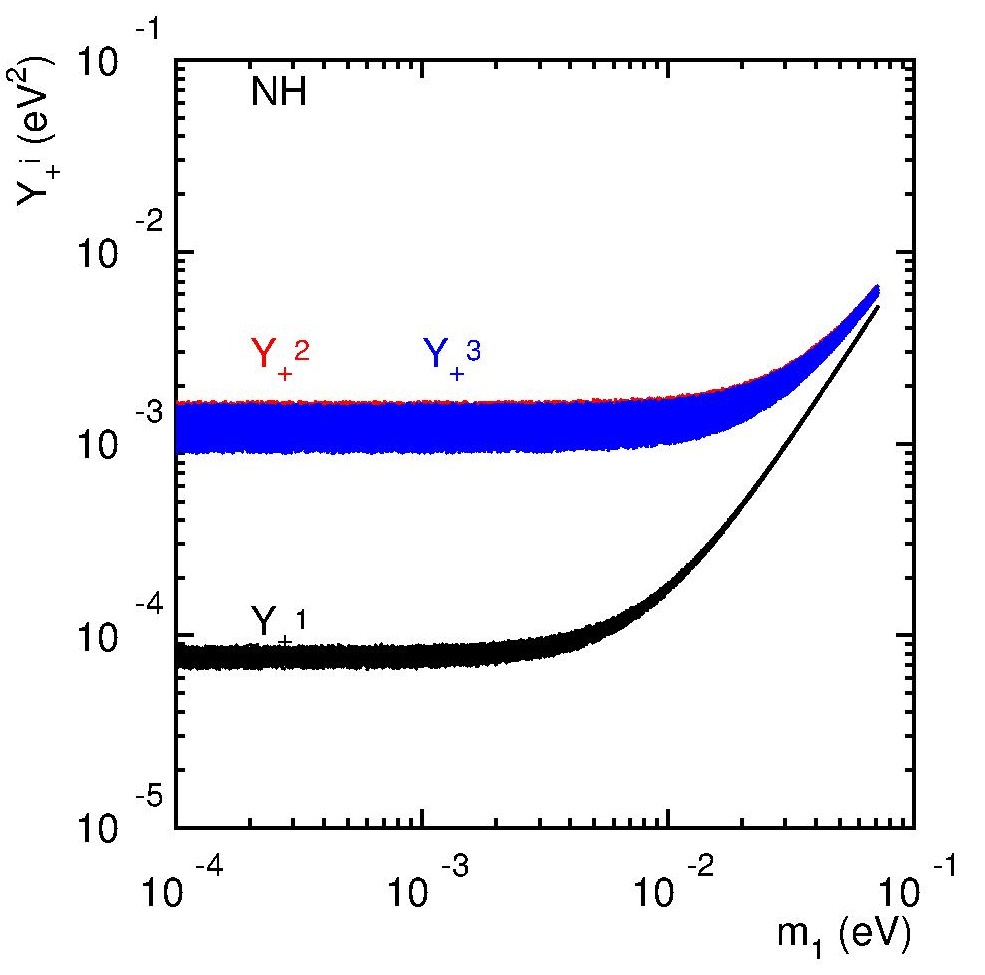}
\minigraph{6cm}{-0.15in}{(b)}{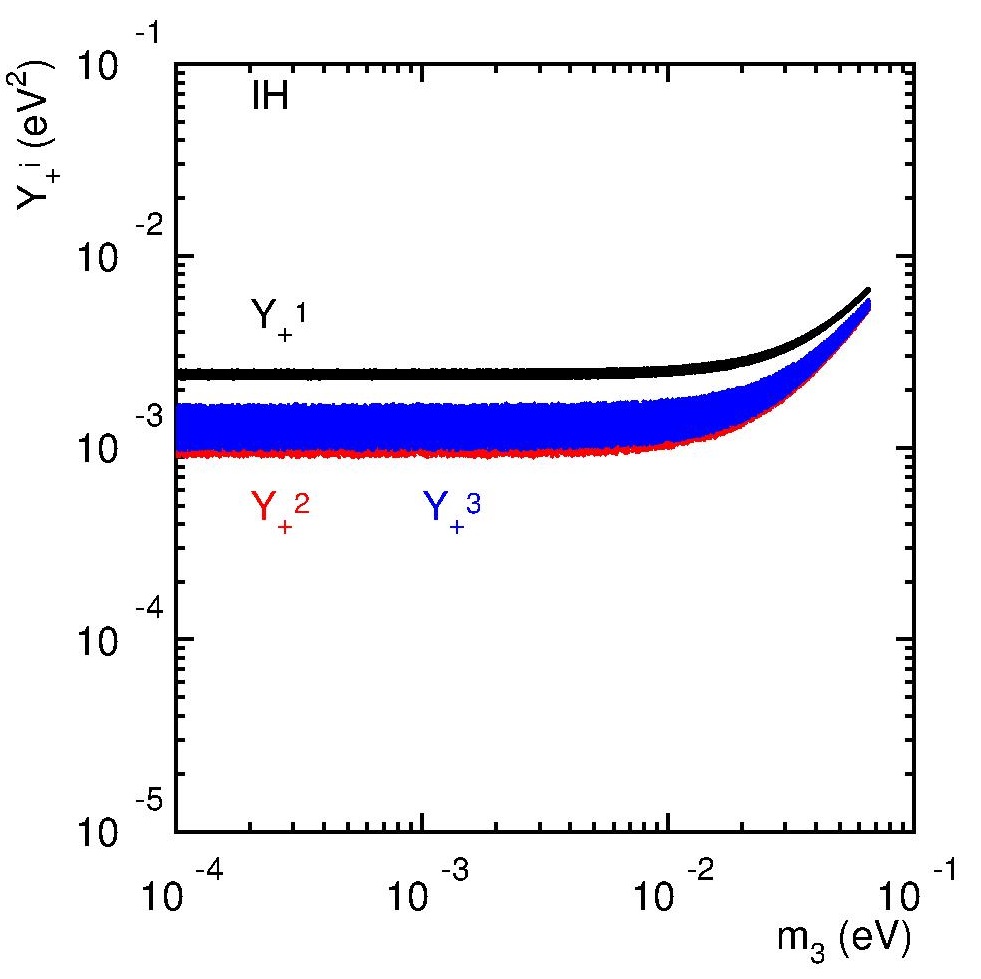}
\end{center}
\caption{Constraints on the squared coupling $Y_+^i\equiv \sum_j
|\Gamma_{+}^{ji}|^2 v^2_{\Delta}$, versus the lowest neutrino mass
for NH (a) and IH (b). } \label{Yi}
\end{figure}

For SM-like Yukawas $Y_\nu \sim 10^{-6} - 1$, one finds that $v_\Delta \sim 0.1\ev - 100\kev$ are needed in order to reproduce $0.1$ eV neutrino masses.
Subsequently, for $\mu \sim M_\Delta$, then $M_\Delta \sim \mu \sim 10^8 -10^{14}$ GeV, and for $\mu \sim v_0$, then  $M_\Delta\sim 10^5 - 10^8$ GeV.
In either case, these scales are too high for present-day experiments.
However, as nonzero $\mu$ is associated with both lepton number and custodial symmetry non-conservation,
one may expect it to be small~\cite{Senjanovic:1978ev} and natural, in the t'Hooft sense~\cite{tHooft:1979rat}. Imposing technical naturalness can have dramatic impact on LHC phenomenology: for example, if $\mu \sim 1$ MeV~(keV), then $M_\Delta\sim 10^2 - 10^5~(10^1-10^4)$ GeV, scales well within the LHC's energy budget. Moreover, this also indicates that proposed future hadron collider experiments~\cite{Arkani-Hamed:2015vfh,Golling:2016gvc}
will be sensitive to MeV-to-GeV values of the scalar-doublet mixing parameter $\mu$, independent of precision Higgs coupling measurements, which are presently at the 10\% level~\cite{Khachatryan:2016vau}. Assuming Higgs coupling deviations of $\mathcal{O}(\mu/M_h)$, this implies the weak 7/8 TeV LHC limit of $\mu\lesssim\mathcal{O}$(10 GeV). While not yet competitive with constraints from EW precision data, improvements on Higgs coupling measurements will be greatly improved over the LHC's lifetime.

After decomposition of leptons into their mass eigenstates,
the Yukawa interactions of the singly and doubly charged Higgses are
\begin{eqnarray}
\nu_L^T \ C \ \Gamma_+ \ H^+ \ \ell_L, 		\quad&:&\quad
\Gamma_+  =  \cos \theta_+ \ \frac{m_\nu^{diag}}{v_{\Delta}} \ U_{PMNS}^\dagger, \ \ \ \theta_+ \approx {\sqrt 2v_\Delta\over v_0}, \\
\ell_L^T \ C \ \Gamma_{++} \ H^{++} \ \ell_L 	\quad&:&\quad
\Gamma_{++} = {M_\nu\over \sqrt{2} v_\Delta} = U_{PMNS}^* \ \frac{m_{\nu}^{diag}}{\sqrt{2} \ v_{\Delta}} \ U_{PMNS}^{\dagger}.
\end{eqnarray}
The constrained neutrino mass matrix $M_\nu=\sqrt{2}v_\Delta \Gamma_{++}$ and squared Yukawa coupling $Y_+^i\equiv \sum_j |\Gamma_{+}^{ji}|^2
v^2_{\Delta}$ with vanishing Majorana phases are shown in Figs.~\ref{mii} and \ref{Yi} respectively.
The results reveal the following mass and Yukawa patterns:
\begin{align}
M_\nu^{22},M_\nu^{33}\gg M_\nu^{11} \quad\text{and}\quad Y_+^2,Y_+^3\gg Y_+^1 \qquad & {\rm for \ NH}; \\
M_\nu^{11}\gg M_\nu^{22},M_\nu^{33} \quad\text{and}\quad Y_+^1\gg Y_+^2,Y_+^3 \qquad & {\rm for \ IH}.
\end{align}

\begin{figure}[tb]
\begin{center}
\minigraph{7.cm}{-0.15in}{(a)}{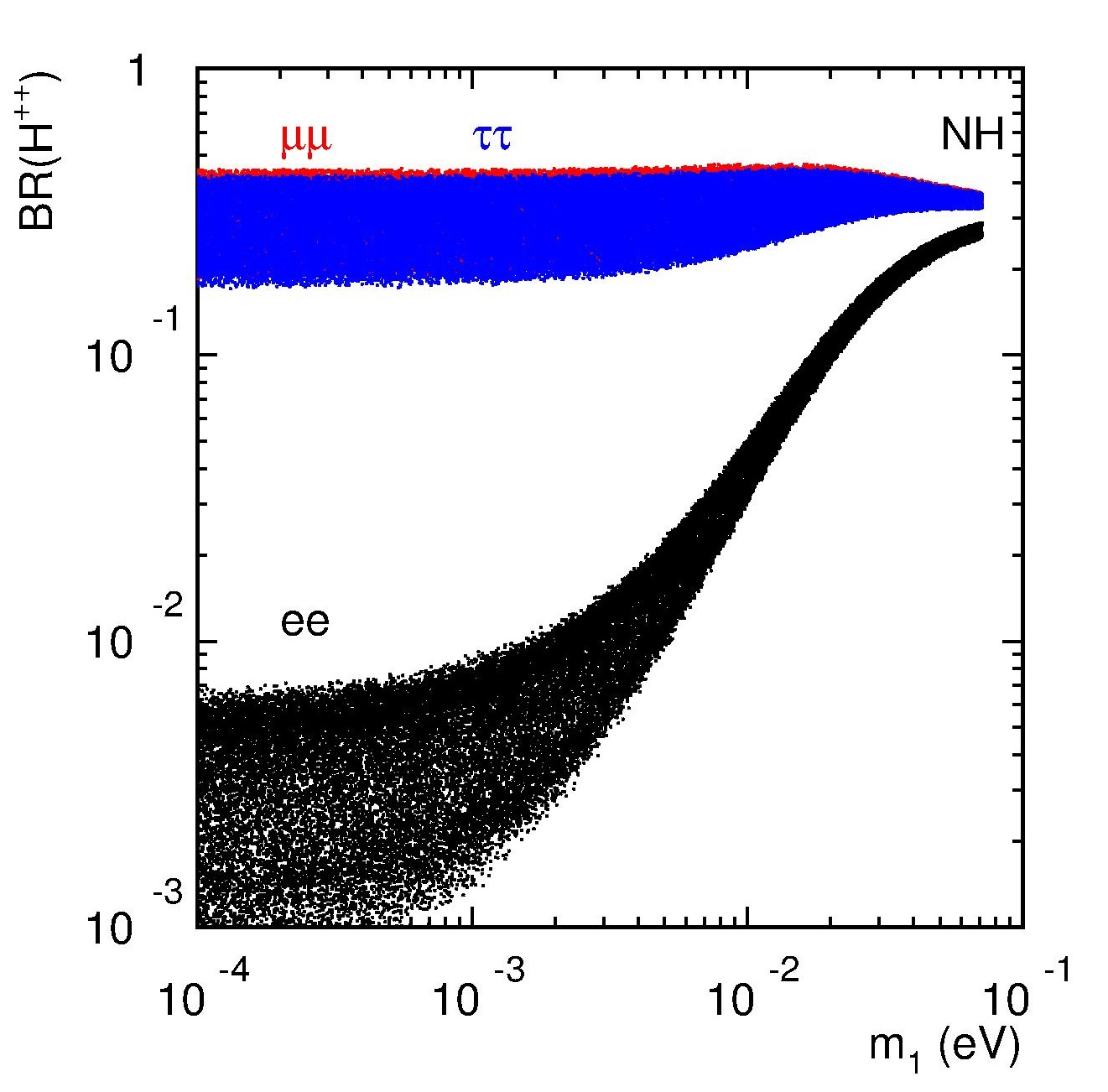}
\minigraph{7.cm}{-0.15in}{(b)}{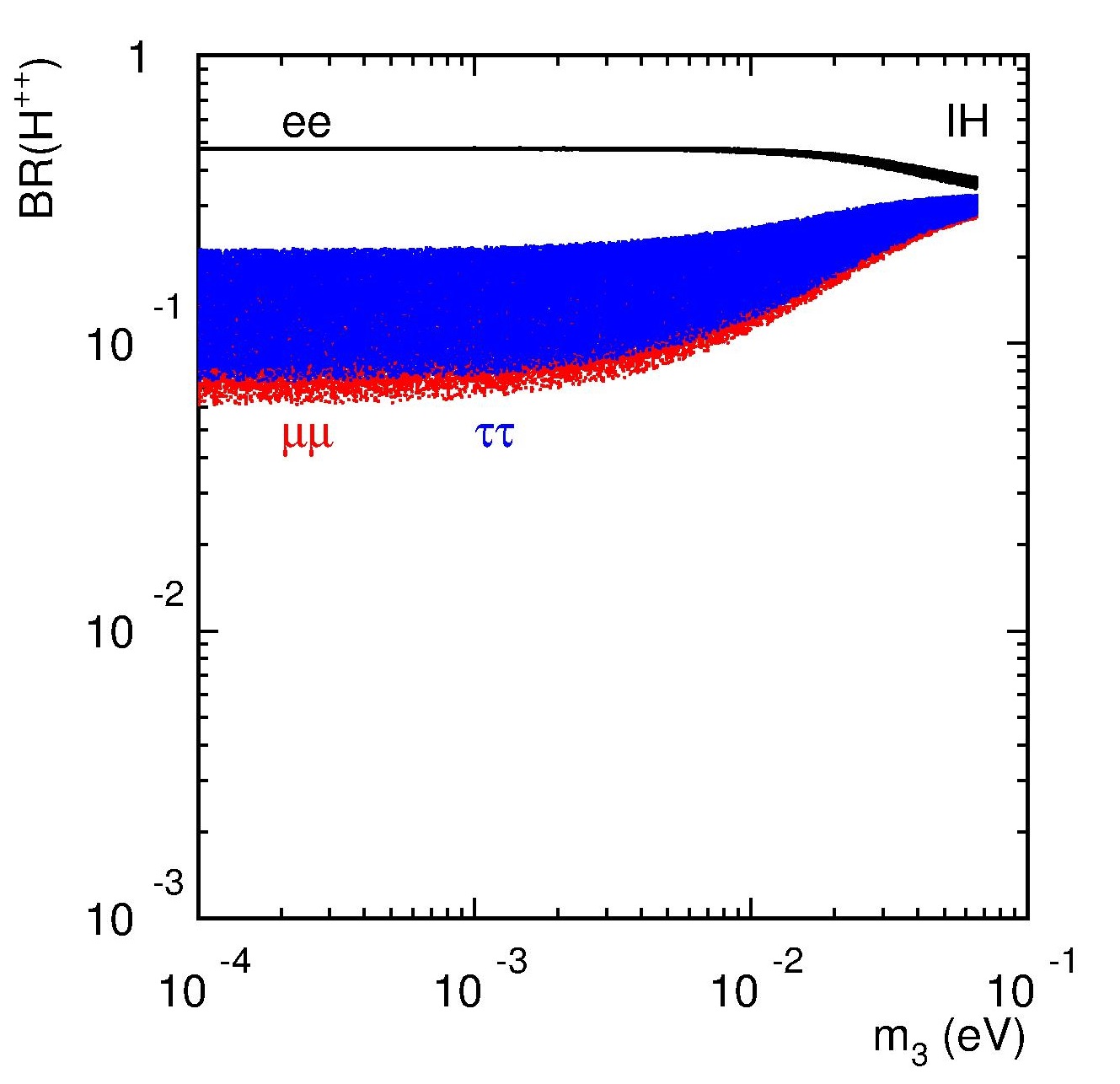}
\end{center}
\caption{Scatter plots for the $H^{++}$ decay branching fractions
 to the flavor-diagonal like-sign dileptons versus the lowest
neutrino mass for NH (a) and IH (b) with $\Phi_1 = \Phi_2 = 0$.}
\label{brii}
\end{figure}

\begin{figure}[tb]
\begin{center}
\minigraph{7.cm}{-0.15in}{(a)}{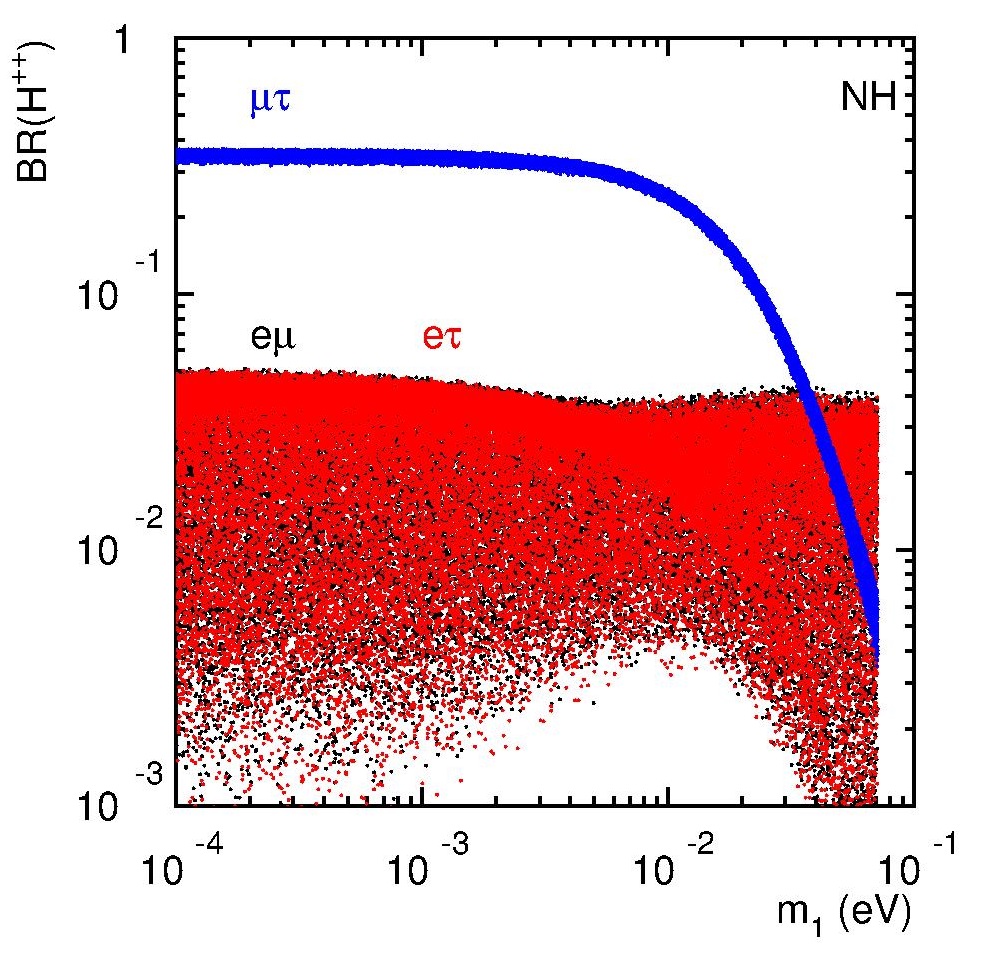}
\minigraph{7.cm}{-0.15in}{(b)}{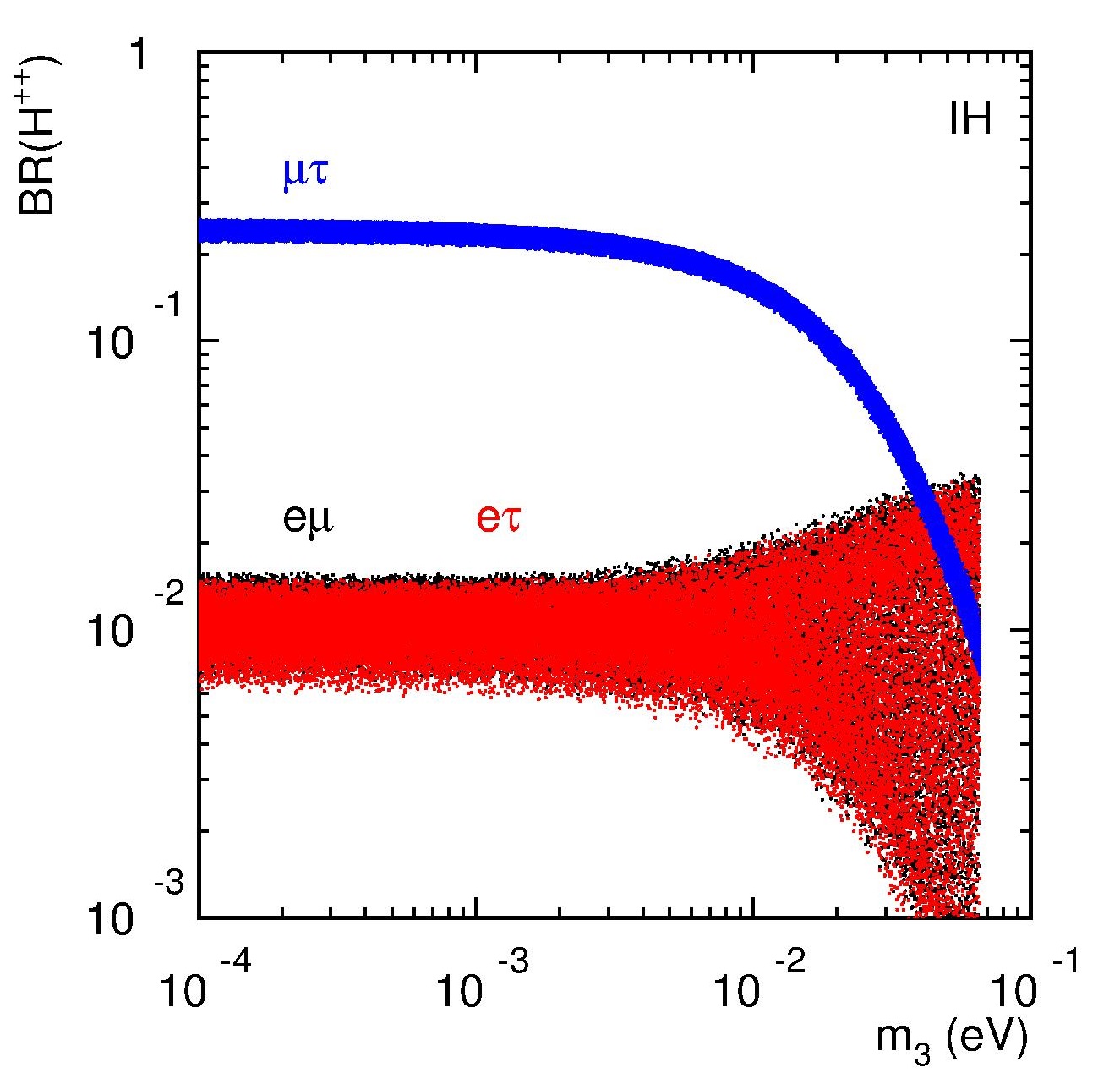}
\end{center}
\caption{$H^{++}$ decay to the flavor-off-diagonal like-sign dileptons
versus the lowest neutrino mass for NH (a) and IH (b) with $\Phi_1 = \Phi_2 = 0$.
}
\label{brij}
\end{figure}

Below $v_{\Delta}\approx 10^{-4}$ GeV, the doubly charged Higgs $H^{\pm\pm}$ decays dominantly to same-sign lepton pairs.
For vanishing Majorana phases $\Phi_1 = \Phi_2 = 0$, we show in Figs.~\ref{brii} and \ref{brij} the branching fraction of the decays into
same-flavor and different-flavor leptonic final states, respectively.
Relations among the branching fractions of the lepton
number violating Higgs decays of both the singly- and doubly-charged Higgs in the NH
and IH, with vanishing Majorana phases, are summarized in Table.~\ref{relationii}.

\begin{table}[t!]
{\small
\begin{center}
\begin{tabular}{|c|c|}
\hline
\text{ } & Relations
\\
\hline	NH	& BR$( H^{++} \to \tau^+ \tau^+ / \mu^+ \mu^+ ) \sim (20-40)\% \gg$ BR$( H^{++} \to e^+ e^+ )\sim (0.1-0.6)\%$\\
		& \qquad BR$( H^{++} \to \mu^+ \tau^+ )\sim (30-40)\% \gg $ BR$(H^{++} \to e^+ \mu^+ / e^+ \tau^+ )\lesssim 5\%$\\
		& BR$( H^{+}  \to \tau^+ \bar{\nu} / \mu^+ \bar{\nu} )\sim (30-60)\% \gg $ BR$(H^{+} \to e^+ \bar{\nu} )\sim (2.5-3)\%$
\\
\hline	IH	& \quad\qquad BR$( H^{++} \to e^+ e^+ )\sim 50\% \ > \ $ BR$( H^{++} \to \mu^+ \mu^+ / \tau^+ \tau^+ )\sim (6-20)\%$\\
		& BR$( H^{++} \to \mu^+ \tau^+ )\sim (20-30)\% \ \gg \ $ BR$( H^{++} \to e^+ \mu^+ / e^+ \tau^+ )\sim (0.1-4)\%$\\
		& \qquad BR$( H^{+} \to e^+ \bar{\nu} )\sim 50\% \ > \ $ BR$( H^{+} \to \mu^+ \bar{\nu} / \bar{\nu} )\sim (20-30)\%$
\\
\hline
\end{tabular}
\caption{Relations among the branching fractions of the lepton
number violating Higgs decays for the neutrino mass patterns of NH
and IH, with vanishing Majorana phases. }
\label{relationii}
\end{center} }
\end{table}

\begin{figure}[tb]
\begin{center}
\minigraph{6cm}{-0.15in}{(a)}{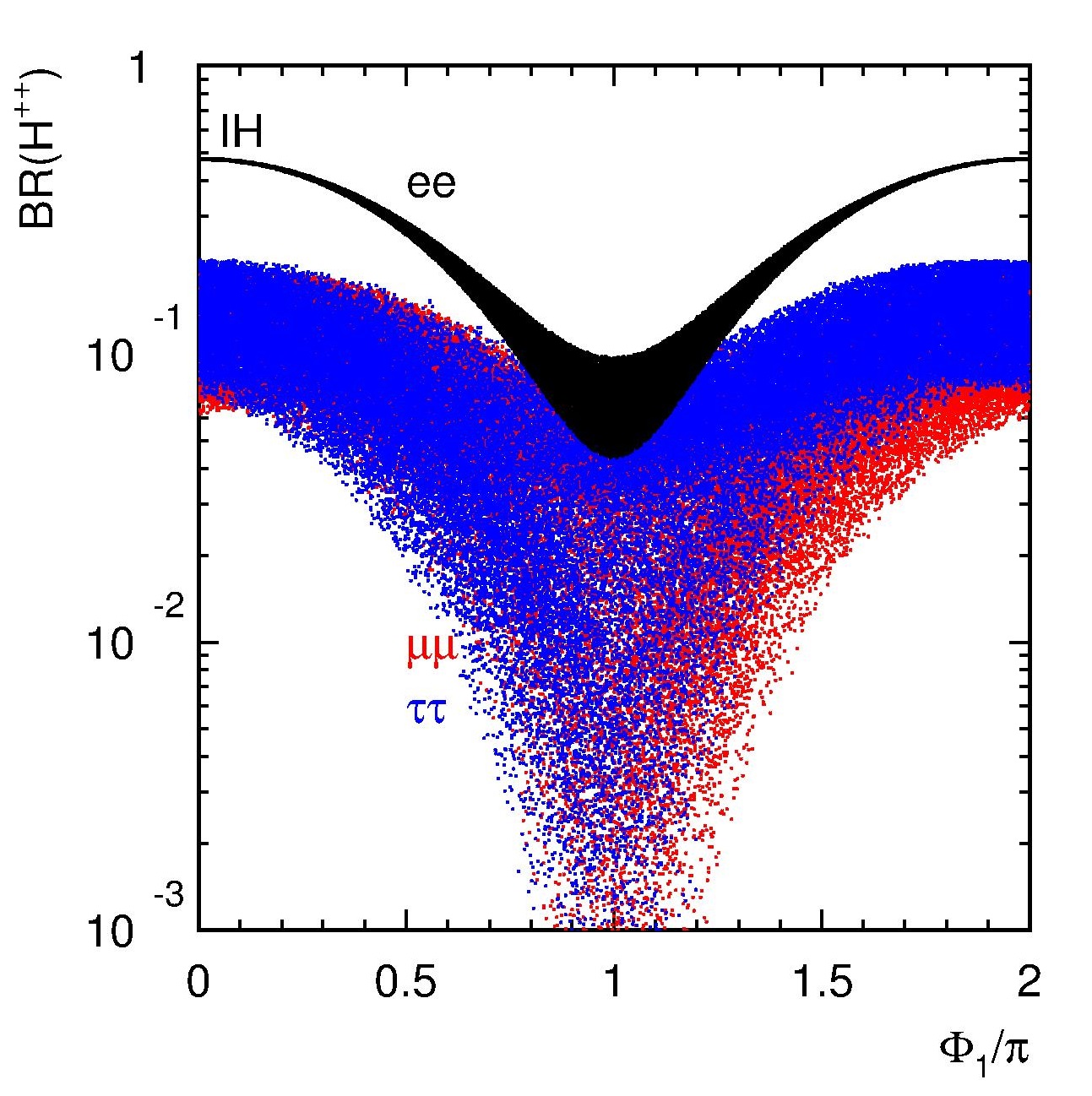}
\minigraph{6cm}{-0.15in}{(b)}{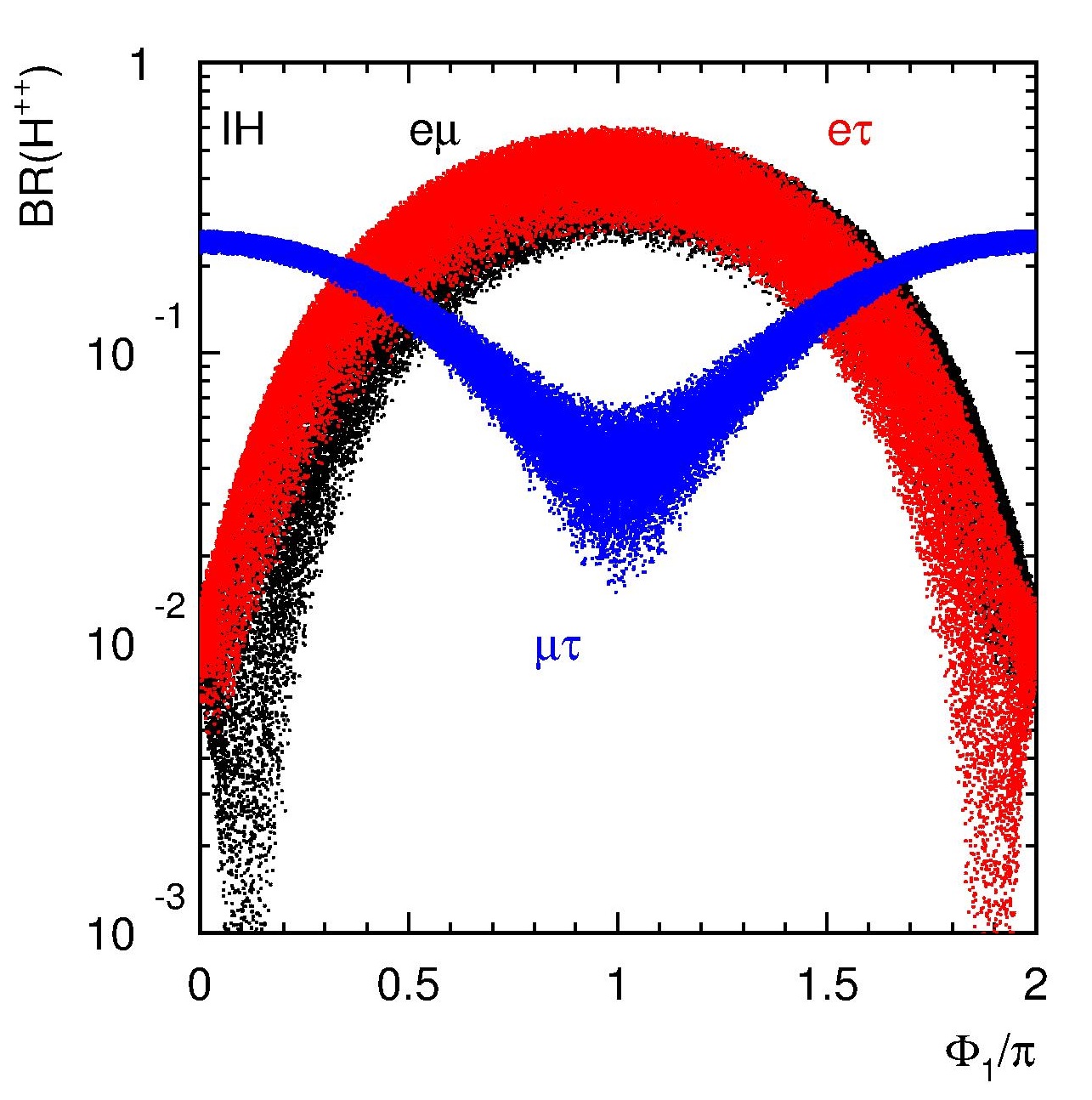}
\end{center}
\caption{Scatter plots of the same (a) and different (b) flavor leptonic
branching fractions for the $H^{++}$ decay versus the Majorana phase
$\Phi_1$ for the IH with $m_3=0$ and $\Phi_2 \in (0, 2\pi)$.}
\label{Majorana2}
\end{figure}

The impact of Majorana phases can be substantial in doubly charged Higgs decays~\cite{Akeroyd:2007zv,Garayoa:2007fw}.
In the case of the IH,  a large cancellation among the relevant channels occurs due to the phase at $\Phi_1=\pi$.
As a result, in this scenario, the dominant channels swap from $H^{++} \to e^+e^+,\ \mu^{+} \tau^+$ when $\Phi_1 \approx 0$
to $H^{++} \to e^+ \mu^+,\ e^{+} \tau^+$ when $\Phi_1 \approx \pi$, as shown in Fig.~\ref{Majorana2}.
Therefore this qualitative change can be made use of to extract the value of the Majorana phase $\Phi_1$.
In the NH case, however, the dependence of the decay branching fractions on the phase
is rather weak because of the lack of a subtle cancellation~\cite{Perez:2008ha}.

The Type II mechanism can be embedded in a number of extended gauge scenarios, for example the LRSM as discussed in Sec.~\ref{sec:hybrid},
as well as GUTs, such as  $(331)$ theories~\cite{Cuypers:1996ia,Tully:2000kk,Fonseca:2016xsy,Cogollo:2008zc}
and the extensions of minimal SU$(5)$~\cite{Georgi:1974sy}.
For $(331)$ models, one finds the presence of bileptons~\cite{Frampton:1992wt,Pisano:1991ee},
\ie, gauge bosons with $L = \pm2$ charges and hence $Q=\pm2$ electric charges. In a realistic extension of the Georgi-Glashow model, a scalar 15-dimensional representation is added~\cite{Dorsner:2005fq} and the scalar triplet stays in the $\textbf{15}$ representation together with scalar leptoquark $\Phi\sim (3,2,1/6)$. The SU$(5)$ symmetry thus indicates that the couplings of the leptoquark to matter gain the same Yukawas $Y_\nu$ responsible for neutrino mass matrix~\cite{FileviezPerez:2008dw}.
Extensions with vector-like leptons in nontrivial SU$(2)_L$ representations are also possible~\cite{Bahrami:2013bsa}.
Unsurprisingly, the phenomenology~\cite{Cuypers:1996ia,Fonseca:2016xsy,Nepomuceno:2016jyr,Corcella:2017dns,Meirose:2011cs}
and direct search constraints~\cite{Nepomuceno:2016jyr,Corcella:2017dns}
for $L$-violating, doubly charged vector bosons are similar to $L$-violating, doubly charged scalar bosons, which we now discuss.

\subsection{Triplet Higgs Scalars at Colliders}\label{sec:type2Collider}

\subsubsection{Triplet Higgs Scalars and the Type II Seesaw at Colliders}

\begin{figure}[!t]
\begin{center}
\includegraphics[scale=1,width=.9\textwidth]{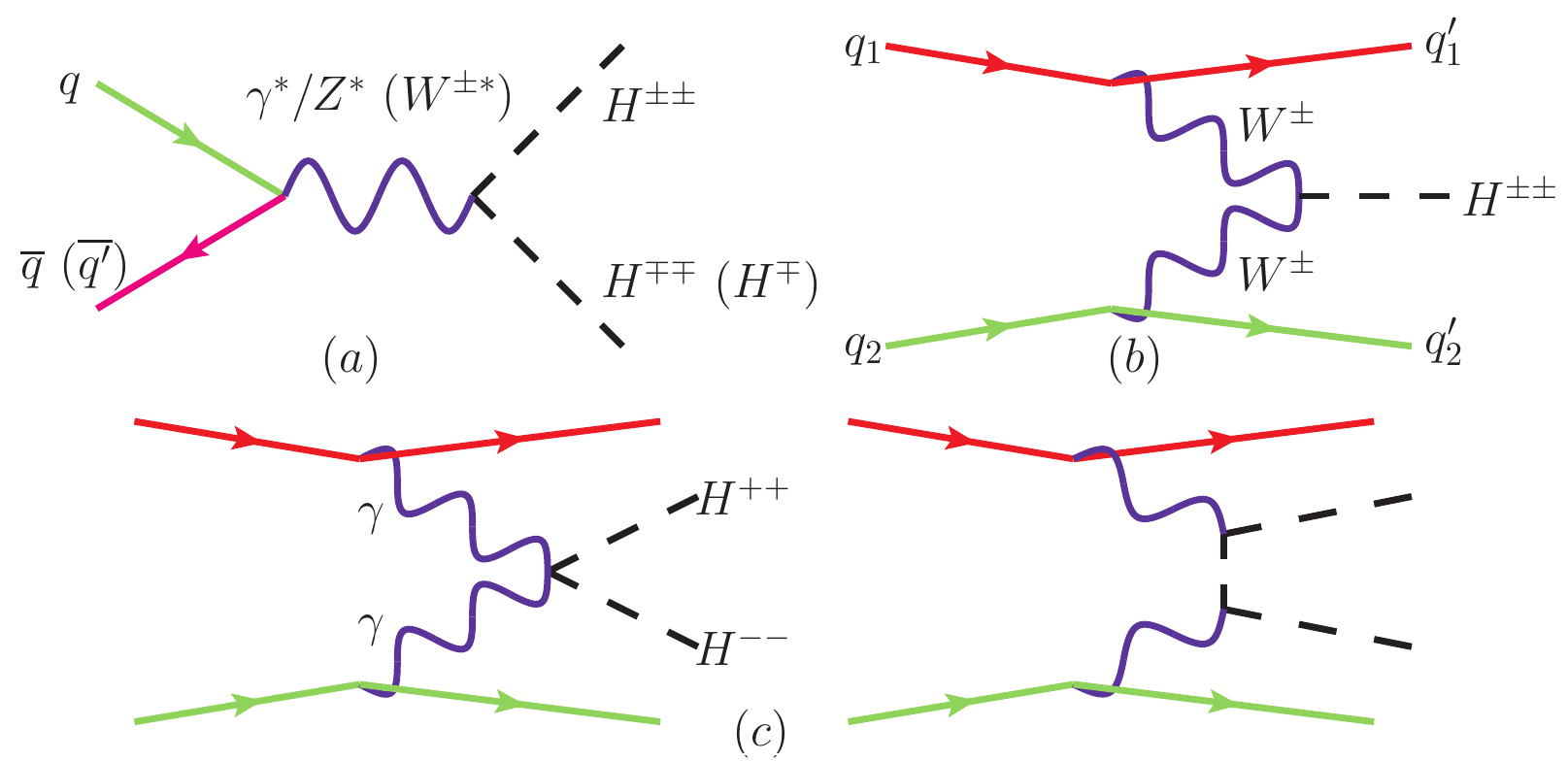} \label{fig:feynmanTypeIIScalars}
\end{center}
\caption{
Born-level diagrams depicting Type II triplet scalar production in $pp$ collisions via (a) the DY mechanism, (b) same-sign $W^\pm W^\pm$ scattering, and (c) $\gamma\gamma$ fusion.
}
\label{fig:feynmanTypeIIScalars}
\end{figure}

\begin{figure}[!t]
\begin{center}
\subfigure[]{\includegraphics[scale=1,width=0.48\textwidth]{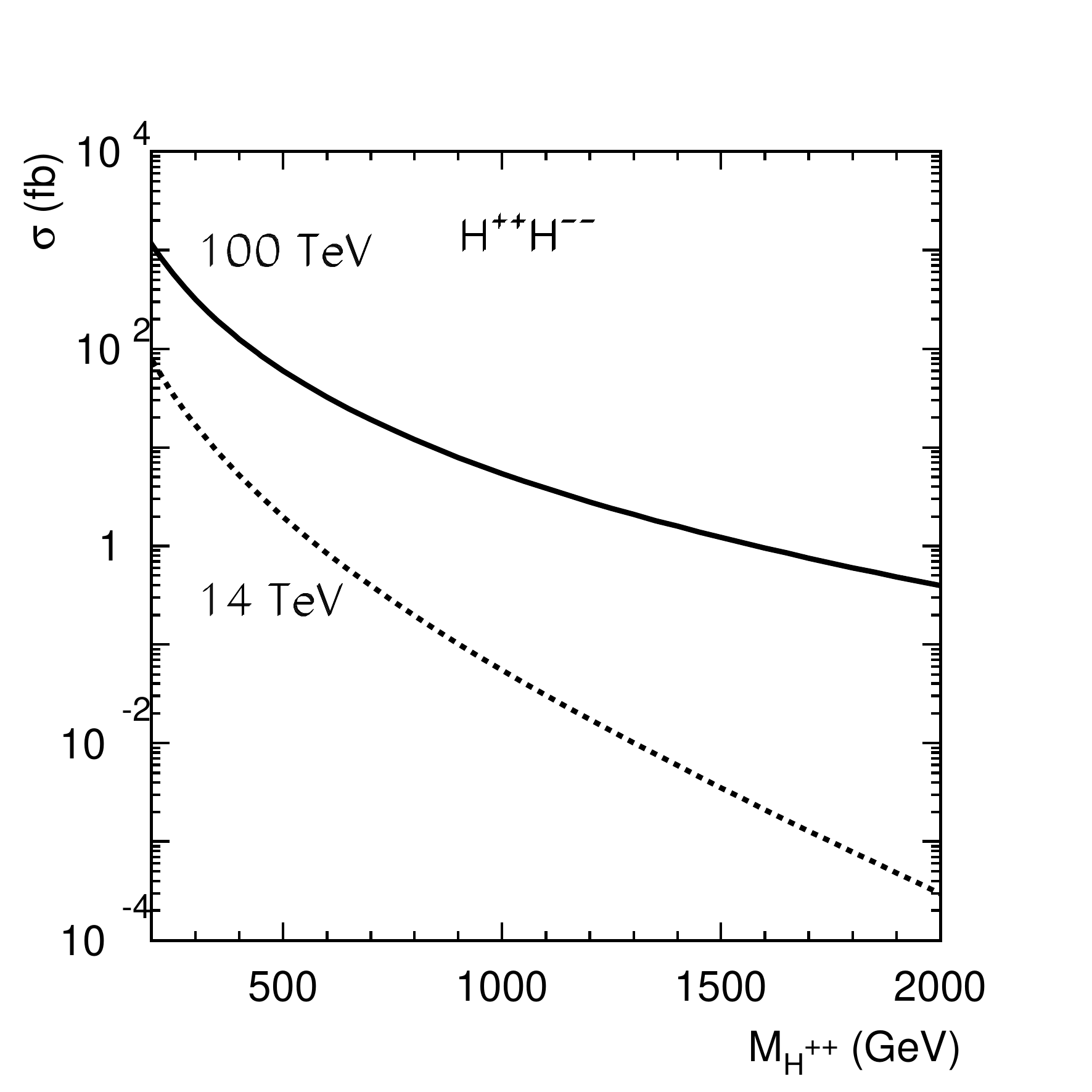} }
\subfigure[]{\includegraphics[scale=1,width=0.48\textwidth]{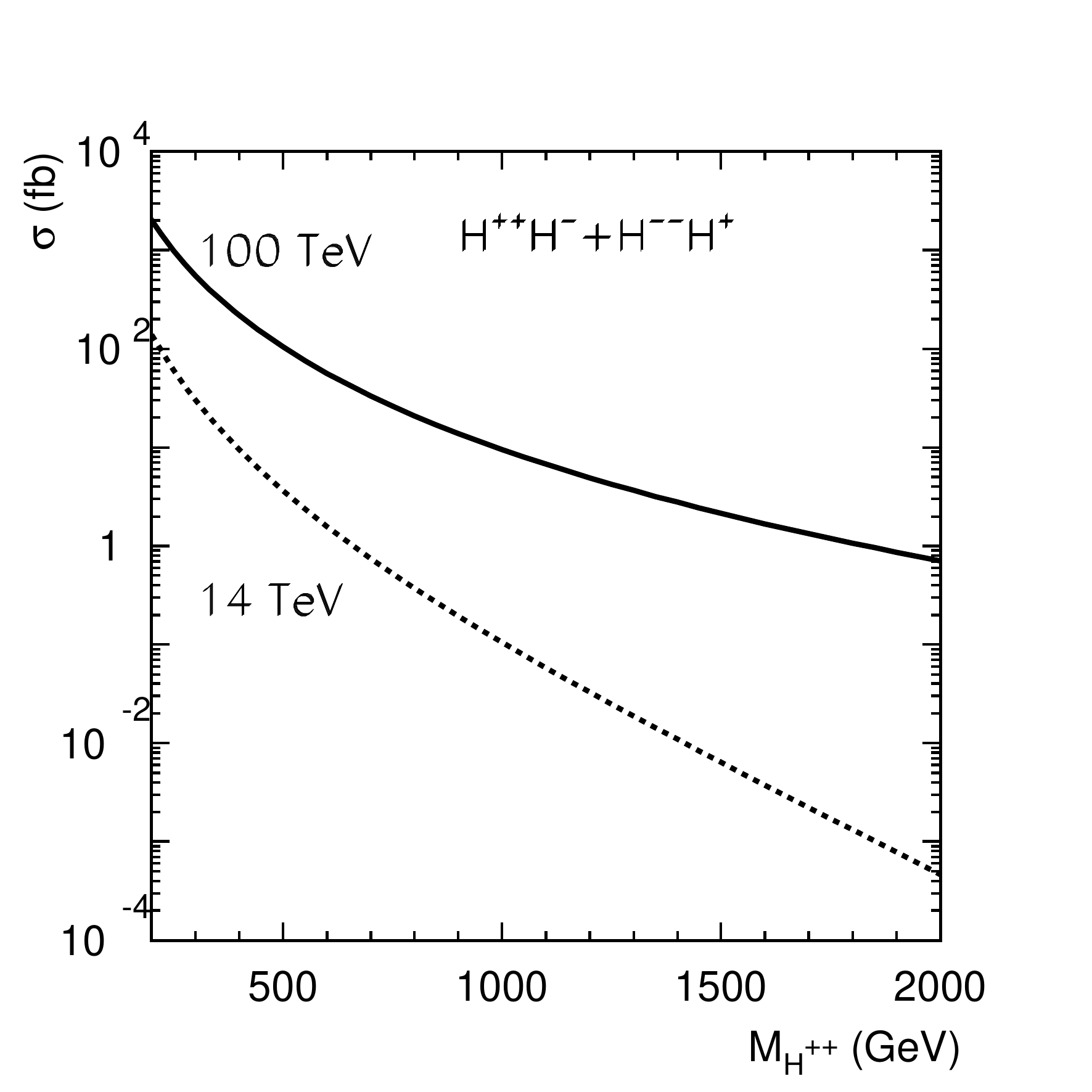} }
\\
\subfigure[]{\includegraphics[scale=1,width=0.48\textwidth]{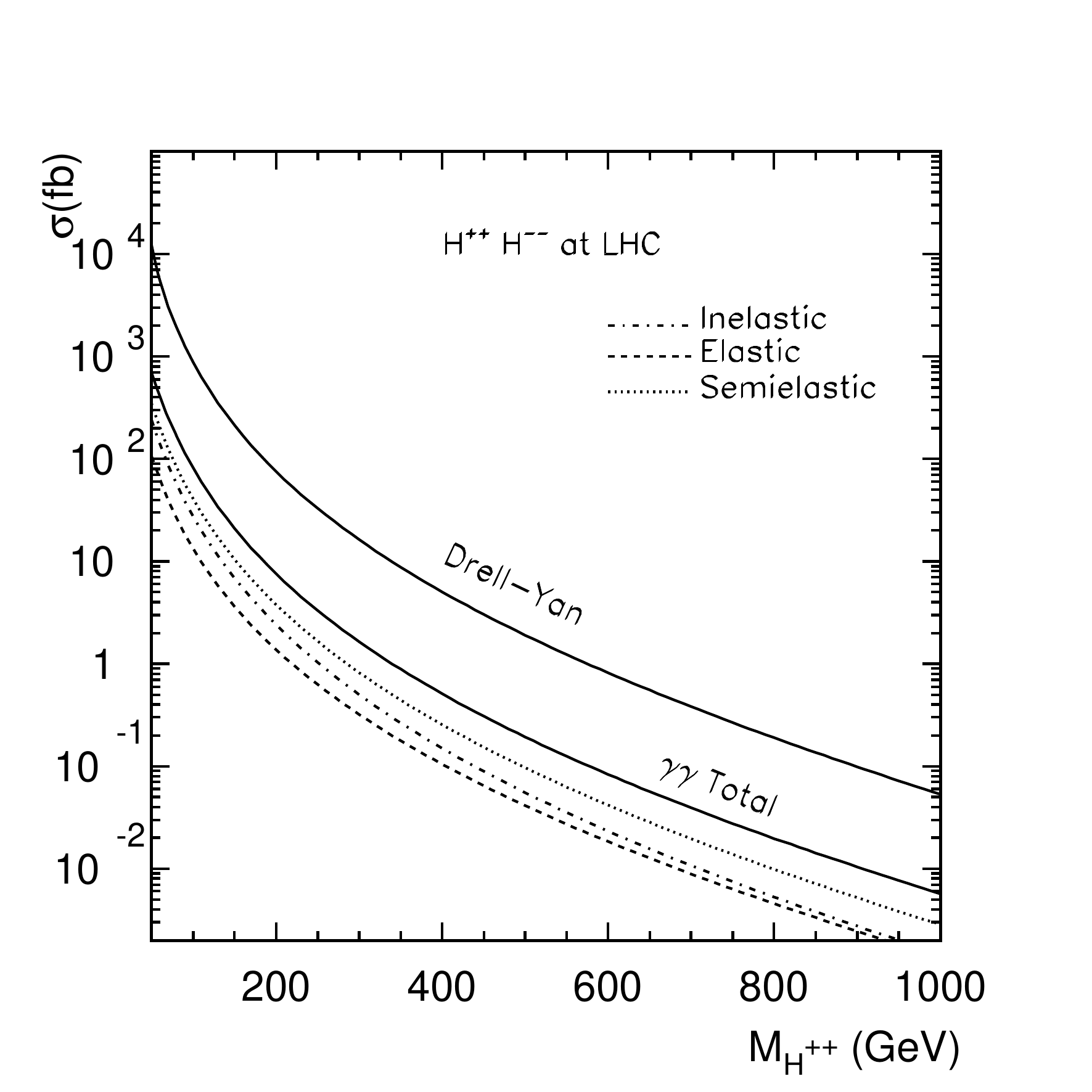}	\label{fig:type2_DblyChargedHiggsXSec}	}
\subfigure[]{\includegraphics[scale=1,width=0.48\textwidth,height=6cm]{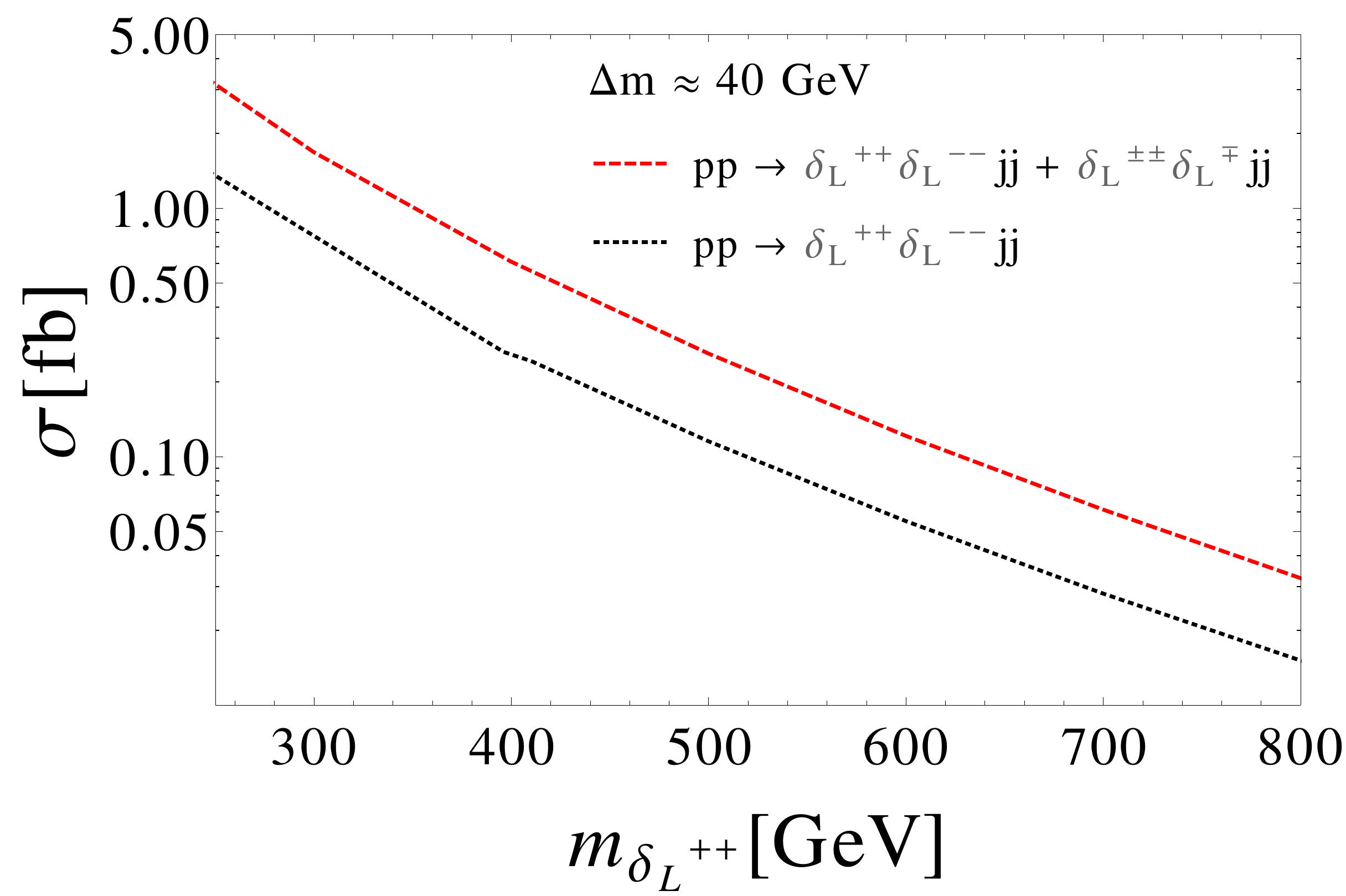}	\label{fig:type2_vbfHiggs_ProdXSec}	}
\end{center}
\caption{Production cross section for (a) $pp\to H^{++} H^{--}$ and (b) $H^{\pm\pm}H^\mp$ at $\sqrt{s} = 14$ and 100 TeV,
as well as for (c) $pp\to H^{++} H^{--} jj$ from $\gamma\gamma$ fusion~\cite{Han:2007bk} and (d) $pp\to H^{++} H^{--} jj + H^{\pm\pm} H^{\mp}$
from VBF at $\sqrt{s}=14$ TeV~\cite{Dutta:2014dba}.
}
\label{fig:typeiiScalarXSec}
\end{figure}

If kinematically accessible, the canonical and well-studied~\cite{Rizzo:1981xx,Akeroyd:2005gt,Perez:2008ha,Bambhaniya:2013wza}
triplet scalars production channels  at hadron colliders
are the neutral and charged current DY processes,  given by
\begin{eqnarray}
pp\to \gamma^*/Z^* \to H^{++} H^{--}, \ \ \ pp\to W^{\pm*} \to H^{\pm\pm} H^{\mp},
\end{eqnarray}
and shown in Fig.~\ref{fig:feynmanTypeIIScalars}(a).
Unlike Type I models, scalars in the Type II Seesaw couple to EW bosons directly via gauge couplings.
Subsequently, their production rates are sizable and can be predicted as a function of mass without additional input.
In Fig.~\ref{fig:typeiiScalarXSec} we show the LO pair production cross section
of triplet scalars via the (a) neutral and (b) charged current DY process at $\sqrt{s} = 14$ and 100 TeV.
NLO in QCD corrections to these processes are well-known~\cite{Muhlleitner:2003me} and span
$K^{\rm NLO } = \sigma^{\rm NLO}/\sigma^{\rm LO} = 1.1 - 1.3$ away from boundaries of collider phase space;
moreover, due to the color-structure of DY-like processes, inclusive kinematics of very heavy scalar triplets are Born-like and
thus na\"ive normalization of kinematics by $K^{\rm NLO }$ gives reliable estimates of both NLO- and NLO+PS-accurate results~\cite{Ruiz:2015zca,Fuks:2017vtl}.
For $M_{H^{\pm\pm}} = 1$ TeV, one finds that the LO pair production rates can reach $\sigma\sim0.1~(10)$ fb at $\sqrt{s}=14$ (100) TeV,
indicating $\mathcal{O}(10^2)~(\mathcal{O}(10^4))$ of events with the ab$^{-1}$-scale data sets expected at the respective collider program.

In addition to the DY channels are:
single production of charged Higgses via weak boson scatter, as shown in Fig.~\ref{fig:feynmanTypeIIScalars}(b) and investigated in~\cite{Han:2005nk,Chen:2008qb};
charged Higgs pair production via $\gamma\gamma$ scattering, as shown in Fig.~\ref{fig:feynmanTypeIIScalars}(c),
studied in~\cite{Drees:1994zx,Han:2007bk,Bambhaniya:2015wna,Dutta:2014dba,Babu:2016rcr},
and computed at $\sqrt{s}=14$ TeV~\cite{Han:2007bk} in Fig.~\ref{fig:type2_DblyChargedHiggsXSec};
as well as pair production through weak boson scattering, as studied in ~\cite{Bambhaniya:2015wna,Dutta:2014dba}
and computed for the 14 TeV LHC~\cite{Dutta:2014dba} in  Fig.~\ref{fig:type2_vbfHiggs_ProdXSec}.
As in the case of $W\gamma$ scattering in heavy $N$ production in Sec.~\ref{sec:type1},
there is renewed interest~\cite{Bambhaniya:2015wna} in the $\gamma\gamma$-mechanisms due to the new availability of photon PDFs that
include both elastic and (deeply) inelastic contributions, \eg, NNPDF 2.3 and 3.0 QED PDF sets~\cite{Ball:2013hta,Ball:2014uwa}.
However, care should be taken in drawing conclusions based on these specific PDF sets due to the (presently) large $\gamma$-PDF uncertainty, particularly at large Bjorken-$x$ where this can reach greater than $100\%$~\cite{Ball:2013hta}. For example:  As shown in Fig.~\ref{fig:type2_DblyChargedHiggsXSec}, $\gamma\gamma$ production is unambiguously sub-leading to the DY mechanism and only contributes about $10\%$
despite recent claims to the contrary~\cite{Babu:2016rcr,Ghosh:2017jbw}. The collinear behavior and the factorization scale dependence of the incoming photons must be treated with great care. As more data is collected and $\gamma$-PDF methodology further matures, one anticipates these uncertainties to greatly shrink;
for further discussions of $\gamma$-PDFs, see Refs.~\cite{Martin:2014nqa,Alva:2014gxa,Harland-Lang:2016kog,Degrande:2016aje,Manohar:2016nzj,Manohar:2017eqh}.
For a list of recommended $\gamma$-PDFs, see the discussion just above Eq.~(\ref{eq:heavyNnj}).

\begin{figure}[!t]
\begin{center}
\subfigure[]{\includegraphics[scale=1,width=0.48\textwidth]{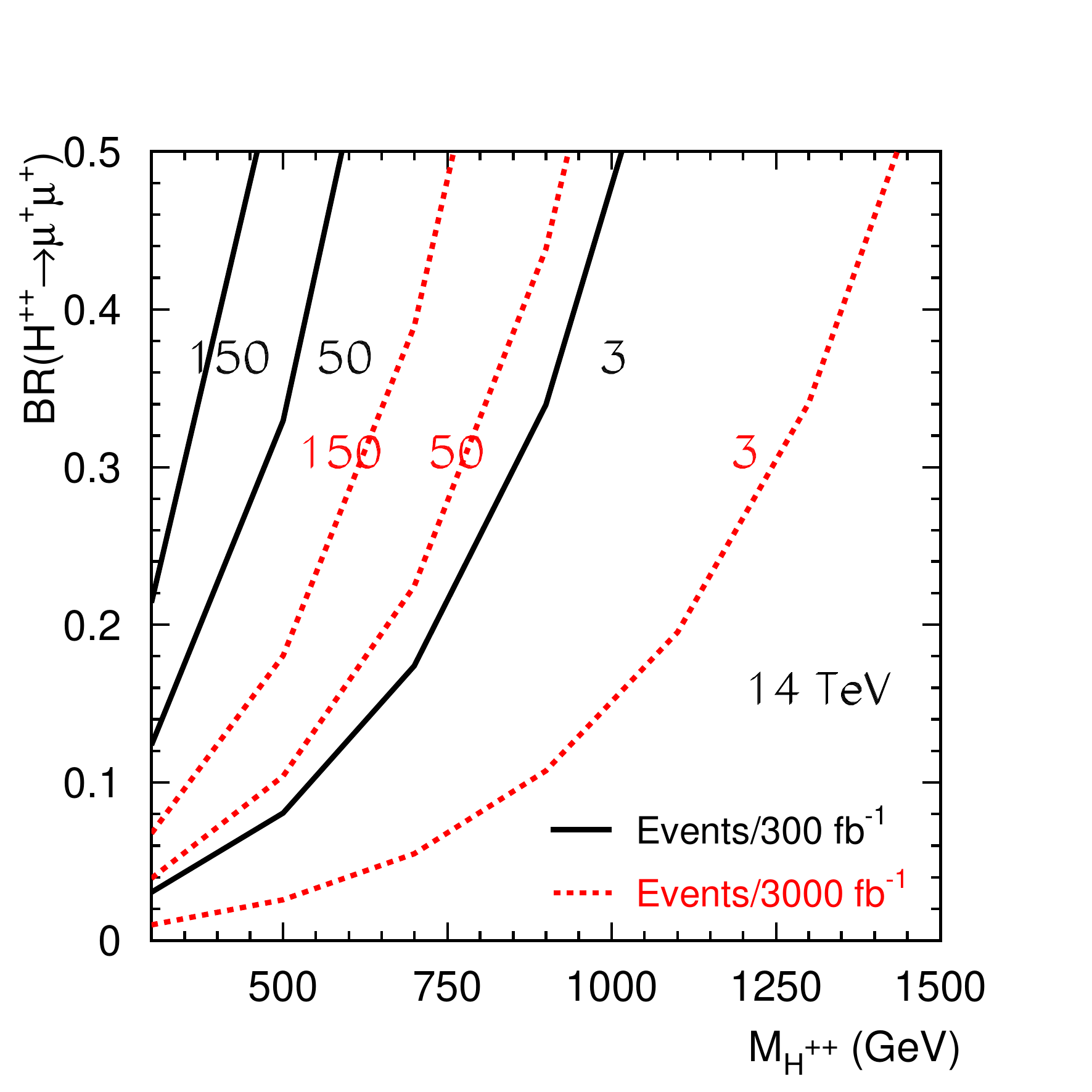}	\label{fig:type2_TripletContour14TeV}	}
\subfigure[]{\includegraphics[scale=1,width=0.48\textwidth]{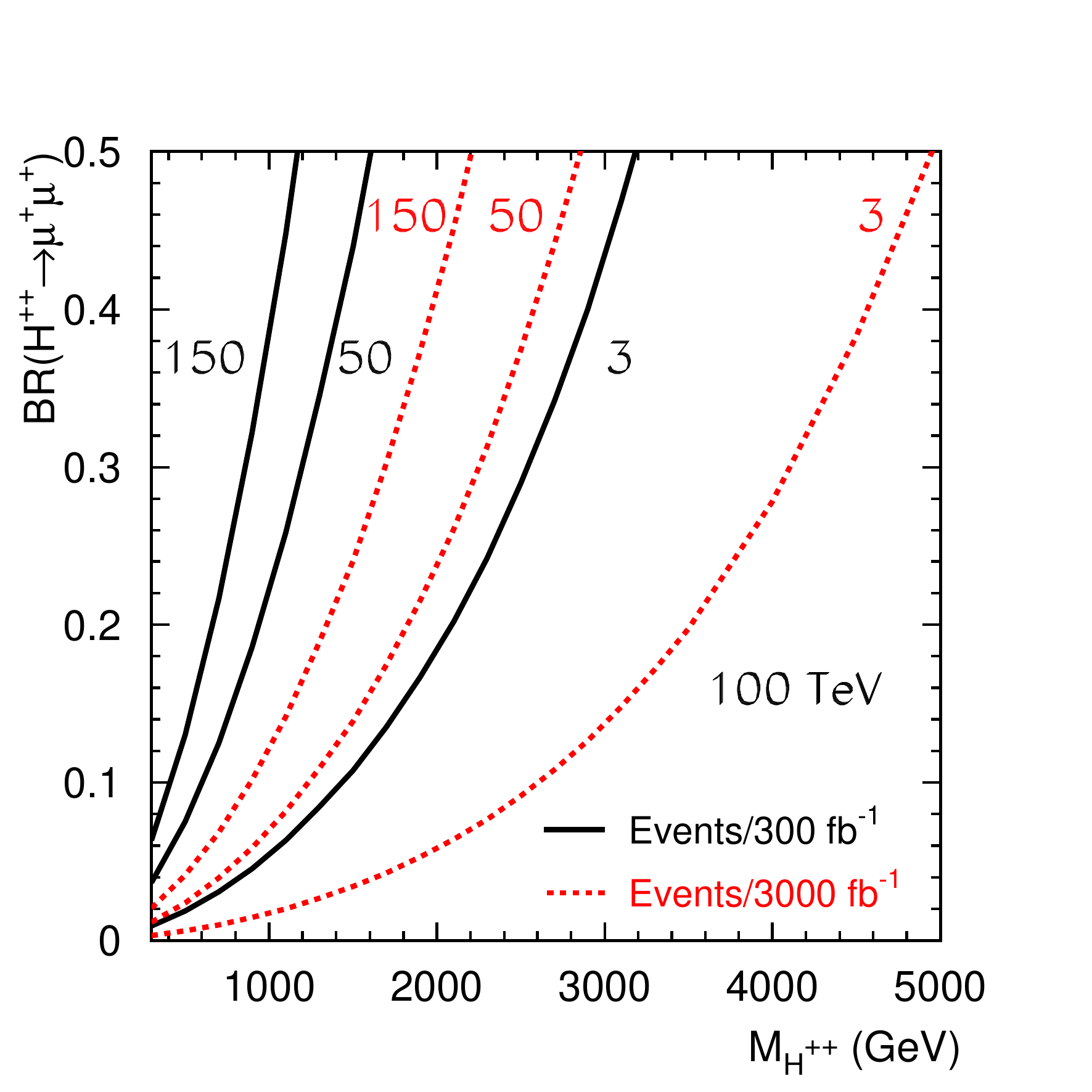}	\label{fig:type2_TripletContour100TeV}	}
\\
\subfigure[]{\includegraphics[scale=1,width=0.48\textwidth]{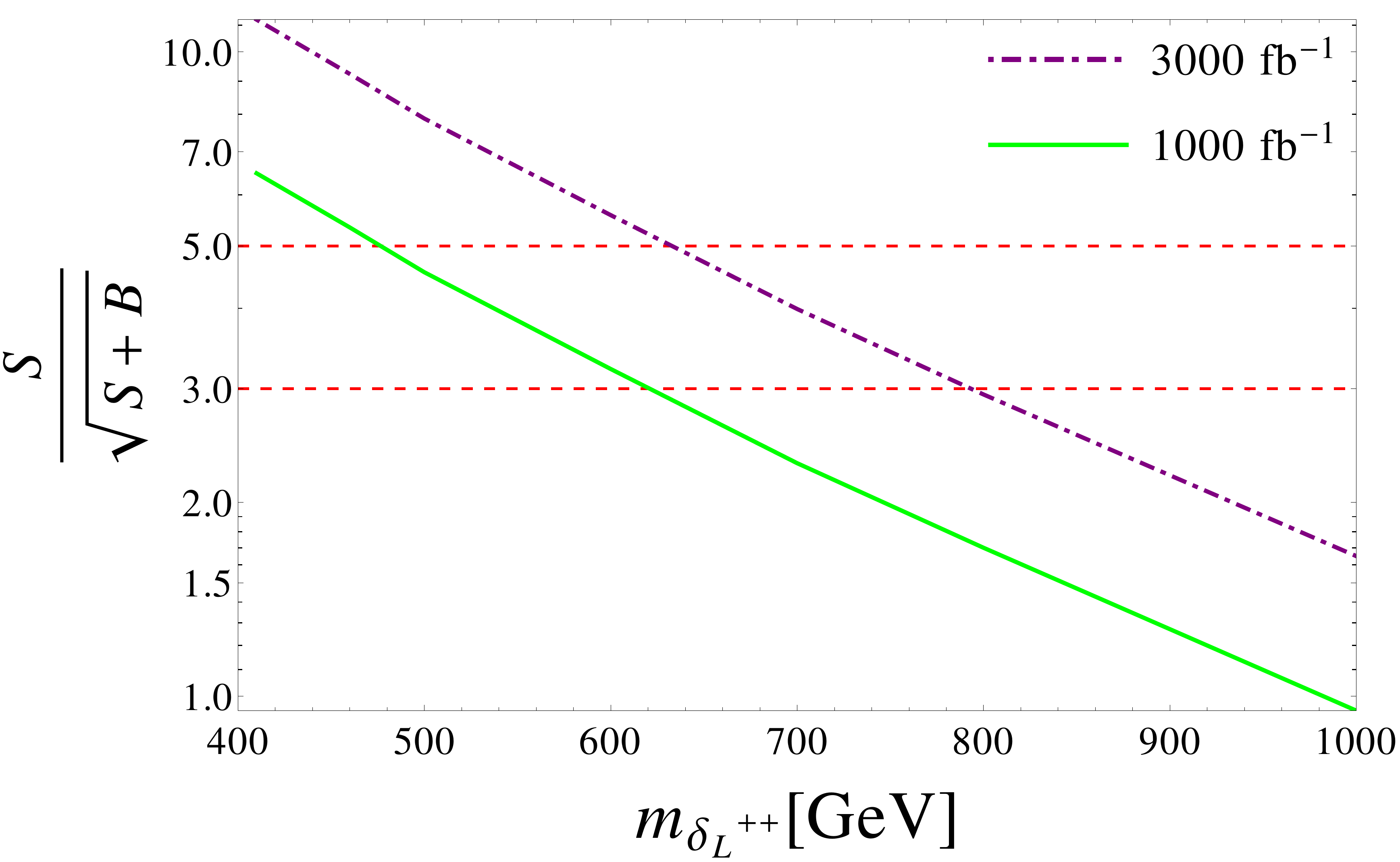}		\label{fig:type2_vbfHiggs_emuSig} }
\subfigure[]{\includegraphics[scale=1,width=0.48\textwidth]{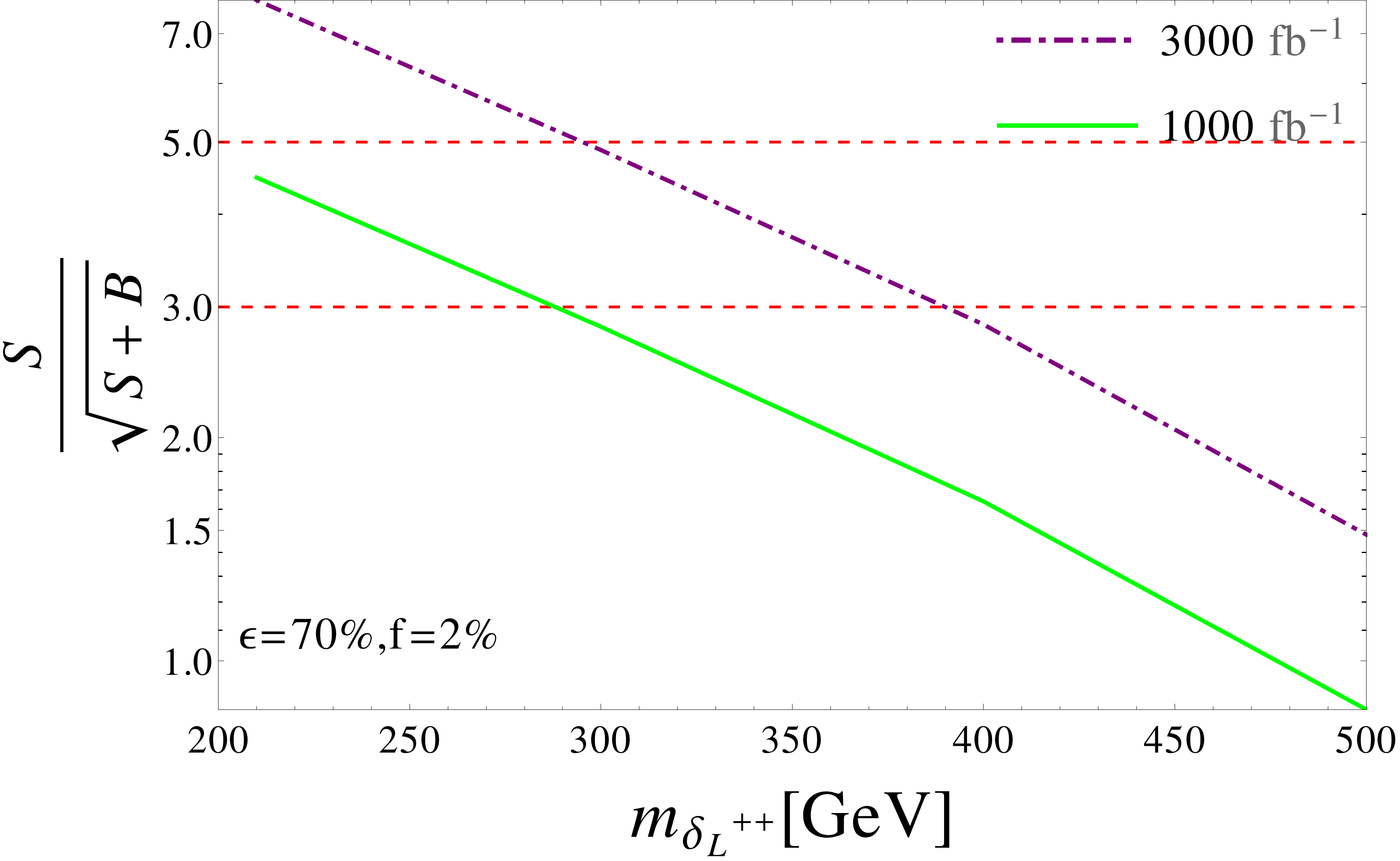}	\label{fig:type2_vbfHiggs_tauSig} }
\end{center}
\caption{Event contour for $H^{++}H^{--}\to \mu^+\mu^+\mu^-\mu^-$ in the ${\rm BR}(H^{++}\to \mu^+\mu^+)$ vs. $M_{H^{++}}$ plane at (a) $\sqrt{s}=$14 TeV and (b) 100 TeV, assuming $\mathcal{L}=$300 fb$^{-1}$ and 3000 fb$^{-1}$, and based on the analysis of Ref.~\cite{Perez:2008ha}.
Signal significance for VBF production of doubly charged Higgs pairs and their decays to (c) $e^\pm\mu^\pm$ and (d) $\tau^\pm\tau^\pm$ final-states, after $\mathcal{L}=1$ and $3\invab$ at the 14 TeV LHC ~\cite{Dutta:2014dba}.
}
\label{fig:typeiiDisc}
\end{figure}

Similar to the $\gamma\gamma$ channel, production of triplet scalars from gluon fusion is sub-leading with respect to DY due to multiple vanishing contributions~\cite{delAguila:1990yw,Hessler:2014ssa} and despite
an expectedly large QCD correction of $K^{\rm N^3LL} = \sigma^{\rm N^3LL}/\sigma^{\rm LO}\sim2.5-3$~\cite{Ruiz:2017yyf}.
If triplet scalar couplings to the SM-like Higgs are not too small and if sufficiently light, then such scalars may appear in pairs as rare decays of the 125 GeV scalar boson~\cite{Nemevsek:2016enw}.
Likewise, if neutral triplet scalars mix appreciably with the SM-like Higgs, then single production via gluon fusion is also possible~\cite{Nemevsek:2016enw};
one should note that in such cases, the QCD $K$-factors calculated in Ref.~\cite{Ruiz:2017yyf} are applicable.

A noteworthy direction of progress in searches for triplet scalars at colliders are the implementation of exotically charged scalars into FeynRules model files.
In particular, lepton number violating scalars are available in the \texttt{LNV-Scalars}~\cite{delAguila:2013mia,FeynRules:LNVScalars} model file
as well as in a full implementation of LRSM at LO accuracy~\cite{Roitgrund:2014zka,FeynRules:LRSMlo};
the Georgi-Machacek model~\cite{Georgi:1985nv} is also available at NLO in QCD accuracy~\cite{Degrande:2015xnm,FeynRules:GMnlo}.
These permit simulation of triplet scalar production in inclusive $\ell\ell/\ell p/pp$ collisions using modern, general-purpose event generators,
such as Herwig~\cite{Bellm:2015jjp}, MadGraph5\_aMC@NLO~\cite{Alwall:2014hca}, and Sherpa~\cite{Gleisberg:2008ta}.

\begin{figure}[!t]
\begin{center}
\subfigure[]{\includegraphics[scale=1,width=0.48\textwidth]{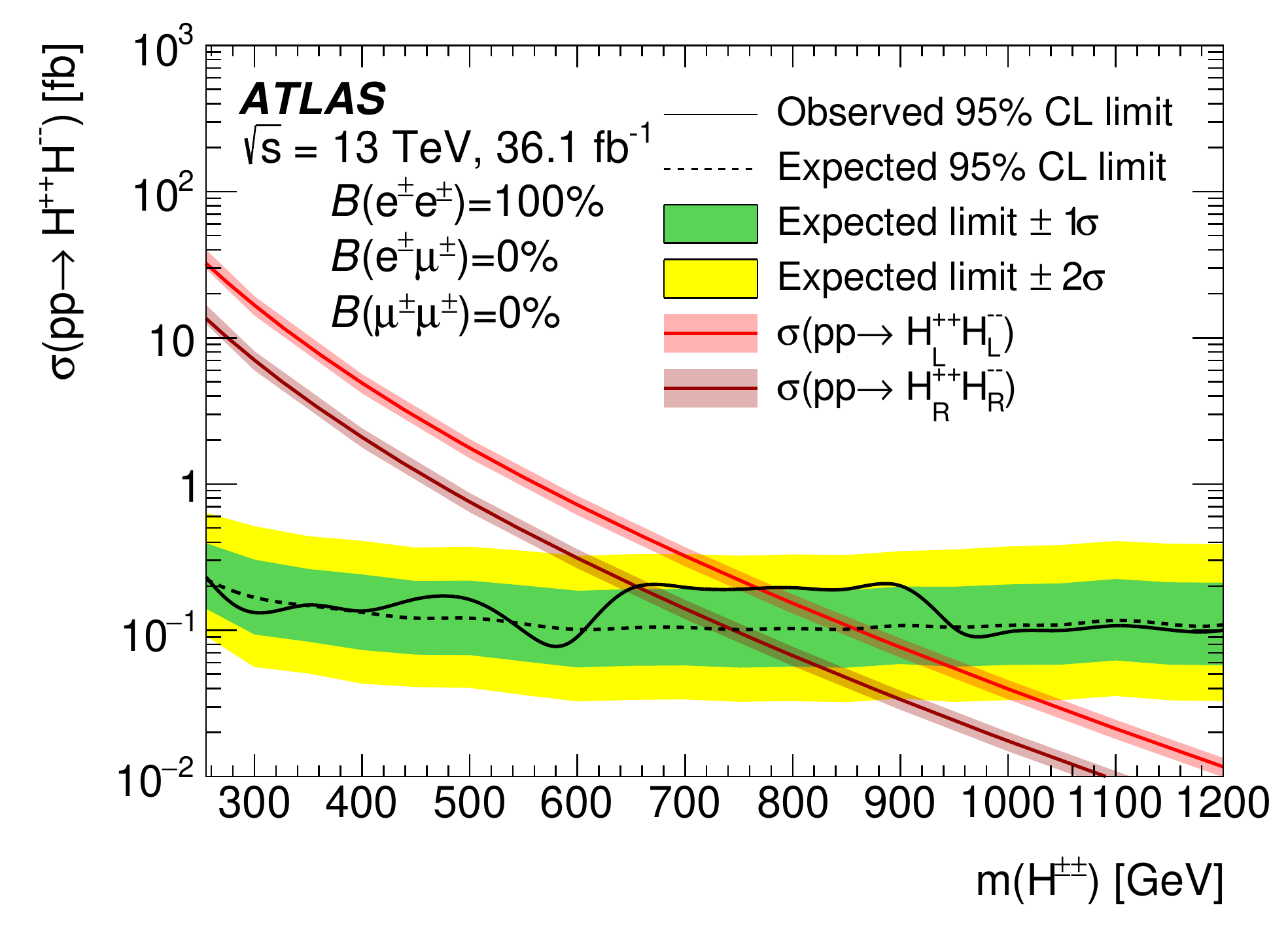} 		\label{fig:type2_SSee_Excl_ATLAS}}
\subfigure[]{\includegraphics[scale=1,width=0.48\textwidth]{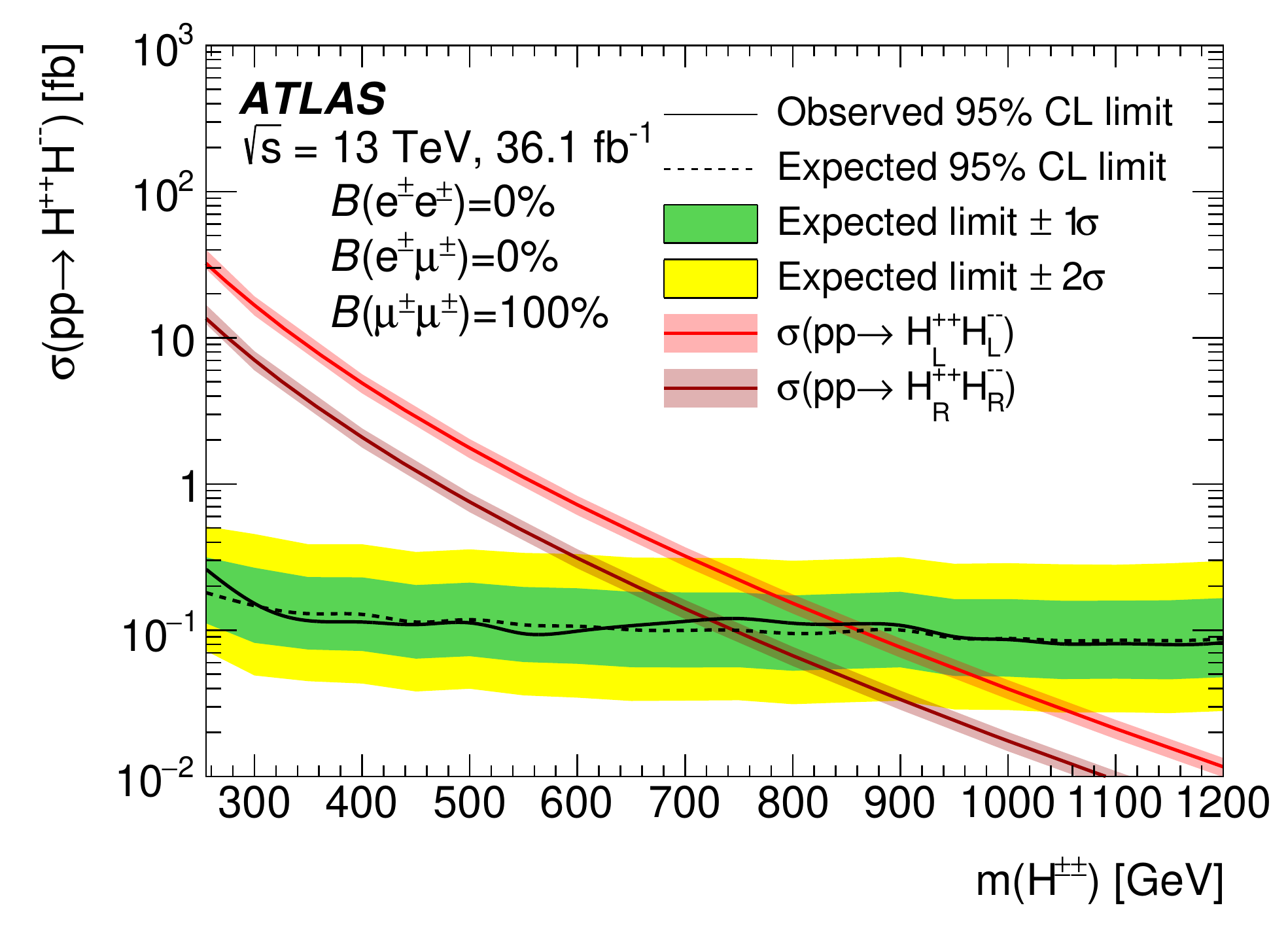}		\label{fig:type2_SSmumu_Excl_ATLAS}}
\\
\subfigure[]{\includegraphics[scale=1,width=0.48\textwidth]{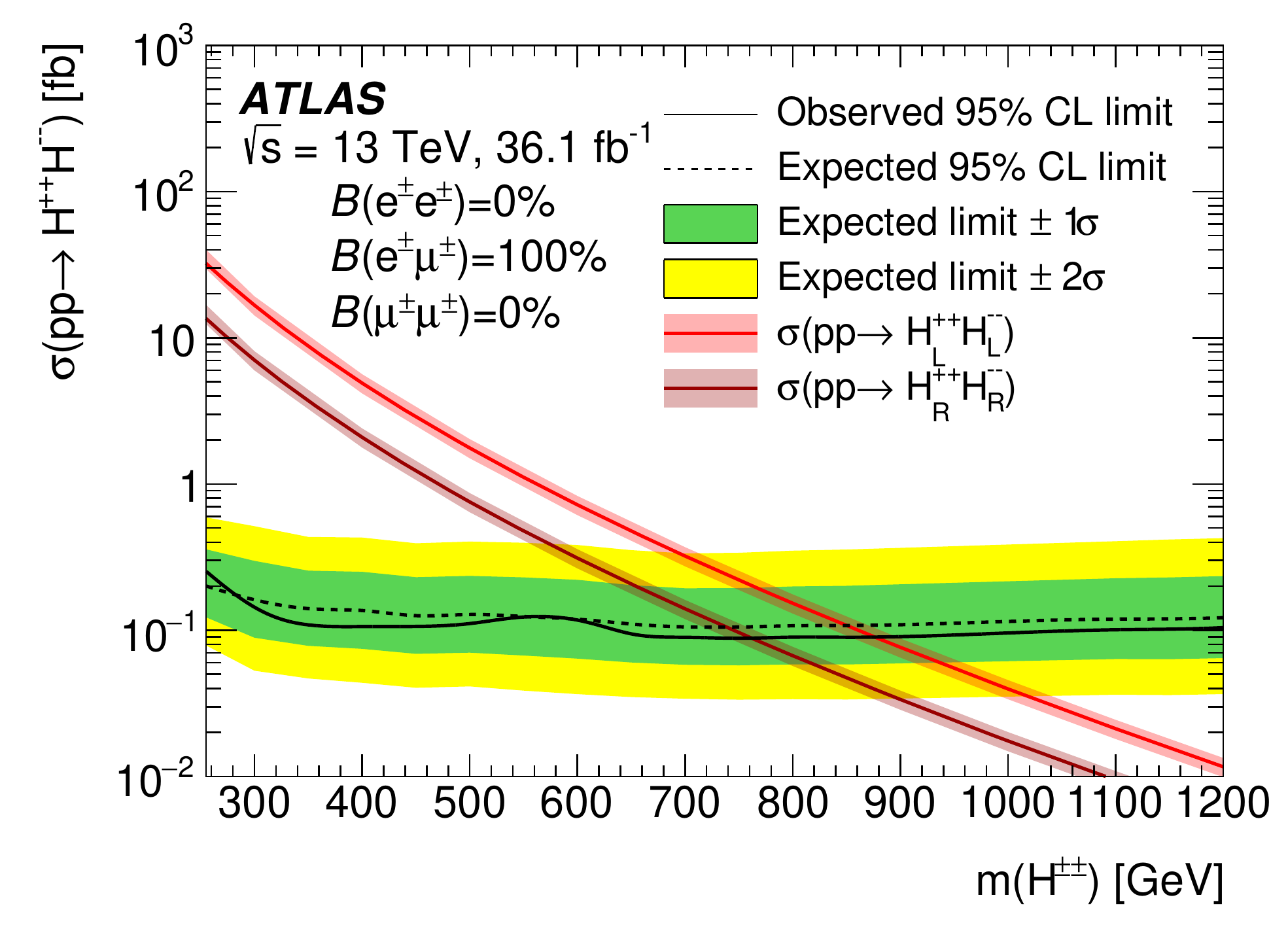} 		\label{fig:type2_SSemu_Excl_ATLAS}}
\subfigure[]{\includegraphics[scale=1,width=0.48\textwidth]{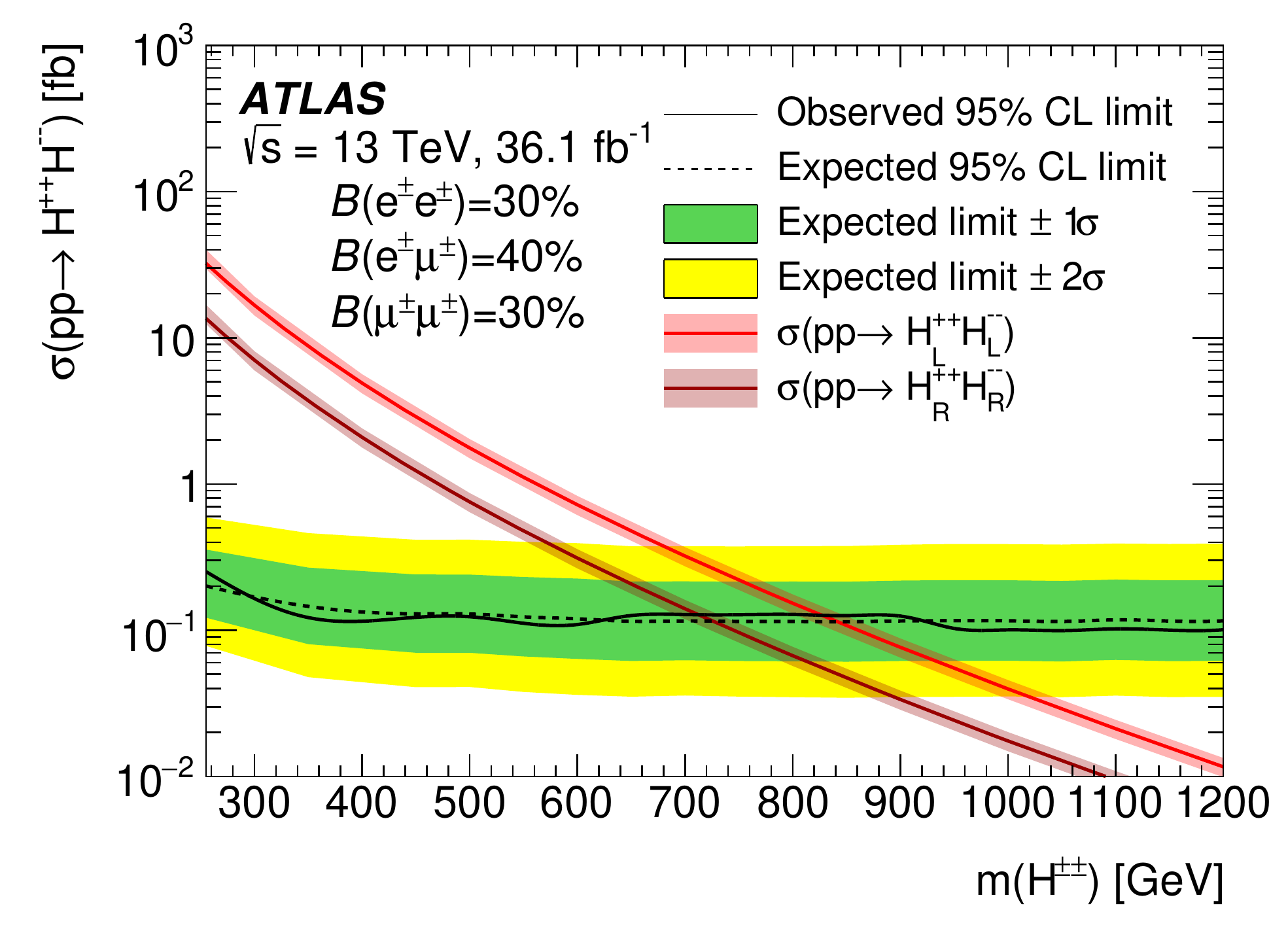}	\label{fig:type2_SSmixed_Excl_ATLAS}}
\end{center}
\caption{
ATLAS 95\% CLs exclusion at 13 TeV after $\mathcal{L}=36\invfb$ on $\sigma(pp\to H^{++}H^{--})$
for various representative branching rates to SM charged leptons
in the (a) pure $e^\pm e^\pm$, (b) pure $\mu^\pm\mu^\pm$, (c) pure $e^\pm\mu^\pm$ and (d) mixed final-states~\cite{ATLAS:2017iqw,Aaboud:2017qph}.
}
\label{fig:type2ExclATLAS}
\end{figure}

Due to the unknown Yukawa structure in Eq.~(\ref{eq:typeIIYukawa}), the decays of the triplet scalars to SM states are much more ambiguous than their production. Subsequently, branching rates of $H^{\pm}\to \ell^\pm \nu$ and  $H^{\pm\pm}\to\ell^\pm_1\ell^\pm_2$ are often taken as phenomenological parameters in analyses and experimental searches. When taking such a model-agnostic approach, it may be necessary to also consider the lifetimes of scalar triplets: In a pure Type II scenario, for $M_{H^{\pm\pm}}<270$ GeV and sub-MeV values of the triplet vev $v_L$, the proper decay length of $H^{\pm\pm}$ can exceed 10~$\mu$m~\cite{Han:2005nk}. As a result, exotically charged triplet scalars may manifest at collider experiments in
searches for long-lived, multi-charged particles such as Refs.~\cite{Aad:2013pqd,Aad:2015oga,CMS:2012aza,Barrie:2017eyd}.

For prompt decays of triplet scalars, the discovery potential at hadron colliders is quantified in Fig.~\ref{fig:typeiiDisc}. In particular, following the analysis of Ref.~\cite{Perez:2008ha}, Figs.~\ref{fig:type2_TripletContour14TeV} and \ref{fig:type2_TripletContour100TeV} show event contours in the ${\rm BR}(H^{++}\to \mu^+\mu^+)$ vs. $M_{H^{\pm\pm}}$ plane after  $\mathcal{L}=300~(3000)\invfb$ of data at $\sqrt{s}=14$ TeV and  100 TeV, respectively. At the $2\sigma$ level, one finds the sensitivity to doubly charged Higgs is about $M_{H^{\pm\pm}}=0.75 ~(1.1)$ TeV at 14 TeV and $M_{H^{\pm\pm}}=2~(3.5)$ TeV at 100 TeV. In Figs.~\ref{fig:type2_vbfHiggs_emuSig} and \ref{fig:type2_vbfHiggs_tauSig}, one similarly has the signal significance $\sigma = S/\sqrt{S+B}$ after $\mathcal{L}=1$ and $3\invab$ at the 14 TeV LHC for VBF production of doubly charged Higgs pairs and their decays to $e^\pm\mu^\pm$ and $\tau^\pm\tau^\pm$ final-states, respectively~\cite{Dutta:2014dba}.
Upon the fortuitous discovery of a doubly charged scalar, however, will require also observing other charged scalars to determine its precise weak isospin and hypercharge quantum numbers~\cite{Bambhaniya:2013wza,delAguila:2013yaa,delAguila:2013mia}.

In light of such sensitivity at hadron colliders, it is unsurprising then that null results from searches at the 7/8/13 TeV LHC~\cite{Chatrchyan:2012ya,ATLAS:2014kca,ATLAS:2017iqw,CMS:2017pet} have placed stringent constraints on EW-scale triplet scalar masses, assuming benchmark branching rates. As seen in Fig.~\ref{fig:type2ExclATLAS}, results from the ATLAS experiment in searches for doubly charged Higgs pairs decaying to leptons, after collecting $\mathcal{L}=36\invfb$ of data at 13 TeV, have ruled out $M_{H^{\pm\pm}} > 600-900$ GeV at 95\% CLs in both the (a) single-flavor and (b) mixed light-lepton final states~\cite{ATLAS:2017iqw}. Comparable limits have been reached by the CMS experiment~\cite{CMS:2017pet}.

At future $e^-e^+$ colliders, triplet scalars can appear in $t$-channel exchanges, inducing charged lepton flavor violation (cLFV) and forward-backward asymmetries~\cite{Nomura:2017abh}; in three-body decays of taus that are absent of light-neutrinos in the final state, \ie, $\tau^\pm \to \ell^\mp H^{\pm\pm *} \to \ell^\mp \mu^\pm \mu^\pm$~\cite{Hays:2017ekz}; and, of course, in pairs via $s$-channel gauge currents~\cite{Frank:1995ex}.
In the event of such observations, the nontrivial conversion of an $e^-e^+$ beam into an $e^-e^-/e^-\mu^-/\mu^-\mu^-$ facility
could provide complimentary information on scalar triplet Yukawa couplings by means of the ``inverse'' $0\nu\beta\beta$ processes,
$\ell^-_i\ell^-_j \to W^-_{L/R} W^-_{L/R}$~\cite{Rizzo:1982kn,Rodejohann:2010jh,Barry:2012ga}.

\subsubsection{Triplet Higgs Scalars and the Left-Right Symmetric Model at Colliders}\label{sec:type2LRSM}

\begin{figure}[!t]
\begin{center}
\subfigure[]{\includegraphics[scale=1,width=0.48\textwidth]{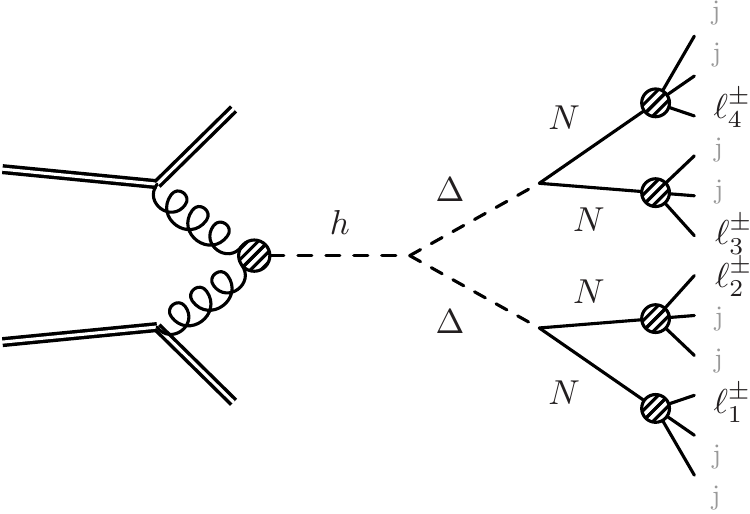}	\label{fig:feynman_LRSM_Higgs4NR}	}
\subfigure[]{\includegraphics[scale=1,width=0.48\textwidth]{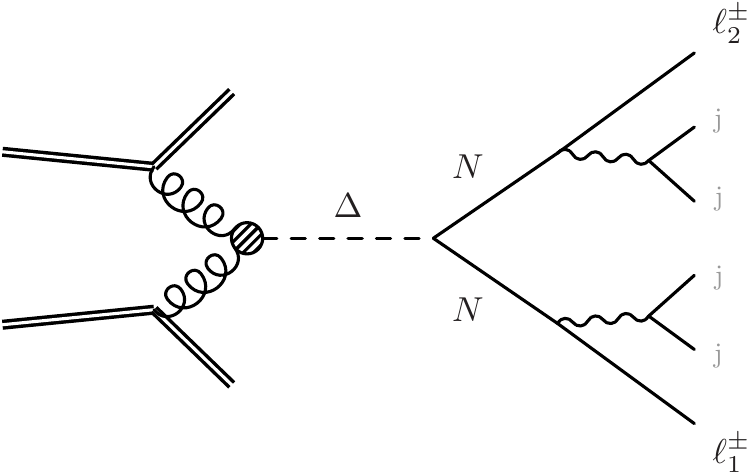}	\label{fig:feynman_LRSM_Triplet2NR}	}
\end{center}
\caption{Feynman diagrams depicting gluon fusion production of Majorana neutrinos via (a) SM Higgs boson $(h)$
and (b) SU$(2)_R$ triplet Higgs $(H)$ through their mixing in $pp$ collisions~\cite{Maiezza:2015lza,Nemevsek:2016enw}.
}
\label{fig:majoranaHiggsDiagrams}
\end{figure}

\begin{figure}[!t]
\begin{center}
\subfigure[]{\includegraphics[scale=1,width=0.48\textwidth]{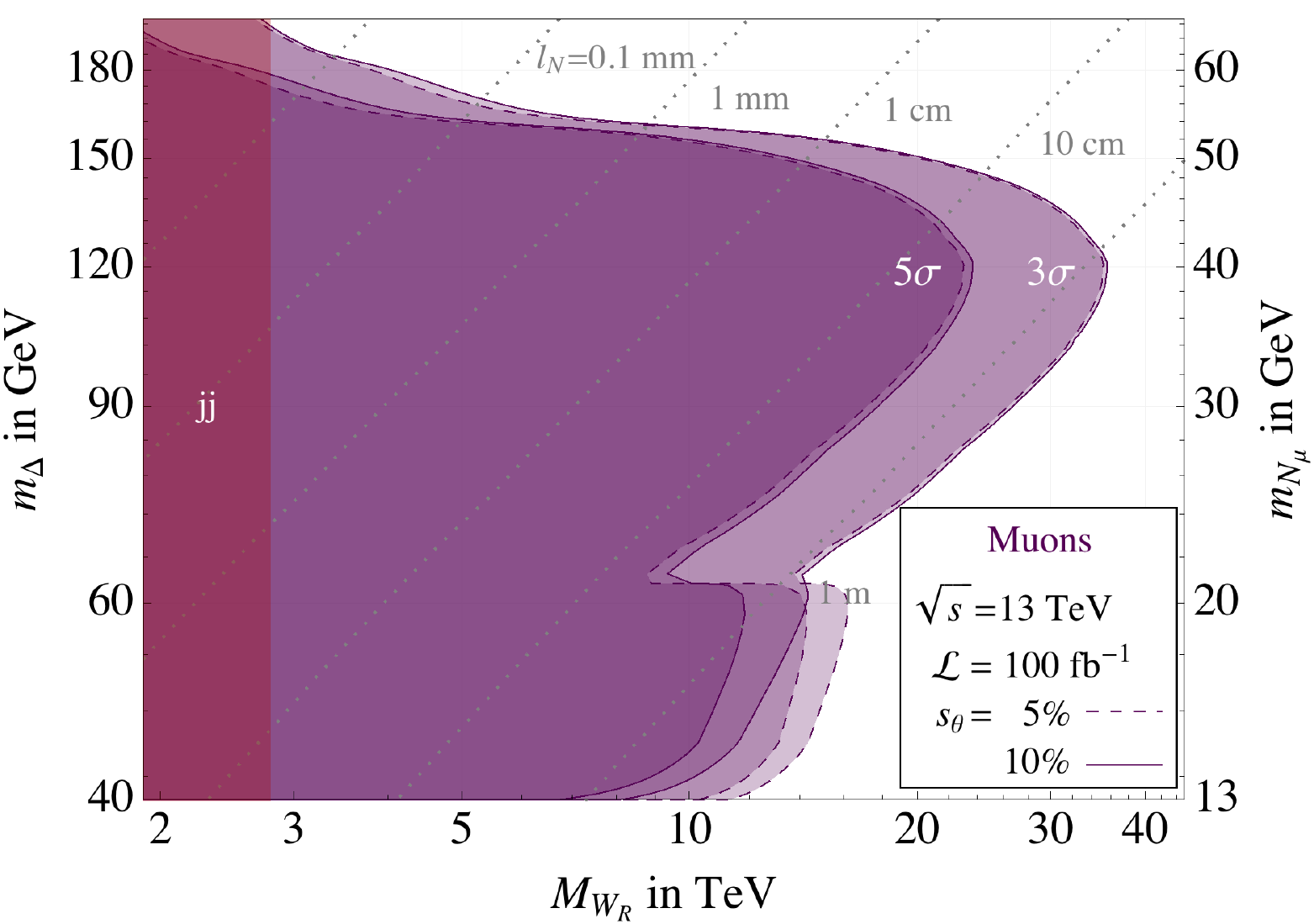}		\label{fig:lrsm_MajorHiggs_Excl_El} }
\subfigure[]{\includegraphics[scale=1,width=0.48\textwidth]{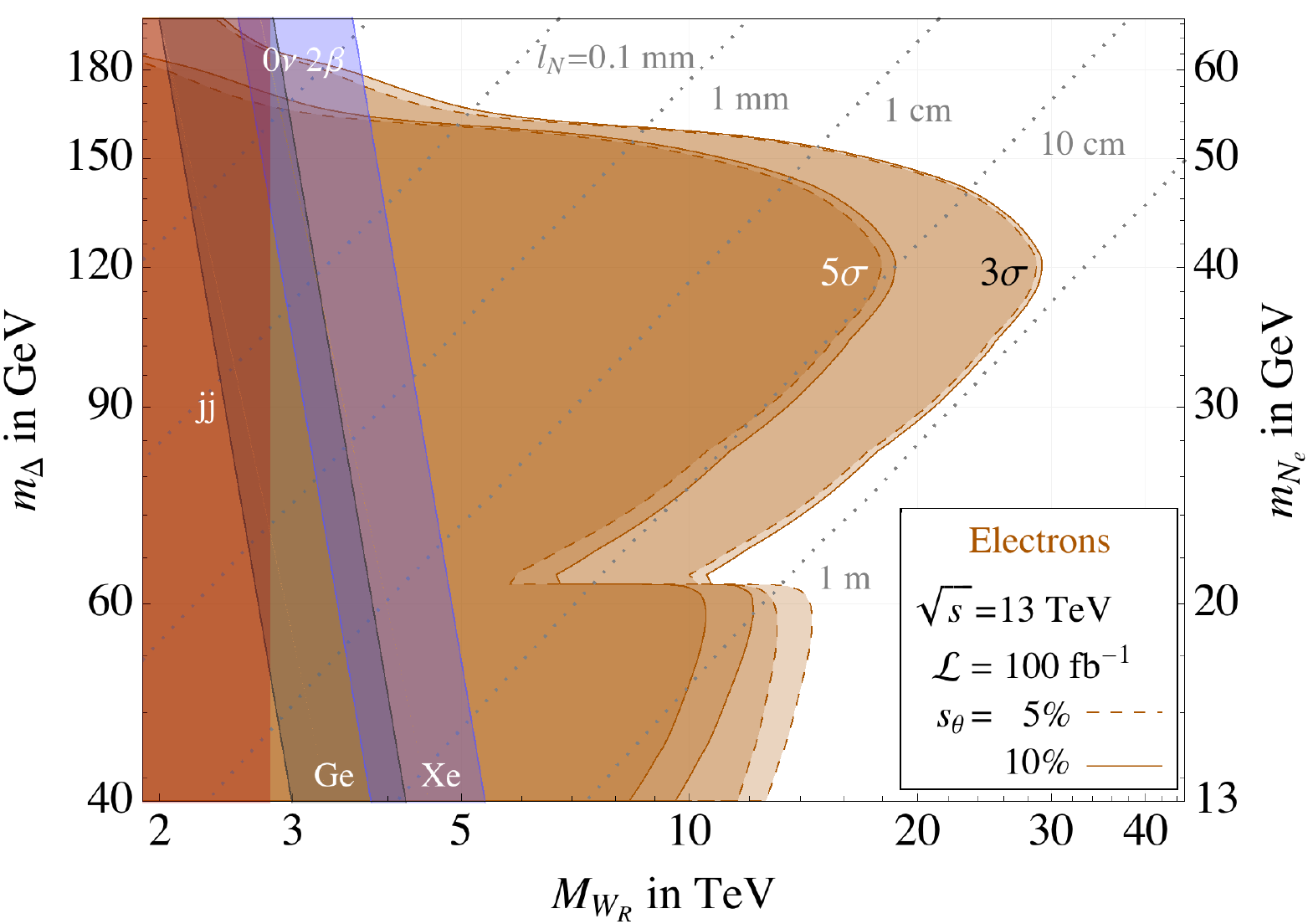}		\label{fig:lrsm_MajorHiggs_Excl_Mu} }
\end{center}
\caption{
(a,b) 13 TeV LHC sensitivity to the LRSM in the $(M_N,M_{W_R})$ plane to the processes shown
in Fig.~\ref{fig:majoranaHiggsDiagrams} after $\mathcal{L}=100\invfb$
~\cite{Maiezza:2015lza,Nemevsek:2016enw}.
}
\label{fig:majoranaHiggs}
\end{figure}

Turning to scalars in the LRSM, as introduced in Sec.~\ref{sec:hybrid},
it was recently  observed~\cite{Maiezza:2015lza,Nemevsek:2016enw} that in a certain class of neutrino mass models,
decays of the SM-like Higgs boson $h(125~{\rm GeV})$ to heavy neutrino pairs, $h\to NN$, may occur much more readily than previously thought.
The significance of this reaction is one's ability to confirm neutrino masses are generated, in part, through EWSB.
It would also indicate sensitivity to the scalar sector responsible for generating RH Majorana masses. 
Interactions between SM particles and $N$ typically proceed through heavy-light neutrino mixing, $\vert V_{\ell N}\vert$, which, is a numerically small quantity.
As $h\to NN$ involves two $N$, the issue is compounded and usually renders the decay rate prohibitively small in a pure Type I scenario. 
For $H\in\{H^0,~H^\pm,~H^{\pm\pm}\}$ predicted in Type I+II Seesaws, and in particular the LRSM, the situation is more interesting:
it may be that $h(125\gev)$ and the RH neutral scalars mix sufficiently that decays to relatively light $(2M_N < 125\gev)$ heavy neutrino pairs are possible~\cite{Maiezza:2015lza}.
This is allowed as $H$ can couple appreciable to $N$ and the mixing between $H^0$ and $h$ is much less constrained.
Subsequently, the na\"ive neutrino mixing suppression is avoided by exploiting that $h\to N N$ decays can proceed instead through $H^0-h$ mixing.
In a similar vein, it may be possible for $h$ to decay to triplet pairs and subsequently to $N$ or same-sign charged leptons,
or for single $H^0$ production to proceed directly~\cite{Nemevsek:2016enw}.
Such processes are shown diagrammatically in Fig.~\ref{fig:majoranaHiggsDiagrams}.
As a result, the $L$-violating Higgs decays,
\begin{eqnarray}
 h(125\gev) &\to& ~N ~N ~\to ~W_R^{\pm *} ~W_R^{\pm *} ~\ell_1^\mp ~\ell_2^\mp ~\to ~\ell_1^\mp ~\ell_2^\mp ~+~ nj,
 \\
 h(125\gev) &\to& ~H^0 ~H^0 ~\to ~4N ~\to ~\ell_1^\pm ~\ell_2^\mp  ~\ell_3^\pm ~\ell_4^\mp ~+~ nj,
 \\
 h(125\gev) &\to&  H^{++} ~H^{--} ~\to  ~\ell_1^\pm ~\ell_2^\pm  ~\ell_3^\mp ~\ell_4^\mp,
\end{eqnarray}
are not only possible, but also provide complementary coverage of low-mass $N$ scenarios that are outside the reach of $0\nu\beta\beta$ experiments and direct searches for $W_R$ at colliders.
The sensitivity of such modes are summarized in Fig.~\ref{fig:majoranaHiggs}~\cite{Maiezza:2015lza,Nemevsek:2016enw}. The associated production channels,
\begin{equation}
 p p \to H^{0,\pm\pm}~W_R^\mp \quad\text{and}\quad pp\to H^0 Z_R,
\end{equation}
are also possible. However, as in the SM, these channels are $s$-channel and phase space suppressed,
which lead to prohibitively small cross sections in light of present mass limits~\cite{Bambhaniya:2013wza}.

Lastly, one should note that the search for such Higgs decays is not limited to hadron colliders.
As presently designed future lepton colliders are aimed at operating as Higgs factories,
searches for such $L$-violating Higgs decays~\cite{Atwood:2007zza,Ren:2008sz,Yue:2010zzb} at such facilities represent an attractive discovery prospect.
In this context, a relatively understudied topic is the possible manifestation of Seesaw in precision measurements of the known SM-like Higgs boson~\cite{BhupalDev:2012zg,Maiezza:2015lza,Banks:2008xt}.
Some related studies also exist in the literature such as for generic pheno~\cite{Han:2007bk,delAguila:2013mia,Han:2007bk};
for little Higgs~\cite{Hektor:2007uu,Han:2005nk};
and for decay ratios and mixing patterns of exotically charged Higgs\cite{Kadastik:2007yd,Chun:2003ej}.

\section{The Type III Seesaw and Lepton Number Violation at Colliders}\label{sec:type3}

We now turn to collider searches for lepton number violation in the context of the Type III Seesaw mechanism~\cite{Foot:1988aq}
as well as its embedding in GUTs and other SM extensions.
In some sense, the Type III model is the fermionic version of the Type II scenario,
namely that Seesaw partner fermions couple to the SM via both weak gauge and Yukawa couplings.
Subsequently, much of the Type III collider phenomenology resembles that of Type I-based models.
However, quantitatively, the presence of gauge couplings lead to a very different outlook and level of sensitivity.
We now summarize the main highlights of the canonical Type III Seesaw (Sec.~\ref{sec:type3Canon}),
Type III-based models (Sec.~\ref{sec:type3Extend}),
and then review their $L$-violating collider phenomenology (Sec.~\ref{sec:type3Collider}).
As with the previous Seesaw scenarios, a discussion of cLFV is outside the scope of this review.
For recent summaries on cLFV in the Type III Seesaw, see Refs.~\cite{Abada:2007ux,Abada:2008ea,Eboli:2011ia,Agostinho:2017biv} and references therein.

\subsection{Type III Seesaw Models}\label{sec:type3Theory}

\subsubsection{The Canonical Type III Seesaw Mechanism}\label{sec:type3Canon}
In addition to the SM field content, the Type III Seesaw~\cite{Foot:1988aq} consists of SU$(2)_L$ triplet~(adjoint) leptons,
\begin{eqnarray}
&&\Sigma_L=  \Sigma^a_L \sigma^a = \left(
  \begin{array}{cc}
    \Sigma_L^0/\sqrt{2} & \Sigma^+_L \\
    \Sigma_L^- & -\Sigma_L^0/\sqrt{2} \\
  \end{array}
\right) \ ,
\quad \Sigma^\pm_L \equiv \frac{\Sigma^1_L \mp i\Sigma^2_L}{\sqrt{2}}, \quad \Sigma_L^0=\Sigma_L^3,
\end{eqnarray}
which transform as $(1,3,0)$ under the SM gauge group.
Here $\Sigma^\pm_L$ have U$(1)_{\rm EM}$ charges $Q=\pm1$, and the $\sigma^{a}$ for $a=1,\dots,3$, are the usual Pauli SU$(2)$ matrices.
The RH conjugate fields are related by
\begin{equation}
 \Sigma_R^c = \begin{pmatrix} \Sigma^{0 c}_R/\sqrt{2} & \Sigma^{-c}_R \\ \Sigma^{+c} & -\Sigma^{0 c}_R/\sqrt{2} \end{pmatrix},
 \quad\text{for}\quad \psi_R^c \equiv (\psi^c)_R = (\psi_L)^c.
\end{equation}
The Type III Lagrangian is given by the sum of the SM Lagrangian, the triplet's kinetic and mass terms,
\begin{equation}
 \mathcal{L}_{T} =
 \frac{1}{2}\Tr\left[\overline{\Sigma_L}i\not\!\!D\Sigma_L\right]
-\left(\frac{M_\Sigma}{2}\overline{\Sigma_L^0}\Sigma_R^{0c} + M_\Sigma\overline{\Sigma^-_L}\Sigma^{+c}_R + \text{H.c.}\right),
\label{eq:typeIIIMassLag}
\end{equation}
and the triplet's Yukawa coupling to the SM LH lepton $(L)$ and Higgs $(H)$ doublet fields,
\begin{equation}
 \mathcal{L}_Y = -Y_\Sigma \overline{L} ~\Sigma_R^c ~i\sigma^2 H^* + \text{H.c.}
 \label{eq:typeIIIYukawa}
\end{equation}
From Eq.~(\ref{eq:typeIIIYukawa}), one can deduce the emergence of a Yukawa coupling between the charged SM leptons and the charged triplet leptons.
This, in turn, induces a mass mixing among charged leptons that is similar to doublet-singlet and doublet-triplet neutrino mass mixing, and represents
one of the more remarkable features of the Type III mechanism. The impact of EW fermion triplets on the SM Higgs,
naturalness in the context of the Type III Seesaw has been discussed in Refs.~\cite{Gogoladze:2008ak,He:2012ub,Gogoladze:2010in}.

After expanding Eqs.~(\ref{eq:typeIIIMassLag})-(\ref{eq:typeIIIYukawa}), the relevant charged lepton and neutrino mass terms are~\cite{Li:2009mw}
\begin{align}
\mathcal{L}^m_{\rm III}= - &\left(
  \begin{array}{cc}
    \overline{l_R} & \overline{\Psi_R} \\
  \end{array}
\right) \left(
  \begin{array}{cc}
    m_l & 0 \\
    Y_\Sigma v_0 & M_\Sigma \\
  \end{array}
\right) \left(
  \begin{array}{c}
    l_L \\
    \Psi_L \\
  \end{array}
\right)
\nonumber\\
& -\left(
  \begin{array}{cc}
    \overline{\nu_L^c} & \overline{\Sigma_L^{0c}} \\
  \end{array}
\right) \left(
  \begin{array}{cc}
    0 & Y_\Sigma^Tv_0/2\sqrt{2} \\
    Y_\Sigma v_0/2\sqrt{2} & M_\Sigma/2 \\
  \end{array}
\right) \left(
  \begin{array}{c}
    \nu_L \\
    \Sigma^0_L \\
  \end{array}
\right) + \text{H.c.},
\label{mass-matrix}
\end{align}
with $\Psi_L\equiv\Sigma_L^-$, $\Psi_R\equiv \Sigma_L^{+c}$, and $\Psi = \Psi_L + \Psi_R$.
After introducing unitarity matrices to transit light doublet and heavy triplet lepton fields as below
\begin{eqnarray}
&&\left(
    \begin{array}{c}
      l_{L,R} \\
      \Psi_{L,R} \\
    \end{array}
  \right)
  =U_{L,R}
  \left(
    \begin{array}{c}
      l_{mL,R} \\
      \Psi_{mL,R} \\
    \end{array}
  \right), \ \ \
\left(
    \begin{array}{c}
      \nu_{L} \\
      \Sigma^0_{L} \\
    \end{array}
  \right)
  =U_0
  \left(
    \begin{array}{c}
      \nu_{mL} \\
      \Sigma^0_{mL} \\
    \end{array}
  \right),\\
&&U_L\equiv
\left(
  \begin{array}{cc}
    U_{Lll} & U_{Ll\Psi} \\
    U_{L\Psi l} & U_{L\Psi\Psi} \\
  \end{array}
\right), \ \ U_R\equiv \left(
  \begin{array}{cc}
    U_{Rll} & U_{Rl\Psi} \\
    U_{R\Psi l} & U_{R\Psi\Psi} \\
  \end{array}
\right), \ \ U_0\equiv \left(
  \begin{array}{cc}
    U_{0\nu\nu} & U_{0\nu\Sigma} \\
    U_{0\Sigma \nu} & U_{0\Sigma\Sigma} \\
  \end{array}
\right),
\end{eqnarray}
one obtains the diagonal mass matrices and mass eigenvalues for neutrinos and charged leptons,
\begin{align}
{\rm diag}(\mathcal{N}) =
U_0^\dagger \left(
  \begin{array}{cc}
    0 & Y_\Sigma^\dagger v_0/\sqrt{2} \\
    Y_\Sigma^\ast v_0/\sqrt{2} & M_\Sigma^\ast \\
  \end{array}
\right) U_0^\ast = \left(
  \begin{array}{cc}
    m_\nu^{diag} & 0 \\
    0 & M_{N}^{diag} \\
  \end{array}
\right),
\\
{\rm diag}(\mathcal{E}) =
\ U_L^\dagger \left(
  \begin{array}{cc}
    m_l^\dagger & Y_\Sigma^\dagger v_0 \\
    0 & M_\Sigma^\dagger\\
  \end{array}
\right) U_R = \left(
  \begin{array}{cc}
    m_l^{diag} & 0 \\
    0 & M_{E}^{diag} \\
  \end{array}
\right).
\end{align}
The light neutrino mass eigenstates are denoted by $\nu_{j}$ for $j=1,\dots,3$; whereas
the heavy neutral and charged leptons are respectively given by $N_{j'}$ and $E^\pm_{k'}$.
In the literature, $N$ and $E^\pm$ are often denoted as $T^0,~T^\pm$ or $\Sigma^0,~\Sigma^\pm$.
However, there is no standard convention as to what set of symbols are used to denote gauge and mass eigenstates.
Where possible, we follow the convention of Ref.~~\cite{Arhrib:2009mz} and generically denote triplet-doublet mixing by $Y_T$ and $\varepsilon_T$.
This means that in the mass basis, triplet gauge states are given by
\begin{eqnarray}
 \Psi^\pm = Y_T~E^\pm 	+ \sqrt{2} \varepsilon_T~ \ell^\pm & \quad\text{and}\quad & \Psi^0   = Y_T~N	+ \varepsilon_T~ \nu_m,
 \nonumber\nonumber\\
 \text{with}\quad \vert Y_T\vert\sim \mathcal{O}(1)	 & \quad\text{and}\quad &   \vert\varepsilon_T\vert\sim {Y_\Sigma v_0\over \sqrt{2}M_\Sigma}\ll1.
\end{eqnarray}
The resulting interaction Lagrangian, in the mass eigenbasis then contains~\cite{Arhrib:2009mz}
\begin{eqnarray}
 \mathcal{L}_{\rm Type~III}^{\rm Mass~Basis} &\ni&
- \overline{E^-_{k'}}\left( e Y_T A_\mu\gamma^\mu  + g\cos\theta_W Y_T Z_\mu\gamma^\mu \right)E^-_{k'}  -g Y_T \overline{E^-_{k'}} W^-_\mu\gamma^\mu N_{j'}
 \nonumber \\
 &-&
 \frac{e}{2s_wc_w}Z_\mu \left(\varepsilon_T \overline{N_{j'}} \gamma^\mu P_R \nu_j +\sqrt{2} \varepsilon_T \overline{E^-_{k'}} \gamma^\mu P_R \ell^-_k
\right)
\nonumber\\
&-&
\frac{e}{s_w}W_\mu^+ \left(\varepsilon_T \overline{\nu_j} \gamma^\mu P_L E^-_{k'} + \frac{1}{\sqrt{2}} \varepsilon_T \overline{N_{j'}} \gamma^\mu P_R \ell^-_k
\right)
+{\rm H.c.}
\label{int-typeiii}
\end{eqnarray}
From this, one sees a second key feature of the Type III Seesaw, that gauge interactions between heavy lepton pairs proceeds largely through pure vector currents
with axial-vector deviations (not shown) suppressed by $\mathcal{O}(\varepsilon_T^2)$ at the Lagrangian level.
This follows from the triplet fermions vector-like nature.
Similarly, the mixing-suppressed gauge couplings between heavy and light leptons proceeds through SM-like currents.

Explicitly, the light and heavy neutrino mass eigenvalues are
\begin{eqnarray}
m_\nu\approx {Y_\Sigma^2v_0^2\over 2M_\Sigma}, \ \ \ M_{N}\approx M_\Sigma,
\end{eqnarray}
and for the charged leptons are
\begin{eqnarray}
m_l-m_l{Y_\Sigma^2 v_0^2\over 2M_\Sigma^2}\approx m_l, \ \ \ M_E\approx M_\Sigma.
\end{eqnarray}
This slight deviation in the light, charged leptons' mass eigenvalues implies a similar variation in the anticipated Higgs coupling to the same charged leptons.
At tree-level, the heavy leptons $N$ and $E^\pm$ are degenerate in mass, a relic of SU$(2)_L$ gauge invariance.
However, after EWSB, and for $M_{\Sigma} \gtrsim 100$ GeV, radiative corrections split this degeneracy by~\cite{Arhrib:2009mz},
\begin{eqnarray}
&& \Delta M_T\equiv M_{E}-M_{N}=\frac{\alpha_W}{2\pi}\frac{M_W^2}{M_\Sigma}
\left[f\left(\frac{M_\Sigma}{M_Z}\right)-f\left(\frac{M_\Sigma}{M_W}\right)\right] \approx 160\MeV, \\
&& \text{where}\quad~f\left(y\right)=\frac{1}{4y^2}\log{y^2}- \left(1+\frac{1}{2y^2}\right)
\sqrt{4y^2-1}\arctan{\sqrt{4y^2-1}},
\end{eqnarray}
and opens the $E^\pm \to N \pi^\pm$ decay mode.
Beyond this are the heavy lepton decays to EW bosons and light leptons that proceed through doublet-triplet lepton mixing.
The mixings are governed by the elements in the unitary matrices $U_{L,R}$ and $U_0$. Expanding $U_{L,R}$ and $U_0$ up to order $Y_\Sigma^2v_0^2M_\Sigma^{-2}$, one gets the following results~\cite{Abada:2008ea,He:2009tf}
\begin{eqnarray}
&&U_{Lll} = 1- \epsilon\;,\;\;U_{Ll\Psi} = Y^\dagger_\Sigma
M^{-1}_\Sigma v_0\;,\;\;\;\;\;\;\;
U_{L\Psi l} = - M^{-1}_\Sigma Y_\Sigma v_0\;,\;\;\;\;\;\;U_{L\Psi\Psi} = 1-\epsilon'\;,\nonumber\\
&&U_{Rll} = 1\;,\;\;\;\;\;\;\;\;U_{Rl\Psi} = m_l Y^\dagger_\Sigma
M^{-2}_\Sigma v_0\;,\;\;\;
U_{R\Psi l} = - M^{-2}_\Sigma Y_\Sigma m_l v_0\;,\;\;U_{R\Psi\Psi} = 1\;,\nonumber\\
&&U_{0\nu\nu} = (1- \epsilon/2)U_{PMNS}\;,\;\; U_{0\nu \Sigma} =
Y^\dagger_\Sigma M^{-1}_\Sigma v_0/\sqrt{2}\;,\;\;
U_{0\Sigma \nu} = - M^{-1}_\Sigma Y_\Sigma U_{0\nu\nu} v_0/\sqrt{2}\;,\nonumber\\
&&U_{0\Sigma\Sigma } =
1-\epsilon'/2\;,\;\;\;\;\;\;\;\; \epsilon =
Y^\dagger_\Sigma M^{-2}_\Sigma Y_\Sigma v_0^2/2\;,\;\;\;\; \epsilon'
= M^{-1}_\Sigma Y_\Sigma Y^\dagger_\Sigma M^{-1}_\Sigma
v_0^2/2\;.\nonumber \label{approx}
\end{eqnarray}
To the order of $Y_\Sigma v_0M_\Sigma^{-1}$, the mixing between the SM charged leptons and triplet leptons, \ie,
$V_{\ell N}=-Y_\Sigma^\dagger v_0 M_\Sigma^{-1}/\sqrt{2}$, follows the same relation as Eq.~(\ref{typei}) in the Type I Seesaw~\cite{Li:2009mw} and the couplings in the interactions in Eq.~(\ref{int-typeiii}) are all given by $V_{\ell N}$~\cite{AguilarSaavedra:2009ik,Li:2009mw}.

\begin{figure}[!t]
\includegraphics[scale=1,width=.95\textwidth]{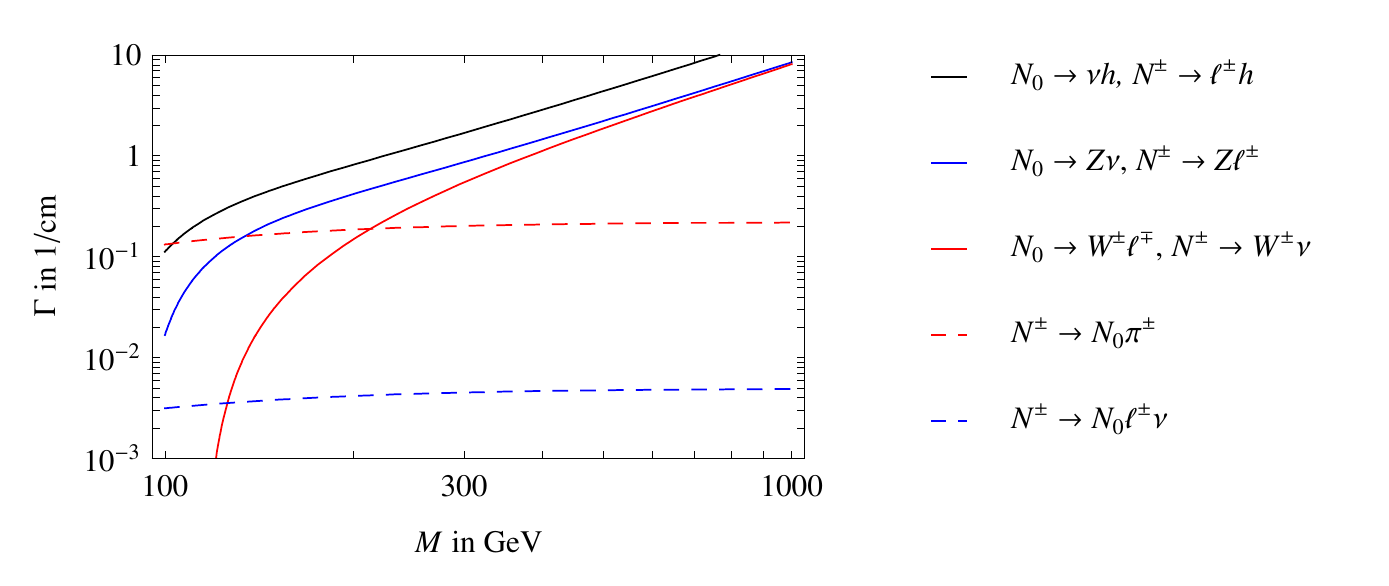}
\caption{Triplet decay widths as function of the triplet mass and assuming $M_{h_{\rm SM}}=115\GeV$~\cite{Franceschini:2008pz}.
}
\label{fig:type3Width}
\end{figure}

Hence, the partial widths for both the heavy charged lepton and heavy neutrino are proportional to $|V_{\ell N}|^2$.
For $M_E\approx M_N\gg M_W, M_Z, M_{h}$, the partial widths behave like~\cite{delAguila:2008cj,AguilarSaavedra:2009ik}
\begin{eqnarray}
&&{1\over 2}\Gamma(N\to \sum_\ell \ell^+ W^-+\ell^-W^+)\approx \Gamma(N\to \sum_\nu\nu Z+\bar{\nu}Z)\approx \Gamma(N\to \sum_\nu \nu h+\bar{\nu}h)\nonumber \\
&\approx &{1\over 2} \Gamma(E^\pm\to \sum_\nu \overset{(-)}\nu W^\pm)\approx \Gamma(E^\pm\to \sum_\ell \ell^\pm Z)\approx \Gamma(E^\pm\to \sum_\ell \ell^\pm h)
\nonumber\\
&\approx & {G_F\over 8\sqrt{2}\pi}\sum_\ell |V_{\ell N}|^2 M_\Sigma^3.
\end{eqnarray}
Thus the heavy lepton branching ratios exhibit
asymptotic behavior consistent with the Goldstone Equivalence Theorem~\cite{Chanowitz:1985hj,Lee:1977yc},
and are given by the relations~\cite{Franceschini:2008pz,delAguila:2008cj,AguilarSaavedra:2009ik,Arhrib:2009mz},
\begin{eqnarray}
&&{1\over 2}{\rm BR}(N\to \sum_\ell \ell^+ W^-+\ell^-W^+)\approx {\rm BR}(N\to \sum_\nu\nu Z+\bar{\nu}Z)\approx {\rm BR}(N\to \sum_\nu \nu h+\bar{\nu}h)\nonumber \\
&\approx &{1\over 2} {\rm BR}(E^\pm\to \sum_\nu \overset{(-)}\nu W^\pm)\approx {\rm BR}(E^\pm\to \sum_\ell \ell^\pm Z)\approx {\rm BR}(E^\pm\to \sum_\ell \ell^\pm h)
\approx {1\over 4}.\quad
\end{eqnarray}
As displayed in Fig.~\ref{fig:type3Width} by Ref.~\cite{Franceschini:2008pz},
as the triplet mass grows, this asymptotic behavior can be seen explicitly in the triplet lepton partial widths.

\subsubsection{Type I+III Hybrid Seesaw in Grand Unified and Extended Gauge Theory}\label{sec:type3Extend}

One plausible possibility to rescue the minimal grand unified theory, \ie, SU$(5)$,
is to introduce an adjoint $24_F$ fermion multiplet in addition to the original $10_F$ and $\bar{5}_F$ fermionic representations~\cite{Ma:1998dn,Bajc:2006ia}.
As the $24_F$ contains both singlet and triplet fermions in this non-supersymmetric SU$(5)$,
the SM gauge couplings unify and neutrino masses can generated through a hybridization of the Types I and III Seesaw mechanisms.
The Yukawa interactions and Majorana masses in this Type I+III Seesaw read~\cite{Arhrib:2009mz}
\begin{eqnarray}
\Delta\mathcal{L}_{\rm I+III}^Y=Y_S LHS + Y_T LHT-{M_S\over 2} SS - {M_T\over 2} TT +{\rm H.c.},
\end{eqnarray}
where $S$ and $T=\left({T^-+T^+\over \sqrt{2}},{T^--T^+\over i\sqrt{2}},T^0\right)$ are the fermionic singlet and triplet fields, respectively,
with masses $M_S$ and $M_T$. In the limit that $M_{S},M_{T}\gg Y_S v_0, Y_Tv_0$,
the light neutrino masses are then given by the sum of the individual Type I and III contributions
\begin{eqnarray}
m_\nu=- (Y_S v_0/\sqrt{2})^2 M_S^{-1} - (Y_Tv_0/\sqrt{2})^2 M_T^{-1},
\end{eqnarray}
The most remarkable prediction of this SU$(5)$ theory is that the unification constraint and the stability of proton require the triplet mass to be small: $M_T\lesssim 1$ TeV~\cite{Bajc:2006ia,Dorsner:2006fx}.
Thus, in  SU$(5)$ scenarios, the triplet leptons of this Type I+III Seesaw are within the LHC's kinematic reach
and can be tested via $L$-violating
collider signatures~\cite{Ma:1998dn,Bajc:2007zf,Perez:2007rm,AristizabalSierra:2010mv,Aguilar-Saavedra:2013twa,Biggio:2010me}.

Other GUT models that can accommodate the Type III Seesaw and potentially lead to collider-scale $L$-violation
include variations of SO$(10)$~\cite{Chakrabortty:2010az} theories.
It is also possible to embed the Type III scenario into extended gauge sectors, including
Left-Right Symmetric theories~\cite{FileviezPerez:2008sr,Duerr:2013opa,Gu:2011yx,Mohapatra:2009fj},
which also represents a Type I+II+III hybrid Seesaw hat trick.
Additionally, Type III-based hybrid Seesaws can be triggered via fermions in other SU$(2)_L\times$U$(1)_Y$
representations~\cite{Delgado:2011iz,Ma:2013tda,Yu:2015pwa,Nomura:2017abu},
The collider phenomenology in many of these cases is very comparable to that of the Type I and II Seesaws,
as discussed in Secs.~\ref{sec:type1} and ~\ref{sec:type2}, or the more traditional Type III scenario, which we now discuss.

\subsection{Heavy Charged Leptons and Neutrinos at Colliders} \label{sec:type3Collider}

\subsubsection{Heavy Charged Leptons and Neutrinos at $pp$ Colliders}

\begin{figure}[!t]
\includegraphics[width=0.95\textwidth]{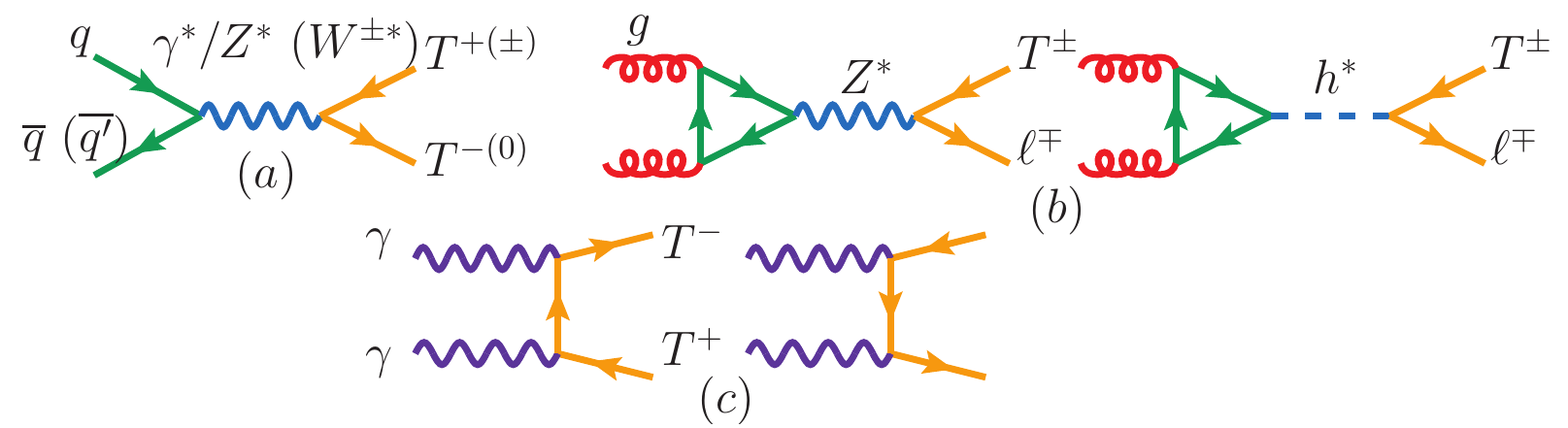}	
\caption{
Born level production of Type III lepton pairs via (a) Drell-Yan, (b) gluon fusion, and (c) photon fusion.}
\label{fig:feynman_Type3_MultiProd}
\end{figure}

\begin{figure}[!t]
\subfigure[]{\includegraphics[width=0.48\textwidth]{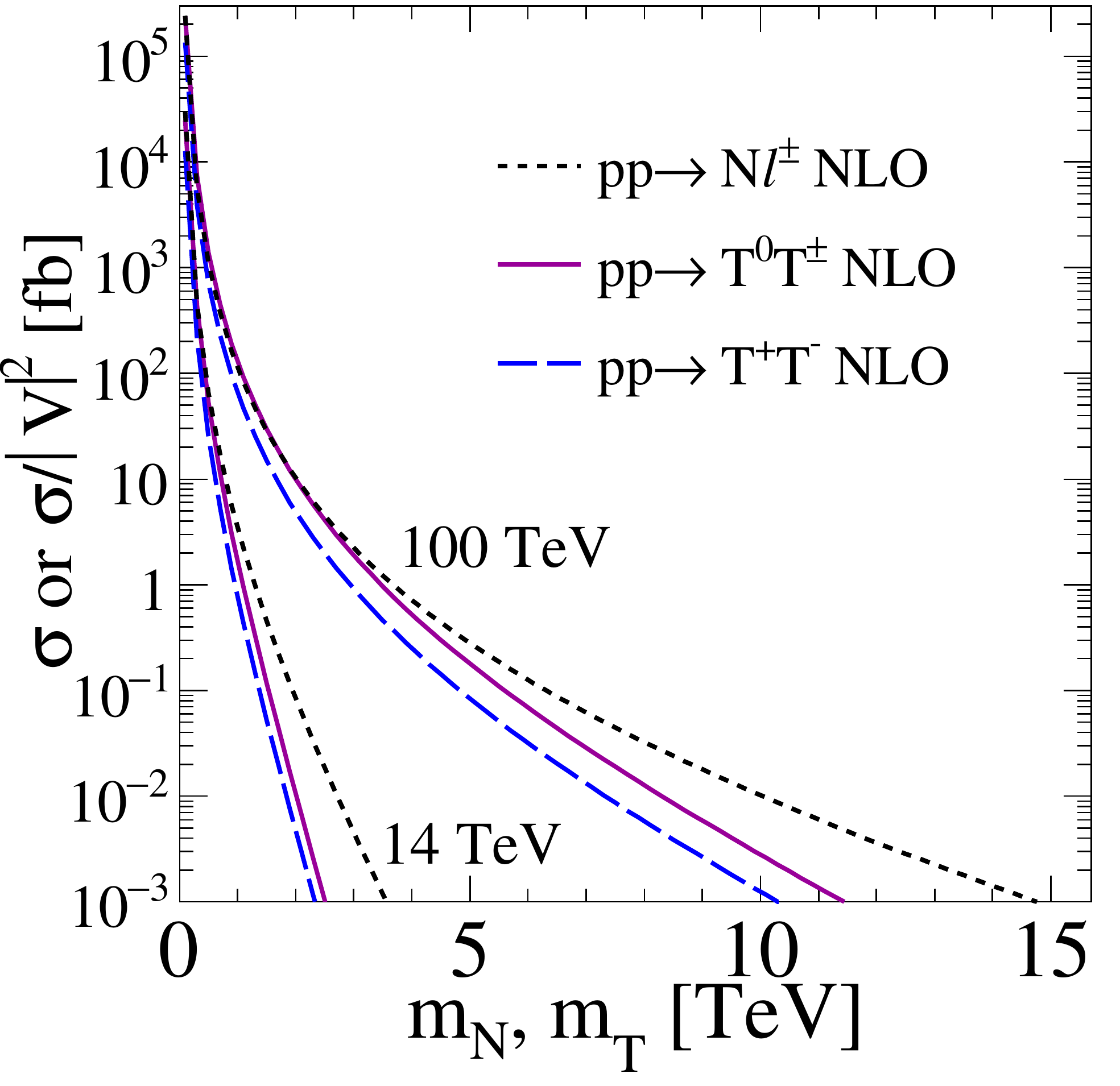}	\label{fig:SeesawTripletXSecMass}}
\subfigure[]{\includegraphics[width=0.48\textwidth]{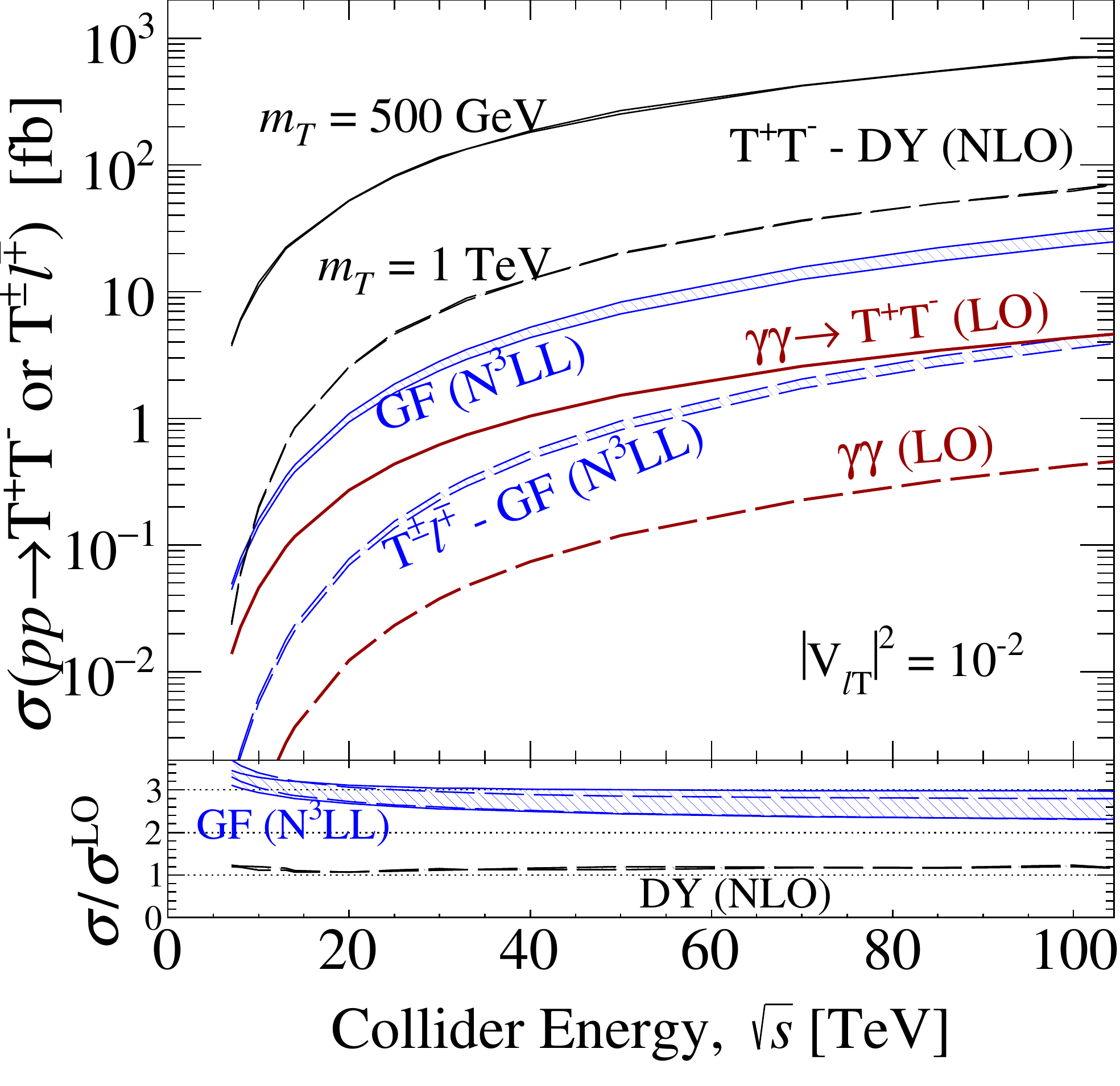}		\label{fig:SeesawTripletXSecBeam}}
\caption{
(a) As a function of mass, the $N\ell^\pm$ (singlet) as well as $T^+T^-$ and $T^\pm T^0$ (triplet)
DY production cross sections at $\sqrt{s}=14$ and 100 TeV.
(b) As a function of collider energy $\sqrt{s}$, the $T^+T^-$ and $T^\pm\ell^\mp$ (assuming benchmark $\vert V_{\ell T}\vert^2 = 10^{-2}$)
production cross sections via various production mechanisms.
}
\label{typeiii-xsec}
\end{figure}

Due to the presence of both gauge and Yukawa couplings to SM fields, the collider phenomenology for triplet leptons is exceedingly rich.
In hadron collisions, for example, pairs of heavy triplet leptons are produced dominantly via charged and neutral Drell-Yan (DY) currents,
given by
\begin{eqnarray}
q\bar{q}'\to W^{\ast\pm}\to T^\pm T^0  	\quad\text{and}\quad 	q\bar{q} \to \gamma^\ast/Z^\ast \to T^+ T^-,
\end{eqnarray}
and shown in Fig.~\ref{fig:feynman_Type3_MultiProd}(a).
For the DY process, the total cross section is now known up to NLO and differentially at NLO+LL in $k_T$ resummation \cite{Ruiz:2015zca}.
As function of mass, the $N\ell^\pm$ (singlet) as well as $T^+T^-$ and $T^\pm T^0$ (triplet) DY production cross sections
at $\sqrt{s}=14$ and 100 TeV are displayed in Fig.~\ref{fig:SeesawTripletXSecMass}.
While the three rates are na\"ively comparable, one should assign a mixing factor of $\vert V_{\ell N}\vert^2 \lesssim 10^{-2}$ to the singlet production
since it proceeds through active-sterile neutrino mixing, \ie,  Yukawa couplings, whereas triplet lepton pair production proceeds through gauge couplings.
Heavy triplet leptons can also be produced singly in the association with light leptons and neutrinos,
\begin{eqnarray}
q\bar{q}'\to W^{\ast\pm}\to T^\pm \nu, ~T^0\ell^\pm 	\quad\text{and}\quad 	q\bar{q} \to \gamma^\ast/Z^\ast \to T^\pm \ell^\mp.
\end{eqnarray}
As single production modes are proportional to the small~\cite{delAguila:2008pw} doublet-triplet mixing, denoted by $|V_{\ell T}|$,
these processes suffer from the same small signal rates at colliders as does singlet production in Type I-based Seesaws (see Sec.~\ref{sec:type1Canon}).
However, as heavy-light lepton vertices also posses axial-vector contributions, new production channels are present,
such as the gluon fusion mechanism~\cite{Willenbrock:1985tj,Dicus:1991wj,Hessler:2014ssa,Ruiz:2017yyf},
shown in Fig.~\ref{fig:feynman_Type3_MultiProd}(b) and given by
\begin{eqnarray}
 g g \to Z^* / h^* \to T^\pm \ell^\mp.
 \label{eq:type3GF}
\end{eqnarray}
It is noteworthy that the partonic expression for gluon fusion channels $gg \to Z^*/h^* \to T^\pm \ell^\mp$
is equal to the Type I analogue $gg\to N \nu_\ell$~\cite{Hessler:2014ssa}, 
and hence so are its QCD corrections~\cite{Ruiz:2017yyf}.
Conversely, heavy triplet pair production through gluon fusion, \ie,  $gg \to T\overline{T}$,
is zero since their couplings to weak bosons are vector-like, and hence vanish
according to Furry's Theorem~\cite{Willenbrock:1985tj,delAguila:1990yw,Dicus:1991wj}.
For $\sqrt{s} = 7-100$ TeV, the N$^3$LL(Threshold) corrections to the Born rates of Eq.~(\ref{eq:type3GF}) span $+160\%$ to $+260\%$~\cite{Ruiz:2017yyf}.
Hence, for singly produced triplet leptons, the gluon fusion mechanism is dominant over the DY channel for $\sqrt{s}\gtrsim 20-25$ TeV,
over a wide range of EW- and TeV-scale triplet masses~\cite{Hessler:2014ssa,Ruiz:2017yyf}.
More exotic production channels also exist, such as the $\gamma\gamma \to T^+ T^-$ VBF channel, shown in Fig.~\ref{fig:feynman_Type3_MultiProd}(c),
as well as permutations involving $W$ and $Z$. However, the $\gamma\gamma$ contributions is sub-leading due to coupling and phase space suppression.

For representative heavy lepton masses of $M_T = 500$ GeV and 1 TeV as well as doublet-triplet mixing of $|V_{\ell T}|^2 = 10^{-2}$,
we display in Fig.~\ref{fig:SeesawTripletXSecBeam}  the
$pp\to T^+T^-$ and $T^\pm\ell^\mp$ production cross sections via various hadronic production mechanisms as a function of collider energy $\sqrt{s}$.
In the figure, the dominance of pair production over single production is unambiguous.
Interestingly, considering that the triplet mass splitting is $\Delta M_T \sim \mathcal{O}(200)$ MeV as stated above,
one should not expect to discover the neutral current single production mode without also observing the charged channel almost simultaneously.
Hence, despite sharing much common phenomenology, experimentally differentiating a Type I scenario from a Type III (or I+III) scenario is straightforward.

Leading order-accurate Monte Carlo simulations for tree-level processes involving Type III leptons are possible
with the \texttt{Type III Seesaw} FeynRules UFO model~\cite{Biggio:2011ja,Abada:2008ea,FeynRules:TypeIII},
as well as a Minimal Lepton Flavor Violation variant \texttt{MLFV Type III Seesaw}~\cite{Eboli:2011ia,Agostinho:2017biv,FeynRules:Type3mlfv}.
The models can be ported into modern, general-purpose event generators,
such at Herwig~\cite{Bellm:2015jjp}, MadGraph5\_aMC@NLO~\cite{Alwall:2014hca}, and Sherpa~\cite{Gleisberg:2008ta}.

\begin{figure}[!t]
\begin{center}
\subfigure[]{\includegraphics[scale=1,width=.48\textwidth,,height=6cm]{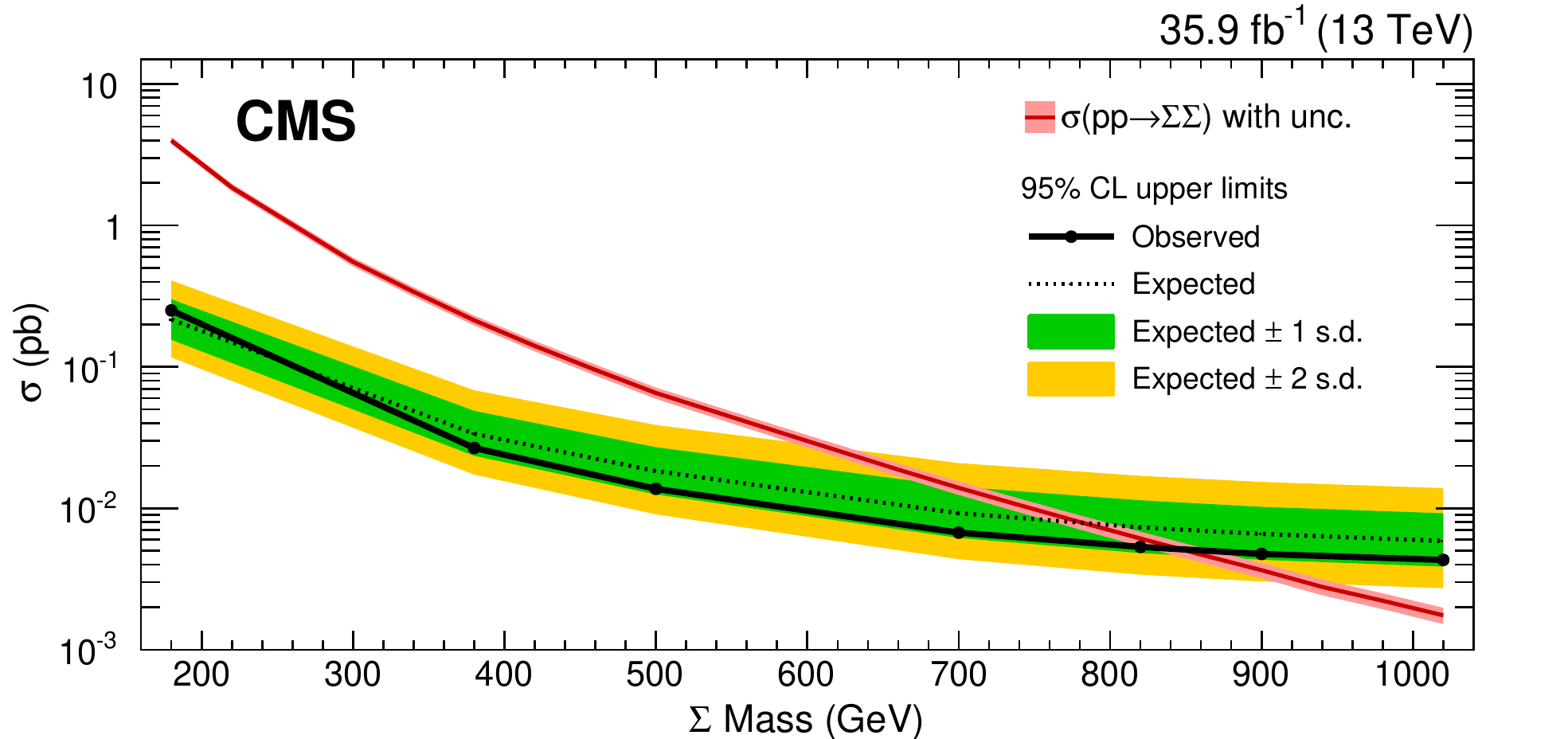}}
\subfigure[]{\includegraphics[scale=1,width=.48\textwidth]{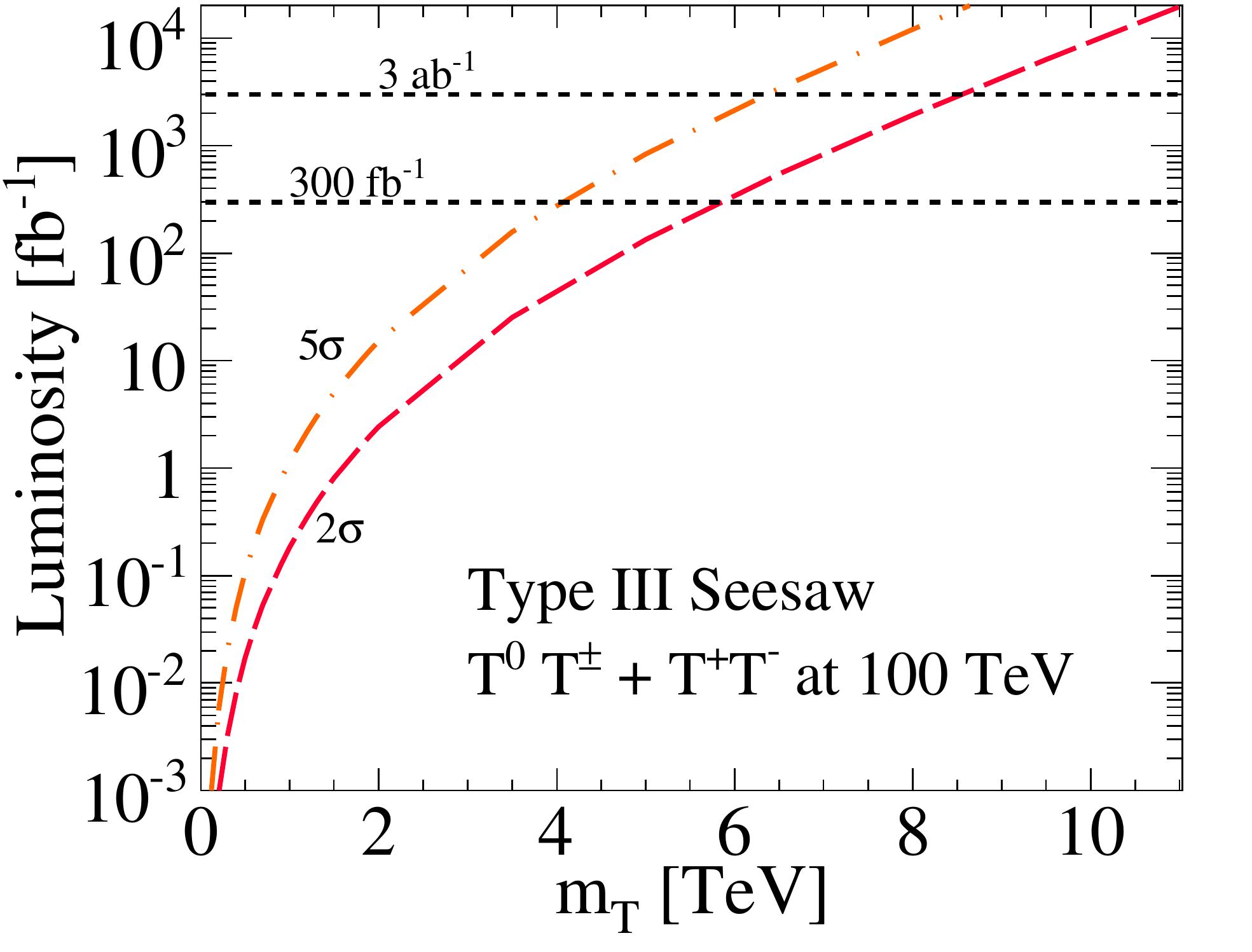}}
\\
\subfigure[]{\includegraphics[scale=1,width=.45\textwidth]{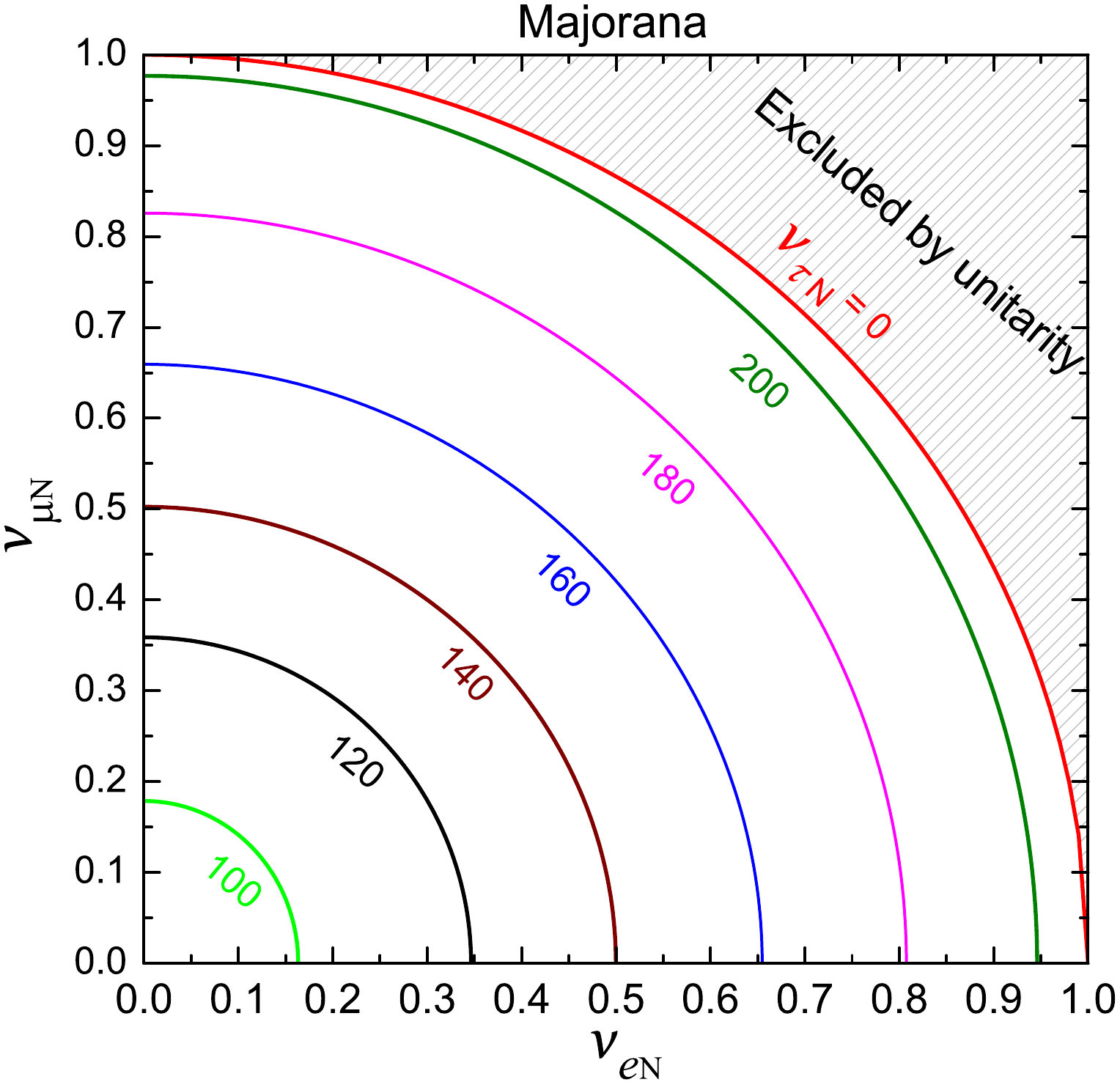}	\label{fig:type3MixingLimits_MuE}}	
\subfigure[]{\includegraphics[scale=1,width=.45\textwidth]{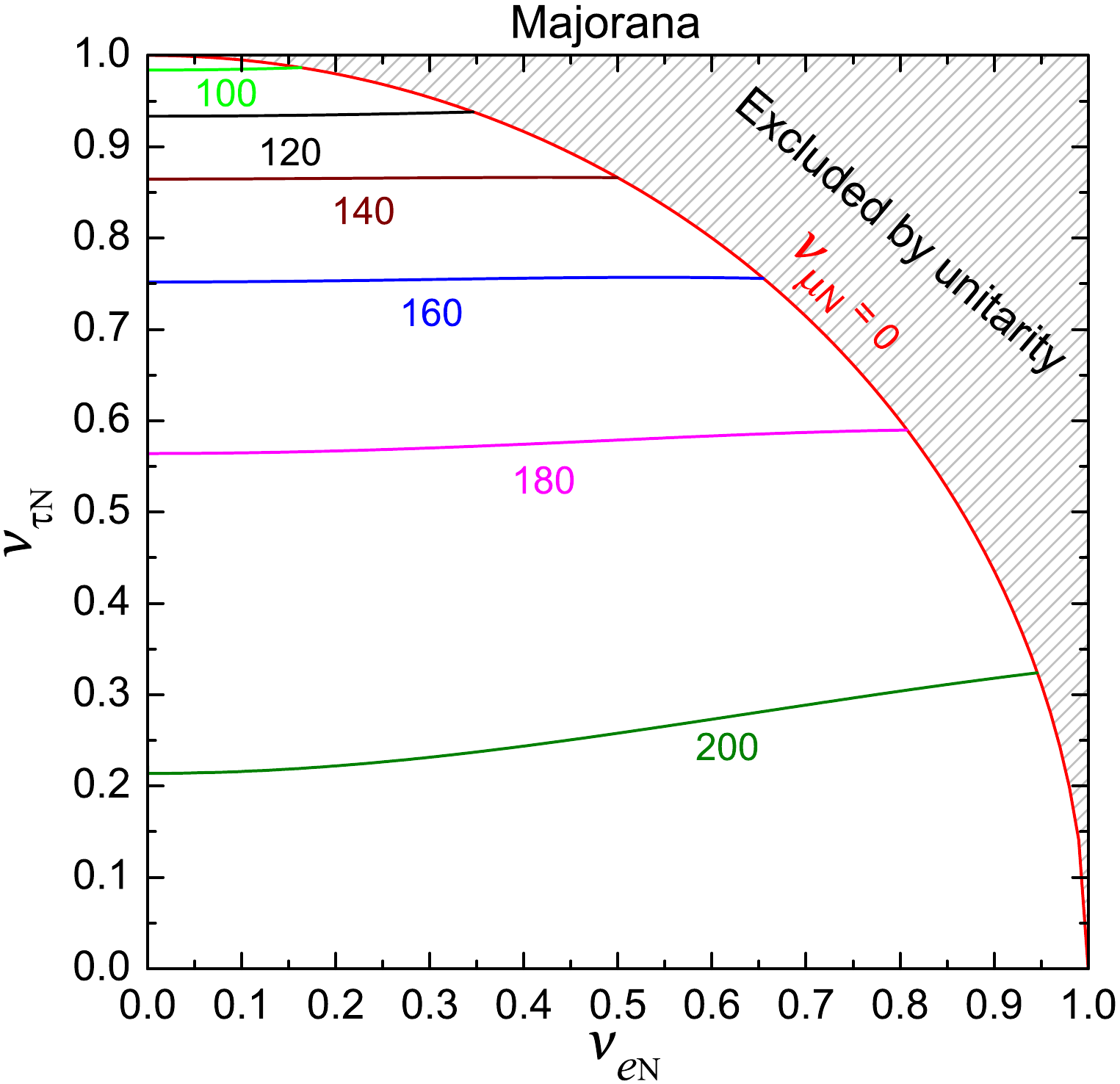}	\label{fig:type3MixingLimits_TaE}}
\end{center}
\caption{
(a) Limits on Type III leptons at $\sqrt{s}=13$ TeV LHC~\cite{CMS:2017wua,Sirunyan:2017qkz};
(b) required luminosity for $2~(5) \sigma$ sensitivity (discovery) with fully reconstructible final states~\cite{Ruiz:2015zca,Golling:2016gvc}.
(c,d) Exclusion contours of doublet-triplet neutrino mixing in $\vert V_{\mu N}\vert-\vert V_{eN}\vert$ and
$\vert V_{\tau N}\vert-\vert V_{eN}\vert$ spaces after $\mathcal{L}=4.9$ fb$^{-1}$ of data at CMS
(labels denote heavy neutral lepton mass in GeV) \cite{Aguilar-Saavedra:2013twa}.
}
\label{fig:type3Searches}
\end{figure}

Hadron collider tests of the Type III Seesaw can be categorized according to the final-state lepton multiplicities, which include:
the $L$-violating, same-sign dilepton and jets final state,
$\ell_1^\pm \ell_2^\pm + nj$~\cite{Biggio:2011ja,Bajc:2006ia,Franceschini:2008pz,Li:2009mw,delAguila:2008cj,Arhrib:2009mz,AguilarSaavedra:2009ik,Choubey:2009wp};
the four-lepton final state,
$\ell_1^\pm \ell_2^\pm \ell_3^\mp \ell_4^\mp + nj$~\cite{Biggio:2011ja,Li:2009mw,delAguila:2008cj,AguilarSaavedra:2009ik,Franceschini:2008pz};
other charged lepton multiplicities~\cite{Biggio:2011ja,delAguila:2008cj,AguilarSaavedra:2009ik,Franceschini:2008pz,Bandyopadhyay:2011aa};
and also displaced charged lepton vertices~\cite{Franceschini:2008pz,He:2009ua}.
Other ``displaced'' signatures, include triplet lepton decays to displaced Higgs bosons~\cite{Bandyopadhyay:2010wp}.
Direct searches for Type III Seesaw partners at the $\sqrt{s}=7/8$ TeV~\cite{CMS:2012ra,Aad:2015cxa,Aad:2015dha}
and $\sqrt{s}=13$ TeV~\cite{CMS:2016hmk,CMS:2017wua,Sirunyan:2017qkz} LHC have yet to show evidence of heavy leptons.
As shown in Fig.~\ref{fig:type3Searches} (a), triplet masses below $M_T\lesssim 800$ GeV have been excluded at 95\% CLs~\cite{CMS:2017wua}.
Figure~\ref{fig:type3Searches} (b) displays the discovery potential of triplet leptons at high-luminosity 100 TeV collider.
One can discover triplet lepton as heavy as 4 (6.5) TeV with 300 (3000) fb$^{-1}$ integrated luminosity.
The absence of triplet leptons in multi-lepton final states can also be interpreted as a constrain on doublet-triplet neutrino mixing.
In Fig.~\ref{fig:type3Searches}(c,d), one sees the exclusion contours of doublet-triplet neutrino mixing in $\vert V_{\mu N}\vert-\vert V_{eN}\vert$ and
$\vert V_{\tau N}\vert-\vert V_{eN}\vert$ spaces after $\mathcal{L}=4.9$ fb$^{-1}$ of data at CMS
(labels denote heavy neutral lepton mass in GeV) \cite{Aguilar-Saavedra:2013twa}.

\subsubsection{Heavy Charged Leptons and Neutrinos at $ee$ and $ep$ Colliders}

\begin{figure}[!t]
\begin{center}
\subfigure[]{\includegraphics[scale=1,width=.43\textwidth]{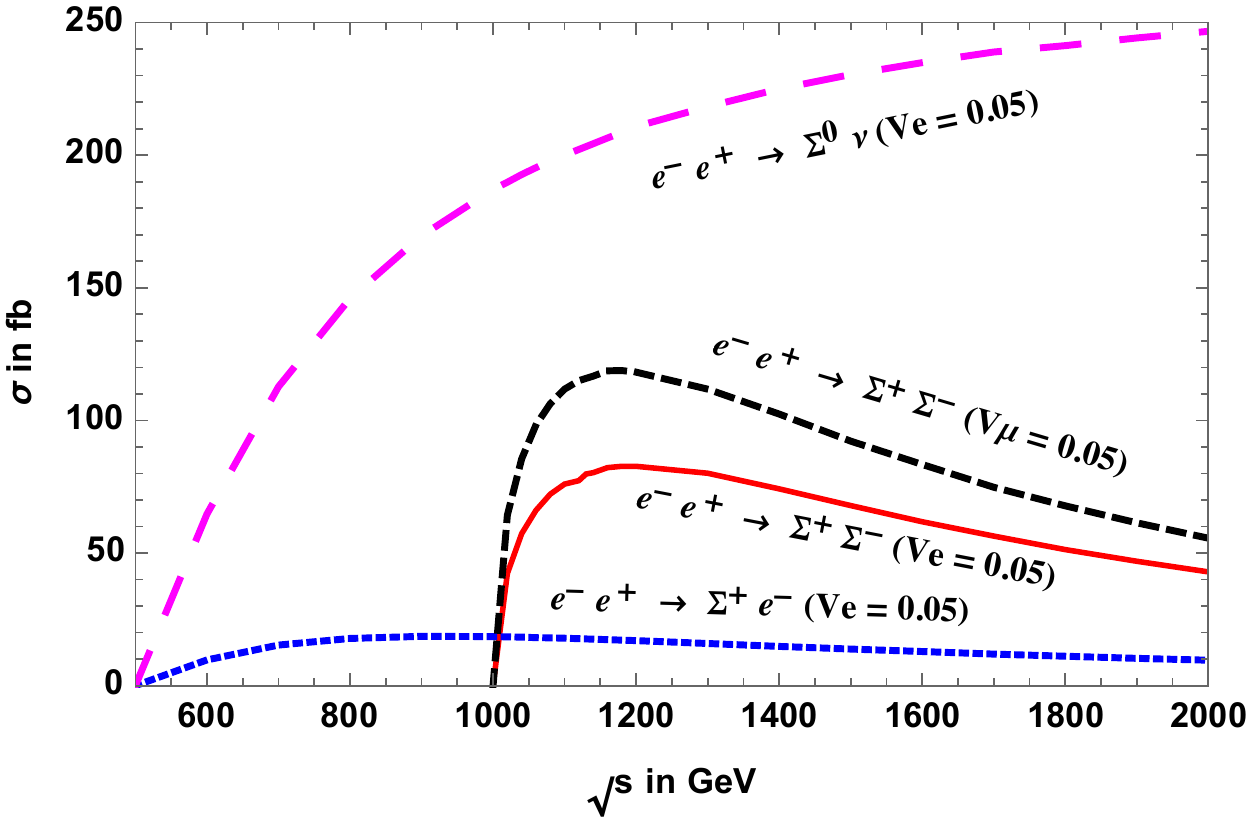}}	
\subfigure[]{\includegraphics[scale=1,width=.53\textwidth]{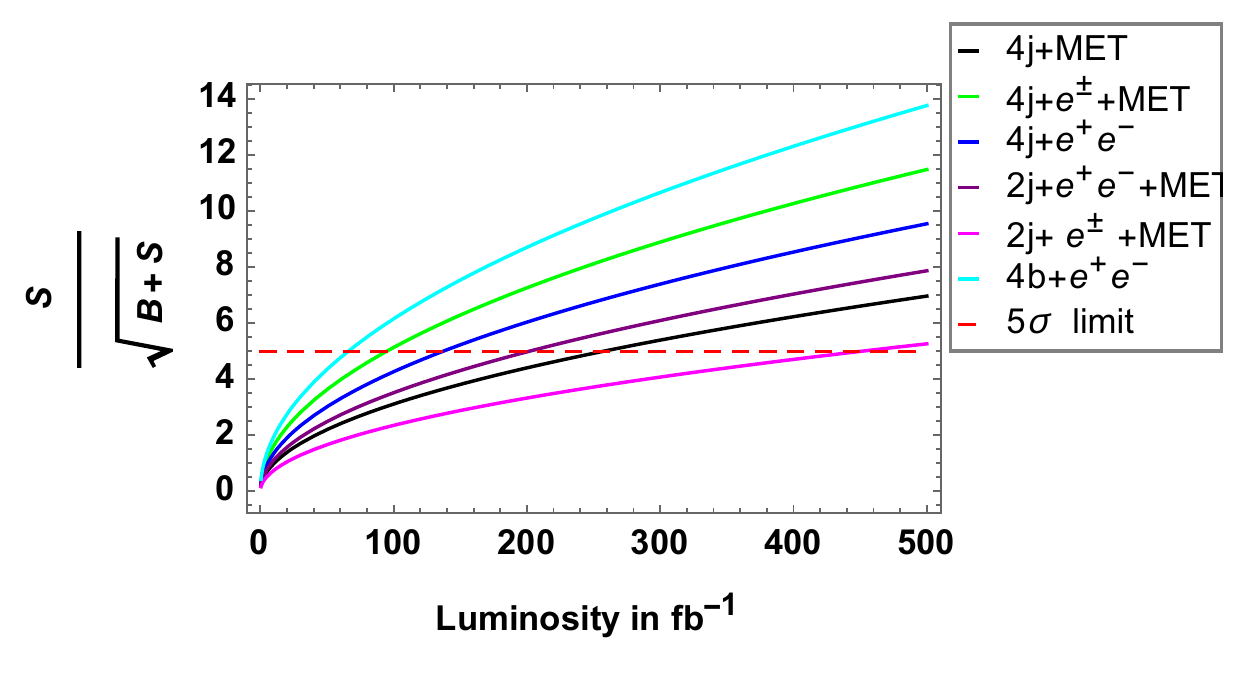}}
\end{center}
\caption{(a) Production cross section of $e^+e^- \to \Sigma^0 \nu, \Sigma^\pm e^\mp, \Sigma^+\Sigma^-$ as a function of the center of mass energy for $e^+e^-$ colliders, with $M_\Sigma=500$ GeV and $V_{eN}=0.05$~\cite{Goswami:2017jqs};
(b) Significance of $\Sigma^+\Sigma^-$ production vs. integrated luminosity at $\sqrt{s}=2$ TeV~\cite{Goswami:2017jqs}.
}
\label{typeiii-eexsec}
\end{figure}

The triplet leptons can also be produced at the leptonic colliders like the ILC and the Compact Linear Collider (CLIC)~\cite{Arhrib:2009mz,Goswami:2017jqs}, and the electron-hadron collider like LHeC~\cite{Liang:2010gm}.
Besides the similar s-channels as hadron colliders, at $e^+e^-$ colliders, the triplet lepton single and pair productions can also happen in $t$-channel via the exchange of $h,~W$, or $Z$ boson. Triplet leptons can also lead to anomalous pair production of SM weak bosons~\cite{Yue:2010zzb}.
Assuming $M_\Sigma=500$ GeV and $V_{eN}=0.05$, the cross sections of triplet lepton single and pair productions are shown in Fig.~\ref{typeiii-eexsec} (a).
For the single production at 1 TeV $e^+e^-$ collider, the triplet lepton with mass up to about 950-980 GeV can be reached with 300 fb$^{-1}$. To discover the heavy charged lepton through $e^+e^-\to \Sigma^+\Sigma^-$ production at $\sqrt{s}=2$ TeV, the luminosity as low (high) as 60 (480) fb$^{-1}$ is needed as shown in Fig.~\ref{typeiii-eexsec} (b).

\section{Radiative Neutrino Mass Models and Lepton Number Violation at Colliders}\label{sec:loop}

A common feature of the Seesaw mechanisms discussed in the previous sessions is that they are
all tree-level, UV completion of the dimension-5 Weinberg operator in of Eq.~(\ref{Weinberg}).
Though economical and elegant, these models
often imply subtle balancing between a Seesaw mass scale at a TeV or below and small Yukawa couplings, 
in the hope for them to be observable at current and near future experiments.
In an altogether different paradigm, it may be the case that small neutrino masses are instead generated radiatively.
In {\it radiative neutrino mass models},
loop and (heavy) mass factors can contribute to the suppression of light neutrino masses and partly  explain their smallness.
A key feature of radiative neutrino mass models is that the Weinberg operator is not generated at tree-level:
For some models, this may be because the particles required to generate tree-level masses, \ie, SM singlet fermions in Type I, triplet scalars in Type II, or triplet leptons in Type III, do not exist in the theory.
For others, it may be the case that the required couplings are forbidden by new symmetries.
Whatever the case, it is necessary that the new field multiplets run in the loops to generate neutrino masses.

At one-loop, such models were first proposed in Refs.~\cite{Zee:1980ai, Hall:1983id},
at two-loop in Refs.~\cite{Cheng:1980qt, Zee:1985id, Babu:1988ki},
and more recently at three-loop order in Ref.~\cite{Krauss:2002px}.
Besides these early works, a plethora of radiative mass models exist
due to the relative ease with which unique loop topologies can be constructed at a given loop order,
as well as the feasibility to accommodate loop contributions from various exotic particles,
including leptoquarks, vector-like leptons and quarks, electrically charged scalars, and EW multiplets.
For a recent, comprehensive review, see Ref.~\cite{Cai:2017jrq}.

However, the diversity of the exotic particles and
interactions in radiative neutrino mass models make it neither feasible nor pragmatic
to develop a simple and unique strategy to test these theories at colliders.
Although some effort has been made to advance approaches to collider tests of radiative neutrino mass models
more systematically~\cite{Cai:2014kra, AristizabalSierra:2007nf},  it remains largely model-dependent.
As a comprehensive summary of the literature for radiative neutrino mass models and their collider study is beyond the scope of this review, in this section, we focus on a small number of representative models with distinctive $L$-violating collider signatures.

It is worth pointing out that some popular radiative neutrino mass models do not predict clear lepton number violation at collider scales.
A prime example are the Scotogenic models~\cite{Ma:2006km}, a class of one-loop radiative neutrino mass scenario with a discrete $Z_2$ symmetry.
Scotogenic models typically contain three SM singlet fermions $N_i$ with Majorana masses and are odd under the $Z_2$, whereas SM fields are even.
The discrete symmetry forbids the mixing between the SM neutrinos and $N_i$ that
one needs to trigger the Type I and III Seesaw mechanisms.
As a result, collider strategies to search for lepton number violation mediated by heavy Majorana neutrinos as presented in Sec.~\ref{sec:type1}
are not applicable to the Scotogenic model. Instead, collider tests of Scotogenic models include, for example,
searches for the additional EW scalars~\cite{Ho:2013hia, Ho:2013spa, Hessler:2016kwm, Diaz:2016udz}
that facilitate lepton number conserving processes. Subsequently, we avoid further discussing radiative models without collider-scale lepton number violation.

Like in the previous sections, we first present in Sec.~\ref{sec:radModels} an overview of representative radiative models.
Then, in Sec.~\ref{sec:radAtColliders}, we review collider searches for lepton number violation associated with radiative neutrino mass models.

\subsection{Selected Radiative Neutrino Mass Models}\label{sec:radModels}
\subsubsection{The Zee-Babu Model}
The first radiative scenario we consider is the well-known Zee-Babu model,
a two-loop radiative neutrino mass model proposed independently by Zee~\cite{Zee:1985id} and Babu~\cite{Babu:1988ki}.
In the model, the SM field content is extended by including one singly-charged scalar ($h^\pm$) and one doubly-charged scalar ($k^{\pm\pm}$).
Both scalars are singlets under $SU(3)_c\times SU(2)_L$, leading to the lepton number violating interaction Lagrangian
\begin{align}
	\Delta \mathcal{L} = \bar{L} Y^\dagger e_R H + \bar{\tilde{L}} f L h^+ + \overline{e_R^c} g e_R k^{++} + \mu_{ZB} h^+h^+ k^{--}  + {\rm H.c.},
\end{align}
where $L~(H)$ is the SM LH lepton (Higgs) doublet.
The $3\times 3$ Yukawa coupling matrices $f$ and $g$ are anti-symmetric and symmetric, respectively.
The trilinear coupling $\mu_{ZB}$ contributes to the masses of the charged scalars at the loop level.
For large values of $(\mu_{ZB}/m_{h^\pm})$ or $(\mu_{ZB}/m_{k^{\pm\pm}})$, where  $m_{h^\pm,k^{\pm\pm}}$ are the masses of $h^\pm$ and $k^{\pm\pm}$,
the scalar potential may have QED-breaking minima. This can be avoided by imposing the condition $\left| \mu_{ZB}\right| \ll 4\pi \; {\rm min}(m_h, m_k)$.

\begin{figure}[t]
\begin{center}
 \subfigure[]{\includegraphics[width=0.45\textwidth]{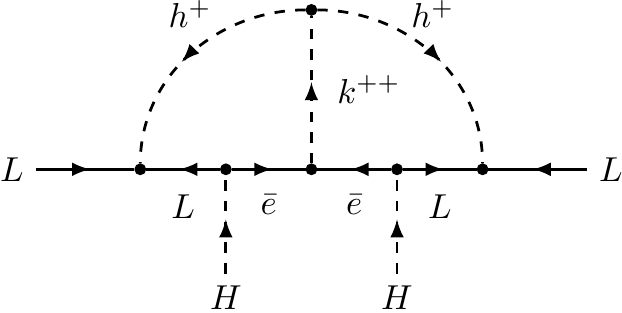}\label{fig:zeebabu}}
 \subfigure[]{\includegraphics[width=0.45\textwidth]{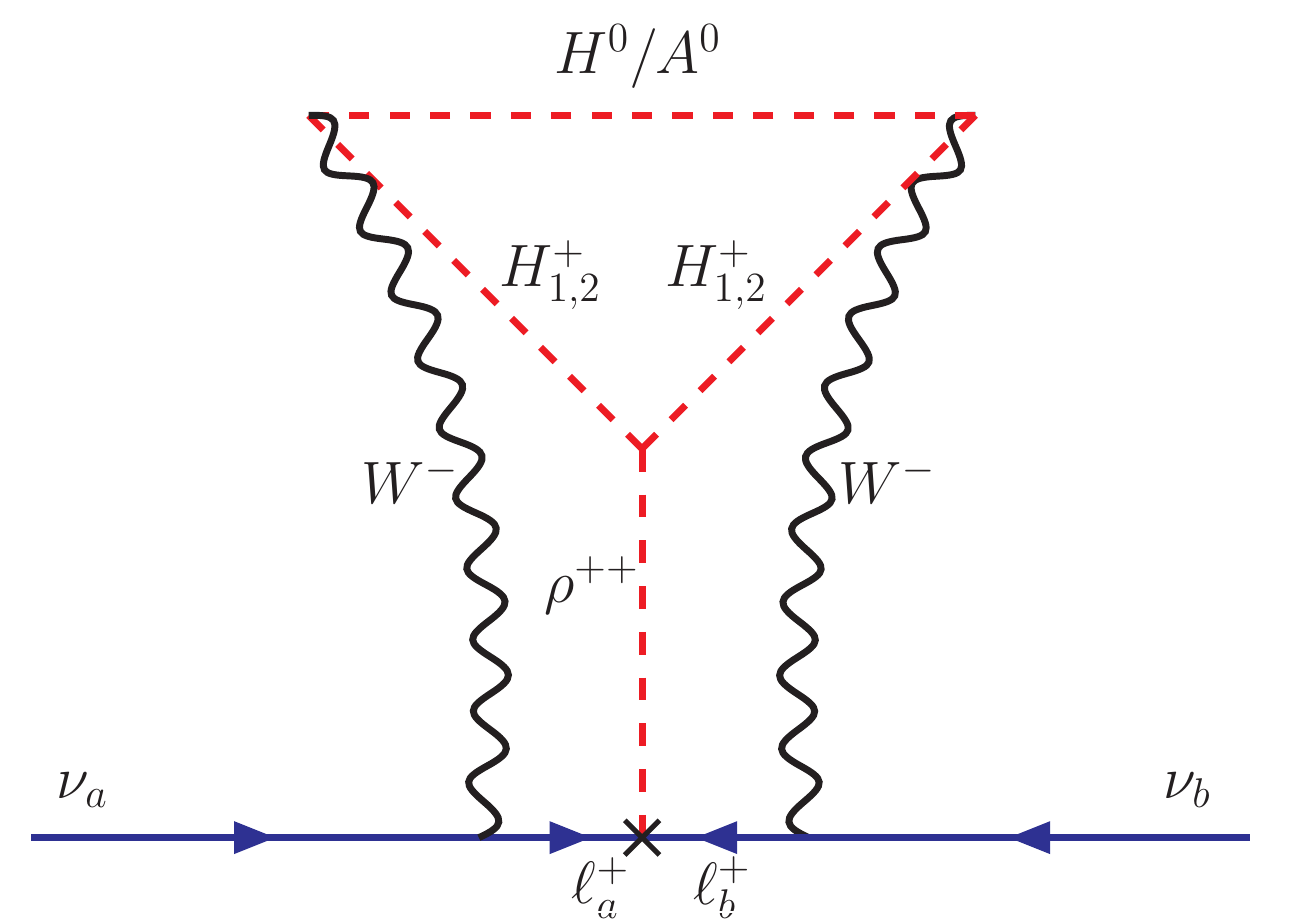}\label{fig:cocktail}}
\end{center}
\caption{(a) Feynman diagram for the generation of neutrino masses at two-loop order in the Zee-Babu model.
(b) Feynman diagram at three-loop order in the cocktail model~\cite{Gustafsson:2012vj}.
}
\label{fig:radiative}
\end{figure}

The combined presence of $Y$, $f$, $g$ and $\mu_{ZB}$ collectively break lepton number and lead to
the generation of a small Majorana neutrino mass.
At lowest order, neutrino masses in the Zee-Babu model arise at two-loop order, as depicted in Fig.~\ref{fig:zeebabu}.
The resulting neutrino mass matrix scales as
\begin{align}
	\mathcal{M}_{\nu} \simeq \left(\frac{v^2 \mu_{ZB}}{96\pi^2 M^2}\right)~ f Y g^\dagger Y^T f^T \; ,
\end{align}
where $M=\max(m_{h^\pm}, m_{k^{\pm\pm}})$ is the heaviest mass in the loop.
Since $f$ is antisymmetric, the determinant of the neutrino mass matrix vanishes, $\det \mathcal{M}_\nu = 0$.
Therefore the Zee-Babu models yields at least one exactly massless neutrino.
An important consequence is that the heaviest neutrino mass is determined by the atmospheric mass difference, which can be estimated as
\begin{align}
	m_\nu \approx 6.6\times 10^{-3} f^2 g \left(\frac{m_\tau^2}{M}\right) \approx 0.05~{\rm eV} \; ,
	\label{eq:ZBRadMass}
\end{align}
where $m_\tau\approx1.778$ GeV is the tau lepton mass.
This implies the product $f^2 g$ can not be arbitrarily small, \eg, for $M\sim 100~{\rm GeV}$, one finds $g^2 f\gtrsim 10^{-7}$.
Subsequently, the parameter space of the Zee-Babu model is constrained by both neutrino oscillation data,
low-energy experiments such as decays mediated $k^{\pm\pm}$ at tree level,
and high-energy searches for direct pair production of $k^{\pm\pm}$.

The study of $h^\pm$ is mostly similar to that of the singly-charged scalar in the Zee model~\cite{Zee:1980ai}, although the lepton number violating effects are not experimentally  observable due to the missing information carried away by the light (Majorana) neutrino in the decay product.
The doubly-charged scalar $k^{\pm\pm}$ can decay to a pair of same-sign leptons,
which manifestly violates lepton number by $\Delta L = \pm2$, with a partial decay width given by
\begin{align}
	\Gamma(k^{\pm\pm}\to \ell_a^\pm \ell_b^\pm) = \frac{\left|g_{ab}\right|^2}{4\pi (1+\delta_{ab})} m_k \;.
\end{align}
If $m_{k^{\pm\pm}}> 2 m_{h^\pm},$ then the $k^{\pm\pm}\to h^\pm h^\pm$ decay mode opens with a partial decay width of
\begin{equation}
	\Gamma(k^{\pm\pm}\to h^\pm h^\pm) =\frac{m_{k^{\pm\pm}}}{8\pi}
		\left(\frac{\mu_{ZB}}{m_{k^{\pm\pm}}}\right)^2  \sqrt{1-\frac{4 m_{h^{\pm}}^2  }{m_{k^{\pm\pm}}^2}} \; .
\end{equation}

Doubly-charged scalars,
appear in many other radiative neutrino mass models,
including the three-loop Cocktail Model~\cite{Gustafsson:2012vj},
whose eponymous mass-generating diagram is shown in the right panel of Fig.~\ref{fig:radiative}.
The doubly-charged scalar couples to the SM lepton doublet and a singly-charged scalar in the same manner as in the Zee-Babu model, and thus again is similar to a Type II scenario.
Radiative Type II Seesaw model~\cite{Ma:2015xla} that generates neutrino mass at one-loop order
contains an $SU(2)_L$ triplet scalar and thus also has similar LHC phenomenology as the tree-level Type II Seesaw mechanism~\cite{Guo:2016dzl}.

\subsubsection{The Colored Zee-Babu Model with Leptoquark}

\begin{figure}[t]
\centering
\includegraphics[width=0.5\textwidth]{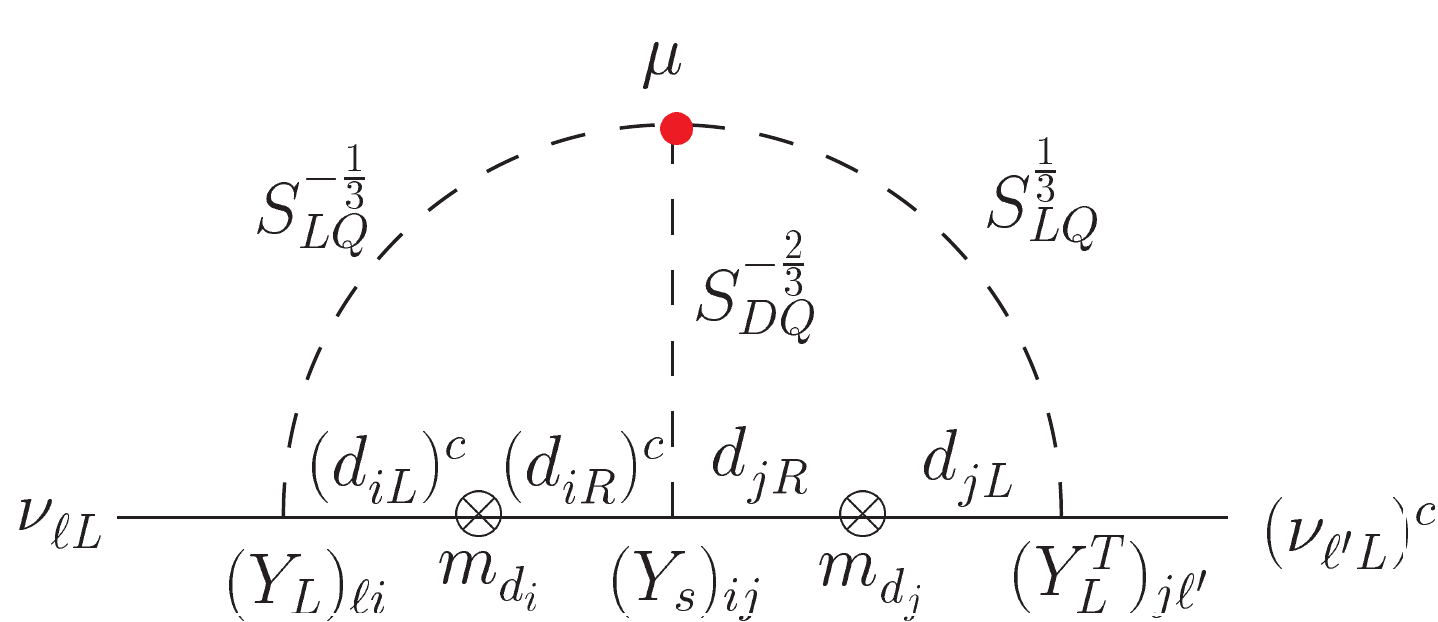}
\caption{Feynman diagram for the generation of neutrino masses at two-loop order in the colored Zee-Babu model~\cite{Kohda:2012sr}.
}
\label{fig:cZBM}
\end{figure}

In a particularly interesting variant of the Zee-Babu model, proposed in Ref.~\cite{Kohda:2012sr}, all particles in the neutrino mass-loop are charged under QCD.
As shown in Fig.~\ref{fig:cZBM}, the lepton doublet in the loop of the Zee-Babu model is replaced with down-type quark while the singly- and doubly-charged scalars are replaced with a leptoquark $S_{LQ}^{-\frac{1}{3}}$ and a diquark $S_{DQ}^{-\frac{2}{3}}$.
Under the SM gauge group, the leptoquark and diquark quantum numbers are
\begin{align}
		S_{LQ}^{-\frac{1}{3}} \quad:\quad (3,1,-\frac{1}{3}) \quad\text{and}\quad  S_{DQ}^{-\frac{2}{3}} \quad:\quad(6, 1, -\frac{2}{3}) \; .
\end{align}
The decay of the diquark $S_{DQ}^{-\frac{2}{3}}$ is analogous to that of the doubly-charged scalar $k^{\pm\pm}$
in that it can decay to a pair of same-sign down-type quarks or a pair of same-sign leptoquarks, if kinematically allowed.

For the models mentioned above, we will only review the collider study with the characteristics
different from the tree-level Seesaws in the following.


\subsection{Radiative Neutrino Mass Models at Colliders}\label{sec:radAtColliders}

\subsubsection{Doubly-charged Scalar at the LHC}
As mentioned above, the Zee-Babu model contains two singlet charged scalars, $h^{\pm}$ and $k^{\pm\pm}$.
Moreover, due to the presence of the doubly-charged scalar decay mode to two same-sign leptons $k^{\pm\pm}\to  \ell^\pm \ell^\pm$ via the coupling $\mu_{ZB}$,
collider searches for $L$-violating effects in the context of the Zee-Babu model are centered on $k^{\pm\pm}$ and its decays.

Like the triplet Higgs in Type II Seesaw, the doubly-charged scalar $k^{\pm\pm}$ can
be pair produced via the Drell-Yan process at the LHC if kinematically accessible and is given by
\begin{align}
	p p \to \gamma^*/Z^* \to k^{++} k^{--}.
\end{align}
This is the same process as shown in Fig.~\ref{fig:feynmanTypeIIScalars}(a).
However, an important distinction is that while $H^{\pm\pm}$ in the Type II Seesaw is an $SU(2)_L$ triplet, the $k^{\pm\pm}$ here is a singlet.
As this quantum-number assignment leads to different $Z$ boson couplings, and hence
different production cross section at colliders, it is a differentiating characteristic of the model.
Note the $\gamma\gamma$ fusion processes, shown in Fig.~\ref{fig:feynmanTypeIIScalars},
also applies to $k^{++}k^{--}$ pair production and leads to the same production cross section.

Since the collider signal for pair produced $k^{\pm\pm}$ is the same as $H^{\pm\pm}$ in the Type II Seesaw,
the search for doubly-charged scalar can be easily performed for both cases
as shown in Fig.~\ref{fig:type2ExclATLAS}.
Obviously, the constraint on the singlet is less stringent due to the absence of weak isospin interactions.
With $36.1$ fb$^{-1}$ data at 13 TeV,
ATLAS has excluded $k^{\pm\pm}$ mass lower than 656-761 GeV for ${\rm BR}(k^{\pm \pm}\to e^\pm e^\pm) + {\rm BR}(k^{\pm\pm} \to \mu^\pm \mu^\pm) = 1$
at 95\% CLs~\cite{ATLAS:2017iqw}.

\begin{figure}[t]
\subfigure[]{\includegraphics[width=0.48\textwidth]{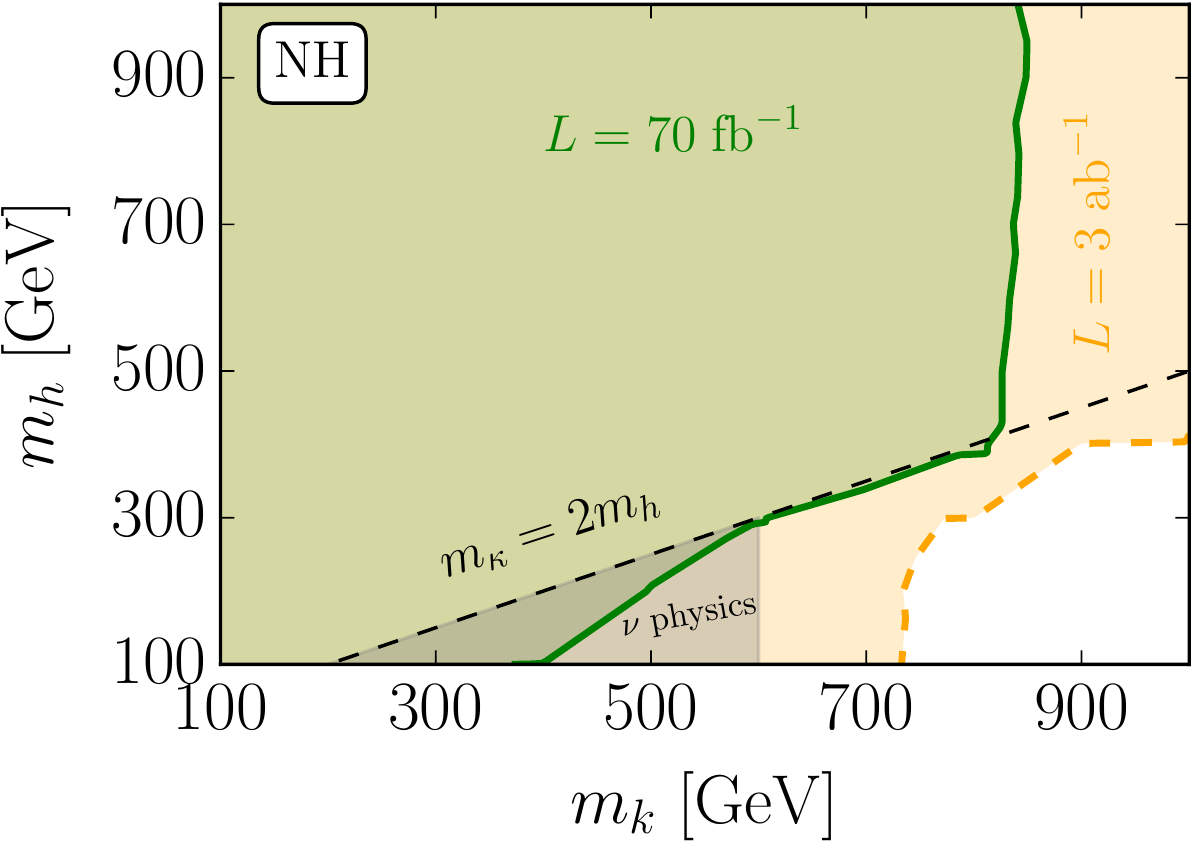}}
\subfigure[]{\includegraphics[width=0.48\textwidth]{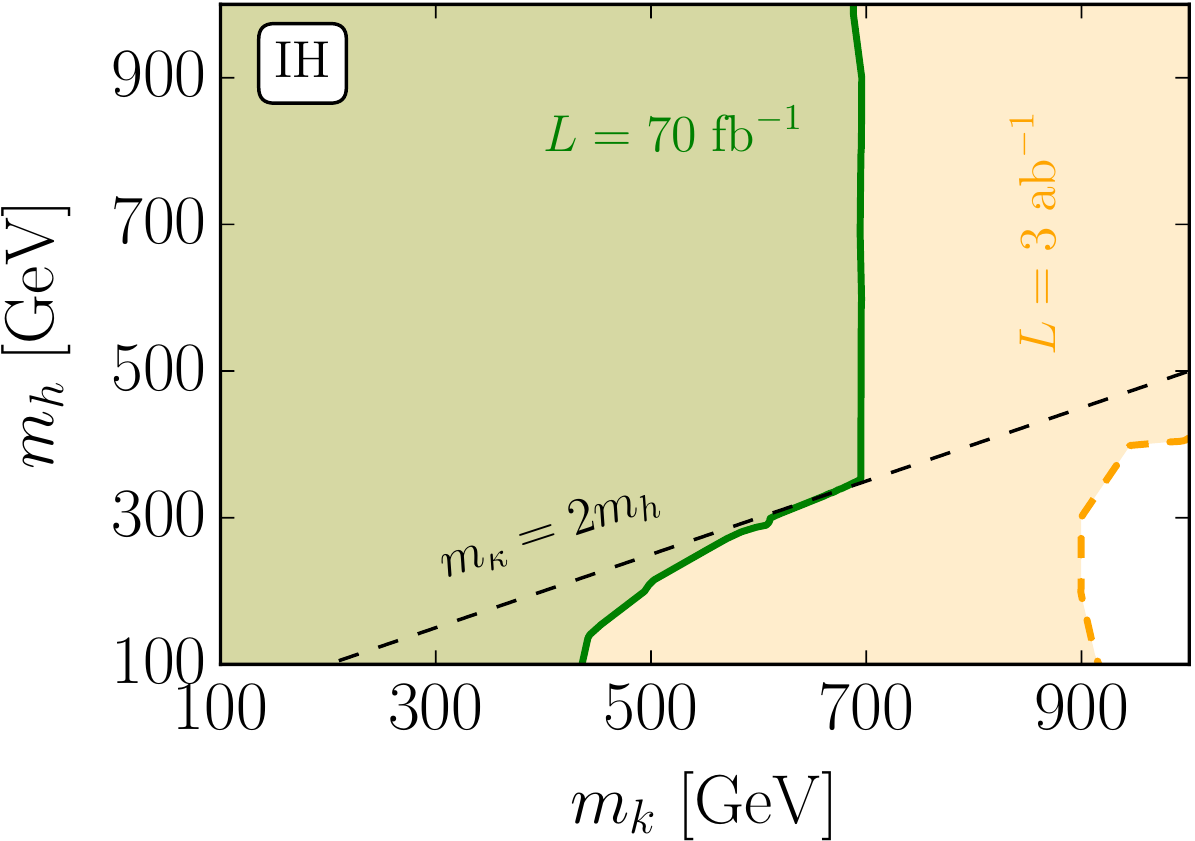}}
	\caption{Projection of sensitivities at the LHC in the $m_{k^{\pm\pm}}$-$m_{h^\pm}$ plane:
	(a) the NH benchmark with $g_{11,22}=0.1$, $g_{12,13,33}=0.001$,
	$f_{12,13}=0.01$ and $f_{23}=0.02$; (b) the IH benchmark with $g_{11,23}=0.1$, $g_{12,22,13,33}=0.0001$, $f_{12}=-f_{13}=0.1$ and $f_{23}=0.01$.
	For both benchmarks, the trilinear coupling is chosen to be $\mu_{ZB}= 5 \; \min(m_{k^{\pm\pm}}, m_{h^{\pm}})$.
	The gray shaded region in the left panel is excluded by low energy experiments. The green and orange regions are
	excluded by future experiments with an integrated luminosity of 70 fb$^{-1}$ and 3 ab$^{-1}$ respectively~\cite{Alcaide:2017dcx}.
	}
\label{fig:limit_zeebabu}
\end{figure}

Low energy LFV experiments, especially $\mu\to e\gamma$, impose very stringent constraints on the parameter space of the Zee-Babu model.
The MEG experiment~\cite{Adam:2013mnn, TheMEG:2016wtm} has placed an upper bound on the decay branching ratio ${\rm BR}(\mu\to e\gamma)<4.2\times 10^{-13}$, which can be roughly translated as~\cite{Herrero-Garcia:2014hfa}
\begin{align}
	\left|f_{13}^* f_{23}\right|^2 \frac{m_{k^{\pm\pm}}^2}{m_{h^\pm}^2} + 16 \left| \sum g_{1i}^* g_{i2} \right|^2 < 1.2\times 10^{-6} \left(\frac{m_k}{{\rm TeV}}\right)^4 \; .
\end{align}
To satisfy LFV constraints, the doubly- and singly-charged scalar masses are pushed well above TeV, with
$m_{k^{\pm\pm}}>1.3~(1.9)$ TeV and $m_{h^{\pm}}>1.3~(2.0)$ TeV for the NH~(IH), assuming $\mu_{ZB}= min(m_{k^{\pm\pm}}, m_{h^{\pm}})$.
This can be very easily relaxed, however, by choosing larger $\mu_{ZB}$ and balancing smaller Yukawa couplings to generate the right neutrino mass spectrum.

A recent study has projected the sensitivities of the LHC with large luminosities by scaling
the cross section bound by $1/\sqrt{\mathcal{L}}$ for two benchmark scenarios: one for NH and one for IH~\cite{Alcaide:2017dcx}.
The projected sensitivities are shown in Fig.~\ref{fig:limit_zeebabu} for model parameters consistant with
neutrino oscillation data.
Note that these benchmarks are chosen to have $\mu_{ZB}=5 \min(m_{k^{\pm\pm}}, m_{h^{\pm}})$ such that the constraints from flavor experiments such as $\mu\to e\gamma$ are much less stringent at the price of a more fine-tuned the scalar potential.
We can see that the NH benchmark is less constrained than the IH one when $m_{k^{\pm\pm}}<2 m_{h^{\pm}}$ because
$k^{\pm\pm}$ has a smaller branching ratio to leptons.

\subsubsection{Leptoquark at the LHC}
\begin{figure}[t]
\centering
\includegraphics[width=0.4\textwidth]{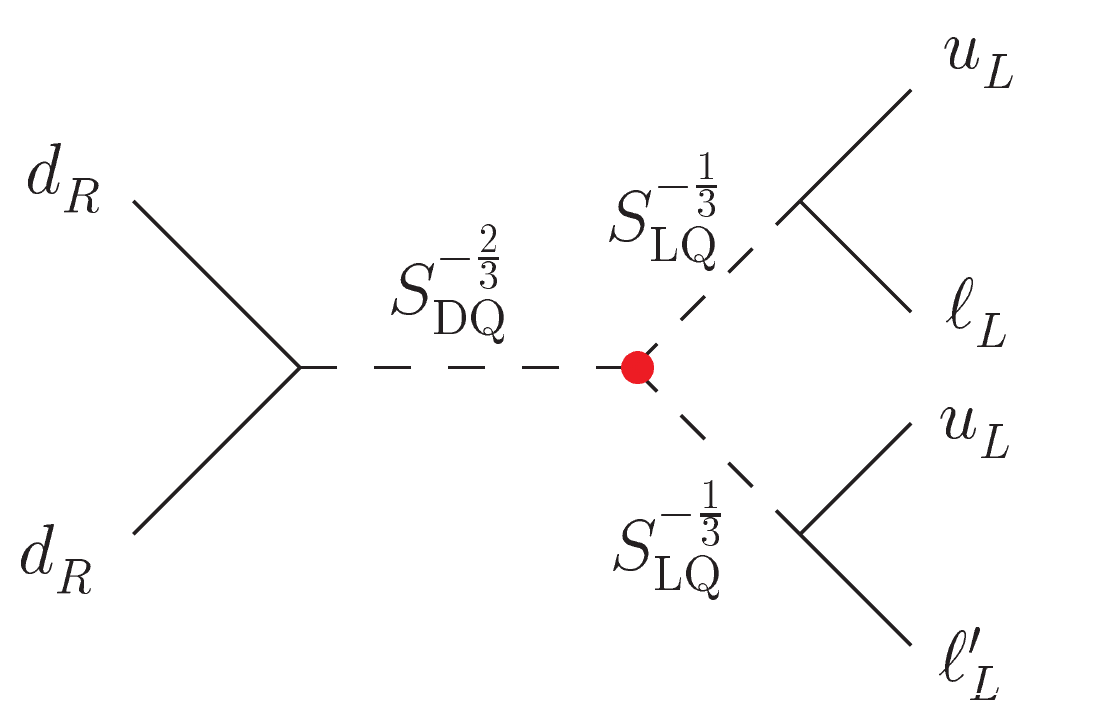}
	\caption{$L$-violating processes at the LHC in the colored Zee-Babu model~\cite{Kohda:2012sr}.}
	\label{fig:cZBM_LHC}
\end{figure}

In the colored Zee-Babu model, $L$-violating signals can be observed in events with pair produced leptoquarks $S_{LQ}^{-\frac{1}{3}}$ via
$s$-channel diquark $S_{DQ}^{-\frac{2}{3}}$, shown in Fig.~\ref{fig:cZBM_LHC}, and given by, 
\begin{align}
	pp\to S_{DQ}^{-\frac{2}{3}*}\to S_{LQ}^{-\frac{1}{3}} S_{LQ}^{-\frac{1}{3}} \to u\ell^- u\ell^{\prime -} .
	\label{eq:radLQlnv}
\end{align}
One benchmark has been briefly studied in Ref.~\cite{Kohda:2012sr}.
For leptoquark mass of 1 TeV and diquark mass of 4 TeV, a benchmark consistent with neutrino oscillation data and low energy experiments,
the $L$-violating process in Eq.~(\ref{eq:radLQlnv}) can proceed with an LHC cross section of $0.18$ fb at $\sqrt{s}=14$ TeV.
So far, there are no dedicated collider study for this model. 
In principle, however, one can recast 
ATLAS or CMS searches for heavy neutrinos, such as Refs.~\cite{Aad:2015xaa, Khachatryan:2014dka}, to derive the limit on the model parameter space.

\begin{figure}[t]
	\centering
\subfigure[]{\includegraphics[width=0.48\textwidth]{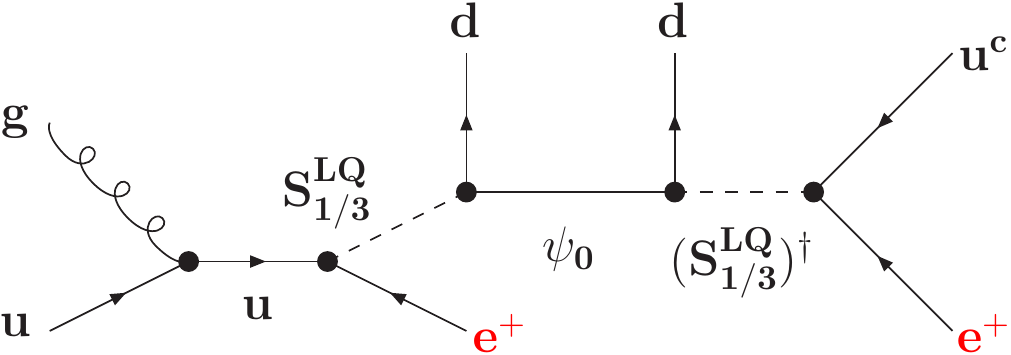}}
\subfigure[]{\includegraphics[width=0.48\textwidth]{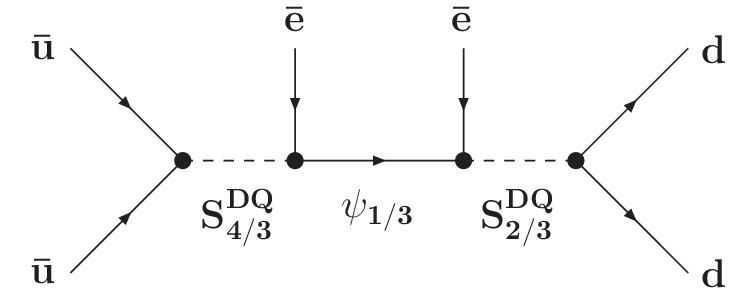}}
	\caption{Example diagrams of $L$ violation processes with (a) leptoquark $S_{1/3}^{\rm LQ}$
	and (b) diquarks $S_{4/3, 2/3}^{\rm DQ}$~\cite{Helo:2013ika, Helo:2013dla}.
	The singlet fermion $\psi_0$ in the left panel leads to Type I Seesaw.
}
 \label{fig:LNVLHC}
\end{figure}

Lepton number violating collider processes, $pp\to\ell^\pm\ell^\pm+nj$,
involving charged scalars, leptoquarks and diquarks have also been studied for the LHC in Refs.~\cite{Peng:2015haa, Helo:2013ika, Helo:2013dla}.
Example diagrams are shown in Fig.~\ref{fig:LNVLHC}.
Even though these studies are performed without a concrete neutrino mass model,
they possess the most important ingredient of Majorana neutrino mass models: $L$ violation by two units,
and therefore radiative neutrino mass models can be constructed from the relevant matter content.
Some processes, however, are realized with a SM singlet fermion (for example the left panel of Fig.~\ref{fig:LNVLHC}),
which implies the existence of a tree-level Seesaw.
Other processes without SM singlet fermions, SU$(2)_L$ triplet scalars, or triplet fermions, such as the one on the right panel of Fig.~\ref{fig:LNVLHC},
can be realized in a radiative neutrino mass model.
Detailed kinematical analyses for resonant mass reconstruction would help to sort out the underlying dynamics.

\subsubsection{Correlation with Lepton Flavor Violation}
In radiative neutrino mass models the breaking of lepton number generally needs the simultaneous presence of multiple couplings.
For example, in the Zee-Babu model, $Y$, $f$, $g$ and $\mu_{ZB}$ together break lepton number.
The observation of pair produced $k^{\pm\pm}$ itself is insufficient to declare $L$ violation.
In order to establish $L$ violation in the theory and thus probe the Majorana nature of the neutrinos, the couplings of $h^\pm$ to SM leptons
and to $k^{\pm\pm}$ have to be studied at the same time.
For the colored Zee-Babu model, the $L$ violation process shown in
Fig.~\ref{fig:cZBM_LHC} involves all couplings except the SM Yukawa
necessary to break the lepton number.
Note, however, the cross section for this process is proportional to the product
of couplings and suppressed by the heavy exotic masses, which both contribute to the smallness of the neutrino masses.
Thus the cross section for this processes must be kinematically suppressed.
For radiative neutrino mass models with dark matter candidates,
probing lepton number violation at colliders alone is generally much more
difficult as the dark matter candidate appears as missing transverse
energy just as neutrinos.
Overall, the study of $L$-violation of radiative neutrino mass models
can be performed either with the combination of different processes that
test different subsets of the couplings
or in a single process that involves all couplings at once whose production cross section
is generally suppressed.

On the contrary, radiative neutrino mass models contain
LFV couplings and exotic particles that can be tested much easier than $L$ violation stated above.
The search strategies for LFV couplings and new particles vary from model to model.
It is definitely impossible to cover all and they are also not the focus of this review.
Thus we will take a few simple examples to illustrate the searches.

The leading LFV signals can be produced in a radiative neutrino mass model from the QCD pair production of the leptoquark $S_{LQ}^{-\frac{1}{3}}$ with its suitable subsequent decays such as
\begin{align}
pp ~\to~ S_{LQ}^{+\frac{1}{3}} S_{LQ}^{-\frac{1}{3}} ~\to~ \bar{t}\ell^+ t\ell^{\prime -}
\end{align}
where $S_{LQ}^{+\frac{1}{3}}=\left(S_{LQ}^{-\frac{1}{3}}\right)^*$ and the top quarks decay hadronically.
Note that the leptoquark pair can also decay to $\bar{b}\nu_{\ell} \ t\ell^{\prime -}$ or $\bar{b}\nu_{\ell} \ b \bar{\nu}_{\ell^\prime}$,
where the LFV effects are not easy to disentangle at colliders due to the invisible neutrinos. 
However, these decay channels can result in  final states $\ell^+\ell^{\prime -}X$, 
inclusive flavour off-diagonal charged lepton pair accompanied by missing transverse energy, jets \etc,  if the quarks decay to appropriate leptons. 
The same final states have been used to search for stop in SUSY theories
and thus the results for stop searches at the LHC can be translated to that of the
leptoquark $S_{LQ}^{-\frac{1}{3}}$, $m({S_{LQ}^{-\frac{1}{3}}})\gtrsim 600$ GeV~\cite{Cai:2014kra} based
on the ATLAS stop search at $\sqrt{s}=8$ TeV~\cite{Aad:2014qaa}~\footnote{There are many dedicated leptoquark searches at the LHC~\cite{Aaboud:2016qeg, Aad:2015caa, CMS:2016imw, CMS:2016qhm, CMS:2016hsa}. However, the leptoquarks searched only couple to one generation of fermions at a time and thus generate no LFV signals.}. 
No recast of stop searches has been performed for 13 TeV run at the time of this work.
Besides leptoquarks, radiative neutrino mass models also comprise exotic particles
such as
vector-like quarks, vector-like leptons, charged scalar singlets (both singly- and doubly-charged)
and higher-dimensional EW multiplets.
For example, disappearing tracks can be used to search for higher-dimensional EW multiplet fermions
whose mass splitting between the neutral and the singly-charged component is around 100 MeV.
The current LHC searches have set a lower mass limit of 430 GeV at $95\%$ CL for a triplet fermion
with a lifetime of about 0.2 ns~\cite{ATLAS:2017bna, Aad:2013yna, CMS:2014gxa}.
We refer the readers to the section about collider tests of radiative neutrino mass model
in Ref.~\cite{Cai:2017jrq} and the references therein for details.

We want to stress, however, that even though $L$ violation in the radiative models is more complicated and challenging
to search for in collider experiments, their observation is essential and conclusive
to establish the Majorana nature of neutrinos.
So once we find signals in either LFV processes or new particles searches,
we should search for $L$ violation in specific radiative neutrino mass models that give these LFV processes or contain these new particles,
in order to ultimately test the generation of neutrino masses.

\section{Summary and Conclusions}\label{sec:con}

Exploring the origin of neutrinos' tiny masses, their large mixing,
and their Dirac or Majorana nature are among the most pressing issues in particle physics today. If one or more neutrino Seesaw mechanisms are realized in nature,
it would be ultimately important to identify the new scales responsible for generating neutrino masses. Neutrino oscillation experiments, however, may not provide such information, and thus complementary pathways, such as collider experiments, are vital to understanding the nature of neutrinos.
Observing lepton number violation at collider experiments would be a conclusive verdict for the existence of neutrino Majorana masses, but also direct evidence of a mass scale qualitatively distinct from those in the SM.

In this context, we have reviewed tests of low-scale neutrino mass models at $pp$, $ep$, and $ee$ colliders,
focusing particularly on searches for lepton number $(L)$ violation:
We begin with summarizing present neutrino oscillation and cosmology data and their impact on the light neutrino mass spectra in Sec.~\ref{sec:nuparameters}.
We then consider several representative scenarios as phenomenological benchmarks, including the characteristic Type I Seesaw in Sec.~\ref{sec:type1},
the Type II Seesaw in Sec.~\ref{sec:type2}, the Type III in Sec.~\ref{sec:type3},
radiative constructions in Sec.~\ref{sec:loop}, as well as extensions and hybridizations of these scenarios.
We summarize the current status of experimental signatures featuring $L$ violation,
and present anticipated coverage in the theory parameter space at current and future colliders.
We emphasize new production and decay channels, their phenomenological relevance and treatment across different collider facilities.
We also summarize available Monte Carlo tools available for studying Seesaw partners in collider environments.

The Type I Seesaw is characterized by new right-handed, SM gauge singlet neutrinos, known also as ``sterile neutrinos,''
which mix with left-handed neutrinos via mass diagonalization.
As this mixing scales with light neutrino masses and elements of the PMNS matrix,
heavy neutrino decays to charged leptons may exhibit some predictable patterns if one adopts some simplifying assumptions for the mixing matrix,
as shown for example in Figs.~\ref{fig:brnlwcase1} and \ref{fig:nibr}, that are correlated with neutrino oscillation data.
The canonical high-scale Type I model, however, predicts tiny active-sterile mixing, with $\vert V_{\ell N}\vert^2 \sim m_\nu/M_N$,
and thus that heavy $N$ decouple from collider experiments.
Subsequently, observing lepton number violation in collider experiments, as discussed in Sec.~\ref{sec:typeIhybrid},
implies a much richer neutrino mass-generation scheme than just the canonical, high-scale Type I Seesaw.
In exploring the phenomenological parameter space, the 14 TeV LHC (and potential 100 TeV successor) and $\mathcal{L}=1$ ab$^{-1}$
integrated luminosity could reach at least $2\sigma$ sensitivity for heavy neutrino masses of $M_N\lesssim 500$ GeV ($1$ TeV)
with a mixing $\vert V_{\ell N}\vert^2 \lesssim 10^{-3}$, as seen in Fig.~\ref{100TeVdiscovery.fig}.
If $N$ is charged under another gauge group that also couples to the SM, as in $B$-$L$ or LR gauge extensions,
then the discovery limit may be extended to $M_N \sim M_{Z'}, M_{W_R}$, when kinematically accessible;
see Secs.~\ref{sec:type1Abelian} and Sec.~\ref{sec:lrsmCollider}.

The Type II Seesaw is characterized by heavy SU$(2)_L$ triplet scalars, which result in new singly- and doubly-charged Higgs bosons.
They can be copiously produced in pairs via SM electroweak gauge interactions if kinematically accessible at collider energies,
and search for the doubly-charged Higgs bosons via the same-sign dilepton channel $H^{\pm\pm} \to \ell^\pm \ell^\pm$
is an on-going effort at the LHC. Current direct searches at 13 TeV bound triplet scalar masses to be above (roughly) 800 GeV.
With anticipated LHC luminosity and energy upgrades, one can expect for the search to go beyond a TeV.
Furthermore, if neutrino masses are dominantly from triplet Yukawa couplings,
then the patterns of the neutrino mixing and mass relations from the oscillation experiments
will correlate with the decays of the triplet Higgs bosons to charged leptons, 
as seen from the branching fraction predictions in Figs.~\ref{brii} and \ref{brij} and in Table \ref{relationii}.
Since a Higgs triplet naturally exists in certain extensions beyond the SM, such as in Little Higgs theory, the LRSM, and GUT theories,
the search for such signals may prove beneficial as discussed in \ref{sec:type2LRSM}.

The Type III Seesaw is characterized by heavy SU$(2)_L$ triplet leptons, which result in vector-like, charged and neutral leptons.
Such multiplets can be realized in realistic GUT theories in hybridization with heavy singlet neutrinos from a Type I Seesaw.
Drell-Yan pair production of heavy charged leptons at hadron colliders is sizable as it is governed by the SM gauge interactions.
They can decay to the SM leptons plus EW bosons, leading to same-sign dilepton events. 
Direct searches for promptly decaying triplet leptons at the LHC set a lower bound on the triplet mass scale of around 800 GeV.
A future 100 TeV $pp$ collider can extend the mass reach to at least several TeV, as seen in Fig.~\ref{fig:type3Searches}.

Finally, neutrino masses can also be generated radiatively,
which provides an attractive explanation for the smallness of neutrino masses with a plausibly low mass scale.
Among the large collection of radiative neutrino mass models,
the Zee-Babu model contains a doubly-charged SU$(2)_L$ singlet scalar
with collider signal akin to the doubly-charged Higgs in the Type II Seesaw.
ATLAS has excluded $k^{\pm\pm}$ mass below $660-760$ GeV assuming
the benchmark decay rate $\sum_{\ell_i=e,\mu} {\rm BR}(k^{\pm\pm} \to \ell_1^\pm \ell_2^\pm )=1$.
The high luminosity LHC is sensitive up to about a TeV for both $k^{\pm\pm}$ and its companion scalar $h^\pm$ in the Zee-Babu model with constraints from neutrino oscillation data and other low energy experiments.
For the colored variant of the Zee-Babu model, a pair of same-sign leptoquark can be produced via an $s$-channel diquark at the LHC.
Their subsequent decay lead to the lepton number violating same-sign dilepton plus jets final state, which still await dedicated studies.

As a final remark, viable low-scale neutrino mass models often generate a rich flavor structure in the charged lepton sector that  predict lepton flavor-violating transitions. Such processes are typically much more easily observable than lepton number violating processes,
in part due to larger production and decay rates, and should be searched for in both high- and low-energy experiments.

\section{Acknowledgements}\label{sec:ack}
We thank Michael A. Schmidt for useful discussions.
Past and present members of the IPPP are thanked for discussions.
The work of TH was supported in part by the U.S. Department of Energy under grant No.~DE-FG02- 95ER40896 and in part by the PITT PACC.
The work of TL was supported in part by the Australian Research Council Centre of Excellence for Particle Physics at the
Tera-scale.
The work of RR was funded in part by the UK Science and Technology Facilities Council (STFC),
the European Union's Horizon 2020 research and innovation programme under the Marie Sklodowska-Curie grant agreements
No 690575 (InvisiblesPlus RISE) and No 674896~(InvisiblesPlus RISE).
Support for the open accessibility of this work is provided by the Research Councils UK, external grant number ST/G000905/1.\\

Republication of the various figures is granted under the terms of the Creative Commons Attribution Licenses,
American Physical Society PRD License Nos.:
~4234250091794,
~4234250774584,
~4234260138469,
~4234270420615,
~4234270625995,
~4234270758210,
~4234280275772,\\
~4234280455112,
~4234291212539,
~4234300360477,
~4234301170893,
~4234310175148,
and PRL License No.~4234301302410.


\bibliography{colliderLNVrefs}

\end{document}